\pdfoutput=1
\RequirePackage{ifpdf}
\ifpdf %We are running pdfTeX in pdf mode
\documentclass[pdftex]{sigma}
\else
\documentclass{sigma}
\fi

\usepackage{tikz}

\numberwithin{equation}{section}
\numberwithin{theorem}{section}

\newtheorem{lem}[theorem]{Lemma}
\newtheorem{prop}[theorem]{Proposition}

{\theoremstyle{definition}
\newtheorem{rem}[theorem]{Remark}
\newtheorem{exam}[theorem]{Example}
}

\newcommand{\cB}{\mathcal{B}}
\newcommand{\cC}{\mathcal{C}}
\newcommand{\cH}{\mathcal{H}}
\newcommand{\cL}{\mathcal{L}}
\newcommand{\cO}{\mathcal{O}}
\newcommand{\cP}{\mathcal{P}}
\newcommand{\cR}{\mathcal{R}}
\newcommand{\cS}{\mathcal{S}}
\newcommand{\cT}{\mathcal{T}}
\newcommand{\cU}{\mathcal{U}}
\newcommand{\bbN}{\mathbb{N}}

\newcommand{\bsy}{\boldsymbol}

\newcommand{\xrarrow}{\xrightarrow{\hspace*{1.6cm}}}

\renewcommand\vec[1]{\overrightarrow{#1}}
\newcommand\cev[1]{\overleftarrow{#1}}

\definecolor{darkgreen}{rgb}{0,0.4,0}
\definecolor{darkred}{rgb}{1, 0.0, 0.0}
\definecolor{brown}{rgb}{0.59, 0.29, 0.0}

\usetikzlibrary{positioning,arrows,decorations.pathmorphing}
\tikzset{
  state/.style={minimum size=6ex},
  arrow/.style={-latex, shorten >=1ex, shorten <=1ex}}

\begin{document}

\newcommand{\arXivNumber}{1409.7855}

\allowdisplaybreaks

\renewcommand{\PaperNumber}{042}

\FirstPageHeading

\ShortArticleName{Simplex and Polygon Equations}

\ArticleName{Simplex and Polygon Equations}

\Author{Aristophanes DIMAKIS~$^\dag$ and Folkert M\"ULLER-HOISSEN~$^\ddag$}

\AuthorNameForHeading{A.~Dimakis and F.~M\"uller-Hoissen}

\Address{$^\dag$~Department of Financial and Management Engineering, University of the Aegean,\\
\hphantom{$^\dag$}~82100 Chios, Greece}
\EmailD{\href{mailto:dimakis@aegean.gr}{dimakis@aegean.gr}}

\Address{$^\ddag$~Max Planck Institute for Dynamics and Self-Organization,
         37077 G\"ottingen, Germany}
\EmailD{\href{mailto:folkert.mueller-hoissen@ds.mpg.de}{folkert.mueller-hoissen@ds.mpg.de}}

\ArticleDates{Received October 23, 2014, in f\/inal form May 26, 2015; Published online June 05, 2015}

\Abstract{It is shown that higher Bruhat orders admit a decomposition into a higher Tamari order,
the corresponding dual Tamari order, and a ``mixed order''. We describe simplex equations
(including the Yang--Baxter equation) as realizations of higher Bruhat orders.
Correspondingly, a family of ``polygon equations'' realizes higher Tamari orders.
They generalize the well-known pentagon equation.
The structure of simplex and polygon equations is visualized in terms of deformations
of maximal chains in posets forming 1-ske\-le\-tons of polyhedra.
The decomposition of higher Bruhat orders induces
a reduction of the $N$-simplex equation to the $(N+1)$-gon equation, its dual, and a
compatibility equation. }


\Classification{06A06; 06A07; 52Bxx; 82B23}
% 06A06 partial order, general
% 06A07 Combinatorics of partially ordered sets
% 52Bxx Polytopes and polyhedra
% 82B23  Exactly solvable models; Bethe ansatz
% For 2010 Mathematics Subject Classification see http://www.ams.org/mathscinet/msc/msc2010.html

\section{Introduction}
The famous (quantum) \emph{Yang--Baxter equation} is
\begin{gather*}
  \hat{\cR}_{\bsy{12}}   \hat{\cR}_{\bsy{13}}   \hat{\cR}_{\bsy{23}}
  = \hat{\cR}_{\bsy{23}}   \hat{\cR}_{\bsy{13}}   \hat{\cR}_{\bsy{12}}   ,
            %\label{YBeq_param}
\end{gather*}
where $\hat{\cR} \in \mathrm{End}(V \otimes V)$, for a vector space $V$, and
boldface indices specify the two factors of a threefold tensor product on which $\hat{\cR}$ acts.
This equation plays an important role in exactly solvable two-dimensional models of statistical mechanics,
in the theory of integrable systems, quantum groups, invariants of knots and three-dimensional
manifolds, and conformal f\/ield theory (see, e.g.,~\cite{BaLuZa99,Cart89,Etin+Lato05,Jimb89,Jimb90}).

A set-theoretical version of the Yang--Baxter equation considers $\hat{\cR}$ as a map
$\hat{\cR}\colon \cU \times \cU \rightarrow \cU \times \cU$,
where $\cU$ is a set (not necessarily supplied with further structure).
Nontrivial examples of ``set-theoretical solutions'' of the Yang--Baxter equation,
for which Veselov later introduced the name \emph{Yang--Baxter maps}~\cite{Vese03,Vese07},
apparently f\/irst appeared in~\cite{Skly88} (cf.~\cite{Vese07}), and Drinfeld's work~\cite{Drin92}
stimulated much interest in this subject.
Meanwhile quite a number of examples and studies of such maps have appeared (see, e.g.,
\cite{ABS04,Etin03,Etin+Sche+Solo99,Gonc+Vese04,Suri+Vese03}).

The Yang--Baxter equation is a member of a family, called \emph{simplex equations}~\cite{Bazh+Stro82}
(also see, e.g.,~\cite{Fren+Moor91,KST14, Mail+Nijh89PLB,Mail+Nijh89GTMP,Mail+Nijh90}).
The $N$-simplex equation is an equation imposed on a map $\hat{\cR}\colon V^{\otimes N} \rightarrow V^{\otimes N}$,
respectively $\hat{\cR}\colon \cU^N \rightarrow \cU^N$ for the set-theoretical version.
The next-to-Yang--Baxter equation, the $3$-simplex equation,
\begin{gather*}
     \hat{\cR}_{\bsy{123}}  \hat{\cR}_{\bsy{145}}  \hat{\cR}_{\bsy{246}}  \hat{\cR}_{\bsy{356}}
   = \hat{\cR}_{\bsy{356}}  \hat{\cR}_{\bsy{246}}  \hat{\cR}_{\bsy{145}}  \hat{\cR}_{\bsy{123}}   ,
\end{gather*}
is also called \emph{tetrahedron equation} or \emph{Zamolodchikov equation}. This equation acts
on $V^{\otimes 6}$.
A set of tetrahedron equations f\/irst appeared as factorization conditions for the $S$-matrix
in Zamolodchikov's $(2+1)$-dimensional scattering theory of straight lines (``straight
strings''), and in a related three-dimensional exactly solvable lattice model~\cite{Zamo80,Zamo81}.
This has been inspired by Baxter's eight-vertex lattice
model~\cite{Baxt72,Baxt78} and stimulated further important work~\cite{Baxt83,Baxt86}, also see
the survey~\cite{Stro97}.
Meanwhile the tetrahedron equation has been the subject of many publications (see, in particular,
\cite{Byts+Volk13,Cart+Sait98,Hiet94,Kore93,KST14,Lawr95,Mail+Nijh89PLA,Serg09book}).
An equation of similar structure as the above $3$-simplex equation, but acting on $V^{\otimes 4}$,
has been proposed in~\cite{Fren+Moor91}.

In a similar way as the $2$-simplex (Yang--Baxter) equation describes a
factorization condition for the scattering matrix of particles in two space-time dimensions
\cite{Cher84, Yang67}, as just mentioned,
the $3$-simplex equation describes a corresponding condition for straight lines
on a plane~\cite{Zamo80,Zamo81}. Manin and Schechtman~\cite{Manin+Schecht86a,Manin+Shekhtman86b}
looked for what could play the role of the permutation group, which acts on the particles in the
Yang--Baxter case, for the higher simplex equations. They were led in this way to introduce the
\emph{higher Bruhat order} $B(N,n)$, with positive integers $n<N$. This is a partial order on the set of
certain equivalence classes
of ``admissible'' permutations of ${[N] \choose n}$, which is the set of $n$-element subsets
of $[N] := \{ 1,2,\ldots,N \}$ (see Section~\ref{subsec:hB}).
The $N$-simplex equation is directly related to the higher Bruhat order $B(N+1,N-1)$.

Let us consider the \emph{local Yang--Baxter equation}\footnote{In very much the same form, the
local Yang--Baxter equation appeared in~\cite{Serg97}, for example. A natural generalization is obtained
by replacing the three appearances of $\hat{\cL}$ by three dif\/ferent maps.}
\begin{gather*}
    \hat{\cL}_{\bsy{12}}(x)   \hat{\cL}_{\bsy{13}}(y)   \hat{\cL}_{\bsy{23}}(z)
    = \hat{\cL}_{\bsy{23}}(z')   \hat{\cL}_{\bsy{13}}(y')   \hat{\cL}_{\bsy{12}}(x')   ,
\end{gather*}
where the $\hat{\cL}_{\bsy{ij}}$ depend on variables in such a way that this equation uniquely determines
a~map $(x,y,z) \mapsto (x',y',z')$.
Then this map turns out to be a set-theoretical solution
of the tetrahedron equation. Here we wrote $\hat{\cL}_{\bsy{ij}}$ instead of $\hat{\cR}_{\bsy{ij}}$
in order to emphasize that such a~``localized'' equation may be regarded as a~``Lax system''
for the tetrahedron equation, i.e., the latter arises as a consistency condition of the system.
This is a familiar concept in integrable systems theory.
 If the variables $x$, $y$, $z$ are elements of a (f\/inite-dimensional, real or complex) vector space,
and if the maps $\hat{\cL}_{\bsy{ij}}$ depend \emph{linearly}
on them, then $\hat{\cL}_{\bsy{ij}}(x) = x_a   \hat{\cL}_{\bsy{ij}}^a$, using the
summation convention and expressing $x = x_a   E^a$ in a basis $E^a$, $a=1,\ldots,m$.
In this case the above equation takes the form
\begin{gather*}
   \hat{\cL}_{\bsy{12}}^a   \hat{\cL}_{\bsy{13}}^b   \hat{\cL}_{\bsy{23}}^c
    = \hat{\cR}^{abc}_{def}   \hat{\cL}_{\bsy{23}}^d   \hat{\cL}_{\bsy{13}}^e
        \hat{\cL}_{\bsy{12}}^f    ,
\end{gather*}
where the coef\/f\/icients $\hat{\cR}^{abc}_{def}$ are def\/ined by
$z_d'   y_e'   x_f' = x_a   y_b   z_c   \hat{\cR}^{abc}_{def}$.
The last system is also known as the tetrahedral Zamolodchikov algebra
(also see~\cite{BaMaOkSe13, Kore93}).
Analogously, there is a~Lax system for the Yang--Baxter equation~\cite{Suri+Vese03},
consisting of $1$-simplex equations,
which is the Zamolodchikov--Faddeev algebra~\cite{Kuli84}, and this structure extends to all
simplex equations~\cite{Mail+Nijh89pre,Mail+Nijh89PLB,Mail+Nijh89GTMP,Mail+Nijh90}.
The underlying idea of relaxing a system of $N$-simplex equations in the above way, by introducing
an object $\hat{\cR}$, such that consistency imposes the $(N+1)$-simplex equation on it,
is the ``obstruction method'' in~\cite{DMH12KPBT,Hiet+Nijh97, Mail+Nijh89pre,Mail+Nijh89PLB,Mail+Nijh89GTMP,Mail+Nijh90,Michi+Nijh93}.
Also see~\cite{Baez+Neuc96, Kapr+Voev94PSPM,Kapr+Voev94JPAA,Kazh+Soib93,Stre95} for a~formulation
in the setting of 2-categories. Indeed, the obstruction method corresponds to the introduction
of laxness (``laxif\/ication''~\cite{Stre95}).

An equation of a similar nature as the Yang--Baxter equation is the
\emph{pentagon equation}
\begin{gather}
      \hat{\cT}_{\bsy{12}}   \hat{\cT}_{\bsy{13}}   \hat{\cT}_{\bsy{23}}
    = \hat{\cT}_{\bsy{23}}   \hat{\cT}_{\bsy{12}}   ,  \label{pentagon_eq_introd}
\end{gather}
which appears as the Biedenharn--Elliott identity for Wigner $6j$-symbols and Racah coef\/f\/icients in the
representation theory of the rotation group~\cite{Bied+Louc81}, as an identity for fusion
matrices in conformal f\/ield theory~\cite{Moor+Seib89}, as a consistency condition for the
\emph{associator} in quasi-Hopf algebras~\cite{Drin89,Drin91} (also see
\cite{AET10,Alek+Toro12,Bar-N98,Etin+Kazh96,Etin+Schi98,Furu10,Furu14}),
as an identity for the Rogers dilogarithm function~\cite{Roge07}
and matrix generalizations~\cite{Kash99TMP}, for the quantum dilogarithm
\cite{And+Kash14,Bazh+Resh95,Byts+Volk13, Fadd+Kash94,Kash+Naka11,Volk97},
and in va\-rious other contexts (see, e.g.,
\cite{Doli+Serg11,Kash96AA, Kash+Resh07,Kash+Serg98,Kore04,Kore+Sait99, Mail94}).
In particular, it is satisf\/ied by the Kac--Takesaki operator $(\hat{\cT} f)(g,g') = f(gg',g')$, $g,g' \in G$,
$G$ a group, where it expresses the associativity of the group operation (see, e.g.,~\cite{Woro96}). A unitary operator acting on $\cH \otimes \cH$,
where~$\cH$ is a Hilbert space, and satisfying the pentagon equation, has been termed a
\emph{multiplicative unitary}~\cite{Baaj+Skan93,Baaj+Skan98,Baaj+Skan03,Izum+Kosa02, Masu+Naka93,Skan91,Woro96,Zakr92}.
It plays an essential role in the development of harmonic analysis on quantum groups.
Under certain additional conditions, such an operator can be used to construct
a quantum group on the $C^\ast$ algebra level~\cite{Baaj+Skan93,Ip13,Skan91,Woro96}.\footnote{See
\cite{Woro+Zakr02} for the example of the Hopf algebra of the quantum plane $ab = q^2 ba$
with~$q$ a root of unity.}
For any locally compact quantum group, a multiplicative unitary can be constructed in terms of the coproduct.
Any f\/inite-dimensional Hopf algebra is characterized by an invertible solution of the pentagon equation~\cite{Mili04}.
The pentagon equation arises as a $3$-cocycle condition in Lie group cohomology
and also in a category-theoretical framework (see, e.g.,~\cite{Stre98}).

A pentagon relation arises from ``laxing'' the associativity law~\cite{MacL63,Stash63, Tamari1951thesis}.
In its most basic form, it describes a partial order (on a set of f\/ive elements), which is the simplest
\emph{Tamari lattice} (also see~\cite{MHPS12}). This is $T(5,3)$ in the notation of this work (also see
\cite{DMH12KPBT}).

We will show that the pentagon equation belongs to an inf\/inite family of equations,
which we call \emph{polygon equations}. They are associated with \emph{higher Tamari orders},
as def\/ined\footnote{This def\/inition
of higher Tamari orders emerged from our exploration of a special class
of line soliton solutions of the Kadomtsev--Petviashvili (KP) equation~\cite{DMH11KPT,DMH12KPBT}. }
in~\cite{DMH12KPBT}, in very much the same way as the simplex equations
are associated with higher Bruhat orders (also see~\cite{DMH12KPBT}).
We believe that these higher Tamari orders coincide with
\emph{higher Stasheff--Tamari orders}, def\/ined in terms of triangulations of cyclic polytopes
\cite{Edel+Rein96,Kapr+Voev91,Ramb+Rein12}.\footnote{More precisely, HST${}_1(n,d)$, as def\/ined, e.g.,
in~\cite{Ramb+Rein12}, is expected to be order isomorphic to $T(n,d+1)$, as def\/ined in Section~\ref{sec:B&T}. }

In Section~\ref{subsec:Bruhat_decomp} we show that any higher Bruhat order can be decomposed
into a corresponding higher Tamari order, its dual (which is the reversed Tamari order),
and a ``mixed order''. A~certain projection of higher Bruhat to higher (Stashef\/f--)Tamari orders appeared in~\cite{Kapr+Voev91} (also see~\cite{Ramb+Rein12} and references cited there).
Whereas this projects, for example,  $B(4,1)$ (permutahedron) to $T(6,3)$ (associahedron), we
describe a projection $B(4,1) \rightarrow T(4,1)$ (tetrahedron), and more generally $B(N,n) \rightarrow T(N,n)$.

The $(N+1)$-simplex equation arises as a consistency
condition of a system of $N$-simplex equations. The higher Bruhat orders are also crucial for
understanding this ``integrability'' of the simplex equations. In the same way,
the higher Tamari orders provide the combinatorial tools to express integrability of
polygon equations.

\begin{figure}
\centering
\includegraphics[width=.8\linewidth]{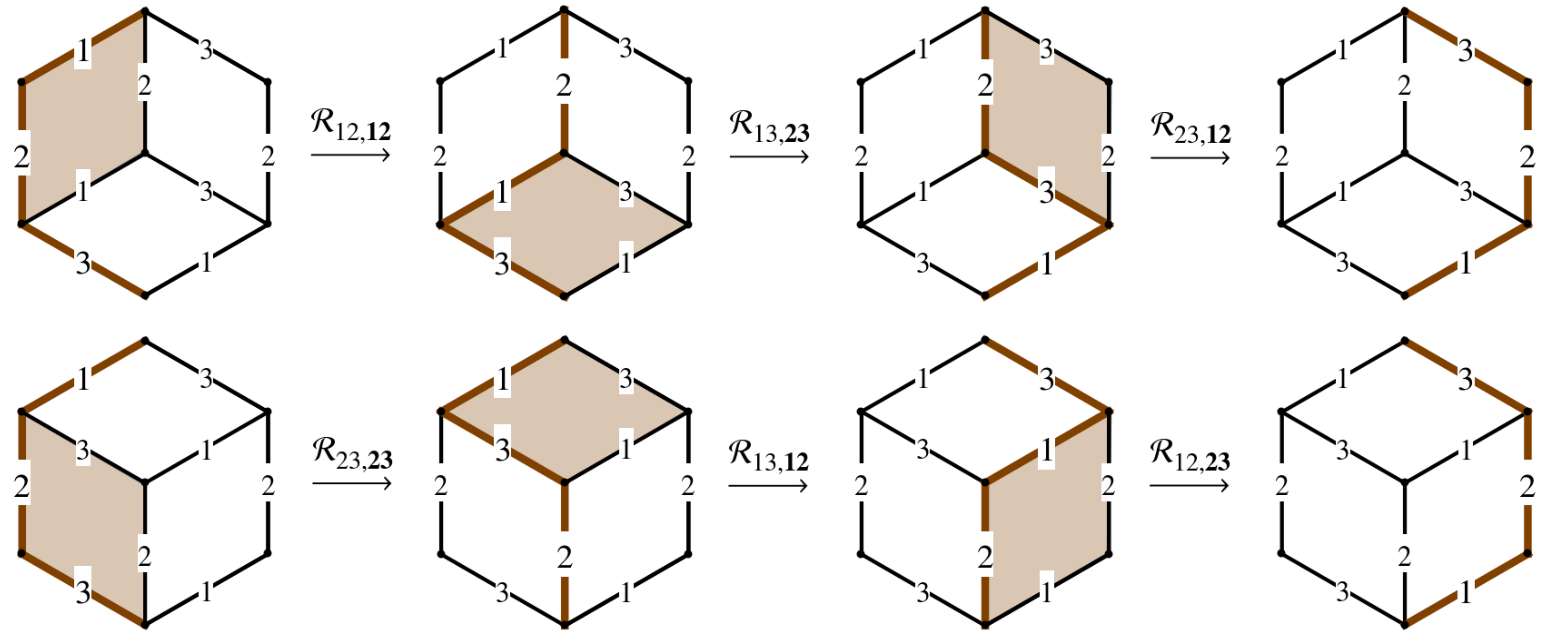}
\caption{The f\/irst row shows a sequence of maximal chains on half of the poset $B(3,0)$, which is
the Boolean lattice on $\{1,2,3\}$.
The second row shows a corresponding sequence on the complementary part.
Glued together along their boundaries, they form
a cube. The two ways of deforming the initial lexicographically ordered maximal chain
to the f\/inal, reverse lexicographically ordered chain, results in a~consistency
condition, which is the Yang--Baxter equation~(\ref{2-simplex_eq_2}).
Edges are associated with spaces~$V_i$, $i=1,2,3$. For example, the second step in the f\/irst row corresponds
to the action of~$\mathrm{id} \otimes \cR_{13}$ on~$V_2 \otimes V_1 \otimes V_3$.
The boldface indices that determine where, in a product of spaces, the map acts, correspond
to the positions of the active edges in the respective (brown) maximal chain, counting from
the top downward.\label{fig:YB_on_cube} }
\end{figure}

Using the transposition map $\cP$ to def\/ine $\cR := \hat{\cR}   \cP$, and generalizing this
to maps $\cR_{ij}\colon V_i \otimes V_j \rightarrow V_j \otimes V_i$, the Yang--Baxter equation
takes the form
\begin{gather}
    \cR_{23,\bsy{1}}   \cR_{13,\bsy{2}}   \cR_{12,\bsy{1}}
  = \cR_{12,\bsy{2}}   \cR_{13,\bsy{1}}   \cR_{23,\bsy{2}}
        \qquad \mbox{on} \quad V_1 \otimes V_2 \otimes V_3  .   \label{2-simplex_eq_2}
\end{gather}
$\cR_{ij,\bsy{a}} := \cR_{ij,\bsy{a},\bsy{a}+1}$ acts on $V_i$ and $V_j$, at positions
$\bsy{a}$ and $\bsy{a}+1$, \looseness=-1
in a product of such spaces. The higher Bruhat orders ensure a correct matching
of the two dif\/ferent types of indices. In fact, the boldface indices are completely determined,
they do not contain independent information. Fig.~\ref{fig:YB_on_cube} shows a
familiar visualization of the Yang--Baxter equation in terms of deformations of chains of edges
on a cube. Supplying the latter with the Bruhat order $B(3,0)$, these are maximal chains.
The information given in the caption of Fig.~\ref{fig:YB_on_cube}
will also be relevant for subsequent f\/igures in this work.

The (weak) Bruhat orders $B(N,1)$, $N >2$, form polytopes called \emph{permutahedra}.
Not all higher Bruhat orders can be realized on polytopes.
The $N$-simplex equation is associated with the Bruhat order $B(N+1,N-1)$, but its structure
is rather visible on $B(N+1,N-2)$. The latter possesses a reduction to the 1-skeleton of
a polyhedron on which the simplex
equation can be visualized in the same way as the Yang--Baxter equation is visualized on $B(3,0)$
(also see~\cite{Aitc10} for a similar view).
This is elaborated in Section~\ref{sec:simplex_eqs}.

\begin{figure}
\centering
\includegraphics[width=.8\linewidth]{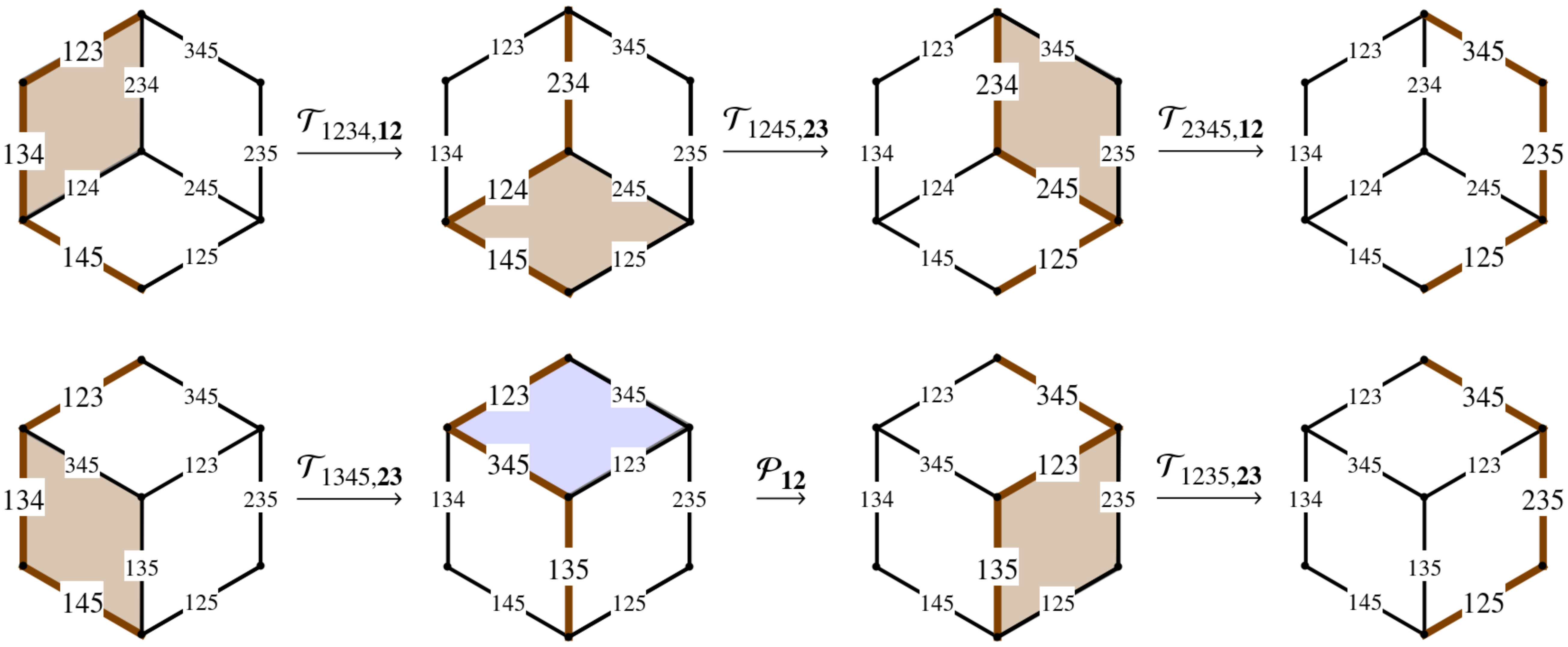}
\caption{ Here the poset is $T(5,2)$, which also forms a cube. Edges are now numbered by
$3$-element subsets of $12345 := \{1,2,3,4,5\}$.
If the edges of a face are labeled by the four $3$-element subsets of $ijkl$,
there is a map $\cT_{ijkl}$ associated with it.
This rule does not apply to the second step in the second row, however, since here the edges
of the active face involve \emph{five} (rather than only four) digits.
The two possibilities of deforming the initial
(lexicographically ordered) maximal chain $(123,134,145)$ into the f\/inal (reverse lexicographically ordered)
maximal chain $(345,235,125)$ result in the pentagon equation~(\ref{pentagon_eq_mod}).
\label{fig:pentagon_eq_on_cube} }
\end{figure}

Also for the polygon equations, proposed in this work, all appearances
of a map, like $\cT$ in the pentagon equation in Fig.~\ref{fig:pentagon_eq_on_cube}, will be treated
as a priori dif\/ferent maps (now on both sides of the equation).
Again we attach to them additional indices that carry combinatorial information, now governed by
higher Tamari orders.

The Tamari orders $T(N,1)$, $N=3,4, \ldots$, $T(N,2)$, $N=4,5, \ldots$, $T(N,3)$, $N=5,6, \ldots$,
as def\/ined in~\cite{DMH12KPBT} and Section~\ref{sec:B&T}, form simplexes, hypercubes and \emph{associahedra}
(Stashef\/f--Tamari polytopes), respectively.
But not all Tamari orders can be realized on polytopes.

The $N$-gon equation is associated with $T(N,N-2)$, but its structure is rather revealed by
$T(N,N-3)$. For small enough $N$, the latter forms a polyhedron. For higher $N$ it admits a~polyhedral reduction.
The structure of the $N$-gon equation can then be visualized in terms of deformations of maximal chains
on the corresponding polyhedron. Fig.~\ref{fig:pentagon_eq_on_cube} shows the example
of the pentagon equation, here obtained in the form
\begin{gather}
    \cT_{2345,\bsy{1}}  \cT_{1245,\bsy{2}}  \cT_{1234,\bsy{1}}
  = \cT_{1235,\bsy{2}}  \cP_{\bsy{1}}  \cT_{1345,\bsy{2}}
                \label{pentagon_eq_mod}
\end{gather}
(also see Section~\ref{sec:polygon_eqs}). This implies (\ref{pentagon_eq_introd}) for $\hat{\cT} = \cT \cP$.

Section~\ref{sec:B&T} f\/irst provides a brief account of higher Bruhat orders
\cite{Manin+Schecht86a,Manin+Shekhtman86b,Manin+Schecht89,Ziegler93}. The main result in this section is
a decomposition of higher Bruhat orders, where higher Tamari orders (in the form
introduced in~\cite{DMH12KPBT}) naturally appear.
Section~\ref{sec:simplex_eqs} explains the relation between higher Bruhat orders and simplex
equations, and how the next higher simplex equation arises as a~consistency condition of
a localized system of simplex equations.
Section~\ref{sec:polygon_eqs} associates in a similar way \emph{polygon equations}, which
generalize the pentagon equation, with higher Tamari orders.
As in the case of simplex equations, the $(N+1)$-gon equation arises as
a consistency condition of a~system of localized $N$-gon equations.
Section~\ref{sec:simplex_polygon_rel} reveals relations between simplex and polygon equations,
in particular providing a deeper explanation for and considerably generalizing a relation between
the pentagon equation and the $4$-simplex equation, f\/irst observed in~\cite{Kash+Serg98}.
Finally, Section~\ref{sec:concl} contains some concluding remarks and Appendix~\ref{app:alg}
supplements all this by expressing some features of simplex and polygon equations via a
more abstract approach.

\section{Higher Bruhat and Tamari orders}
\label{sec:B&T}
In the f\/irst subsection we recall some material about higher Bruhat orders from
\cite{Manin+Schecht86a,Manin+Shekhtman86b,Manin+Schecht89,Ziegler93}. The second subsection
introduces a decomposition of higher Bruhat orders that includes higher Tamari orders, in
the form we def\/ined them in~\cite{DMH12KPBT}.

\subsection{Higher Bruhat orders}
\label{subsec:hB}
For $N \in \bbN$, let $[N]$ denote the set $\{ 1,2,\ldots,N \}$.
The \emph{packet} $P(K)$ of $K \subset [N]$ is the set of all subsets of $K$
of cardinality one less than that of $K$.
For $K = \{k_1, \ldots, k_{n+1}\}$ in natural order,
i.e., $k_1 < \cdots < k_{n+1}$, we set
\begin{gather*}
   \vec{P}(K) := ( K \setminus \{k_{n+1}\} , K \setminus \{k_n\} , \ldots ,  K \setminus \{k_1\} )   ,
     \\ \cev {P}(K) := ( K \setminus \{k_1\} , K \setminus \{k_2\} , \ldots ,  K \setminus \{k_{n+1}\} )
        .
\end{gather*}
The f\/irst displays the packet of $K$ in lexicographical order ($<_{\mathrm{lex}}$).
The second displays $P(K)$ in reverse lexicographical order.

Let ${[N] \choose n}$,
$0 \leq n \leq N$, denote the set of all subsets of $[N]$ of cardinality $n$. Its cardinality is
\begin{gather*}
      c(N,n) := {N \choose n}   .
\end{gather*}
A linear (or total) order $\rho$ on ${[N] \choose n}$ can be written as a
sequence $\rho = (J_1,\ldots,J_{c(N,n)})$ with $J_a \in {[N] \choose n}$. It is called \emph{admissible} if, for
all $K \in {[N] \choose n+1}$, $\rho$ induces on $P(K)$ either the lexicographical or the reverse
lexicographical order, i.e., either $\vec{P}(K)$ or~$\cev{P}(K)$ is a subsequence of~$\rho$.
Let $A(N,n)$ denote the set of admissible linear orders of ${[N] \choose n}$.

The \emph{envelope} $E(J)$ of $J \in {[N] \choose n}$ is the set of $K \in {[N] \choose n+1}$
such that $J \in P(K)$.
An equivalence relation on $A(N,n)$ is obtained by setting $\rho \sim \rho'$
if $\rho$ and $\rho'$ only dif\/fer by a sequence of exchanges of neighboring elements $J$, $J'$
with $E(J) \cap E(J') = \varnothing$. We set
\begin{gather*}
      B(N,n) := A(N,n)/{\sim }  .
\end{gather*}

\begin{exam}
${[4] \choose 2} = \{ 12, 13, 14, 23, 24, 34 \}$, where $ij := \{i,j\}$, allows $6! = 720$
linear orders, but only $14$ are admissible. For example, $\rho = (12,34,14,13,24,23) \in A(4,2)$,
since it contains the packets of the four elements of ${[4] \choose 3}$ in the orders
$\vec{P}(123), \vec{P}(124), \cev{P}(134), \cev{P}(234)$. We have
$\rho \sim (34,12,14,13,24,23) \sim (34,12,14,24,13,23)$. $B(4,2) = A(4,2)/{\sim}$ has $8$ elements.
\end{exam}

The \emph{inversion set} $\mathrm{inv}[\rho]$ of $\rho \in A(N,n)$ is the set of all
$K \in {[N] \choose n+1}$ such that $P(K)$ is contained in $\rho$ in \emph{reverse} lexicographical order.
All members of the equivalence class $[\rho] \in B(N,n)$ have the same inversion set.
Next we introduce the inversion operation
\begin{gather*}
       I_K\colon \ \vec{P}(K) \mapsto \cev{P}(K)   .
\end{gather*}
If $\vec{P}(K)$ appears in $\rho \in A(N,n)$ at consecutive positions, let $I_K\rho$ be the
linear order obtained by inversion of $\vec{P}(K)$ in $\rho$. Then\footnote{Since two dif\/ferent packets
have at most a single member in common, such an inversion does not change the order of other
packets in $\rho$ than that of $K$.}
$I_K\rho \in A(N,n)$ and $\mathrm{inv}[I_K\rho] = \mathrm{inv}[\rho] \cup \{K\}$.
This corresponds to the covering relation
\begin{gather*}
     [\rho] \stackrel{K}{\rightarrow} [I_K\rho]   ,  %\label{hB_inversion_on_rho}
\end{gather*}
which determines the \emph{higher Bruhat order} on the set $B(N,n)$
\cite{Manin+Schecht86a,Manin+Schecht89}.
In the following, we will mostly drop the adjective ``higher''.
$B(N,n)$ has a unique minimal element $[\alpha]$ that contains the lexicographically ordered
set ${[N] \choose n}$, hence $\mathrm{inv}[\alpha] = \varnothing$, and
a unique maximal element $[\omega]$ that contains the reverse lexicographically ordered
set ${[N] \choose n}$, hence $\mathrm{inv}[\omega] = {[N] \choose n+1}$.
The Bruhat orders are naturally extended by def\/ining $B(N,0)$ as the Boolean lattice
on $[N]$, which corresponds to the 1-skeleton of the $N$-cube,
with edges directed from a f\/ixed vertex toward the opposite vertex.

\begin{rem}
\label{rem:A(N,n+1)-B(N,n)}
There is a natural correspondence between the elements of $A(N,n+1)$ and the maximal chains of $B(N,n)$
\cite{Manin+Schecht89}.
Associated with $\sigma = (K_1, \ldots, K_{c(N,n+1)}) \in A(N,n+1)$ is the maximal chain
\begin{gather*}
   [\alpha] \stackrel{K_1}{\longrightarrow} [\rho_1] \stackrel{K_2}{\longrightarrow} [\rho_2]
                    \stackrel{K_3}{\longrightarrow}  \cdots
             \stackrel{K_{c(N,n+1)}}{\longrightarrow} [\omega]
                       ,   %\label{hB_max_chain}
\end{gather*}
where
$\mathrm{inv}[\rho_r] = \{ K_1, \ldots, K_r \}$. This allows to construct $B(N,n)$ from $B(N,n+1)$.
As a~consequence, all Bruhat orders $B(N,n)$, $n<N-1$, can be constructed recursively from the
highest non-trivial, which is $B(N,N-1)$.
\end{rem}

\begin{exam}
$B(N,N-1)$ is simply $\vec{P}([N]) \stackrel{[N]}{\rightarrow} \cev{P}([N])$. From the two
admissible linear orders $\vec{P}([N]) = (\hat{N}, \ldots, \hat{2}, \hat{1})$ and
$\cev{P}([N]) = (\hat{1}, \hat{2}, \ldots, \hat{N})$, where $\hat{k} := [N] \setminus \{k\}$
(``complementary notation''),
we can construct the two maximal chains of $B(N,N-2)$:
\begin{gather}
   [\alpha] \stackrel{\hat{N}}{\longrightarrow} [\rho_1] \stackrel{\widehat{N-1}}{\longrightarrow} [\rho_2]
      \longrightarrow \cdots \longrightarrow [\rho_{N-1}] \stackrel{\hat{1}}{\longrightarrow} [\omega]
              , \nonumber \\
  [\alpha] \stackrel{\hat{1}}{\longrightarrow} [\sigma_1] \stackrel{\hat{2}}{\longrightarrow} [\sigma_2]
      \longrightarrow \cdots \longrightarrow [\sigma_{N-1}] \stackrel{\hat{N}}{\longrightarrow} [\omega]   .
             \label{B(N,N-2)_chains}
\end{gather}
The example $B(4,2)$ is displayed below in (\ref{B(4,2)-chains}).
\end{exam}

\begin{rem}
\label{rem:consistent_sets}
$U \subset {[N] \choose n+1}$ is called a \emph{consistent set} if, for all $L \in {[N] \choose n+2}$,
$U \cap P(L)$ can be ordered in such a way that it becomes a \emph{beginning segment} either of $\vec{P}(L)$
or of $\cev{P}(L)$.\footnote{A \emph{beginning segment} of a sequence is a subsequence that starts with the
f\/irst member of the sequence and contains all its members up to a f\/inal one.
Also the empty sequence and the full sequence are beginning segments.}
Consistent sets are in bijective correspondence with inversion sets~\cite{Ziegler93}.
\end{rem}

\begin{rem}
\label{rem:Bruhat_reduction}
 For f\/ixed $k \in [N+1]$, we def\/ine an equivalence relation in $A(N+1,n+1)$ as follows.
Let $\rho \stackrel{k}{\sim} \rho'$ if $\rho$ and $\rho'$ only dif\/fer in the order of elements
$K \in {[N+1] \choose n+1}$ with $k \notin K$.
 For $\rho = (K_1,\ldots, K_{c(N+1,n+1)}) \in A(N+1,n+1)$, the equivalence class
$\rho_{(k)} \in A(N+1,n+1)/{\stackrel{k}{\sim}}$ is then completely characterized by
the subsequence $(K_{i_1},\ldots, K_{i_{c(N,n)}})$ consisting of only those~$K_i$
that contain $k$. Hence we can identify $\rho_{(k)}$ with this subsequence. As a~consequence,
there is an obvious bijection between $A(N+1,n+1)/{\stackrel{k}{\sim}}$ and $A(N,n)$.
Clearly, inversions of pac\-kets~$P(L)$, $L \in {[N+1] \choose n+2}$, with $k \notin L$, have no
ef\/fect on the equivalence classes. If $k \in L$, and if~$\vec{P}(L)$ appears in $\rho$
at consecutive positions, then $(I_L\rho)_{(k)}$ is obtained from $\rho_{(k)}$ by
the inversion
$\vec{P}(L) \setminus \{L \setminus \{k\}\} \rightarrow \cev{P}(L) \setminus \{L \setminus \{k\}\}$.
Since the latter naturally corresponds to $\vec{P}(L \setminus \{k\}) \rightarrow \cev{P}(L \setminus \{k\})$,
the bijection $A(N+1,n+1)/\!\!\stackrel{k}{\sim} \, \rightarrow A(N,n)$ is monotone, i.e., order-preserving.
Since the equivalence relation $\sim$ is compatible with $\stackrel{k}{\sim}$, the bijection induces
a corresponding monotone bijection $B(N+1,n+1)/{\stackrel{k}{\sim}} \rightarrow B(N,n)$.
We will use this \emph{projection} in Section~\ref{subsec:simplex_red}.
Examples are shown in Figs.~\ref{fig:B41_red} and \ref{fig:B52_red}.
\begin{figure}[t]\centering
\includegraphics[width=.4\linewidth]{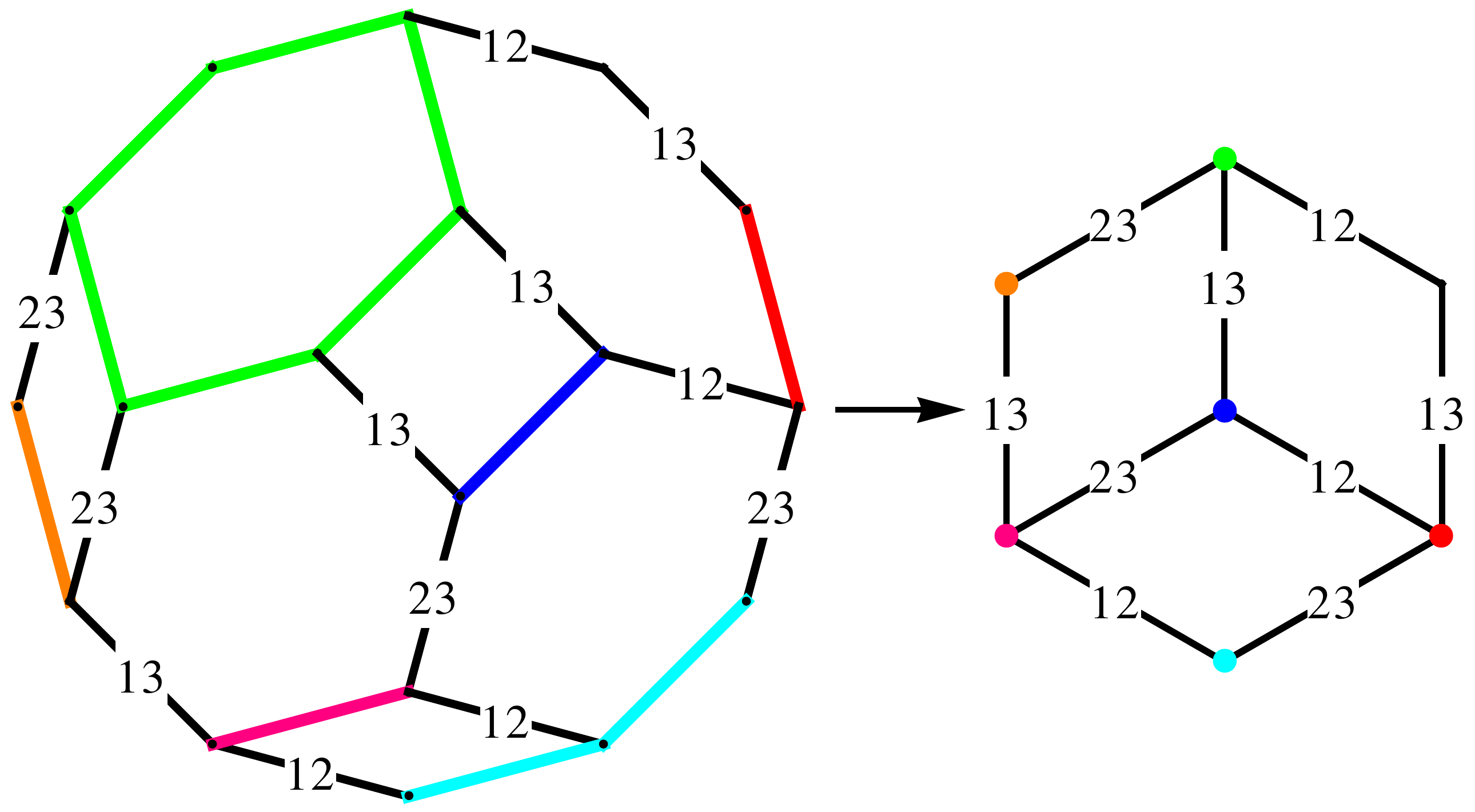}
\hspace{1.cm}
\includegraphics[width=.4\linewidth]{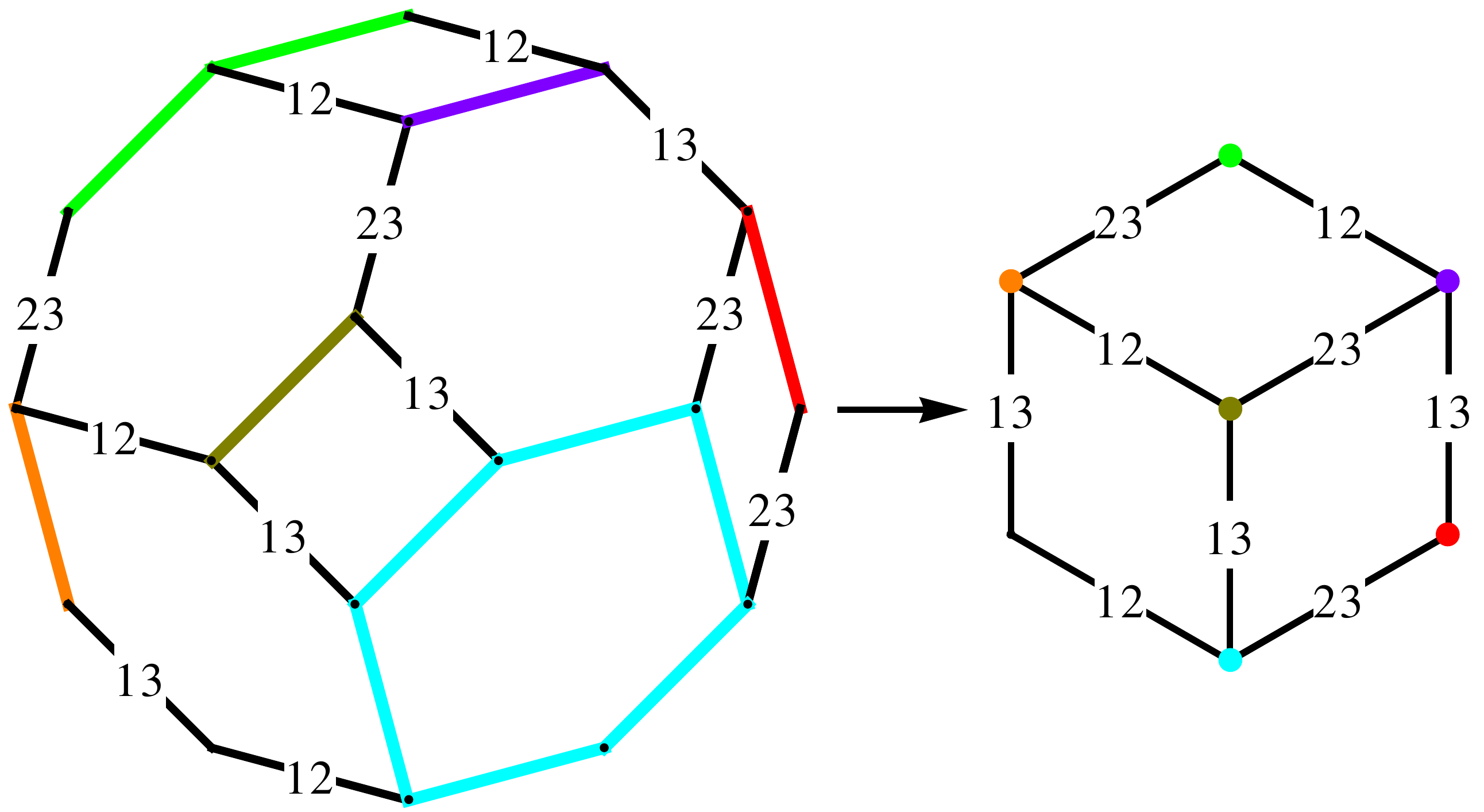}
\caption{Projection of $B(4,1)$ (permutahedron) to $B(3,0)$ (cube), each split into two
complementary parts. Here we chose $k=4$
in Remark~\ref{rem:Bruhat_reduction} and use complementary notation for the labels, but with hats omitted.
The coloring marks those parts in the two Bruhat orders that are related by the projection.
\label{fig:B41_red} }
\end{figure}
\begin{figure}[t]
\centering
\includegraphics[width=.44\linewidth]{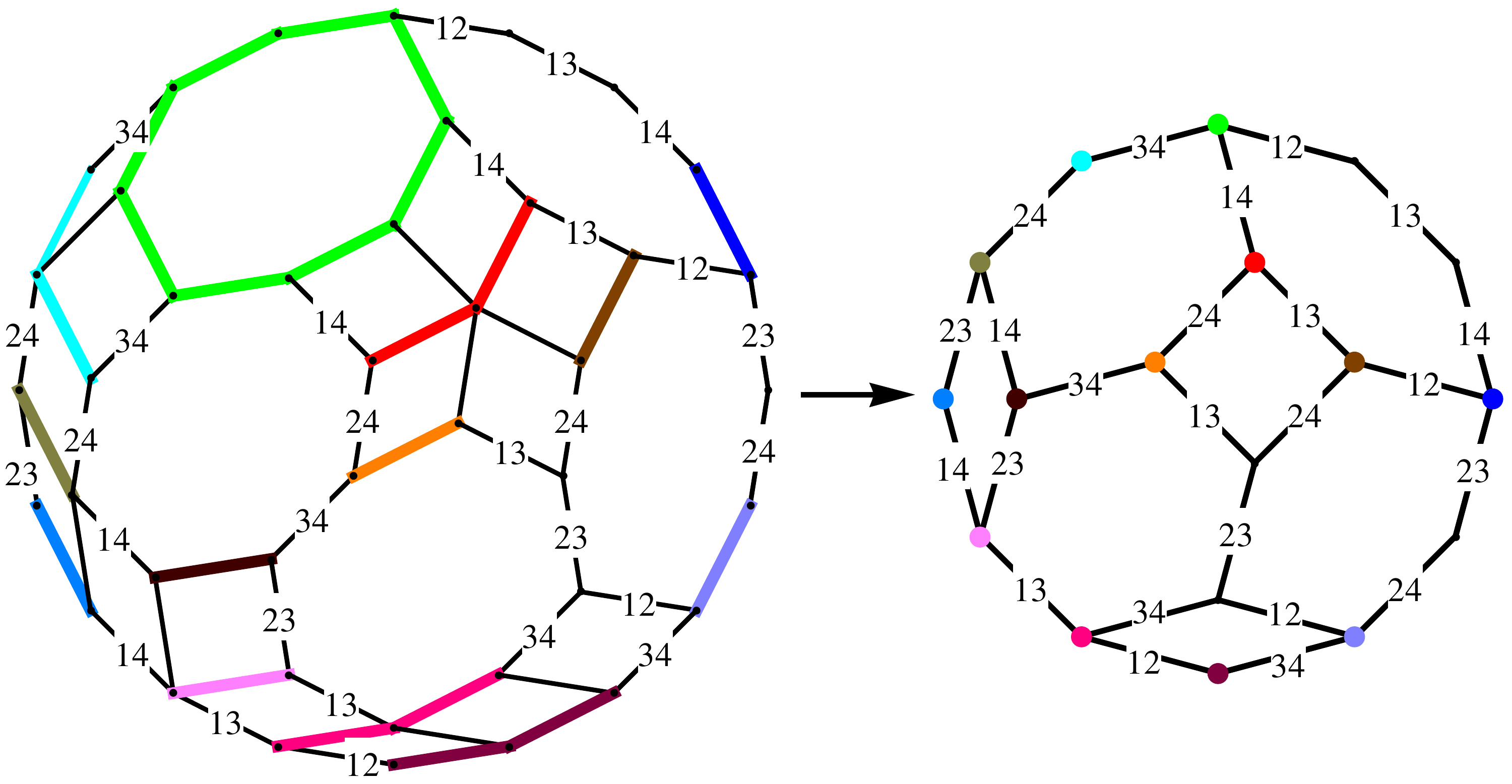}
\hspace{.5cm}
\includegraphics[width=.44\linewidth]{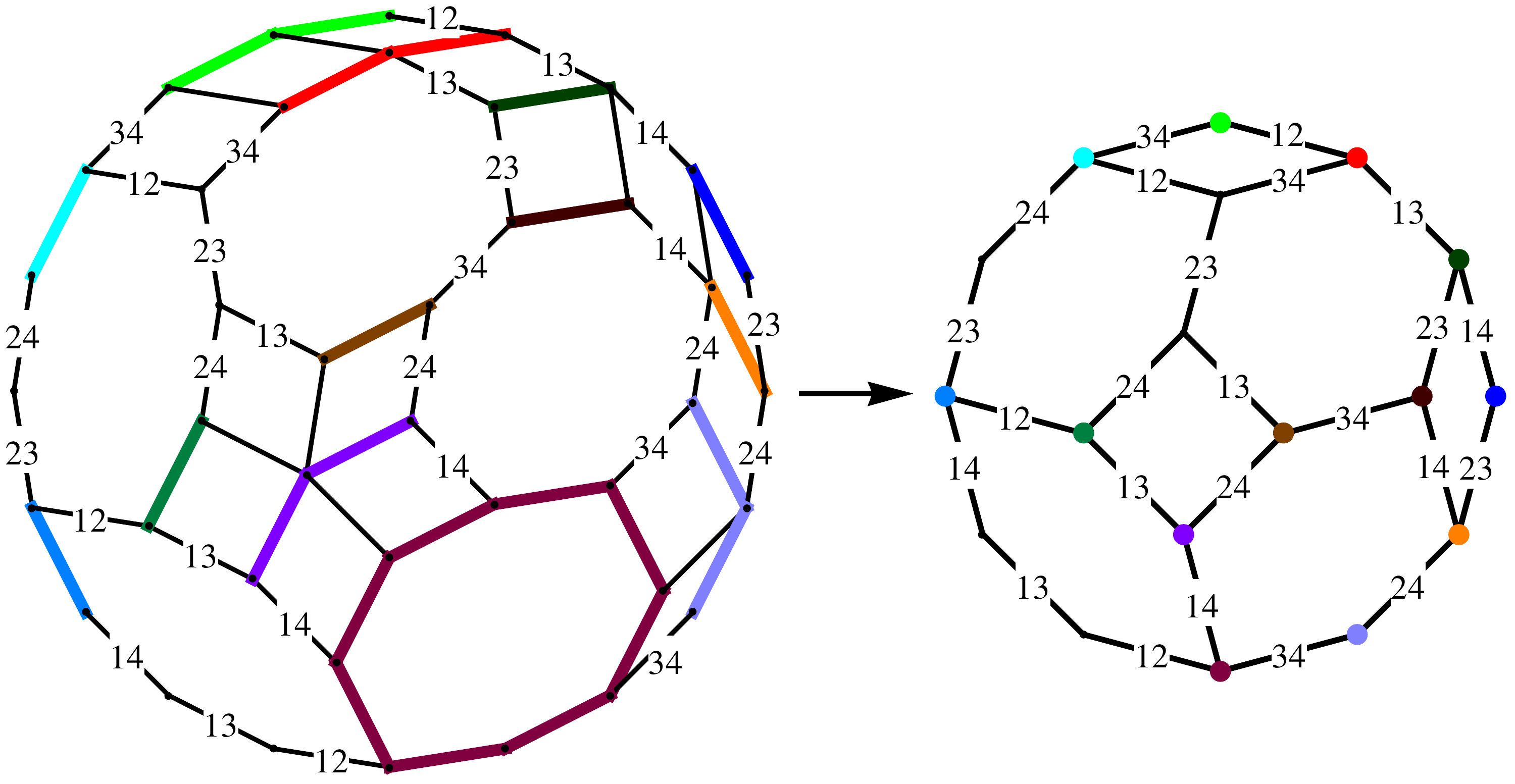}
\caption{Projection of $B(5,2)$ (Felsner--Ziegler polyhedron~\cite{Fels+Zieg01})
to $B(4,1)$ (permutahedron).
Here we chose $k=5$ in Remark~\ref{rem:Bruhat_reduction}. Again, we use complementary labeling,
and the coloring marks the parts related by the projection.
\label{fig:B52_red} }
\end{figure}
\end{rem}

\begin{exam}
$\alpha = (123,124,134,234,125,135,235,145,245,345)$ represents the minimal
element $[\alpha]$ of $B(5,3)$. Let $k=5$. Then $\alpha_{(5)}$ is
represented by $(125,135,235,145,245,345)$. This corresponds to $(12,13,23,14,24,34)$, which
represents the minimal element of $B(4,2)$. Inversion of the packet of $L = \{ i,j,m, 5\}$,
$1 \leq i < j < m < 5$, corresponds to inversion of the packet of $\{ i,j,m\}$, which
def\/ines an edge in $B(4,2)$.
\end{exam}

\subsection{Three color decomposition of higher Bruhat orders}
\label{subsec:Bruhat_decomp}
 For $K \in {[N] \choose n+1}$, let $P_o(K)$, respectively $P_e(K)$, denote the \emph{half-packet}
of elements of $P(K)$ with odd, respectively even, position in the lexicographical order.
We assign colors to elements of~$\vec{P}(K)$, respectively $\cev{P}(K)$, as follows.
An element of $P_o(K)$ is \emph{blue} in $\vec{P}(K)$ and \emph{red} in $\cev{P}(K)$,
and an element of $P_e(K)$ is \emph{red} in $\vec{P}(K)$ and \emph{blue} in $\cev{P}(K)$.

\begin{exam}
For $K = \{1,2,3,4,5\} = 12345$, we have $P_o(K) = \{1234,1245,2345\}$ and
$P_e(K) = \{1235,1345\}$. Hence
\begin{gather*}
    \vec{P}(12345)
   = ( {\color{blue} 1234}, {\color{darkred} 1235}, {\color{blue}1245},
         {\color{darkred} 1345}, {\color{blue}2345} )   , \qquad
    \cev{P}(12345)
   =  ( {\color{darkred} 2345}, {\color{blue} 1345}, {\color{darkred} 1245},
         {\color{blue} 1235}, {\color{darkred} 1234} )   .
\end{gather*}
\end{exam}

We say $J \in {[N] \choose n}$ is \emph{blue (red) in} $\rho \in A(N,n)$ if,
for all $K \in E(J)$, $J$ is blue (red) in $\vec{P}(K)$, respectively $\cev{P}(K)$, depending
in which order $P(K)$ appears in~$\rho$.\footnote{This means that $J$ is blue (red)
in $\rho$ if $J \in P_o(K)$ ($J\in P_e(K)$) for all $K \in E(J) \setminus \mathrm{inv}[\rho]$,
and $J \in P_e(K)$ ($J \in P_o(K)$) for all $K \in E(J) \cap \mathrm{inv}[\rho]$. \label{foot:color_def}}
$J$ is called \emph{green in} $\rho$, if there are $K,K' \in E(J)$, such that $J$
is blue with respect to $K$ and red with respect to $K'$.

\begin{exam}
The following element of $A(5,3)$ has empty inversion set,
\begin{gather*}
  \alpha = ({\color{blue} 123},{\color{darkgreen}124},{\color{darkred}125},{\color{blue} 134},
           {\color{darkgreen} 135},{\color{blue} 145},
           {\color{darkgreen}234},{\color{darkred}235},{\color{darkgreen}245},{\color{darkred}345})   .
\end{gather*}
For example, we have $E(124) = \{ 1234,1245 \}$, and $\alpha$ contains
$\vec{P_o}(1245) = (124,145)$ and $\vec{P_e}(1234) = (124,234)$ as subsequences.
This shows that $124$ is blue in $\vec{P}(1245)$ and red in $\vec{P}(1234)$, therefore
green in~$\alpha$.
\end{exam}

For each $c \in \{b,r,g\}$ (where $b,r,g$ stands for blue, red and green, respectively), we def\/ine
an equivalence relation on $A(N,n)\colon \rho \sim_c   \rho'$ if $\rho$ and $\rho'$ have the same
elements with color $c$ in the same order.
Let $\rho^{(c)}$ denote the corresponding equivalence class, and
\begin{gather*}
       A^{(c)}(N,n) := A(N,n)/{\sim_c}, \qquad c \in \{b,r,g\}   .
\end{gather*}
$\rho^{(c)}$ can be identif\/ied with the subsequence of elements in $\rho$ having color $c$.

The def\/inition of the color of an element $J$ of a linear order $\rho$ only involves the inversion
set of $\rho$, but not $\rho$ itself (also see footnote~\ref{foot:color_def}).
Hence, if $J$ has color $c$ in $\rho$, then it has the same
color in any element of $[\rho]$. As a~consequence,
for each $c$, the equivalence relation $\sim_c$ is compatible with $\sim$. Def\/ining
\begin{gather*}
       B^{(c)}(N,n) := A^{(c)}(N,n)/{\sim} = (A(N,n)/{\sim})/{\sim_c}, \qquad c \in \{b,r,g\}   ,
\end{gather*}
we thus obtain a projection $B(N,n) \rightarrow B^{(c)}(N,n)$ via $[\rho] \mapsto [\rho^{(c)}]$.
We will show that the resulting single-colored sets inherit a partial order from the respective
Bruhat order.

\begin{lem}
\label{lem:color}
Let $K \in E(J) \cap P_o(L)$ $($respectively, $K \in E(J) \cap P_e(L))$ for some
$J \in {[N] \choose n}$ and $L \in {[N] \choose n+2}$, where $n<N-1$.
Let $K' \in E(J) \cap P(L)$, $K' \neq K$.

If $K <_{\mathrm{lex}} K'$, then $J \in P_o(K')$ $($respectively, $J \in P_e(K'))$.

If $K' <_{\mathrm{lex}} K$, then $J \in P_e(K')$ $($respectively, $J \in P_o(K'))$.
\end{lem}

\begin{proof}
Since $K,K' \in E(J)$ and $K \neq K'$, we can write $K=J \cup \{k\}$ and $K' = J \cup \{k'\}$, with
$k,k' \notin J$, $k \neq k'$, and $L = K \cup \{k'\} = K' \cup \{k\}$.
$K <_{\mathrm{lex}} K'$ is equivalent to $k < k'$.
Let us write $\vec{P}(L) = (L \setminus \{\ell_{n+2}\},\ldots,L \setminus \{\ell_1\})$ with
$\ell_1 < \ell_2 < \cdots < \ell_{n+2}$.

$K \in P_o(L)$ ($K \in P_e(L)$) means that $K = L \setminus \{k'\}$ has an odd (even) position in $\vec{P}(L)$,
hence~$k'$ has an odd (even) position in $(\ell_{n+2},\ldots,\ell_1)$.
If $k < k'$, then removal of $k$ from $(\ell_{n+2},\ldots,\ell_1)$ does not change this,
so that $k'$ also has an odd (even) position in $(\ell_{n+2},\ldots,\check{k},\ldots,\ell_1)$, where~$\check{ }$ indicates an omission. It follows that $J = K' \setminus \{k'\} \in P_o(K')$ ($J \in P_e(K')$).
If $k' < k$, then the position of $k'$ in $(\ell_{n+2},\ldots,\check{k},\ldots,\ell_1)$ is even (odd),
hence $J = K \setminus \{k'\} \in P_e(K')$ ($J \in P_o(K')$).
\end{proof}

In view of the bijection between $A(N,n+1)$ and the set of maximal chains of $B(N,n)$,
it is natural to say that $[\rho] \stackrel{K}{\rightarrow} [I_K \rho]$ \emph{has color} $c$
in some maximal chain of $B(N,n)$ if $K$ has this color in the associated element of $A(N,n+1)$.
An equivalent statement, formulated next, in particular shows that the color of
$[\rho] \stackrel{K}{\rightarrow} [I_K \rho]$ is the same in any maximal chain in which it
appears, hence we can speak about $[\rho] \stackrel{K}{\rightarrow} [I_K \rho]$ having color $c$
in $B(N,n)$.

If $\vec{P}(L) \cap \mathrm{inv}[\rho]$ is a beginning segment, we will say that $\vec{P}(L)$
\emph{has beginning segment with respect to} $[\rho]$.
A corresponding formulation applies with $\vec{P}(L)$ replaced by $\cev{P}(L)$. Let $n<N-1$.
Then $[\rho] \stackrel{K}{\rightarrow} [I_K \rho]$ is \emph{blue}
(\emph{red}) if, for all $L \in E(K)$, either $\vec{P}(L)$ has beginning segment w.r.t.~$[\rho]$
and also w.r.t.~$[I_K\rho]$,
and $K \in P_o(L)$ ($K \in P_e(L)$), or $\cev{P}(L)$ has beginning segments w.r.t.~$[\rho]$
and $[I_K\rho]$,
and $K \in P_e(L)$ ($K \in P_o(L)$).\footnote{Both conditions covered by
``has beginning segments with respect to $[\rho]$ and $[I_K\rho]$'' are necessary in order
to avoid ambiguities that would
otherwise arise if $P(L) \cap \mathrm{inv}[\rho]$ is empty or if
$P(L) \cap \mathrm{inv}[I_K \rho]$ is the full packet. We are grateful to one of the
referees for pointing this out. }
Otherwise $[\rho] \stackrel{K}{\rightarrow} [I_K \rho]$ is \emph{green}.

\begin{prop}
\label{prop:color}
Let $n<N-1$, $\rho \in A(N,n)$ and $K \in {[N] \choose n+1}$.
\begin{enumerate}\itemsep=0pt
\item[$(a)$] If $[\rho] \stackrel{K}{\rightarrow} [I_K \rho]$ is blue, then the elements of $P_o(K)$
are blue in $[\rho]$ and green in $[I_K \rho]$, and the elements of $P_e(K)$
are blue in $[I_K \rho]$ and green in $[\rho]$.
\item[$(b)$] If $[\rho] \stackrel{K}{\rightarrow} [I_K \rho]$ is red, then the elements of $P_e(K)$
are red in $[\rho]$ and green in $[I_K \rho]$, and the elements of $P_o(K)$
are red in $[I_K \rho]$ and green in $[\rho]$.
\item[$(c)$] If $[\rho] \stackrel{K}{\rightarrow} [I_K \rho]$ is green, then all elements of
$P(K)$ are green in both, $[\rho]$ and $[I_K \rho]$.
\end{enumerate}
\end{prop}
\begin{proof} (a)
If $[\rho] \stackrel{K}{\rightarrow} [I_K \rho]$ is blue, this means that for all $L \in E(K)$
either (i) $\vec{P}(L)$ has beginning segments w.r.t.~$[\rho]$ and $[I_K\rho]$,
and $K \in P_o(L)$, or (ii) $\cev{P}(L)$ has beginning segments w.r.t.~$[\rho]$ and $[I_K\rho]$,
and $K \in P_e(L)$. In case (i), let $J \in P(K)$ and $K' \in P(L) \cap E(J)$.
If $K' \notin \mathrm{inv}[I_K \rho]$,
then $K <_{\mathrm{lex}} K'$ and thus $J \in P_o(K')$ by Lemma~\ref{lem:color}.
If $K' \in \mathrm{inv}[I_K \rho]$, then $K' <_{\mathrm{lex}} K$, hence $J \in P_e(K')$ according to
Lemma~\ref{lem:color}.
In both cases we can conclude that, if $J \in P_o(K)$, then $J$ is blue in $[\rho]$ and
green in $[I_K\rho]$, and if $J \in P_e(K)$, then $J$ is blue in $[I_K\rho]$ and green in $[\rho]$.
The case (ii) is treated correspondingly.

(b) is proved in the same way.

(c) $[\rho] \stackrel{K}{\rightarrow} [I_K \rho]$ green means that there are $L_1,L_2 \in E(K)$, $L_1 \neq L_2$,
such that one of the following three cases holds:
\begin{enumerate}\itemsep=0pt
\item[(i)] $\vec{P}(L_1)$ and $\vec{P}(L_2)$ have beginning segments w.r.t.~$[\rho]$ and $[I_K\rho]$, and
    $K \in P_o(L_1) \cap P_e(L_2)$,
\item[(ii)] $\vec{P}(L_1)$ and $\cev{P}(L_2)$ have beginning segments w.r.t.~$[\rho]$ and $[I_K\rho]$, and
    $K \in P_o(L_1) \cap P_o(L_2)$ or $K \in P_e(L_1) \cap P_e(L_2)$,
\item[(iii)] $\cev{P}(L_1)$ and $\cev{P}(L_2)$ have beginning segments w.r.t.~$[\rho]$ and $[I_K\rho]$, and
    $K \in P_e(L_1) \cap P_o(L_2)$.
    \end{enumerate}

In case (i), let $J \in P(K)$ and $K_1 \in P(L_1) \cap E(J)$, $K_2 \in P(L_2) \cap E(J)$, $K_1 \neq K_2$.
If $K <_{\mathrm{lex}} K_1,K_2$, then $K_1,K_2 \notin \mathrm{inv}[I_K\rho]$,
hence $J \in P_o(K_1) \cap P_e(K_2)$ by Lemma~\ref{lem:color}.
If $K_1 <_{\mathrm{lex}} K <_{\mathrm{lex}} K_2$, then $K_1 \in \mathrm{inv}[I_K \rho]$ and
$K_2 \notin \mathrm{inv}[I_K\rho]$, so that $J \in P_e(K_1) \cap P_e(K_2)$
by Lemma~\ref{lem:color}.
If $K_2 <_{\mathrm{lex}} K <_{\mathrm{lex}} K_1$, then $K_1 \notin \mathrm{inv}[I_K \rho]$
and $K_2 \in \mathrm{inv}[I_K \rho]$, hence $J \in P_o(K_1) \cap P_o(K_2)$.
Finally, if $K_1,K_2 <_{\mathrm{lex}} K$, then $K_1,K_2 \in \mathrm{inv}[I_K \rho]$,
hence $J \in P_e(K_1) \cap P_o(K_2)$ by Lemma~\ref{lem:color}.
In all these cases $J$ is green in both, $[\rho]$ and $[I_K \rho]$.
The cases~(ii) and~(iii) can be treated in a~similar way.
\end{proof}

The preceding proposition in particular shows that blue (red) elements of $[\rho]$ are not af\/fected by
red (blue) and green inversions.

\begin{prop}
\label{prop:no_reoccurence_of_a_color}
Let $[\rho] \stackrel{K}{\rightarrow} [I_K\rho]$ be blue $($red$)$ in $B(N,n)$, $n<N$.
Then any $J \in P(K)$ that is blue $($red$)$ in $[\rho]$ is  not  blue $($red$)$ in any subsequent
element of $B(N,n)$.
\end{prop}

\begin{proof}
This is obvious if $n=N-1$. Let $n < N-1$, $[\rho] \stackrel{K}{\rightarrow} [I_K\rho]$ and
$J \in P(K)$ blue.
According to Proposition~\ref{prop:color}, $J$ is green in $[\rho_1] = [I_K\rho]$.
Let us assume that $J$ becomes blue again in some subsequent $[\rho_r]$. Then
$[\rho_{r-1}] \stackrel{K_r}{\rightarrow} [\rho_r]$ has to be blue and $J \in P_e(K_r)$.
As a~consequence, $K, K_r \in E(J)$, $K \neq K_r$, and $L = K \cup K_r$ has
cardinality $n+2$.
If $\vec{P}(L)$ has beginning segment w.r.t.~$[\rho_{r-1}]$, then also w.r.t.~$[\rho_1]$.
Since $K \in P_o(L)$ (because $[\rho] \stackrel{K}{\rightarrow} [\rho_1]$ is blue) and
$K <_{\mathrm{lex}} K_r$, Lemma~\ref{lem:color} yields $J \in P_o(K_r)$ and thus
a contradiction. If $\cev{P}(L)$ has beginning segment w.r.t.~$[\rho_{r-1}]$,
then $K \in P_e(L)$ and $K_r <_{\mathrm{lex}} K$, hence $J \in P_o(K_r)$ according to
Lemma~\ref{lem:color}, so we have a contradiction.
The red case is treated in the same way.
\end{proof}

\begin{prop}
\label{prop:alpha_omega_colors}
Let $[\alpha]$ and $[\omega]$ be the minimal and the maximal element of $B(N,n)$, respectively.
Then the blue $($red$)$ elements of $[\alpha]$ are the red $($blue$)$ elements of $[\omega]$. Furthermore,
$[\alpha]$~and~$[\omega]$ share the same green elements.
\end{prop}

\begin{proof}
Let $J$ be blue in $[\alpha]$. Since $\mathrm{inv}[\alpha] = \varnothing$,
$J$ is blue in $\vec{P}(K)$ for all $K \in E(J)$. Since $\mathrm{inv}[\omega]$
contains all $K$, $J$ is red in $[\omega]$. Correspondingly, a red $J$
in $[\alpha]$ is blue in $[\omega]$. The last statement of the proposition is then obvious.
\end{proof}

Let us recall that we also use $\rho^{(c)}$ to denote the subsequence of $\rho \in A(N,n)$ of color $c$.
Now any $\rho \in A(N,n)$ can be decomposed into three subsequences,
$\rho^{(b)}$, $\rho^{(r)}$ and $\rho^{(g)}$, and this decomposition is carried over to $B(N,n)$.

For $K \in {[N] \choose n+1}$, let us introduce the \emph{half-packet inversions}
\begin{gather*}
     I^{(b)}_K \colon \  \vec{P_o}(K) \rightarrow \cev{P_e}(K)   , \qquad
     I^{(r)}_K \colon \ \vec{P_e}(K) \rightarrow \cev{P_o}(K)   .
\end{gather*}
Let $\rho \stackrel{K}{\rightarrow} I_K \rho$ be an inversion in $A(N,n)$. From the above
propositions we conclude:
\begin{itemize}\itemsep=0pt
\item if the inversion is blue, then $(I_K\rho)^{(b)} = I^{(b)}_K \rho^{(b)}$,
$(I_K\rho)^{(r)} = \rho^{(r)}$, and $(I_K\rho)^{(g)} = I^{(r)}_K \rho^{(g)}$,
\item if the inversion is red, then $(I_K\rho)^{(b)} = \rho^{(b)}$,
$(I_K\rho)^{(r)} = I^{(r)}_K \rho^{(r)}$, and $(I_K\rho)^{(g)} = I^{(b)}_K \rho^{(g)}$,
\item if the inversion is green, then $(I_K\rho)^{(b)} = \rho^{(b)}$,
$(I_K\rho)^{(r)} = \rho^{(r)}$, and $(I_K\rho)^{(g)} = I_K \rho^{(g)}$.
\end{itemize}
In the following, $B^{(c)}(N,n)$ shall denote the corresponding set supplied
with the induced partial order.
$B^{(b)}(N,n)$ is the (\emph{higher}) \emph{Tamari order}
$T(N,n)$ (see~\cite{DMH12KPBT} for an equivalent def\/inition).
$B^{(r)}(N,n)$ is the dual of the (higher) Tamari order $T(N,n)$.
$B^{(g)}(N,n)$ will be called \emph{mixed order}. The latter involves all the three inversions.
It should be noted that a red (blue) half-packet inversion in $B^{(g)}(N,n)$ stems from a
blue (red) inversion $I_K$.

\begin{rem}
$B^{(b)}(N,N-1)$ is $\vec{P_o}([N]) \stackrel{[N]}{\rightarrow} \cev{P_e}([N])$ and
$B^{(r)}(N,N-1)$ is $\vec{P_e}([N]) \stackrel{[N]}{\rightarrow} \cev{P_o}([N])$.
$B^{(g)}(N,N-1)$ is empty.
$B(N,N-2)$ consists of a pair of maximal chains, see (\ref{B(N,N-2)_chains}).
We set $m := N \, \mathrm{mod}\, 2$.
The two blue subchains
\begin{gather*}
   [\alpha^{(b)}] \stackrel{\hat{N}}{\longrightarrow} [\rho_1^{(b)}]
      \stackrel{\widehat{N-2}}{\longrightarrow} [\rho_3^{(b)}] \longrightarrow
    \cdots \longrightarrow [\rho_{N+m-5}^{(b)}] \stackrel{\widehat{4-m}}{\longrightarrow}
      [\rho_{N+m-3}^{(b)}] \stackrel{\widehat{2-m}}{\longrightarrow} [\omega^{(b)}]   , \\
  [\alpha^{(b)}] \stackrel{\widehat{m+1}}{\longrightarrow} [\sigma_{m+1}^{(b)}]
      \stackrel{\widehat{m+3}}{\longrightarrow} [\sigma_{m+3}^{(b)}] \longrightarrow
    \cdots \longrightarrow [\sigma_{N-5}^{(b)}]
    \stackrel{\widehat{N-3}}{\longrightarrow} [\sigma_{N-3}^{(b)}]
    \stackrel{\widehat{N-1}}{\longrightarrow}[\omega^{(b)}]   ,
\end{gather*}
constitute $B^{(b)}(N,N-2)$. All inversions are blue, of course.
Correspondingly, the two red chains
\begin{gather*}
  [\alpha^{(r)}] \stackrel{\widehat{N-1}}{\longrightarrow} [\rho_2^{(r)}]
      \stackrel{\widehat{N-3}}{\longrightarrow} [\rho_4^{(r)}]
      \longrightarrow \cdots \longrightarrow
      [\rho_{N-m-4}^{(r)}] \stackrel{\widehat{m+3}}{\longrightarrow} [\rho_{N-m-2}^{(r)}]
      \stackrel{\widehat{m+1}}{\longrightarrow} [\omega^{(r)}]   , \\
   [\alpha^{(r)}] \stackrel{\widehat{2-m}}{\longrightarrow} [ \sigma_{2-m}^{(r)}]
     \stackrel{\widehat{4-m}}{\longrightarrow} [\sigma_{4-m}^{(r)}] \longrightarrow
    \cdots \longrightarrow [\sigma_{N-4}^{(r)}]
    \stackrel{\widehat{N-2}}{\longrightarrow} [\sigma_{N-2}^{(r)}]
    \stackrel{\widehat{N}}{\longrightarrow} [\omega^{(r)}]   ,
\end{gather*}
form $B^{(r)}(N,N-2)$. All inversions are red.
One maximal chain of $B^{(g)}(N,N-2)$ is
\begin{gather*}
   [\alpha^{(g)}] \stackrel{\hat{N}}{\longrightarrow} [\rho_1^{(g)}]
      \stackrel{\widehat{N-1}}{\longrightarrow} [\rho_2^{(g)}] \longrightarrow \cdots \longrightarrow
      [\rho_{N-2}^{(g)}] \stackrel{\hat{2}}{\longrightarrow} [\rho_{N-1}^{(g)}]
      \stackrel{\hat{1}}{\longrightarrow} [\omega^{(g)}]   ,
\end{gather*}
where now an inversion $\stackrel{\hat{k}}{\rightarrow}$ is blue
if $\hat{k} \in P_e([N])$ and red if $\hat{k} \in P_o([N])$. The second chain is
\begin{gather*}
    [\alpha^{(g)}] \stackrel{\hat{1}}{\longrightarrow} [\sigma_1^{(g)}] \stackrel{\hat{2}}{\longrightarrow}
    [\sigma_2^{(g)}] \longrightarrow \cdots \longrightarrow
    [\sigma_{N-2}^{(g)}] \stackrel{\widehat{N-1}}{\longrightarrow}  [\sigma_{N-1}^{(g)}]
      \stackrel{\hat{N}}{\longrightarrow}[\omega^{(b)}]   ,
\end{gather*}
where $\stackrel{\hat{k}}{\rightarrow}$ is blue
if $\hat{k} \in P_o([N])$ and red if $\hat{k} \in P_e([N])$. There are no green inversions in this case.
\end{rem}

\begin{exam}
\label{ex:B(4,2)_decomp}
$B(4,2)$ consists of the two maximal chains
\begin{gather}
 \begin{array}{@{}c@{\;}c@{\;}c@{\;}c@{\;}c@{\;}c@{\;}c@{\;}c@{\;}c@{\;}c@{\;}c@{\;}c@{\;}c@{}}
  \begin{minipage}{.3cm} {\color{blue}12} \\ {\color{darkgreen}13} \\ {\color{darkred}14} \\
     {\color{blue}23} \\ {\color{darkgreen}24} \\ {\color{blue}34} \end{minipage}
               & \stackrel{ \bsy{\sim} }{\rightarrow} &
  \begin{minipage}{.3cm} {\color{blue}12} \\ {\color{darkgreen}13} \\ {\color{blue}23} \\
     {\color{darkred}14} \\ {\color{darkgreen}24} \\ {\color{blue}34} \end{minipage}
               & \stackrel{ \color{blue} 123 }{\rightarrow} &
  \begin{minipage}{.3cm} {\color{darkgreen}23} \\ {\color{blue}13} \\ {\color{darkgreen}12} \\
     {\color{darkred}14} \\ {\color{darkgreen}24} \\ {\color{blue}34} \end{minipage}
         & \stackrel{ \color{darkred} 124 }{\rightarrow} &
  \begin{minipage}{.3cm} {\color{darkgreen}23} \\ {\color{blue}13} \\ {\color{darkred}24} \\
     {\color{darkgreen}14} \\ {\color{darkred}12} \\ {\color{blue}34} \end{minipage}
         & \stackrel{ \bsy{\sim} }{\rightarrow} &
  \begin{minipage}{.3cm} {\color{darkgreen}23} \\ {\color{darkred}24} \\ {\color{blue}13} \\
     {\color{darkgreen}14} \\ {\color{blue}34} \\ {\color{darkred}12} \end{minipage}
         & \stackrel{ \color{blue} 134 }{\rightarrow} &
  \begin{minipage}{.3cm} {\color{darkgreen}23} \\ {\color{darkred}24} \\ {\color{darkgreen}34} \\
     {\color{blue}14} \\ {\color{darkgreen}13} \\ {\color{darkred}12} \end{minipage}
         & \stackrel{ \color{darkred} 234 }{\rightarrow} &
  \begin{minipage}{.3cm} {\color{darkred}34} \\ {\color{darkgreen}24} \\ {\color{darkred}23} \\
     {\color{blue}14} \\ {\color{darkgreen}13} \\ {\color{darkred}12} \end{minipage}
 \end{array}
    \qquad
 \begin{array}{@{}c@{\;\,}c@{\;\,}c@{\;\,}c@{\;\,}c@{\;\,}c@{\;\,}c@{\;\,}c@{\;\,}c@{\;\,}c@{\;\,}c@{\;\,}c@{\;\,}c@{}}
   \begin{minipage}{.3cm} {\color{blue}12} \\ {\color{darkgreen}13} \\ {\color{darkred}14} \\
     {\color{blue}23} \\ {\color{darkgreen}24} \\ {\color{blue}34} \end{minipage}
               & \stackrel{ \color{blue} 234  }{\rightarrow} &
  \begin{minipage}{.3cm} {\color{blue}12} \\ {\color{darkgreen}13} \\ {\color{darkred}14} \\
     {\color{darkgreen}34} \\  {\color{blue}24} \\ {\color{darkgreen}23} \end{minipage}
               & \stackrel{ \color{darkred} 134 }{\rightarrow} &
  \begin{minipage}{.3cm} {\color{blue}12} \\ {\color{darkred}34} \\ {\color{darkgreen}14} \\
     {\color{darkred}13} \\  {\color{blue}24} \\ {\color{darkgreen}23} \end{minipage}
          & \stackrel{ \bsy{\sim} }{\rightarrow} &
  \begin{minipage}{.3cm} {\color{darkred}34} \\ {\color{blue}12} \\ {\color{darkgreen}14} \\
     {\color{blue}24} \\ {\color{darkred}13} \\  {\color{darkgreen}23} \end{minipage}
          & \stackrel{ \color{blue} 124 }{\rightarrow} &
  \begin{minipage}{.3cm} {\color{darkred}34} \\ {\color{darkgreen}24} \\ {\color{blue}14} \\
     {\color{darkgreen}12} \\  {\color{darkred}13} \\ {\color{darkgreen}23} \end{minipage}
          & \stackrel{ \color{darkred} 123 }{\rightarrow} &
  \begin{minipage}{.3cm} {\color{darkred}34} \\ {\color{darkgreen}24} \\ {\color{blue}14} \\
     {\color{darkred}23} \\  {\color{darkgreen}13} \\ {\color{darkred}12} \end{minipage}
         & \stackrel{ \bsy{\sim} }{\rightarrow} &
  \begin{minipage}{.3cm} {\color{darkred}34} \\ {\color{darkgreen}24} \\ {\color{darkred}23} \\
     {\color{blue}14} \\  {\color{darkgreen}13} \\ {\color{darkred}12} \end{minipage}
 \end{array}   \label{B(4,2)-chains}
\end{gather}
Here they are resolved into admissible linear orders, i.e., elements of $A(4,2)$.
These are the maximal chains of $B(4,1)$ (forming a permutahedron).
The blue and red subchains of (\ref{B(4,2)-chains}), forming $B^{(b)}(4,2)$
and $B^{(r)}(4,2)$, respectively, are
\begin{gather*}
 \begin{array}{@{}c@{\;\,}c@{\;\,}c@{\;\,}c@{\;\,}c@{}}
  \begin{minipage}{.3cm} {\color{blue}12} \\ {\color{blue}23} \\ {\color{blue}34} \end{minipage}
            & \stackrel{ \color{blue} 123 }{\rightarrow} &
  \begin{minipage}{.3cm} {\color{blue}13} \\ {\color{blue}34} \end{minipage}
            & \stackrel{ \color{blue} 134 }{\rightarrow} &
  \begin{minipage}{.3cm} {\color{blue}14} \end{minipage}
 \end{array}
 \qquad
 \begin{array}{@{}c@{\;\,}c@{\;\,}c@{\;\,}c@{\;\,}c@{}}
  \begin{minipage}{.3cm} {\color{blue}12} \\ {\color{blue}23} \\ {\color{blue}34} \end{minipage}
            & \stackrel{ \color{blue} 234 }{\rightarrow} &
  \begin{minipage}{.3cm} {\color{blue}12} \\ {\color{blue}24} \end{minipage}
            & \stackrel{ \color{blue} 124 }{\rightarrow} &
  \begin{minipage}{.3cm} {\color{blue}14} \end{minipage}
 \end{array}
  \qquad
  \begin{array}{@{}c@{\;\,}c@{\;\,}c@{\;\,}c@{\;\,}c@{}}
   \begin{minipage}{.3cm} {\color{darkred}14} \end{minipage}
     & \stackrel{ \color{darkred} 124 }{\rightarrow} &
   \begin{minipage}{.3cm} {\color{darkred}24} \\ {\color{darkred}12} \end{minipage}
     & \stackrel{ \color{darkred} 234 }{\rightarrow} &
   \begin{minipage}{.3cm} {\color{darkred}34} \\ {\color{darkred}23} \\ {\color{darkred}12} \end{minipage}
  \end{array}
  \qquad
  \begin{array}{@{}c@{\;\,}c@{\;\,}c@{\;\,}c@{\;\,}c@{}}
   \begin{minipage}{.3cm} {\color{darkred}14} \end{minipage}
     & \stackrel{ \color{darkred} 134 }{\rightarrow} &
   \begin{minipage}{.3cm} {\color{darkred}34} \\ {\color{darkred}13} \end{minipage}
     & \stackrel{ \color{darkred} 123 }{\rightarrow} &
   \begin{minipage}{.3cm} {\color{darkred}34} \\ {\color{darkred}23} \\ {\color{darkred}12} \end{minipage}
  \end{array}
\end{gather*}
$B^{(g)}(4,2)$ is given by
\begin{gather*}
    \begin{array}{@{}c@{\;\,}c@{\;\,}c@{\;\,}c@{\;\,}c@{\;\,}c@{\;\,}c@{\;\,}c@{\;\,}c@{}}
  \begin{minipage}{.3cm} {\color{darkgreen}13} \\ {\color{darkgreen}24} \end{minipage}
              & \stackrel{ \color{darkred} 123 }{\rightarrow} &
  \begin{minipage}{.3cm} {\color{darkgreen}23} \\ {\color{darkgreen}12} \\
     {\color{darkgreen}24}  \end{minipage}
         & \stackrel{ \color{blue} 124 }{\rightarrow} &
  \begin{minipage}{.3cm} {\color{darkgreen}23} \\ {\color{darkgreen}14} \end{minipage}
         & \stackrel{ \color{darkred} 134 }{\rightarrow} &
  \begin{minipage}{.3cm} {\color{darkgreen}23} \\ {\color{darkgreen}34} \\
     {\color{darkgreen}13} \end{minipage}
         & \stackrel{ \color{blue} 234 }{\rightarrow} &
  \begin{minipage}{.3cm} {\color{darkgreen}24} \\ {\color{darkgreen}13} \end{minipage}
 \end{array}
    \qquad
 \begin{array}{@{}c@{\;\,}c@{\;\,}c@{\;\,}c@{\;\,}c@{\;\,}c@{\;\,}c@{\;\,}c@{\;\,}c@{}}
  \begin{minipage}{.3cm} {\color{darkgreen}13} \\ {\color{darkgreen}24}  \end{minipage}
               & \stackrel{ \color{darkred} 234  }{\rightarrow} &
  \begin{minipage}{.3cm} {\color{darkgreen}13} \\ {\color{darkgreen}34} \\ {\color{darkgreen}23} \end{minipage}
               & \stackrel{ \color{blue} 134 }{\rightarrow} &
  \begin{minipage}{.3cm} {\color{darkgreen}14} \\ {\color{darkgreen}23} \end{minipage}
          & \stackrel{ \color{darkred} 124 }{\rightarrow} &
  \begin{minipage}{.3cm} {\color{darkgreen}24} \\ {\color{darkgreen}12} \\ {\color{darkgreen}23} \end{minipage}
          & \stackrel{ \color{blue} 123 }{\rightarrow} &
  \begin{minipage}{.3cm} {\color{darkgreen}24} \\ {\color{darkgreen}13} \end{minipage}
 \end{array}
\end{gather*}
Fig.~\ref{fig:B42_line_diagram} displays the structure of the two maximal chains
(\ref{B(4,2)-chains}) of $B(4,2)$.\footnote{In the context of Soergel bimodules, a corresponding
diagrammatic equation (also see Fig.~\ref{fig:3simplex_decomp_line_diagram}) appeared in~\cite{Elia+Will13} as the $A_3$ Zamolodchikov relation. }
\begin{figure}[t]
\centering
\includegraphics[width=.5\linewidth]{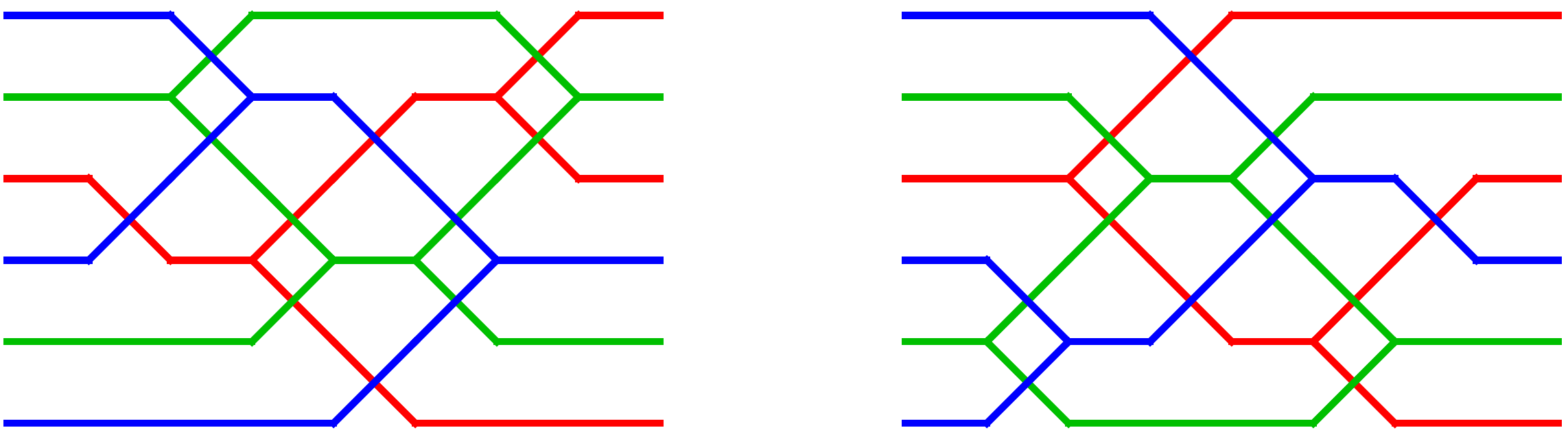}

\caption{Structure of the chains (\ref{B(4,2)-chains}). A rhombus with two green
and two blue (red) edges corresponds to a blue (red) inversion in $B(4,2)$. Such a rhombus
corresponds to a hexagon in $B(4,1)$ (see Remark~\ref{rem:line_diagram_from_polyhedron}).
\label{fig:B42_line_diagram} }
\end{figure}
\end{exam}

\begin{exam}
In the following, we display one of the maximal chains of $B(5,2)$, resolved into linear orders (elements
of $A(5,2)$), and its single-colored subsequences
%\footnotesize
\begin{gather*}
 \begin{array}{@{}c@{\,}c@{\,}c@{\,}c@{\,}c@{\,}c@{\,}c@{\,}c@{\,}c@{\,}c@{\,}c@{\,}c@{\,}c@{\,}c@{\,}c@{\,}c@{\,}c@{\,}c@{\,}c@{\,}
          c@{\,}c@{\,}c@{\,}c@{\,}c@{\,}c@{\,}c@{\,}c@{\,}c@{\,}c@{\,}c@{\,}c@{\,}c@{\,}c@{\,}c@{\,}c@{}}
 \begin{minipage}{.3cm} {\color{blue}12} \\ {\color{darkgreen}13} \\ {\color{darkgreen}14} \\ {\color{darkred}15} \\ {\color{blue}23} \\ {\color{darkgreen}24} \\
    {\color{darkgreen}25} \\ {\color{blue}34} \\ {\color{darkgreen}35} \\ {\color{blue}45} \end{minipage}
               & \stackrel{ \bsy{\sim} }{\,\rightarrow} &
 \begin{minipage}{.3cm} {\color{blue}12} \\ {\color{darkgreen}13} \\ {\color{blue}23} \\ {\color{darkgreen}14} \\ {\color{darkred}15} \\ {\color{darkgreen}24} \\
    {\color{darkgreen}25} \\ {\color{blue}34} \\ {\color{darkgreen}35} \\ {\color{blue}45}  \end{minipage}
               & \stackrel{\,{\color{blue}123}}{\rightarrow} &
 \begin{minipage}{.3cm} {\color{darkgreen}23} \\ {\color{blue}13} \\ {\color{darkgreen}12} \\ {\color{darkgreen}14} \\ {\color{darkred}15} \\ {\color{darkgreen}24} \\
     {\color{darkgreen}25} \\ {\color{blue}34} \\ {\color{darkgreen}35} \\ {\color{blue}45} \end{minipage}
               & \stackrel{ \bsy{\sim} }{\,\rightarrow} &
 \begin{minipage}{.3cm} {\color{darkgreen}23} \\ {\color{blue}13} \\ {\color{darkgreen}12} \\ {\color{darkgreen}14} \\ {\color{darkgreen}24} \\ {\color{darkred}15} \\ {\color{darkgreen}25} \\
     {\color{blue}34} \\ {\color{darkgreen}35} \\ {\color{blue}45} \end{minipage}
               &\stackrel{\,{\color{darkgreen}124}}{\rightarrow}&
  \begin{minipage}{.3cm} {\color{darkgreen}23} \\ {\color{blue}13} \\ {\color{darkgreen}24} \\ {\color{darkgreen}14} \\ {\color{darkgreen}12} \\ {\color{darkred}15} \\ {\color{darkgreen}25} \\
     {\color{blue}34} \\ {\color{darkgreen}35} \\  {\color{blue}45} \end{minipage}
               &\stackrel{\,{\color{darkred}125}}{\rightarrow}&
  \begin{minipage}{.3cm} {\color{darkgreen}23} \\ {\color{blue}13} \\ {\color{darkgreen}24} \\ {\color{darkgreen}14} \\ {\color{darkred}25} \\ {\color{darkgreen}15} \\ {\color{darkred}12} \\
     {\color{blue}34} \\ {\color{darkgreen}35} \\  {\color{blue}45} \end{minipage}
               & \stackrel{ \bsy{\sim} }{\,\rightarrow} &
  \begin{minipage}{.3cm} {\color{darkgreen}23} \\ {\color{darkgreen}24} \\ {\color{blue}13} \\ {\color{darkgreen}14} \\ {\color{blue}34} \\ {\color{darkred}25} \\ {\color{darkgreen}15} \\
     {\color{darkred}12} \\ {\color{darkgreen}35} \\  {\color{blue}45} \end{minipage}
               &\stackrel{\,{\color{blue}134}}{\rightarrow}&
  \begin{minipage}{.3cm} {\color{darkgreen}23} \\ {\color{darkgreen}24} \\ {\color{darkgreen}34} \\ {\color{blue}14} \\ {\color{darkgreen}13} \\ {\color{darkred}25} \\ {\color{darkgreen}15} \\
     {\color{darkred}12} \\ {\color{darkgreen}35} \\  {\color{blue}45} \end{minipage}
               & \stackrel{ \bsy{\sim} }{\,\rightarrow} &
  \begin{minipage}{.3cm} {\color{darkgreen}23} \\ {\color{darkgreen}24} \\ {\color{darkgreen}34} \\ {\color{blue}14} \\ {\color{darkred}25} \\ {\color{darkgreen}13} \\ {\color{darkgreen}15} \\
     {\color{darkgreen}35} \\ {\color{darkred}12} \\  {\color{blue}45} \end{minipage}
               &\stackrel{\,{\color{darkgreen}135}}{\rightarrow}&
  \begin{minipage}{.3cm} {\color{darkgreen}23} \\ {\color{darkgreen}24} \\ {\color{darkgreen}34} \\ {\color{blue}14} \\ {\color{darkred}25} \\ {\color{darkgreen}35} \\ {\color{darkgreen}15} \\
     {\color{darkgreen}13} \\ {\color{darkred}12} \\  {\color{blue}45} \end{minipage}
               & \stackrel{ \bsy{\sim} }{\,\rightarrow} &
  \begin{minipage}{.3cm} {\color{darkgreen}23} \\ {\color{darkgreen}24} \\ {\color{darkgreen}34} \\ {\color{darkred}25} \\ {\color{darkgreen}35} \\ {\color{blue}14} \\ {\color{darkgreen}15} \\
     {\color{blue}45} \\ {\color{darkgreen}13} \\  {\color{darkred}12} \end{minipage}
               &\stackrel{\,{\color{blue}145}}{\rightarrow}&
  \begin{minipage}{.3cm} {\color{darkgreen}23} \\ {\color{darkgreen}24} \\ {\color{darkgreen}34} \\ {\color{darkred}25} \\ {\color{darkgreen}35} \\ {\color{darkgreen}45} \\ {\color{blue}15} \\
     {\color{darkgreen}14} \\ {\color{darkgreen}13} \\  {\color{darkred}12} \end{minipage}
               &\stackrel{{\,\color{darkgreen}234}}{\rightarrow}&
  \begin{minipage}{.3cm} {\color{darkgreen}34} \\ {\color{darkgreen}24} \\ {\color{darkgreen}23} \\ {\color{darkred}25} \\ {\color{darkgreen}35} \\ {\color{darkgreen}45} \\ {\color{blue}15} \\
     {\color{darkgreen}14} \\ {\color{darkgreen}13} \\  {\color{darkred}12} \end{minipage}
               &\stackrel{\,{\color{darkred}235}}{\rightarrow}&
  \begin{minipage}{.3cm} {\color{darkgreen}34} \\ {\color{darkgreen}24} \\ {\color{darkred}35} \\ {\color{darkgreen}25} \\ {\color{darkred}23} \\ {\color{darkgreen}45} \\ {\color{blue}15} \\
     {\color{darkgreen}14} \\ {\color{darkgreen}13} \\  {\color{darkred}12} \end{minipage}
               & \stackrel{ \bsy{\sim} }{\,\rightarrow} &
  \begin{minipage}{.3cm} {\color{darkgreen}34} \\ {\color{darkred}35} \\ {\color{darkgreen}24} \\ {\color{darkgreen}25} \\ {\color{darkgreen}45} \\ {\color{darkred}23} \\ {\color{blue}15} \\
     {\color{darkgreen}14} \\ {\color{darkgreen}13} \\  {\color{darkred}12} \end{minipage}
               &\stackrel{\,{\color{darkgreen}245}}{\rightarrow}&
  \begin{minipage}{.3cm} {\color{darkgreen}34} \\ {\color{darkred}35} \\ {\color{darkgreen}45} \\ {\color{darkgreen}25} \\ {\color{darkgreen}24} \\ {\color{darkred}23} \\ {\color{blue}15} \\
     {\color{darkgreen}14} \\ {\color{darkgreen}13} \\  {\color{darkred}12} \end{minipage}
               &\stackrel{\,{\color{darkred}345}}{\rightarrow}&
  \begin{minipage}{.3cm} {\color{darkred}45} \\ {\color{darkgreen}35} \\ {\color{darkred}34} \\ {\color{darkgreen}25} \\ {\color{darkgreen}24} \\ {\color{darkred}23} \\ {\color{blue}15} \\
     {\color{darkgreen}14} \\ {\color{darkgreen}13} \\  {\color{darkred}12} \end{minipage}
  \end{array}
\end{gather*}
%\normalsize
The three subsequences collapse to
\begin{gather*}
 \begin{array}{@{}c@{\;\;}c@{\;\;}c@{\;\;}c@{\;\;}c@{\;\;}c@{\;\;}c@{\;\;}c@{\;\;}c@{\;\;}c@{\;\;}c@{\;\;}c@{\;\;}c@{\;\;}c@{\;\;}c@{\;\;}c@{\;\;}c@{\;\;}c@{\;\;}c@{\;\;}
          c@{\;\;}c@{\;\;}c@{\;\;}c@{\;\;}c@{\;\;}c@{\;\;}c@{\;\;}c@{\;\;}c@{\;\;}c@{\;\;}c@{\;\;}c@{\;\;}c@{\;\;}c@{\;\;}c@{\;\;}c@{}}
 \begin{minipage}{.3cm} {\color{blue}12} \\ {\color{blue}23} \\ {\color{blue}34} \\ {\color{blue}45} \end{minipage}
               & \stackrel{{\color{blue}123}}{\rightarrow} &
 \begin{minipage}{.3cm} {\color{blue}13} \\ {\color{blue}34} \\ {\color{blue}45} \end{minipage}
               &\stackrel{{\color{blue}134}}{\rightarrow}&
  \begin{minipage}{.3cm} {\color{blue}14} \\ {\color{blue}45} \end{minipage}
               &\stackrel{{\color{blue}145}}{\rightarrow}&
  \begin{minipage}{.3cm} {\color{blue}15} \end{minipage}
  \end{array}
\qquad
 \begin{array}{@{}c@{\;\;}c@{\;\;}c@{\;\;}c@{\;\;}c@{\;\;}c@{\;\;}c@{\;\;}c@{\;\;}c@{\;\;}c@{\;\;}c@{\;\;}c@{\;\;}c@{\;\;}c@{\;\;}c@{\;\;}c@{\;\;}c@{\;\;}c@{\;\;}c@{\;\;}
          c@{\;\;}c@{\;\;}c@{\;\;}c@{\;\;}c@{\;\;}c@{\;\;}c@{\;\;}c@{\;\;}c@{\;\;}c@{\;\;}c@{\;\;}c@{\;\;}c@{\;\;}c@{\;\;}c@{\;\;}c@{}}
 \begin{minipage}{.3cm} {\color{darkred}15} \end{minipage}
               &\stackrel{{\color{darkred}125}}{\rightarrow}&
  \begin{minipage}{.3cm} {\color{darkred}25} \\ {\color{darkred}12} \end{minipage}
               &\stackrel{{\color{darkred}235}}{\rightarrow}&
  \begin{minipage}{.3cm} {\color{darkred}35} \\ {\color{darkred}23} \\ {\color{darkred}12} \end{minipage}
               &\stackrel{{\color{darkred}345}}{\rightarrow}&
  \begin{minipage}{.3cm} {\color{darkred}45} \\ {\color{darkred}34} \\ {\color{darkred}23} \\ {\color{darkred}12} \end{minipage}
  \end{array}
\\
 \begin{array}{@{}c@{\;\;}c@{\;\;}c@{\;\;}c@{\;\;}c@{\;\;}c@{\;\;}c@{\;\;}c@{\;\;}c@{\;\;}c@{\;\;}c@{\;\;}c@{\;\;}c@{\;\;}c@{\;\;}c@{\;\;}c@{\;\;}c@{\;\;}c@{\;\;}c@{\;\;}
          c@{\;\;}c@{\;\;}c@{\;\;}c@{\;\;}c@{\;\;}c@{\;\;}c@{\;\;}c@{\;\;}c@{\;\;}c@{\;\;}c@{\;\;}c@{\;\;}c@{\;\;}c@{\;\;}c@{\;\;}c@{}}
 \begin{minipage}{.3cm} {\color{darkgreen}13} \\ {\color{darkgreen}14} \\ {\color{darkgreen}24} \\
    {\color{darkgreen}25} \\ {\color{darkgreen}35} \end{minipage}
               & \stackrel{{\color{darkred}123}}{\rightarrow} &
 \begin{minipage}{.3cm} {\color{darkgreen}23} \\ {\color{darkgreen}12} \\ {\color{darkgreen}14} \\ {\color{darkgreen}24} \\ {\color{darkgreen}25} \\ {\color{darkgreen}35} \end{minipage}
               &\stackrel{{\color{darkgreen}124}}{\rightarrow}&
  \begin{minipage}{.3cm} {\color{darkgreen}23} \\ {\color{darkgreen}24} \\ {\color{darkgreen}14} \\ {\color{darkgreen}12} \\ {\color{darkgreen}25} \\ {\color{darkgreen}35} \end{minipage}
               &\stackrel{{\color{blue}125}}{\rightarrow}&
  \begin{minipage}{.3cm} {\color{darkgreen}23} \\ {\color{darkgreen}24} \\ {\color{darkgreen}14} \\ {\color{darkgreen}15} \\ {\color{darkgreen}35} \end{minipage}
               &\stackrel{{\color{darkred}134}}{\rightarrow}&
  \begin{minipage}{.3cm} {\color{darkgreen}23} \\ {\color{darkgreen}24} \\ {\color{darkgreen}34} \\ {\color{darkgreen}13} \\ {\color{darkgreen}15} \\ {\color{darkgreen}35} \end{minipage}
               &\stackrel{{\color{darkgreen}135}}{\rightarrow}&
  \begin{minipage}{.3cm} {\color{darkgreen}23} \\ {\color{darkgreen}24} \\ {\color{darkgreen}34} \\ {\color{darkgreen}35} \\ {\color{darkgreen}15} \\ {\color{darkgreen}13} \end{minipage}
               &\stackrel{{\color{darkred}145}}{\rightarrow}&
  \begin{minipage}{.3cm} {\color{darkgreen}23} \\ {\color{darkgreen}24} \\ {\color{darkgreen}34} \\ {\color{darkgreen}35} \\ {\color{darkgreen}45} \\ {\color{darkgreen}14} \\ {\color{darkgreen}13}
                   \end{minipage}
               &\stackrel{{\color{darkgreen}234}}{\rightarrow}&
  \begin{minipage}{.3cm} {\color{darkgreen}34} \\ {\color{darkgreen}24} \\ {\color{darkgreen}23} \\ {\color{darkgreen}35} \\ {\color{darkgreen}45} \\
                 {\color{darkgreen}14} \\ {\color{darkgreen}13} \end{minipage}
               &\stackrel{{\color{blue}235}}{\rightarrow}&
  \begin{minipage}{.3cm} {\color{darkgreen}34} \\ {\color{darkgreen}24} \\ {\color{darkgreen}25} \\ {\color{darkgreen}45} \\ {\color{darkgreen}14} \\ {\color{darkgreen}13} \end{minipage}
               &\stackrel{{\color{darkgreen}245}}{\rightarrow}&
  \begin{minipage}{.3cm} {\color{darkgreen}34} \\ {\color{darkgreen}45} \\ {\color{darkgreen}25} \\ {\color{darkgreen}24} \\ {\color{darkgreen}14} \\ {\color{darkgreen}13} \end{minipage}
               &\stackrel{{\color{blue}345}}{\rightarrow}&
  \begin{minipage}{.3cm} {\color{darkgreen}35} \\ {\color{darkgreen}25} \\ {\color{darkgreen}24} \\ {\color{darkgreen}14} \\ {\color{darkgreen}13} \end{minipage}
  \end{array}
\end{gather*}
Here the blue order is ruled by $I^{(b)}_{ijk}\colon (ij,jk) \mapsto (ik)$, the red order by
$I^{(r)}_{ijk}\colon (ik) \mapsto (jk,ij)$, where $1 \leq i<j<k \leq 5$.
The green order involves these two and in addition $I_{ijk}\colon (ij,ik,jk) \mapsto (jk,ik,ij)$.
\end{exam}

\begin{rem}
\label{rem:A(N,n+1)-B(N,n)_for_b,r}
Inherited from the Bruhat orders, for $c \in \{b,r\}$, there is a one-to-one correspondence between
elements of $A^{(c)}(N,n+1)$
and maximal chains of $B^{(c)}(N,n)$. No such relation exists for the mixed order, but
elements of $A^{(g)}(N,n+1)$ are in one-to-one correspondence with the green inversion subsequences of
maximal chains of $B^{(g)}(N,n)$.
\end{rem}

\begin{rem}
\label{rem:Tamari_red}
In Remark~\ref{rem:Bruhat_reduction} we def\/ined, for each $k \in [N+1]$, a projection
$B(N+1,n+1) \rightarrow B(N,n)$, via an equivalence relation $\stackrel{k}{\sim}$.
If $k \in [N+1] \setminus \{1,N+1\}$, these projections do \emph{not} respect the above three color
decomposition. The reason is that if $L \in {[N+1] \choose n+2}$ contains $k$, then
$P_o(L) \setminus \{L \setminus \{k\}\}$ and $P_e(L) \setminus \{L \setminus \{k\}\}$
cannot be brought into natural correspondence with the half-packets
$P_o(L \setminus \{k\})$ and $P_e(L \setminus \{k\})$, respectively.
For example, if $L=1234$ and $k=2$, then $L \setminus \{k\} = 134$, $P_o(L) = \{ 123,134 \}$,
$P_e(L) = \{ 124,234 \}$, hence $P_o(L) \setminus \{L \setminus \{k\}\} = \{123\}$,
$P_e(L) \setminus \{L \setminus \{k\}\} = \{124,234\}$, while $P_o(L \setminus \{k\}) = \{13,34\}$
and $P_e(L \setminus \{k\}) = \{14\}$.
But if $k=1$, then $L \setminus \{1\}$ is the last element of $\vec{P}(L)$ and its elimination
thus does not inf\/luence the positions of the remaining elements. For $k=1$ we therefore
obtain monotone projections $B^{(c)}(N+1,n+1) \rightarrow B^{(c)}(N,n)$.
The other exception is $k=N+1$. Then $L \setminus \{N+1\}$ is the f\/irst element of $\vec{P}(L)$ and
its elimination turns odd into even elements, and vice versa. In this case we obtain
monotone projections $B^{(b)}(N+1,n+1) \rightarrow B^{(r)}(N,n)$,
$B^{(r)}(N+1,n+1) \rightarrow B^{(b)}(N,n)$ and $B^{(g)}(N+1,n+1) \rightarrow B^{(g)}(N,n)$.
We will use the projection with $k=1$ in Section~\ref{subsec:polygon_red}.
See, in particular, Figs.~\ref{fig:T74_red} and \ref{fig:T85_red}.
\end{rem}

\section{Simplex equations}
\label{sec:simplex_eqs}
In this section we consider realizations of Bruhat orders in terms of sets and maps between
Cartesian\footnote{Alternatively, we may as well consider tensor products or direct sums,
assuming that the sets carry the necessary additional structure.}
products of the sets. The $N$-simplex equation is directly associated with $B(N+1,N-1)$, but
its structure is fully displayed as a polyhedral reduction of $B(N+1,N-2)$.
Section~\ref{subsec:res_of_B(N+1,N-1)} explains the relation with polyhedra and prepares the
stage for the def\/inition of simplex equations in Section~\ref{subsec:simplex_eqs}, which
contains explicit expressions up to the $7$-simplex equation, and the associated polyhedra.
Section~\ref{subsec:simplex_Lax} discusses the integrability of simplex equations.
The reduction of the Bruhat order $B(N+2,N)$ to $B(N+1,N-1)$ induces a reduction of the
$(N+1)$-simplex equation to the $N$-simplex equation. This is the subject of
Section~\ref{subsec:simplex_red}.

\subsection[Resolutions of $B(N+1,N-1)$ and polyhedra]{Resolutions of $\boldsymbol{B(N+1,N-1)}$ and polyhedra}
\label{subsec:res_of_B(N+1,N-1)}
Let $s_a$ denote the operation of exchange of elements $J_a,J_{a+1}$ of
$\rho = (J_1,\ldots,J_{c(N+1,n)})\in A(N+1,n)$, which is applicable (only)
if $E(J_a) \cap E(J_{a+1}) = \varnothing$.
For any $\beta,\beta' \in [\rho]$, there is a minimal number $m$ of exchange operations
$s_{a_1},\ldots, s_{a_m}$, such that
$\beta' = s_{\beta',\beta}   \beta$, where $s_{\beta',\beta} := s_{a_m} \cdots s_{a_1}$.
The sequence $\beta_0,\beta_1,\ldots,\beta_m$,
where $\beta_0 = \beta$, $\beta_i = s_{a_i} \beta_{i-1}$, $i=1,\ldots,m$,
is called a \emph{resolution} of~$[\rho]$ from $\beta$ to $\beta'$. It is unique up to potential
applications of the identities
\begin{gather}
  s_{a}   s_{b} = s_{b}   s_{a} \qquad \mbox{if} \quad |a - b| > 1   ,
   \qquad
   s_{a}   s_{a+1}  s_{a} = s_{a+1}   s_{a}   s_{a+1}   .
           \label{exchange_operation_identities}
\end{gather}
Let $\cC\colon [\rho_0] \stackrel{K_1}{\longrightarrow} [\rho_1] \stackrel{K_2}{\longrightarrow} \cdots
 \stackrel{K_k}{\longrightarrow} [\rho_k]$ be a chain in $B(N+1,n)$. A \emph{resolution}
$\tilde{\cC}$ of $\cC$
is a sequence of resolutions of all $[\rho_i]$, such that the initial element of the resolution
of $[\rho_{i+1}]$ is obtained by application of $I_{K_{i+1}}$ to the f\/inal element of the resolution of $[\rho_i]$,
for $i=0,\ldots,k-1$.

Denoting by $\iota_{a}$ an inversion, acting at positions
$a,a+1,\ldots,a+n$, of some element of $A(N+1,n)$, the resolution $\tilde{\cC}$ uniquely corresponds
to a composition of exchange and inversion operations,
\begin{gather}
  \cO_{\tilde{\cC}} := s_{\beta_k',\beta_k}   \iota_{a_k} \cdots s_{\beta_1',\beta_1}
       \iota_{a_1}   s_{\beta_0',\beta_0}    , \label{operator_rep_of_resolution}
\end{gather}
where $\beta_i$ is the initial and $\beta_i'$ the f\/inal element of the resolution of $[\rho_i]$, and
$\beta_{i+1} = \iota_{a_{i+1}}   \beta_i'$.

\begin{rem}
\label{rem:s_iota_rels}
The operations $s_{a}$ and $\iota_{b}$ satisfy the following identities,
\begin{gather}
  s_{a}   \iota_{b} = \iota_{b}   s_{a} \qquad \mbox{if} \quad
     a < b - 1   \quad \mbox{or} \quad a > b + n    , \nonumber \\
  \iota_a   \iota_b = \iota_b   \iota_a  \qquad \mbox{if} \quad |b-a| > n   , \nonumber \\
  \iota_{a}   s_{a+n} \cdots s_{a+1} s_{a}
     = s_{a+n} \cdots s_{a+1} s_{a}   \iota_{a+1}   .   \label{s_iota_rels}
\end{gather}
In the last identity, $s_{a+n} \cdots s_{a+1} s_{a}$ exchanges the element at position $a$ with the block
of elements at positions $a+1,\ldots,a+n+1$.
The identities~(\ref{s_iota_rels}) take care of the fact that the above def\/inition of a resolution
of a chain in $B(N+1,n)$ does not in general f\/ix all the f\/inal elements of the resolutions of
the $[\rho_i]$.
Using $s_{a}^2 = \mathrm{id}$, in the last of the above relations one can
move exchange operations from one side to the other. Since the relations~(\ref{exchange_operation_identities})
and~(\ref{s_iota_rels}) are homogeneous, all resolutions with the same
initial and the same f\/inal element have the same length.
\end{rem}

The Bruhat order $B(N+1,N-1)$ consists of the two maximal chains\footnote{Here ``$\mathrm{lex}$''
and ``$\mathrm{rev}$'' stand for ``lexicographically ordered'' and ``reverse lexicographically ordered'',
respectively. In these chains we should better use complementary notation,
$\hat{l} := [N+1] \setminus \{l\}$, and we will do this mostly in the following. }
\begin{gather*}
    \cC_{\mathrm{lex}} \colon \  [\alpha] \stackrel{[N+1] \setminus \{N+1\}}{\xrarrow} [\rho_1]
            \stackrel{[N+1] \setminus \{N\}}{\xrarrow} \cdots
            \stackrel{[N+1] \setminus \{2\}}{\xrarrow} [\rho_N]
            \stackrel{[N+1] \setminus \{1\}}{\xrarrow} [\omega]   , \\
    \cC_{\mathrm{rev}} \colon \   [\alpha] \stackrel{[N+1] \setminus \{1\}}{\xrarrow} [\sigma_1]
            \stackrel{[N+1] \setminus \{2\}}{\xrarrow} \cdots
            \stackrel{[N+1] \setminus \{N\}}{\xrarrow} [\sigma_N]
            \stackrel{[N+1] \setminus \{N+1\}}{\xrarrow} [\omega]  .
\end{gather*}
This implies that there are resolutions $\tilde{\cC}_{\mathrm{lex}}$ and $\tilde{\cC}_{\mathrm{rev}}$ of
$\cC_{\mathrm{lex}}$ and $\cC_{\mathrm{rev}}$, respectively, both starting with $\alpha$ and
both ending with $\omega$,
\begin{gather*}
\begin{tikzpicture}
  \node (A) at (-2,0) {$\alpha$};
  \node (B) at (2,0) {$\omega$};
  \path[->,font=\scriptsize,>=angle 90]
  (A) edge [bend left] node[above] {$\tilde{\cC}_{\mathrm{lex}}$} (B)
      edge [bend right] node[below] {$\tilde{\cC}_{\mathrm{rev}}$} (B);
\end{tikzpicture}
\end{gather*}

Via the correspondence between elements of $A(N+1,N-1)$ and maximal chains of $B(N+1,N-2)$ (see
Remark~\ref{rem:A(N,n+1)-B(N,n)}),
each of the two resolutions
corresponds to a sequence of maximal chains of $B(N+1,N-2)$.
For $\rho \in A(N+1,N-1)$, let $\cC_\rho$ be the corresponding maximal chain of $B(N+1,N-2)$.
$\tilde{\cC}_{\mathrm{lex}}$, respectively $\tilde{\cC}_{\mathrm{rev}}$, is then a
rule for deforming $\cC_\alpha$ stepwise into $\cC_\omega$.

Moreover, a resolution of $B(N+1,N-1)$, represented by the above diagram, contains a rule
to construct a \emph{polyhedron}.
Starting from a common vertex, we represent the elements of $\alpha$ and
$\omega = \mathrm{rev}(\alpha)$ ($\alpha$ in reversed order) from top
to bottom as the edges of the left, respectively right side of a regular $N(N+1)$-gon.
Then we deform the left side (corresponding to $\alpha$) stepwise, following the resolution
$\tilde{\cC}_{\mathrm{lex}}$ and ending in the right side (corresponding to $\omega$) of the polygon.
For any appearance of an exchange operation $s$ we insert a rhombus, and for any inversion $\iota$
a~$2N$-gon. This is done in such a way
that opposite edges are parallel and have equal length, so the inserted polygons are \emph{zonogons}.
We proceed in the same way with the resolution $\tilde{\cC}_{\mathrm{rev}}$. The resulting two zonotiles
constitute complementary sides of a zonohedron. Up to ``small cubes'' (see the following
remark), it represents~$B(N+1,N-2)$.

\begin{rem}
\label{rem:small_cubes}
 For the f\/irst few values of $N$, $B(N+1,N-2)$ forms a polyhedron. This is no longer so for
higher values, because ``small cubes'' appear~\cite{Fels+Zieg01}. Fig.~\ref{fig:5simplex_small_cubes}
displays them for $B(6,3)$.
\begin{figure}[t]
\centering
\includegraphics[width=.7\linewidth]{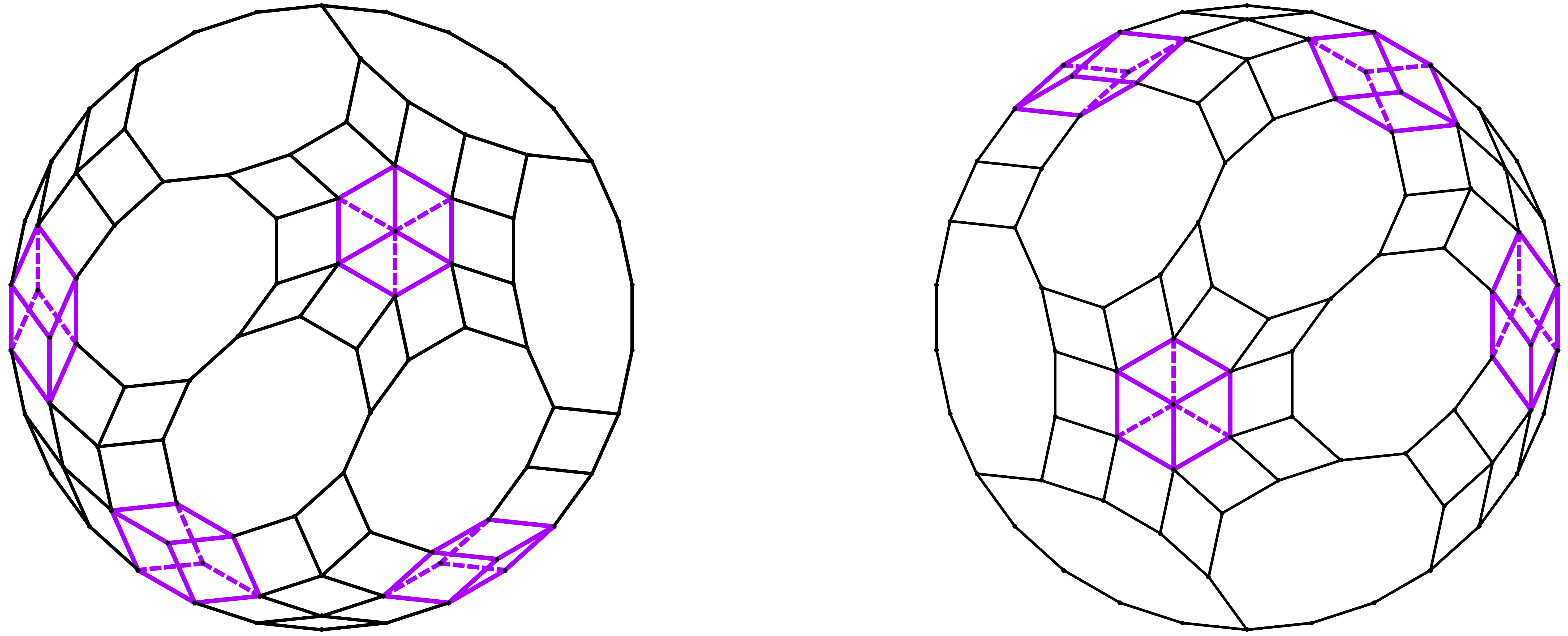}
\caption{Complementary sides of $B(6,3)$.
Because of the ``small cubes'' (here marked purple), $B(6,3)$ is not polyhedral.
\label{fig:5simplex_small_cubes} }
\end{figure}
A small cube is present in the Bruhat order $B(N+1,N-2)$ whenever there are two dif\/ferent resolutions of an
element of $B(N+1,N-1)$, which are identical except that one of them contains a subsequence
$\beta, s_a \beta, s_{a+1} s_a \beta, \beta'$ and the other
$\beta, s_{a+1} \beta, s_a s_{a+1} \beta, \beta'$ instead,
where $\beta' = s_a   s_{a+1}   s_a \beta \equiv s_{a+1}   s_a   s_{a+1} \beta$.
The six members of the two subsequences determine six maximal chains of $B(N+1,N-2)$, which
enclose a cube (similarly as in Fig.~\ref{fig:YB_on_cube}).
The process of deformations of maximal chains described above keeps only half of any small cube.
The polyhedron, constructed in the way described above, is a \emph{polyhedral reduction} of $B(N+1,N-2)$.
\end{rem}

\subsection{Simplex equations and associated polyhedra}
\label{subsec:simplex_eqs}
With each
$J \in {[N+1] \choose n}$, we associate a set $\cU_J$. With $\rho \in A(N+1,n)$,
$\rho = (J_1,\ldots, J_{c(N+1,n)})$, we then associate the Cartesian product
\begin{gather*}
    \cU_\rho := \cU_{J_1} \times \cU_{J_2} \times \cdots \times \cU_{J_{c(N+1,n)}}   .
\end{gather*}
Furthermore, for each $K \in {[N+1] \choose n+1}$, let there be a map
\begin{gather*}
     \cR_K \colon \  \cU_{\vec{P}(K)} \rightarrow \cU_{\cev{P}(K)}   .
\end{gather*}
If $\cU_{\vec{P}(K)}$ appears in $\cU_\rho$ at consecutive positions, starting at position~$\bsy{a}$,
we extend $\cR_K$ to a map $\cR_{K,\bsy{a}}\colon \cU_\rho \rightarrow \cU_{\rho'}$, where it acts
non-trivially only on the sets labeled by the elements of $P(K)$. $\cR_{K,\bsy{a}}$ then represents
the inversion operation $\iota_a$. The exchange operation $s_a$ will be represented by
the transposition map $\cP_{\bsy{a}}$ (where $\cP\colon (u,v) \mapsto (v,u)$), which
acts at positions $\bsy{a}$ and~$\bsy{a}+1$ of~$\cU_\rho$.\footnote{We use boldface ``position'' numbers
in order to distinguish them more clearly from the numbers specifying some $K \in {[N+1] \choose n+1}$. }
In this way, the resolution $\tilde{\cC}$ of the chain $\cC$ considered in Section~\ref{subsec:res_of_B(N+1,N-1)}
translates, via~(\ref{operator_rep_of_resolution}), to a composition of maps,
\begin{gather*}
  \cR_{\tilde{\cC}} := \cP_{\beta_k',\beta_k}   \cR_{K_k,\bsy{a}_k} \cdots
         \cP_{\beta_1',\beta_1}  \cR_{K_1,\bsy{a}_1}   \cP_{\beta_0',\beta_0}   ,
\end{gather*}
where $\cP_{\beta_i',\beta_i} = \cP_{\bsy{a}_{i,m_i}} \cdots \cP_{\bsy{a}_{i,1}}$.

Let us now turn to $B(N+1,N-1)$. Choosing $\alpha$ as the lexicographically
ordered set ${[N+1] \choose N-1}$, and
$\omega$ as $\alpha$ in reverse order, we def\/ine the $N$-\emph{simplex equation} as
\begin{gather}
      \cR_{\tilde{\cC}_{\mathrm{lex}}} = \cR_{\tilde{\cC}_{\mathrm{rev}}}   ,  \label{simplex_eq}
\end{gather}
where $\tilde{\cC}_{\mathrm{lex}}$ and $\tilde{\cC}_{\mathrm{rev}}$ are resolutions of
$\cC_{\mathrm{lex}}$ and $\cC_{\mathrm{rev}}$, respectively, with initial element $\alpha$ and
f\/inal element $\omega$, and $\cR_{\tilde{\cC}_{\mathrm{lex}}}$,
$\cR_{\tilde{\cC}_{\mathrm{rev}}}$ are the corresponding compositions of maps $\cR_{K,\bsy{a}}$, $\cP_{\bsy{b}}$.
(\ref{simplex_eq})~is independent
of the choices of the resolutions $\tilde{\cC}_{\mathrm{lex}}$ and $\tilde{\cC}_{\mathrm{rev}}$,
since $\cP_{\bsy{a}}$ and $\cR_{K,\bsy{b}}$ clearly satisfy all the relations that $s_a$ and
$\iota_b$ fulf\/ill (see Section~\ref{subsec:res_of_B(N+1,N-1)}). Since we have the freedom to move
a~transposition in leftmost or rightmost position from one side of~(\ref{simplex_eq}) to the other,
the above choice of~$\alpha$ is no restriction.

\begin{rem}
\label{rem:simplex_eq_structure}
Let
\begin{gather*}
     \cP_K \colon \ \cU_{\cev{P}(K)} \rightarrow \cU_{\vec{P}(K)}
\end{gather*}
be a composition of transposition maps $\cP_{\bsy{a}}$ corresponding to a \emph{reversion}.
The maps $\hat{\cR}_K$ that we will encounter in this section are related to the
respective maps $\cR_K$ via
\begin{gather*}
    \hat{\cR}_K = \cR_K   \cP_K   ,
\end{gather*}
and they are endomorphisms
\begin{gather*}
   \hat{\cR}_K \colon \  \cU_{\cev{P}(K)} \longrightarrow \cU_{\cev{P}(K)}   .
\end{gather*}
$\cR_K$ acts on $\cU_\rho$ only if $\vec{P}(K)$ appears at consecutive positions in $\rho$, and it
changes the order of the factors of~$\cU_\rho$. In contrast, $\hat{\cR}_K$ only acts on $\cU_\omega$
(not necessarily at consecutive positions).
It does \emph{not} change the order of~$\cU$'s.

In complementary notation, the reverse lexicographical order $\omega$ on ${[N+1] \choose N-1}$ reads
\begin{gather*}
   \omega = \big(\widehat{12},\widehat{13},\ldots,\widehat{1(N+1)},\widehat{23},\widehat{24},\ldots,
             \widehat{2(N+1)},\ldots, \widehat{N(N+1)}\big)   .
\end{gather*}
The $N$-simplex equation has the form
\begin{gather*}
     \hat{\cR}_{\hat{1},\bsy{A}_1}   \hat{\cR}_{\hat{2},\bsy{A}_2}   \cdots   \hat{\cR}_{\widehat{N+1},\bsy{A}_{N+1}}
   = \hat{\cR}_{\widehat{N+1},\bsy{A}_{N+1}}   \hat{\cR}_{\hat{N},\bsy{A}_N}   \cdots
     \hat{\cR}_{\hat{1},\bsy{A}_1}   ,
\end{gather*}
where both sides are maps $\cU_\omega \rightarrow \cU_\omega$. One has to determine the positions,
given by the multi-index $\bsy{A}_k$, of the factors of $\cU_{\omega}$, on which the map
$\hat{\cR}_{\hat{k}}$ acts.
For the examples in this section, it is given by $\bsy{A}_k = ( \bsy{a}_{k,1}, \ldots, \bsy{a}_{k,N+1} )$,
where the integers $\bsy{a}_{k,j}$ are determined by
\begin{gather*}
    \bsy{a}_{k,j} =   \begin{cases} \frac{1}{2} (2 n -k)(k-1)+j & \text{if} \ \ k \leq j,\\
                    \bsy{a}_{j,k-1} &
                    \text{if} \ \   k > j.\end{cases}
\end{gather*}
\end{rem}

\paragraph{1-simplex equation.}
In case of $B(2,0)$ we consider maps $\cR_1,\cR_2\colon \cU_\varnothing \longrightarrow \cU_\varnothing$
subject to
\begin{gather*}
     \cR_2   \cR_1 = \cR_1   \cR_2   ,
\end{gather*}
which is the $1$-simplex equation.

\paragraph{2-simplex equation and the cube.}
Associated with the two maximal chains of $B(3,1)$ is the $2$-simplex, or Yang--Baxter equation,
\begin{gather*}
      \cR_{23,\bsy{1}}   \cR_{13,\bsy{2}}   \cR_{12,\bsy{1}}
    = \cR_{12,\bsy{2}}   \cR_{13,\bsy{1}}   \cR_{23,\bsy{2}}   ,
\end{gather*}
for maps $\cR_{ij}\colon \cU_i \times \cU_j \rightarrow \cU_j \times \cU_i$, $i<j$.
The two sides of this equation correspond to sequences of maximal chains on two complementary sides of the cube,
formed by $B(3,0)$, see Fig.~\ref{fig:YB_on_cube}.
In complementary notation, $23 = \hat{1}$, $13 = \hat{2}$ and $12 = \hat{3}$, the
Yang--Baxter equation reads
\begin{gather*}
      \cR_{\hat{1},\bsy{1}}   \cR_{\hat{2},\bsy{2}}   \cR_{\hat{3},\bsy{1}}
    = \cR_{\hat{3},\bsy{2}}   \cR_{\hat{2},\bsy{1}}   \cR_{\hat{1},\bsy{2}}   .
      %\label{2-simplex_eq_cn}
\end{gather*}
In terms of $\hat{\cR}_{\hat{k}} := \cR_{\hat{k}}   \cP$, it takes the form
\begin{gather*}
    \hat{\cR}_{\hat{1},\bsy{12}}  \hat{\cR}_{\hat{2},\bsy{13}}   \hat{\cR}_{\hat{3},\bsy{23}}
  = \hat{\cR}_{\hat{3},\bsy{23}}  \hat{\cR}_{\hat{2},\bsy{13}}   \hat{\cR}_{\hat{1},\bsy{12}}
      .
\end{gather*}

\begin{figure}[t]
\centering
\includegraphics[width=.74\linewidth]{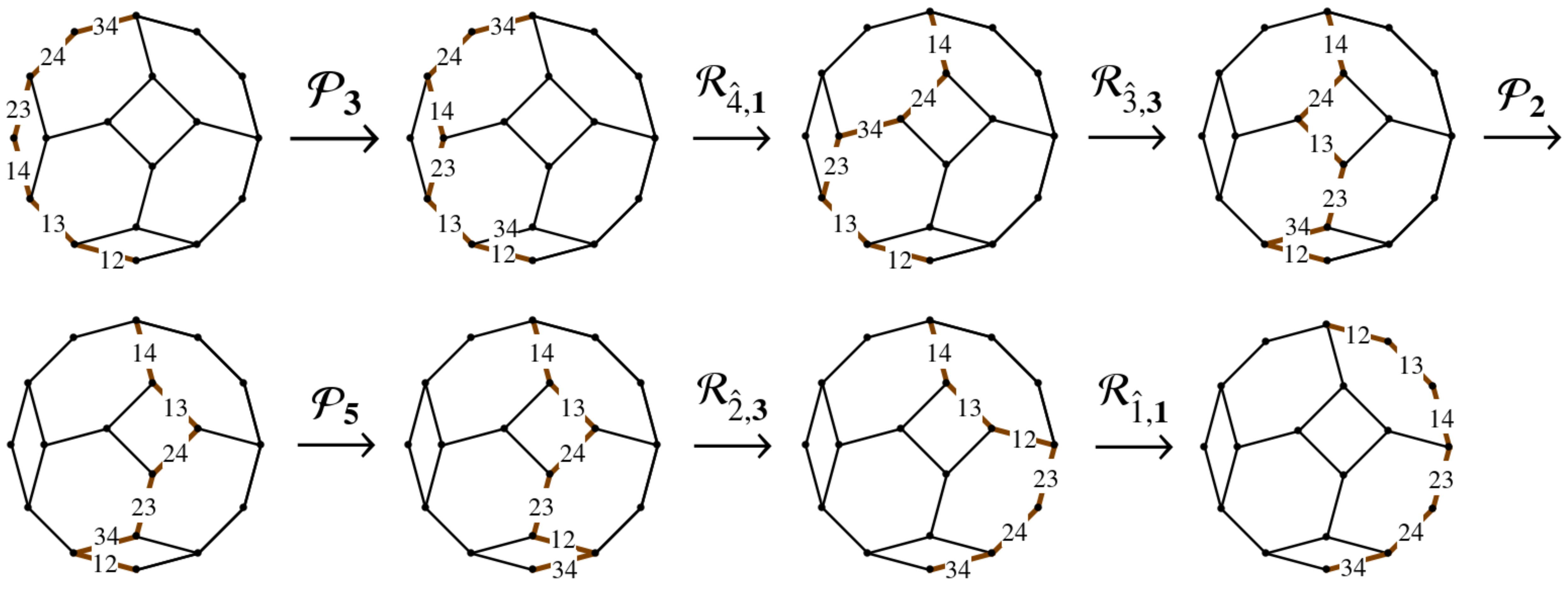}
\caption{The left-hand side of the $3$-simplex equation (\ref{3-simplex_eq_cn}) corresponds to a
sequence of maximal chains of $B(4,1)$. This sequence forms one side of the permutahedron in
three dimensions. Here and in the following f\/igures, if not stated otherwise,
edge labels in graphs will be in complementary notation, but with hats omitted.
\label{fig:left_3simplex_on_permutahedron} }
\end{figure}

\begin{figure}[t]
\centering
\includegraphics[width=.74\linewidth]{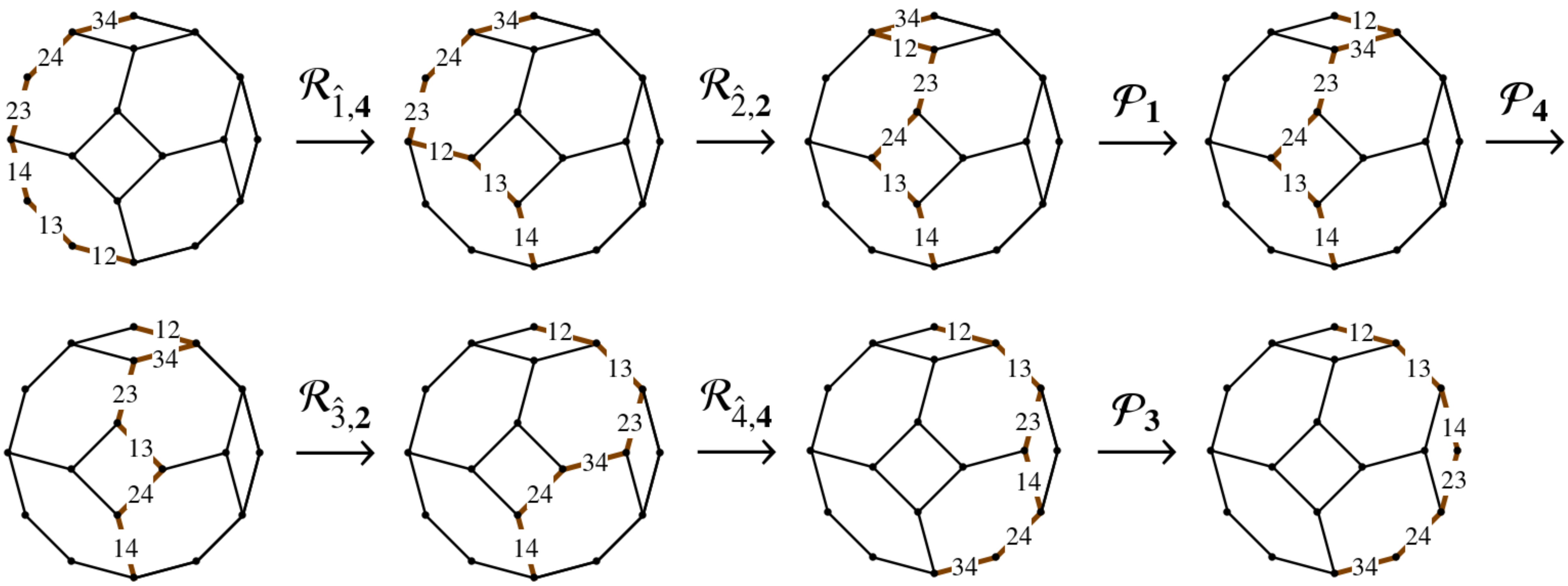}
\caption{The right-hand side of the $3$-simplex equation (\ref{3-simplex_eq_cn}) corresponds to
a sequence of maximal chains of $B(4,1)$, forming the side of the permutahedron complementary to that in
Fig.~\ref{fig:left_3simplex_on_permutahedron}.
\label{fig:right_3simplex_on_permutahedron} }
\end{figure}

\begin{figure}[t]
\centering
\includegraphics[width=.56\linewidth]{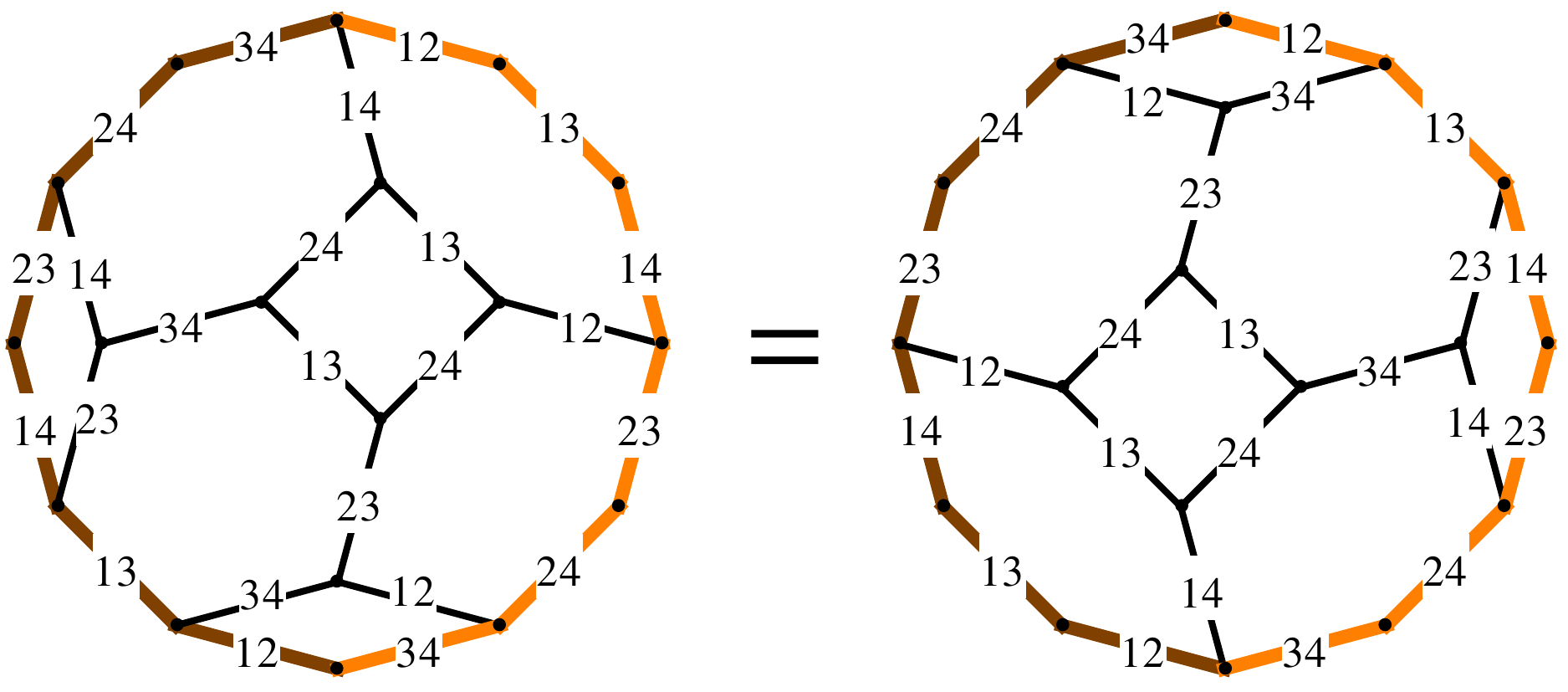}
\caption{Graphical representation of the $3$-simplex equation.
\label{fig:3simplex_eq} }
\end{figure}

\paragraph{3-simplex equation and the permutahedron.}
The two maximal chains of $B(4,2)$ are
\begin{gather*}
    \cC_{\mathrm{lex}}\colon \  [\alpha] \stackrel{123}{\rightarrow} [\rho_1]
            \stackrel{124}{\rightarrow} [\rho_2]
            \stackrel{134}{\rightarrow} [\rho_3]
            \stackrel{234}{\rightarrow} [\omega]   , \qquad
    \cC_{\mathrm{rev}}\colon \ [\alpha] \stackrel{234}{\rightarrow} [\sigma_1] \stackrel{134}{\rightarrow} [\sigma_2]
             \stackrel{124}{\rightarrow} [\sigma_3]  \stackrel{123}{\rightarrow} [\omega]   .
\end{gather*}
Let us start with the lexicographical linear order $\alpha = (12,13,14,23,24,34)$. The minimal element
$[\alpha] \in B(4,2)$ also contains $(12,13,23,14,24,34)$. $\omega$ is $\alpha$ in
reverse order. We already dis\-played~$\tilde{\cC}_{\mathrm{lex}}$ and~$\tilde{\cC}_{\mathrm{rev}}$ in
(\ref{B(4,2)-chains}). From them we read of\/f
\begin{gather*}
    \cR_{\tilde{\cC}_{\mathrm{lex}}}
  = \cR_{234,\bsy{1}}   \cR_{134,\bsy{3}}   \cP_{\bsy{5}}   \cP_{\bsy{2}}
    \cR_{124,\bsy{3}}   \cR_{123,\bsy{1}}   \cP_{\bsy{3}}   , \qquad
    \cR_{\tilde{\cC}_{\mathrm{rev}}}
  = \cP_{\bsy{3}}   \cR_{123,\bsy{4}}   \cR_{124,\bsy{2}}
    \cP_{\bsy{4}}   \cP_{\bsy{1}}   \cR_{134,\bsy{2}}   \cR_{234,\bsy{4}}   ,
\end{gather*}
for maps $\cR_{ijk}\colon \cU_{ij} \times \cU_{ik} \times \cU_{jk} \rightarrow \cU_{jk} \times \cU_{ik} \times \cU_{ij}$, $i<j<k$.
This determines the $3$-simplex equation
\begin{gather*}
   \cR_{234,\bsy{1}}  \cR_{134,\bsy{3}}  \cP_{\bsy{5}}  \cP_{\bsy{2}}
    \cR_{124,\bsy{3}}  \cR_{123,\bsy{1}}  \cP_{\bsy{3}}
 = \cP_{\bsy{3}}  \cR_{123,\bsy{4}}  \cR_{124,\bsy{2}}  \cP_{\bsy{4}}  \cP_{\bsy{1}}
   \cR_{134,\bsy{2}}  \cR_{234,\bsy{4}}  .
\end{gather*}
In complementary notation, $234 = \hat{1}$, $134 = \hat{2}$, etc., it reads
\begin{gather}
   \cR_{\hat{1},\bsy{1}}  \cR_{\hat{2},\bsy{3}}  \cP_{\bsy{5}}  \cP_{\bsy{2}}
   \cR_{\hat{3},\bsy{3}}  \cR_{\hat{4},\bsy{1}}  \cP_{\bsy{3}}
 = \cP_{\bsy{3}} \cR_{\hat{4},\bsy{4}}  \cR_{\hat{3},\bsy{2}}  \cP_{\bsy{4}} \cP_{\bsy{1}}
    \cR_{\hat{2},\bsy{2}}  \cR_{\hat{1},\bsy{4}}  ,
  \label{3-simplex_eq_cn}
\end{gather}
where, for example,
$\cR_{\hat{1}}  \colon \cU_{\widehat{14}} \times \cU_{\widehat{13}} \times \cU_{\widehat{12}}
\rightarrow \cU_{\widehat{12}} \times \cU_{\widehat{13}} \times \cU_{\widehat{14}}$ and
$\cR_{\hat{2}}\colon \cU_{\widehat{24}} \times \cU_{\widehat{23}} \times \cU_{\widehat{12}}
\rightarrow \cU_{\widehat{12}} \times \cU_{\widehat{23}} \times \cU_{\widehat{24}}$.
Left- and right-hand side of (\ref{3-simplex_eq_cn}) correspond, respectively, to
Figs.~\ref{fig:left_3simplex_on_permutahedron} and~\ref{fig:right_3simplex_on_permutahedron}.
Collapsing the sequences of graphs in these f\/igures, we can represent the equation as in
Fig.~\ref{fig:3simplex_eq}. Disregarding the indices associated with the underlying Bruhat order,
this is Fig.~17 in~\cite{Lawr95} and Fig.~5
in Chapter~6 of~\cite{Cart+Sait98}, where the $3$-simplex equation has been called ``permutohedron equation''.

In terms of $\hat{\cR} := \cR  \cP_{\bsy{1} \bsy{3}}$, where $\cP_{\bsy{a} \bsy{b}}$ is the
transposition map acting at positions $\bsy{a}$ and $\bsy{b}$,
the $3$-simplex equation takes the form
\begin{gather*}
    \hat{\cR}_{\hat{1},\bsy{123}}   \hat{\cR}_{\hat{2},\bsy{145}}   \hat{\cR}_{\hat{3},\bsy{246}}
    \hat{\cR}_{\hat{4},\bsy{356}} = \hat{\cR}_{\hat{4},\bsy{356}}   \hat{\cR}_{\hat{3},\bsy{246}}
    \hat{\cR}_{\hat{2},\bsy{145}}   \hat{\cR}_{\hat{1},\bsy{123}}   ,
     %\label{3-simplex_eq_mod}
\end{gather*}
which is also known as the \emph{tetrahedron} or \emph{Zamolodchikov equation}. Ignoring
the boldface indices and interpreting the others as ``position indices'', we formally obtain the
Frenkel--Moore version~\cite{Fren+Moor91}
\begin{gather*}
    \hat{\cR}_{\bsy{234}}   \hat{\cR}_{\bsy{134}}  \hat{\cR}_{\bsy{124}}
    \hat{\cR}_{\bsy{123}} = \hat{\cR}_{\bsy{123}}   \hat{\cR}_{\bsy{124}}
    \hat{\cR}_{\bsy{134}}   \hat{\cR}_{\bsy{234}}   .
\end{gather*}
With a dif\/ferent interpretation of the indices, this equation appeared, for example, in~\cite{Byts+Volk13}.

\begin{figure}[t]
\centering
\includegraphics[width=.85\linewidth]{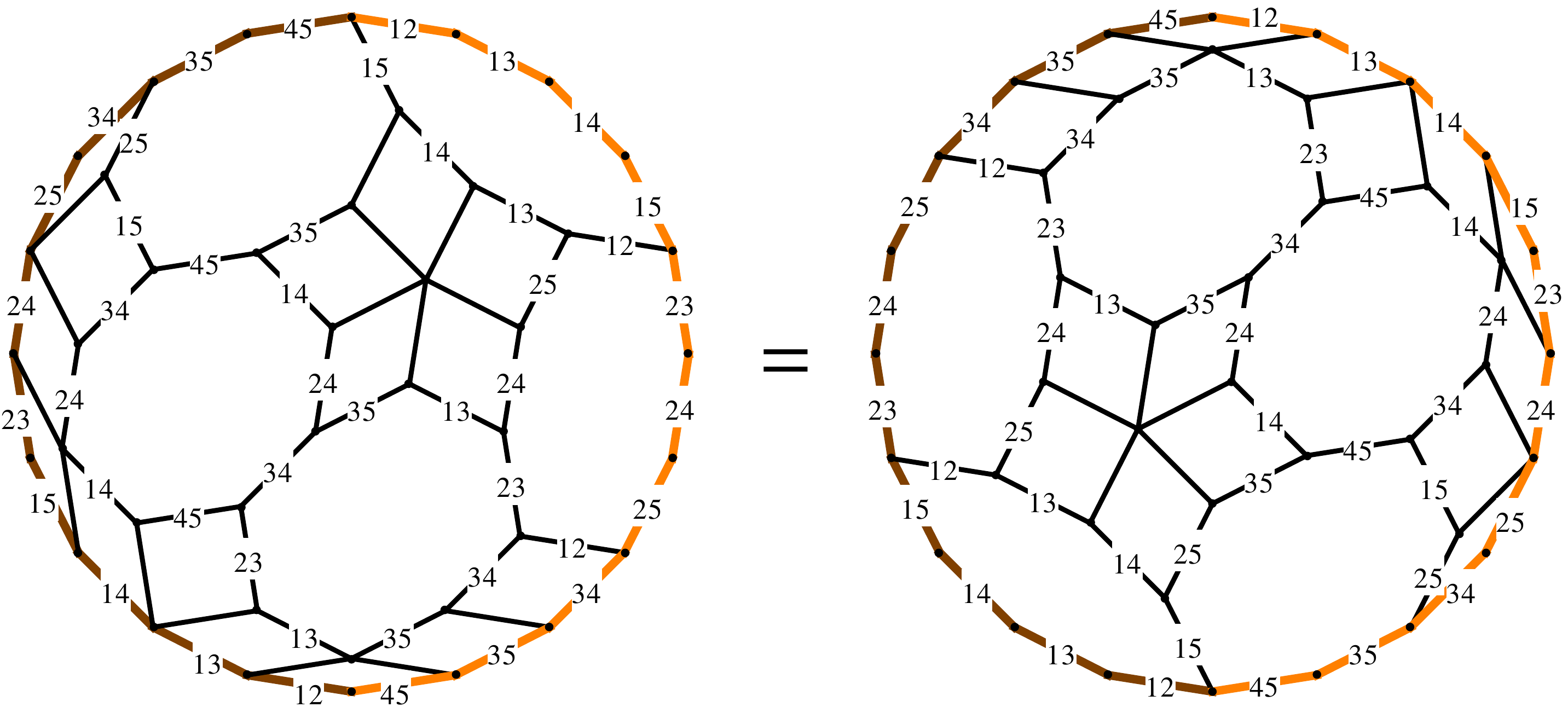}
\caption{The two sides of the $4$-simplex equation correspond to sequences of
maximal chains on two complementary sides (left and right f\/igure)
of the Felsner--Ziegler polyhedron, which carries the partial order~$B(5,2)$. Edge labels are
in complementary notation, but with hats omitted.
\label{fig:4simplex_eq_FZ_polyhedron} }
\end{figure}

\paragraph{4-simplex equation and the Felsner--Ziegler polyhedron.}
In case of $B(5,3)$ we consider maps
$\cR_{ijkl}\colon \cU_{ijk} \times \cU_{ijl} \times \cU_{ikl} \times \cU_{jkl} \rightarrow
\cU_{jkl} \times \cU_{ikl} \times \cU_{ijl} \times \cU_{ijk}$, $i<j<k<l$. Turning to
complementary notation, we have, for example, $\cR_{2345}  = \cR_{\hat{1}} \colon
\cU_{\widehat{15}} \times \cU_{\widehat{14}} \times
\cU_{\widehat{13}} \times \cU_{\widehat{12}} \rightarrow \cU_{\widehat{12}} \times
\cU_{\widehat{13}} \times \cU_{\widehat{14}} \times \cU_{\widehat{15}}$. The maps
are subject to the $4$-simplex equation
\begin{gather}
  \cR_{\hat{1},\bsy{1}}   \cR_{\hat{2},\bsy{4}}   \cP_{\bsy{7}}
      \cP_{\bsy{8}}   \cP_{\bsy{9}}   \cP_{\bsy{3}}
      \cP_{\bsy{2}}   \cP_{\bsy{4}}  \cR_{\hat{3},\bsy{5}}
      \cP_{\bsy{8}}   \cP_{\bsy{7}}   \cP_{\bsy{4}}   \cP_{\bsy{3}}
      \cR_{\hat{4},\bsy{4}}   \cP_{\bsy{7}}   \cR_{\hat{5},\bsy{1}}
      \cP_{\bsy{4}}   \cP_{\bsy{5}}    \cP_{\bsy{6}}
     \cP_{\bsy{3}}               \nonumber \\
\qquad {} =  \cP_{\bsy{7}}   \cP_{\bsy{4}}   \cP_{\bsy{5}}
     \cP_{\bsy{6}}   \cP_{\bsy{3}}   \cR_{\hat{5},\bsy{7}}
     \cR_{\hat{4},\bsy{4}}   \cP_{\bsy{7}}   \cP_{\bsy{8}}
     \cP_{\bsy{6}}   \cP_{\bsy{3}}   \cP_{\bsy{2}}
   \cP_{\bsy{1}}   \cR_{\hat{3},\bsy{3}}   \cP_{\bsy{6}}
     \cP_{\bsy{7}}   \cP_{\bsy{2}}   \cP_{\bsy{3}}
   \cR_{\hat{2},\bsy{4}}   \cR_{\hat{1},\bsy{7}}   .  \label{4-simplex_eq}
\end{gather}
This can be read of\/f from $B(5,2)$, which forms the Felsner--Ziegler polyhedron
($G_5$ in~\cite{Fels+Zieg01}), see Fig.~\ref{fig:4simplex_eq_FZ_polyhedron}.
In terms of $\hat{\cR}_{\hat{k}} := \cR_{\hat{k}} \, \cP_{\bsy{2} \bsy{3}} \, \cP_{\bsy{1} \bsy{4}}$,
the $4$-simplex equation takes the more concise form
\begin{gather}
   \hat{\cR}_{\hat{1},\bsy{1},\bsy{2},\bsy{3},\bsy{4}}   \hat{\cR}_{\hat{2},\bsy{1},\bsy{5},\bsy{6},\bsy{7}}
      \hat{\cR}_{\hat{3},\bsy{2},\bsy{5},\bsy{8},\bsy{9}}
   \hat{\cR}_{\hat{4},\bsy{3},\bsy{6},\bsy{8},\bsy{10}}
   \hat{\cR}_{\hat{5},\bsy{4},\bsy{7},\bsy{9},\bsy{10}} \nonumber  \\
 \qquad{}   = \hat{\cR}_{\hat{5},\bsy{4},\bsy{7},\bsy{9},\bsy{10}}
    \hat{\cR}_{\hat{4},\bsy{3},\bsy{6},\bsy{8},\bsy{10}}
    \hat{\cR}_{\hat{3},\bsy{2},\bsy{5},\bsy{8},\bsy{9}}   \hat{\cR}_{\hat{2},\bsy{1},\bsy{5},\bsy{6},\bsy{7}}
      \hat{\cR}_{\hat{1},\bsy{1},\bsy{2},\bsy{3},\bsy{4}}   .   \label{4-simplex_eq_hatR_cn}
\end{gather}
It is also known as the \emph{Bazhanov--Stroganov equation} (see, e.g.,~\cite{Mail+Nijh89PLA}).

\begin{figure}[t]\centering
\includegraphics[width=.75\linewidth]{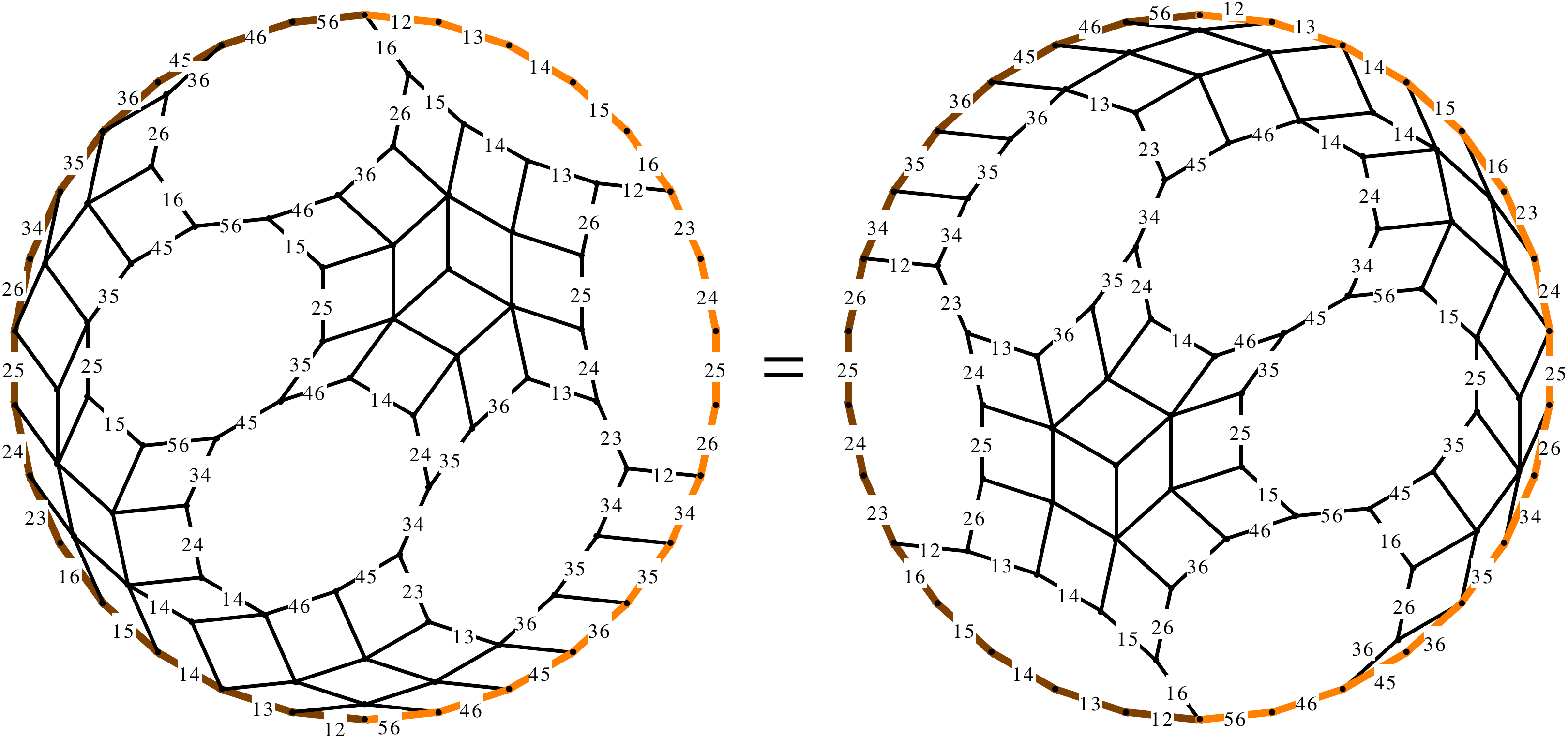}
\caption{The two sides of the $5$-simplex equation correspond to sequences of
maximal chains on two complementary sides (left and right f\/igure)
of a polyhedral reduction of~$B(6,3)$. Parallel edges carry the same label.\label{fig:5simplex_eq_polyhedron} }
\end{figure}

\paragraph{5-simplex equation.}
Turning to $B(6,4)$, we are dealing with maps
\begin{gather*}
 \cR_{ijklm} \colon \  \cU_{ijkl} \times \cU_{ijkm} \times \cU_{ijlm} \times \cU_{iklm} \times \cU_{jklm}
 \longrightarrow \cU_{jklm} \times \cU_{iklm} \times \cU_{ijlm} \times \cU_{ijkm} \times \cU_{ijkl}   ,
\end{gather*}
where $i<j<k<l<m$. In complementary notation, for example,
\begin{gather*}
  \cR_{23456} = \cR_{\hat{1}} \colon \  \cU_{\widehat{16}} \times \cU_{\widehat{15}} \times \cU_{\widehat{14}}
\times \cU_{\widehat{13}} \times \cU_{\widehat{12}} \longrightarrow \cU_{\widehat{12}} \times
\cU_{\widehat{13}} \times \cU_{\widehat{14}} \times \cU_{\widehat{15}} \times \cU_{\widehat{16}}   .
\end{gather*}
These maps have to satisfy the $5$-simplex equation
\begin{gather}
   \cR_{\hat{1},\bsy{1}}   \cR_{\hat{2},\bsy{5}}   \cP_{\bsy{9}}
    \cP_{\bsy{10}}   \cP_{\bsy{11}}   \cP_{\bsy{12}}
    \cP_{\bsy{13}}   \cP_{\bsy{14}}
    \cP_{\bsy{4}}   \cP_{\bsy{3}}   \cP_{\bsy{2}}   \cP_{\bsy{5}}
    \cP_{\bsy{4}}   \cP_{\bsy{6}}   \cR_{\hat{3},\bsy{7}}
    \cP_{\bsy{11}}   \cP_{\bsy{12}}   \cP_{\bsy{13}}   \cP_{\bsy{10}}
       \nonumber \\
  \cP_{\bsy{11}}   \cP_{\bsy{12}}   \cP_{\bsy{6}}   \cP_{\bsy{5}}
    \cP_{\bsy{4}}   \cP_{\bsy{3}}
    \cP_{\bsy{7}}   \cP_{\bsy{6}}   \cR_{\hat{4},\bsy{7}}
    \cP_{\bsy{11}}   \cP_{\bsy{12}}   \cP_{\bsy{10}}   \cP_{\bsy{9}}
    \cP_{\bsy{6}}   \cP_{\bsy{5}}   \cP_{\bsy{4}}
    \cR_{\hat{5},\bsy{5}}   \cP_{\bsy{9}}   \cP_{\bsy{10}}
    \cP_{\bsy{11}}   \cP_{\bsy{8}}   \nonumber \\
  \cR_{\hat{6},\bsy{1}}
    \cP_{\bsy{5}}   \cP_{\bsy{6}}   \cP_{\bsy{7}}
    \cP_{\bsy{8}}   \cP_{\bsy{9}}   \cP_{\bsy{10}}   \cP_{\bsy{4}}
    \cP_{\bsy{5}}    \cP_{\bsy{6}}   \cP_{\bsy{3}}  \nonumber \\
\qquad {} =   \cP_{\bsy{12}}   \cP_{\bsy{9}}   \cP_{\bsy{10}}   \cP_{\bsy{11}}
       \cP_{\bsy{8}}   \cP_{\bsy{5}}   \cP_{\bsy{6}}   \cP_{\bsy{7}}
       \cP_{\bsy{8}}   \cP_{\bsy{9}}   \cP_{\bsy{10}}   \cP_{\bsy{4}}
       \cP_{\bsy{5}}   \cP_{\bsy{6}}   \cP_{\bsy{3}}
     \cR_{\hat{6},\bsy{11}}   \cR_{\hat{5},\bsy{7}}   \cP_{\bsy{11}}
     \cP_{\bsy{12}}   \cP_{\bsy{13}}   \nonumber \\
\qquad \quad \  \cP_{\bsy{10}}
     \cP_{\bsy{11}}   \cP_{\bsy{9}}   \cP_{\bsy{6}}
     \cP_{\bsy{5}}   \cP_{\bsy{4}}   \cP_{\bsy{3}}   \cP_{\bsy{2}}
     \cP_{\bsy{1}}   \cR_{\hat{4},\bsy{5}}   \cP_{\bsy{9}}
     \cP_{\bsy{10}}   \cP_{\bsy{11}}   \cP_{\bsy{12}}
     \cP_{\bsy{8}}   \cP_{\bsy{9}}   \cP_{\bsy{4}}   \cP_{\bsy{3}}
     \cP_{\bsy{2}}   \cP_{\bsy{5}}   \cP_{\bsy{4}}   \cP_{\bsy{3}}
         \nonumber \\
\qquad \quad \   \cR_{\hat{3},\bsy{5}}   \cP_{\bsy{9}}   \cP_{\bsy{10}}
     \cP_{\bsy{11}}   \cP_{\bsy{4}}   \cP_{\bsy{5}}   \cP_{\bsy{6}}
     \cR_{\hat{2},\bsy{7}}   \cR_{\hat{1},\bsy{11}}   ,
         \label{5simplex_eq}
\end{gather}
see Fig.~\ref{fig:5simplex_eq_polyhedron}.
The Bruhat order $B(6,3)$ is not polyhedral~\cite{Fels+Zieg01}, because of the existence of ``small cubes'',
see Remark~\ref{rem:small_cubes} and Fig.~\ref{fig:5simplex_small_cubes}.
We can rewrite the $5$-simplex equation as follows to display the respective appearances (in brackets),
\begin{gather*}
  \cR_{\hat{1},\bsy{1}}   \cR_{\hat{2},\bsy{5}}   \cP_{\bsy{9}}    \cP_{\bsy{10}}
    ( \cP_{\bsy{12}}   \cP_{\bsy{11}}   \cP_{\bsy{12}} )
    \cP_{\bsy{13}}   \cP_{\bsy{14}}   \cP_{\bsy{4}}   \cP_{\bsy{3}}   \cP_{\bsy{2}}
       \cP_{\bsy{5}}   \cP_{\bsy{6}}
    \cR_{\hat{3},\bsy{7}}   \cP_{\bsy{11}}
    \cP_{\bsy{12}}   \cP_{\bsy{13}}   \cP_{\bsy{10}}   \cP_{\bsy{11}}   \nonumber \\
  \cP_{\bsy{6}}
    ( \cP_{\bsy{4}}   \cP_{\bsy{5}}   \cP_{\bsy{4}} )
    \cP_{\bsy{3}}   \cP_{\bsy{7}}   \cP_{\bsy{6}}
    \cR_{\hat{4},\bsy{7}}
    ( \cP_{\bsy{12}}   \cP_{\bsy{11}}   \cP_{\bsy{12}} )
    \cP_{\bsy{10}}   \cP_{\bsy{9}}   \cP_{\bsy{6}}   \cP_{\bsy{5}}   \cP_{\bsy{4}}
    \cR_{\hat{5},\bsy{5}}
    \cP_{\bsy{9}}   \cP_{\bsy{10}}   \cP_{\bsy{11}}
    \cR_{\hat{6},\bsy{1}}  \nonumber \\
  \cP_{\bsy{5}}   \cP_{\bsy{6}}
    ( \cP_{\bsy{8}}   \cP_{\bsy{7}}   \cP_{\bsy{8}} )
    \cP_{\bsy{9}}   \cP_{\bsy{10}}   \cP_{\bsy{4}}
    \cP_{\bsy{5}}    \cP_{\bsy{6}}   \nonumber \\
\qquad {} =  \cP_{\bsy{9}}   \cP_{\bsy{10}}   \cP_{\bsy{11}}   \cP_{\bsy{5}}   \cP_{\bsy{6}}
     ( \cP_{\bsy{8}}   \cP_{\bsy{7}}   \cP_{\bsy{8}} )
     \cP_{\bsy{9}}   \cP_{\bsy{10}}   \cP_{\bsy{4}}   \cP_{\bsy{5}}   \cP_{\bsy{6}}
     \cR_{\hat{6},\bsy{11}}   \cR_{\hat{5},\bsy{7}}   \cP_{\bsy{11}}
     \cP_{\bsy{12}}   \cP_{\bsy{13}}   \cP_{\bsy{10}}  \nonumber \\
\qquad \quad \  \cP_{\bsy{9}}   \cP_{\bsy{6}}   \cP_{\bsy{5}}
     ( \cP_{\bsy{3}}   \cP_{\bsy{4}}   \cP_{\bsy{3}} )
     \cP_{\bsy{2}}   \cP_{\bsy{1}}   \cR_{\hat{4},\bsy{5}}   \cP_{\bsy{9}}
     ( \cP_{\bsy{11}}   \cP_{\bsy{10}}   \cP_{\bsy{11}} )
     \cP_{\bsy{12}}   \cP_{\bsy{8}}   \cP_{\bsy{9}}   \cP_{\bsy{4}}
     \cP_{\bsy{3}}   \cP_{\bsy{2}}   \cP_{\bsy{5}}   \cP_{\bsy{4}}
              \nonumber \\
\qquad \quad \  \cR_{\hat{3},\bsy{5}}   \cP_{\bsy{9}}   \cP_{\bsy{10}}
     \cP_{\bsy{11}}
     ( \cP_{\bsy{3}}   \cP_{\bsy{4}}   \cP_{\bsy{3}} )
     \cP_{\bsy{5}}   \cP_{\bsy{6}}
     \cR_{\hat{2},\bsy{7}}   \cR_{\hat{1},\bsy{11}}   .
\end{gather*}
In terms of $\hat{\cR} := \cR   \cP_{\bsy{24}}   \cP_{\bsy{15}}$,
the $5$-simplex equation takes the form
\begin{gather*}
    \hat{\cR}_{\hat{1},\bsy{1},\bsy{2},\bsy{3},\bsy{4},\bsy{5}}   \hat{\cR}_{\hat{2},\bsy{1},\bsy{6},\bsy{7},\bsy{8},\bsy{9}}
    \hat{\cR}_{\hat{3},\bsy{2}, \bsy{6}, \bsy{10}, \bsy{11}, \bsy{12}}
    \hat{\cR}_{\hat{4},\bsy{3}, \bsy{7}, \bsy{10}, \bsy{13}, \bsy{14}}
    \hat{\cR}_{\hat{5},\bsy{4}, \bsy{8}, \bsy{11}, \bsy{13}, \bsy{15}}
    \hat{\cR}_{\hat{6},\bsy{5}, \bsy{9}, \bsy{12}, \bsy{14}, \bsy{15}}  \\
  \qquad = \hat{\cR}_{\hat{6},\bsy{5}, \bsy{9},\bsy{ 12}, \bsy{14}, \bsy{15}}
    \hat{\cR}_{\hat{5},\bsy{4}, \bsy{8}, \bsy{11}, \bsy{13}, \bsy{15}}
    \hat{\cR}_{\hat{4},\bsy{3}, \bsy{7}, \bsy{10}, \bsy{13},\bsy{ 14}}
    \hat{\cR}_{\hat{3},\bsy{2}, \bsy{6}, \bsy{10},\bsy{ 11}, \bsy{12}}
    \hat{\cR}_{\hat{2},\bsy{1}, \bsy{6}, \bsy{7}, \bsy{8}, \bsy{9}}
    \hat{\cR}_{\hat{1},\bsy{1},\bsy{2},\bsy{3},\bsy{4},\bsy{5}}   .
\end{gather*}

\begin{figure}[t]
\centering
\includegraphics[width=1.\linewidth]{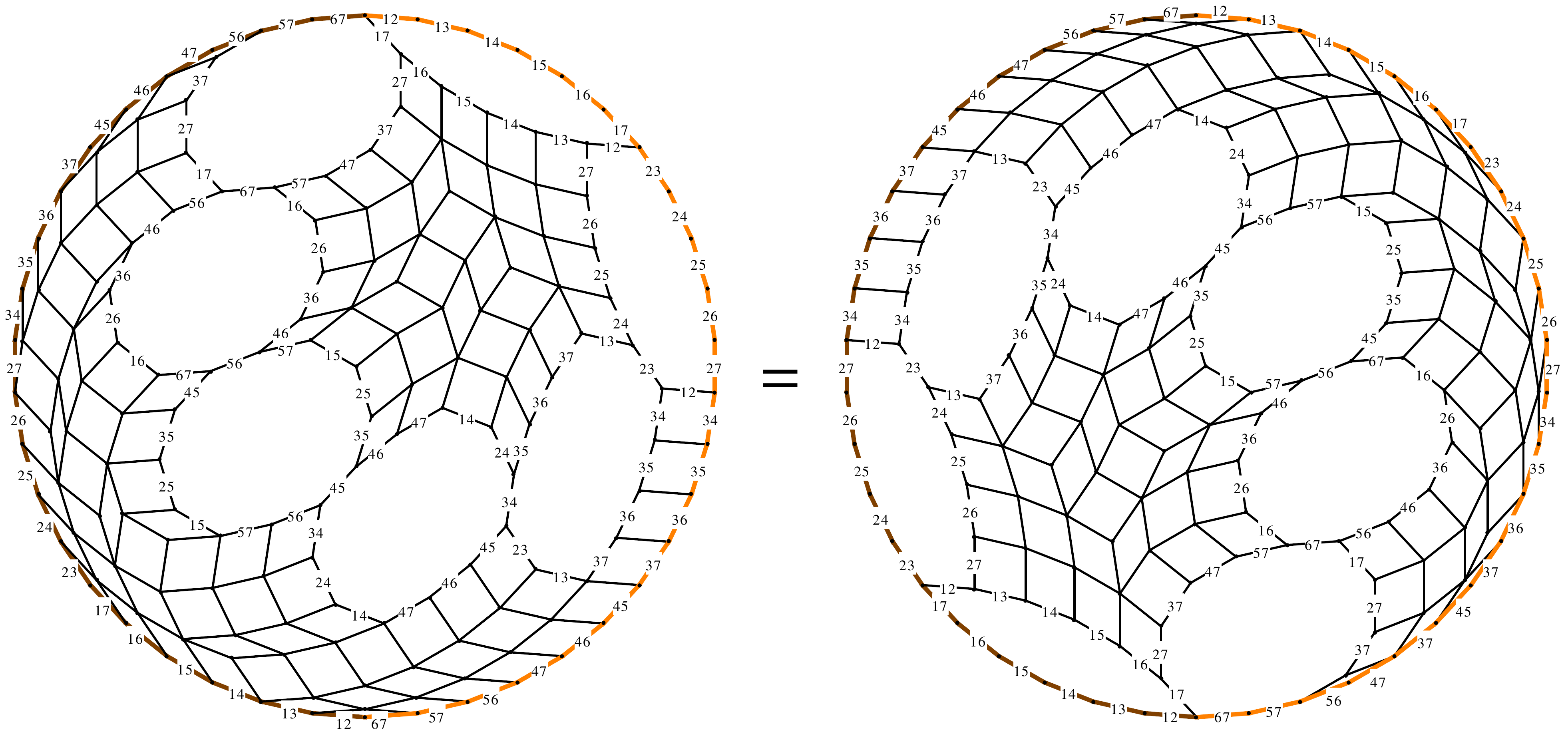}
\caption{The two sides of the $6$-simplex equation correspond to sequences of
maximal chains on two complementary sides (left and right f\/igure)
of a polyhedral reduction of $B(7,4)$.
\label{fig:6simplex_eq_polyhedron} }
\end{figure}

\paragraph{6-simplex equation.}
In case of $B(7,5)$ we consider maps
\begin{gather*}
 \cR_{ijklmn} \colon \  \cU_{ijklm} \times \cU_{ijkln} \times \cU_{ijkmn} \times \cU_{ijlmn}
 \times \cU_{ilmn} \times \cU_{jklmn}  \\
\hphantom{\cR_{ijklmn} \colon}{} \     \longrightarrow   \
 \cU_{jklmn} \times \cU_{iklmn} \times \cU_{ijlmn} \times \cU_{ijkmn} \times \cU_{ijkln}
 \times \cU_{ijklm}   ,
\end{gather*}
where $i<j<k<l<m<n$.
Turning to complementary notation, these maps have to satisfy the $6$-simplex equation
\begin{gather*}
  {\color{brown} \cR_{\hat{1},\bsy{1}}   \cR_{\hat{2},\bsy{6}} }
     ( \cP_{\bsy{11}}   \cP_{\bsy{12}}   \cP_{\bsy{13}}   \cP_{\bsy{14}}
       \cP_{\bsy{15}}   \cP_{\bsy{16}}   \cP_{\bsy{17}}   \cP_{\bsy{18}}
       \cP_{\bsy{19}}   \cP_{\bsy{20}} )   (\cP_{\bsy{5}}   \cP_{\bsy{4}}
       \cP_{\bsy{3}}   \cP_{\bsy{2}} )
     ( \cP_{\bsy{6}}   \cP_{\bsy{5}}   \cP_{\bsy{4}} )
     ( \cP_{\bsy{7}}   \cP_{\bsy{6}} )         \nonumber \\
    \cP_{\bsy{8}}
    {\color{brown} \cR_{\hat{3},\bsy{9}} }
     ( \cP_{\bsy{14}}   \cP_{\bsy{15}}   \cP_{\bsy{16}}
       \cP_{\bsy{17}}   \cP_{\bsy{18}}   \cP_{\bsy{19}})
     ( \cP_{\bsy{13}}   \cP_{\bsy{14}}   \cP_{\bsy{15}}   \cP_{\bsy{16}}
       \cP_{\bsy{17}}   \cP_{\bsy{18}} )
     ( \cP_{\bsy{8}}   \cP_{\bsy{7}}
       \cP_{\bsy{6}}   \cP_{\bsy{5}}   \cP_{\bsy{4}}   \cP_{\bsy{3}})
                      \nonumber \\
  (\cP_{\bsy{9}}   \cP_{\bsy{8}}   \cP_{\bsy{7}}   \cP_{\bsy{6}})
       (\cP_{\bsy{10}}   \cP_{\bsy{9}} )
    {\color{brown} \cR_{\hat{4},\bsy{10}} }
       ( \cP_{\bsy{15}}   \cP_{\bsy{16}}   \cP_{\bsy{17}}   \cP_{\bsy{18}} )
       ( \cP_{\bsy{14}}   \cP_{\bsy{15}}   \cP_{\bsy{16}} )
       ( \cP_{\bsy{13}}   \cP_{\bsy{14}}   \cP_{\bsy{15}} )   \nonumber \\
    ( \cP_{\bsy{9}}   \cP_{\bsy{8}}   \cP_{\bsy{7}}   \cP_{\bsy{6}}
         \cP_{\bsy{5}}   \cP_{\bsy{4}} )
       ( \cP_{\bsy{10}}   \cP_{\bsy{9}}   \cP_{\bsy{8}} )
   {\color{brown} \cR_{\hat{5},\bsy{9}} }
     ( \cP_{\bsy{14}}   \cP_{\bsy{15}}   \cP_{\bsy{16}}   \cP_{\bsy{17}} )
     ( \cP_{\bsy{13}}   \cP_{\bsy{14}} )
     ( \cP_{\bsy{12}}   \cP_{\bsy{11}} )    \nonumber \\
  ( \cP_{\bsy{8}}   \cP_{\bsy{7}}   \cP_{\bsy{6}}   \cP_{\bsy{5}} )
   {\color{brown} \cR_{\hat{6},\bsy{6}} }
    ( \cP_{\bsy{11}}   \cP_{\bsy{12}}   \cP_{\bsy{13}}   \cP_{\bsy{14}}
       \cP_{\bsy{15}}   \cP_{\bsy{16}} )
    ( \cP_{\bsy{10}}   \cP_{\bsy{11}}   \cP_{\bsy{12}} )
      \cP_{\bsy{9}}
   {\color{brown} \cR_{\hat{7},\bsy{1}} }      \nonumber \\
  ( \cP_{\bsy{6}}   \cP_{\bsy{7}}   \cP_{\bsy{8}}   \cP_{\bsy{9}}
      \cP_{\bsy{10}}   \cP_{\bsy{11}}   \cP_{\bsy{12}}   \cP_{\bsy{13}}
      \cP_{\bsy{14}}   \cP_{\bsy{15}} )
    ( \cP_{\bsy{5}}   \cP_{\bsy{6}}   \cP_{\bsy{7}}   \cP_{\bsy{8}}
      \cP_{\bsy{9}}   \cP_{\bsy{10}} )
    ( \cP_{\bsy{4}}   \cP_{\bsy{5}}   \cP_{\bsy{6}} )
      \cP_{\bsy{3}}     \nonumber \\
\qquad {}  =  \cP_{\bsy{18}}
    ( \cP_{\bsy{15}}   \cP_{\bsy{16}}   \cP_{\bsy{17}} )
      \cP_{\bsy{14}}
    ( \cP_{\bsy{11}}   \cP_{\bsy{12}}   \cP_{\bsy{13}}
      \cP_{\bsy{14}}   \cP_{\bsy{15}}   \cP_{\bsy{16}} )
    ( \cP_{\bsy{10}}   \cP_{\bsy{11}}   \cP_{\bsy{12}} )
      \cP_{\bsy{9}}   \nonumber \\
 \qquad \quad \  ( \cP_{\bsy{6}}   \cP_{\bsy{7}}   \cP_{\bsy{8}}
      \cP_{\bsy{9}}   \cP_{\bsy{10}}   \cP_{\bsy{11}}   \cP_{\bsy{12}}
      \cP_{\bsy{13}}   \cP_{\bsy{14}}   \cP_{\bsy{15}} )
    ( \cP_{\bsy{5}}   \cP_{\bsy{6}}   \cP_{\bsy{7}}   \cP_{\bsy{8}}
      \cP_{\bsy{9}}   \cP_{\bsy{10}} )
    ( \cP_{\bsy{4}}   \cP_{\bsy{5}}   \cP_{\bsy{6}} )
      \cP_{\bsy{3}}
   {\color{brown} \cR_{\hat{7},\bsy{16}}   \cR_{\hat{6},\bsy{11}} }   \nonumber \\
\qquad \quad \  ( \cP_{\bsy{16}}   \cP_{\bsy{17}}   \cP_{\bsy{18}}   \cP_{\bsy{19}} )
    ( \cP_{\bsy{15}}   \cP_{\bsy{16}}   \cP_{\bsy{17}} )
    ( \cP_{\bsy{14}}   \cP_{\bsy{15}} )
      \cP_{\bsy{13}}
    ( \cP_{\bsy{10}}   \cP_{\bsy{9}}   \cP_{\bsy{8}}   \cP_{\bsy{7}}
      \cP_{\bsy{6}}   \cP_{\bsy{5}}   \cP_{\bsy{4}}   \cP_{\bsy{3}}
      \cP_{\bsy{2}}   \cP_{\bsy{1}} )
   {\color{brown} \cR_{\hat{5},\bsy{8}} }      \nonumber \\
\qquad \quad \  ( \cP_{\bsy{13}}   \cP_{\bsy{14}}   \cP_{\bsy{15}}
      \cP_{\bsy{16}}   \cP_{\bsy{17}}   \cP_{\bsy{18}} )
    ( \cP_{\bsy{12}}   \cP_{\bsy{13}}   \cP_{\bsy{14}}   \cP_{\bsy{15}} )
    ( \cP_{\bsy{11}}   \cP_{\bsy{12}} )
    ( \cP_{\bsy{7}}   \cP_{\bsy{6}}   \cP_{\bsy{5}}   \cP_{\bsy{4}}
      \cP_{\bsy{3}}   \cP_{\bsy{2}} )       \nonumber \\
\qquad \quad \  ( \cP_{\bsy{8}}   \cP_{\bsy{7}}   \cP_{\bsy{6}}   \cP_{\bsy{5}}
      \cP_{\bsy{4}}   \cP_{\bsy{3}} )
   {\color{brown} \cR_{\hat{4},\bsy{7}} }
    ( \cP_{\bsy{12}}   \cP_{\bsy{13}}   \cP_{\bsy{14}}   \cP_{\bsy{15}}
      \cP_{\bsy{16}}   \cP_{\bsy{17}} )
    ( \cP_{\bsy{11}}   \cP_{\bsy{12}}   \cP_{\bsy{13}} )
    ( \cP_{\bsy{6}}   \cP_{\bsy{5}}   \cP_{\bsy{4}} )
    ( \cP_{\bsy{7}}   \cP_{\bsy{6}}   \cP_{\bsy{5}} )      \nonumber \\
\qquad \quad \  ( \cP_{\bsy{8}}   \cP_{\bsy{7}}   \cP_{\bsy{6}})
   {\color{brown} \cR_{\hat{3},\bsy{8}} }
    ( \cP_{\bsy{13}}   \cP_{\bsy{14}}   \cP_{\bsy{15}}   \cP_{\bsy{16}} )
      \cP_{\bsy{7}}   \cP_{\bsy{8}}   \cP_{\bsy{9}}   \cP_{\bsy{10}}
   {\color{brown} \cR_{\hat{2},\bsy{11}}   \cR_{\hat{1},\bsy{16}} }   .
\end{gather*}
The two sides of this equation correspond to sequences of maximal chains on complementary
sides of a polyhedron, see Fig.~\ref{fig:6simplex_eq_polyhedron}.
In terms of $\hat{\cR} = \cR   \cP_{\bsy{34}}   \cP_{\bsy{25}}   \cP_{\bsy{16}}$, the $6$-simplex equation
takes the form
\begin{gather}
   \hat{\cR}_{\hat{1},\bsy{1},\bsy{2},\bsy{3},\bsy{4},
    \bsy{5},\bsy{6} }
   \hat{\cR}_{\hat{2},\bsy{1},\bsy{7},\bsy{8},\bsy{9},\bsy{10},
    \bsy{11} }
   \hat{\cR}_{\hat{3},\bsy{2},\bsy{7},\bsy{12},\bsy{13},\bsy{14},
    \bsy{15} }
   \hat{\cR}_{\hat{4},\bsy{3},\bsy{8},\bsy{12},\bsy{16},\bsy{17},
    \bsy{18} }
   \hat{\cR}_{\hat{5},\bsy{4},\bsy{9},\bsy{13},\bsy{16},\bsy{19},
    \bsy{20} }               \nonumber \\
   \hat{\cR}_{\hat{6},\bsy{5},\bsy{10},\bsy{14},\bsy{17},\bsy{19},
    \bsy{21} }
   \hat{\cR}_{\hat{7},\bsy{6},\bsy{11},\bsy{15},\bsy{18},\bsy{20},
    \bsy{21} }                \nonumber \\
\qquad {} =  \hat{\cR}_{\hat{7},\bsy{6},\bsy{11},\bsy{15},\bsy{18},\bsy{20},
    \bsy{21} }
   \hat{\cR}_{\hat{6},\bsy{5},\bsy{10},\bsy{14},\bsy{17},\bsy{19},
    \bsy{21} }
   \hat{\cR}_{\hat{5},\bsy{4},\bsy{9},\bsy{13},\bsy{16},\bsy{19},
    \bsy{20} }
   \hat{\cR}_{\hat{4},\bsy{3},\bsy{8},\bsy{12},\bsy{16},\bsy{17},
    \bsy{18} }
\nonumber \\
\qquad \quad \ \hat{\cR}_{\hat{3},\bsy{2},\bsy{7},\bsy{12},\bsy{13},\bsy{14},
    \bsy{15} } \hat{\cR}_{\hat{2},\bsy{1},\bsy{7},\bsy{8},\bsy{9},\bsy{10},
    \bsy{11} }
   \hat{\cR}_{\hat{1},\bsy{1},\bsy{2},\bsy{3},\bsy{4},
    \bsy{5},\bsy{6} }   .     \label{6simplex_eq}
\end{gather}

\begin{figure}[t]
\centering
\includegraphics[width=1.\linewidth]{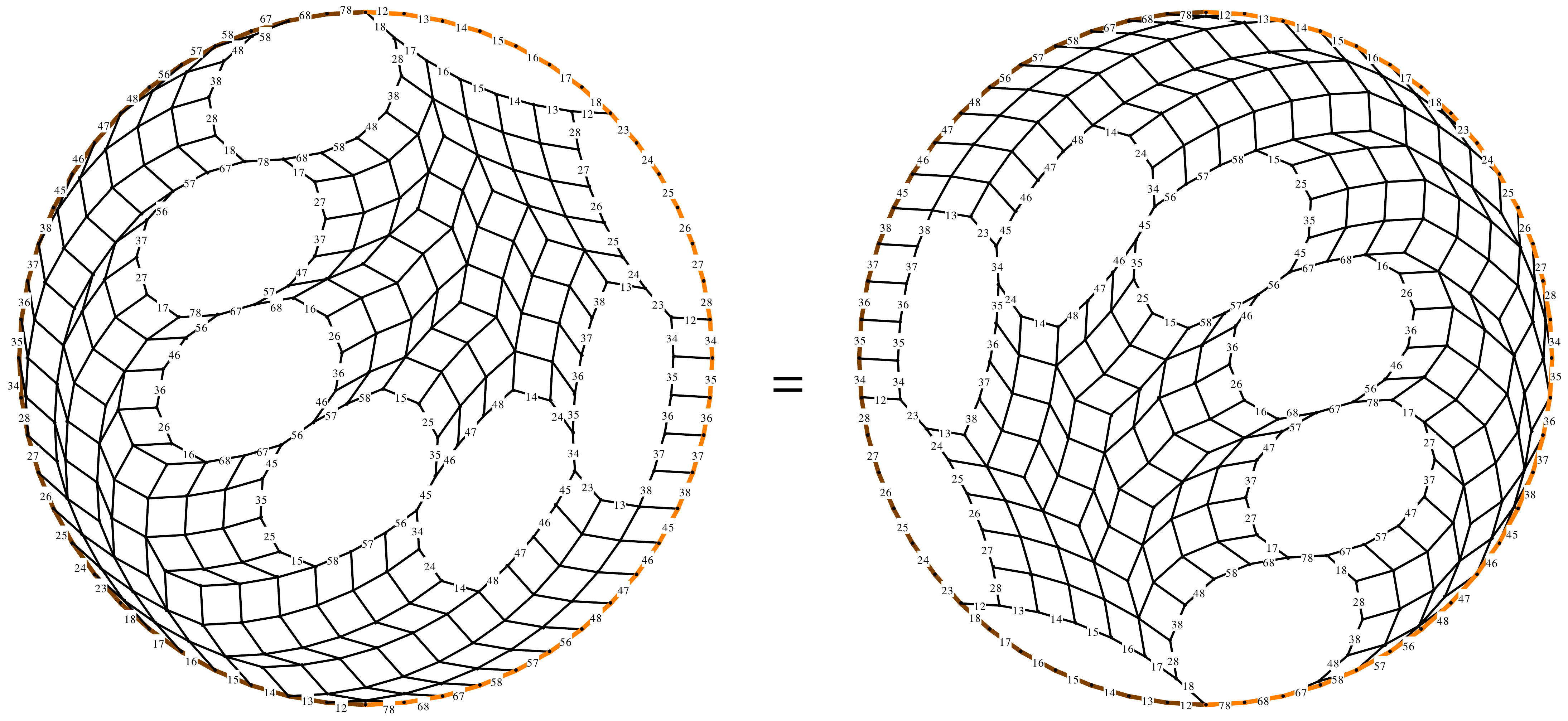}
\caption{Two complementary sides of a polyhedral reduction of $B(8,5)$. The $7$-simplex equation corresponds
to sequences of maximal chains on them.\label{fig:7simplex_eq_polyhedron} }
\end{figure}

\paragraph{7-simplex equation.}
Here we consider maps
\begin{gather*}
 \cR_{ijklmnp} \colon \  \cU_{ijklmn} \times \cU_{ijklmp} \times \cU_{ijknp} \times \cU_{ijkmnp} \times \cU_{ijlmnp}
 \times \cU_{ilmnp} \times \cU_{jklmnp}  \\
\hphantom{\cR_{ijklmnp} \colon}{} \ \longrightarrow  \
 \cU_{jklmnp} \times \cU_{iklmnp} \times \cU_{ijlmnp} \times \cU_{ijkmnp} \times \cU_{ijklnp} \times \cU_{ijklmp}
 \times \cU_{ijklmn}   ,
\end{gather*}
where $i<j<k<l<m<n<p$. The $7$-simplex equation reads
\begin{gather*}
    { \color{brown} \cR_{\hat{1},\bsy{1}}   \cR_{\hat{2},\bsy{7}} }
       \cP_{\bsy{13}}   \cP_{\bsy{14}}   \cP_{\bsy{15}}   \cP_{\bsy{16}}   \cP_{\bsy{17}}   \cP_{\bsy{18}}
       \cP_{\bsy{19}}   \cP_{\bsy{20}}   \cP_{\bsy{21}}   \cP_{\bsy{22}}   \cP_{\bsy{23}}   \cP_{\bsy{24}}
       \cP_{\bsy{25}}   \cP_{\bsy{26}}   \cP_{\bsy{27}}   \cP_{\bsy{6}}   \cP_{\bsy{5}}   \cP_{\bsy{4}}
       \cP_{\bsy{3}}    \\
     \cP_{\bsy{2}}   \cP_{\bsy{7}}   \cP_{\bsy{6}}   \cP_{\bsy{5}}   \cP_{\bsy{4}}   \cP_{\bsy{8}}
       \cP_{\bsy{7}}   \cP_{\bsy{6}}   \cP_{\bsy{9}}   \cP_{\bsy{8}}    \cP_{\bsy{10}}
       { \color{brown} \cR_{\hat{3},\bsy{11}} }
       \cP_{\bsy{17}}   \cP_{\bsy{18}}   \cP_{\bsy{19}}   \cP_{\bsy{20}}   \cP_{\bsy{21}}   \cP_{\bsy{22}}
       \cP_{\bsy{23}}   \cP_{\bsy{24}}   \cP_{\bsy{25}}   \cP_{\bsy{26}}   \\
    \cP_{\bsy{16}}   \cP_{\bsy{17}}   \cP_{\bsy{18}}   \cP_{\bsy{19}}   \cP_{\bsy{20}}   \cP_{\bsy{21}}
       \cP_{\bsy{22}}   \cP_{\bsy{23}}   \cP_{\bsy{24}}   \cP_{\bsy{25}}
       \cP_{\bsy{10}}   \cP_{\bsy{9}}   \cP_{\bsy{8}}   \cP_{\bsy{7}}   \cP_{\bsy{6}}   \cP_{\bsy{5}}
       \cP_{\bsy{4}}   \cP_{\bsy{3}}   \cP_{\bsy{11}}   \cP_{\bsy{10}}   \cP_{\bsy{9}}   \cP_{\bsy{8}}   \\
     \cP_{\bsy{7}}   \cP_{\bsy{6}}   \cP_{\bsy{12}}   \cP_{\bsy{11}}   \cP_{\bsy{10}}   \cP_{\bsy{9}}
       \cP_{\bsy{13}}   \cP_{\bsy{12}}
       { \color{brown} \cR_{\hat{4},\bsy{13}} }
       \cP_{\bsy{19}}   \cP_{\bsy{20}}   \cP_{\bsy{21}}   \cP_{\bsy{22}}   \cP_{\bsy{23}}   \cP_{\bsy{24}}
       \cP_{\bsy{25}}   \cP_{\bsy{18}}   \cP_{\bsy{19}}   \cP_{\bsy{20}}   \cP_{\bsy{21}}   \\
     \cP_{\bsy{22}}   \cP_{\bsy{23}}   \cP_{\bsy{17}}   \cP_{\bsy{18}}   \cP_{\bsy{19}}   \cP_{\bsy{20}}
       \cP_{\bsy{21}}   \cP_{\bsy{22}}   \cP_{\bsy{12}}   \cP_{\bsy{11}}   \cP_{\bsy{10}}   \cP_{\bsy{9}}
       \cP_{\bsy{8}}   \cP_{\bsy{7}}   \cP_{\bsy{6}}   \cP_{\bsy{5}}   \cP_{\bsy{4}}   \cP_{\bsy{13}}
       \cP_{\bsy{12}}   \cP_{\bsy{11}}   \cP_{\bsy{10}}    \\
     \cP_{\bsy{9}}   \cP_{\bsy{8}}   \cP_{\bsy{14}}   \cP_{\bsy{13}}   \cP_{\bsy{12}}
       { \color{brown} \cR_{\hat{5},\bsy{13}} }
       \cP_{\bsy{19}}   \cP_{\bsy{20}}   \cP_{\bsy{21}}   \cP_{\bsy{22}}   \cP_{\bsy{23}}   \cP_{\bsy{24}}
       \cP_{\bsy{18}}   \cP_{\bsy{19}}   \cP_{\bsy{20}}   \cP_{\bsy{21}}   \cP_{\bsy{17}}   \cP_{\bsy{18}}
       \cP_{\bsy{19}}   \cP_{\bsy{16}}     \\
     \cP_{\bsy{17}}   \cP_{\bsy{18}}   \cP_{\bsy{12}}   \cP_{\bsy{11}}   \cP_{\bsy{10}}   \cP_{\bsy{9}}
       \cP_{\bsy{8}}   \cP_{\bsy{7}}   \cP_{\bsy{6}}   \cP_{\bsy{5}}   \cP_{\bsy{13}}   \cP_{\bsy{12}}
       \cP_{\bsy{11}}   \cP_{\bsy{10}}
       { \color{brown} \cR_{\hat{6},\bsy{11}} }
       \cP_{\bsy{17}}   \cP_{\bsy{18}}   \cP_{\bsy{19}}   \cP_{\bsy{20}}   \cP_{\bsy{21}}   \cP_{\bsy{22}}   \\
     \cP_{\bsy{23}}   \cP_{\bsy{16}}   \cP_{\bsy{17}}   \cP_{\bsy{18}}   \cP_{\bsy{19}}   \cP_{\bsy{15}}
       \cP_{\bsy{16}}    \cP_{\bsy{14}}    \cP_{\bsy{13}}   \cP_{\bsy{10}}   \cP_{\bsy{9}}   \cP_{\bsy{8}}
       \cP_{\bsy{7}}   \cP_{\bsy{6}}
       { \color{brown} \cR_{\hat{7},\bsy{7}} }
       \cP_{\bsy{13}}   \cP_{\bsy{14}}   \cP_{\bsy{15}}   \cP_{\bsy{16}}   \cP_{\bsy{17}}   \cP_{\bsy{18}}   \\
     \cP_{\bsy{19}}   \cP_{\bsy{20}}   \cP_{\bsy{21}}   \cP_{\bsy{22}}   \cP_{\bsy{12}}   \cP_{\bsy{13}}
       \cP_{\bsy{14}}   \cP_{\bsy{15}}   \cP_{\bsy{16}}   \cP_{\bsy{17}}   \cP_{\bsy{11}}   \cP_{\bsy{12}}
       \cP_{\bsy{13}}   \cP_{\bsy{10}}
        { \color{brown} \cR_{\hat{8},\bsy{1}} }
       \cP_{\bsy{7}}   \cP_{\bsy{8}}   \cP_{\bsy{9}}   \cP_{\bsy{10}}   \cP_{\bsy{11}}   \cP_{\bsy{12}}   \\
     \cP_{\bsy{13}}   \cP_{\bsy{14}}   \cP_{\bsy{15}}   \cP_{\bsy{16}}   \cP_{\bsy{17}}   \cP_{\bsy{18}}
       \cP_{\bsy{19}}   \cP_{\bsy{20}}   \cP_{\bsy{21}}   \cP_{\bsy{6}}   \cP_{\bsy{7}}   \cP_{\bsy{8}}
       \cP_{\bsy{9}}   \cP_{\bsy{10}}   \cP_{\bsy{11}}   \cP_{\bsy{12}}   \cP_{\bsy{13}}   \cP_{\bsy{14}}
       \cP_{\bsy{15}}   \cP_{\bsy{5}}   \cP_{\bsy{6}}   \cP_{\bsy{7}}   \\
     \cP_{\bsy{8}}   \cP_{\bsy{9}}   \cP_{\bsy{10}}   \cP_{\bsy{4}}   \cP_{\bsy{5}}   \cP_{\bsy{6}}   \cP_{\bsy{3}}
               \\
\qquad {} =  \cP_{\bsy{25}}   \cP_{\bsy{22}}   \cP_{\bsy{23}}   \cP_{\bsy{24}}
       \cP_{\bsy{21}}   \cP_{\bsy{18}}   \cP_{\bsy{19}}   \cP_{\bsy{20}}
       \cP_{\bsy{21}}   \cP_{\bsy{22}}    \cP_{\bsy{23}}   \cP_{\bsy{17}}
       \cP_{\bsy{18}}   \cP_{\bsy{19}}   \cP_{\bsy{16}}   \cP_{\bsy{13}}   \cP_{\bsy{14}}   \cP_{\bsy{15}}
       \cP_{\bsy{16}}   \cP_{\bsy{17}}    \\
\qquad \quad \   \cP_{\bsy{18}}   \cP_{\bsy{19}}
       \cP_{\bsy{20}}   \cP_{\bsy{21}}   \cP_{\bsy{22}}   \cP_{\bsy{12}}
       \cP_{\bsy{13}}   \cP_{\bsy{14}}
       \cP_{\bsy{15}}   \cP_{\bsy{16}}   \cP_{\bsy{17}}   \cP_{\bsy{11}}
       \cP_{\bsy{12}}   \cP_{\bsy{13}}   \cP_{\bsy{10}}   \cP_{\bsy{7}}
       \cP_{\bsy{8}}   \cP_{\bsy{9}}   \cP_{\bsy{10}}   \cP_{\bsy{11}}   \\
\qquad \quad \    \cP_{\bsy{12}}   \cP_{\bsy{13}}
       \cP_{\bsy{14}}   \cP_{\bsy{15}}   \cP_{\bsy{16}}   \cP_{\bsy{17}}
       \cP_{\bsy{18}}   \cP_{\bsy{19}}   \cP_{\bsy{20}}   \cP_{\bsy{21}}
       \cP_{\bsy{6}}   \cP_{\bsy{7}}
       \cP_{\bsy{8}}   \cP_{\bsy{9}}   \cP_{\bsy{10}}   \cP_{\bsy{11}}
       \cP_{\bsy{12}}   \cP_{\bsy{13}}   \cP_{\bsy{14}}   \cP_{\bsy{15}}
       \cP_{\bsy{5}}   \\
\qquad \quad \    \cP_{\bsy{6}}   \cP_{\bsy{7}}   \cP_{\bsy{8}}   \cP_{\bsy{9}}   \cP_{\bsy{10}}
       \cP_{\bsy{4}}   \cP_{\bsy{5}}   \cP_{\bsy{6}}   \cP_{\bsy{3}}
       { \color{brown} \cR_{\hat{8},\bsy{22}}   \cR_{\hat{7},\bsy{16}} }
       \cP_{\bsy{22}}   \cP_{\bsy{23}}   \cP_{\bsy{24}}   \cP_{\bsy{25}}   \cP_{\bsy{26}}
       \cP_{\bsy{21}}   \cP_{\bsy{22}}   \cP_{\bsy{23}}   \cP_{\bsy{24}}   \cP_{\bsy{20}}   \\
\qquad \quad \     \cP_{\bsy{21}}
       \cP_{\bsy{22}}   \cP_{\bsy{19}}   \cP_{\bsy{20}}   \cP_{\bsy{18}}   \cP_{\bsy{15}}   \cP_{\bsy{14}}
       \cP_{\bsy{13}}   \cP_{\bsy{12}}   \cP_{\bsy{11}}   \cP_{\bsy{10}}
       \cP_{\bsy{9}}   \cP_{\bsy{8}}   \cP_{\bsy{7}}   \cP_{\bsy{6}}
       \cP_{\bsy{5}}   \cP_{\bsy{4}}   \cP_{\bsy{3}}   \cP_{\bsy{2}}   \cP_{\bsy{1}}
       { \color{brown} \cR_{\hat{6},\bsy{12}} }
       \cP_{\bsy{18}}   \\
\qquad \quad \     \cP_{\bsy{19}}   \cP_{\bsy{20}}   \cP_{\bsy{21}}   \cP_{\bsy{22}}
       \cP_{\bsy{23}}   \cP_{\bsy{24}}   \cP_{\bsy{25}}
       \cP_{\bsy{17}}   \cP_{\bsy{18}}   \cP_{\bsy{19}}   \cP_{\bsy{20}}   \cP_{\bsy{21}}   \cP_{\bsy{22}}
       \cP_{\bsy{16}}   \cP_{\bsy{17}}   \cP_{\bsy{18}}   \cP_{\bsy{19}}   \cP_{\bsy{15}}   \cP_{\bsy{16}}
       \cP_{\bsy{11}}   \\
\qquad \quad \     \cP_{\bsy{10}}   \cP_{\bsy{9}}   \cP_{\bsy{8}}   \cP_{\bsy{7}}   \cP_{\bsy{6}}
       \cP_{\bsy{5}}   \cP_{\bsy{4}}   \cP_{\bsy{3}}   \cP_{\bsy{2}}
       \cP_{\bsy{12}}   \cP_{\bsy{11}}
       \cP_{\bsy{10}}   \cP_{\bsy{9}}   \cP_{\bsy{8}}   \cP_{\bsy{7}}   \cP_{\bsy{6}}   \cP_{\bsy{5}}
       \cP_{\bsy{4}}   \cP_{\bsy{3}}
       { \color{brown} \cR_{\hat{5},\bsy{10}} }
       \cP_{\bsy{16}}   \cP_{\bsy{17}}   \cP_{\bsy{18}}   \\
\qquad \quad \       \cP_{\bsy{19}}   \cP_{\bsy{20}}   \cP_{\bsy{21}}
       \cP_{\bsy{22}}   \cP_{\bsy{23}}   \cP_{\bsy{24}}   \cP_{\bsy{15}}   \cP_{\bsy{16}}
       \cP_{\bsy{17}}   \cP_{\bsy{18}}   \cP_{\bsy{19}}   \cP_{\bsy{20}}   \cP_{\bsy{14}}   \cP_{\bsy{15}}   \cP_{\bsy{16}}
       \cP_{\bsy{9}}   \cP_{\bsy{8}}   \cP_{\bsy{7}}   \cP_{\bsy{6}}   \cP_{\bsy{5}}   \cP_{\bsy{4}}   \\
\qquad \quad \     \cP_{\bsy{10}}   \cP_{\bsy{9}}   \cP_{\bsy{8}}   \cP_{\bsy{7}}   \cP_{\bsy{6}}   \cP_{\bsy{5}}
       \cP_{\bsy{11}}   \cP_{\bsy{10}}   \cP_{\bsy{9}}   \cP_{\bsy{8}}   \cP_{\bsy{7}}
       \cP_{\bsy{6}}
       { \color{brown} \cR_{\hat{4},\bsy{10}} }
       \cP_{\bsy{16}}   \cP_{\bsy{17}}   \cP_{\bsy{18}}   \cP_{\bsy{19}}   \cP_{\bsy{20}}   \cP_{\bsy{21}}
       \cP_{\bsy{22}}   \cP_{\bsy{23}}   \\
\qquad \quad \     \cP_{\bsy{15}}   \cP_{\bsy{16}}   \cP_{\bsy{17}}   \cP_{\bsy{18}}
       \cP_{\bsy{9}}   \cP_{\bsy{8}}   \cP_{\bsy{7}}   \cP_{\bsy{10}}   \cP_{\bsy{9}}   \cP_{\bsy{8}}
       \cP_{\bsy{11}}   \cP_{\bsy{10}}   \cP_{\bsy{9}}   \cP_{\bsy{12}}   \cP_{\bsy{11}}   \cP_{\bsy{10}}
       { \color{brown} \cR_{\hat{3},\bsy{12}} }
       \cP_{\bsy{18}}   \cP_{\bsy{19}}   \cP_{\bsy{20}}   \cP_{\bsy{21}}   \\
\qquad \quad \     \cP_{\bsy{22}}   \cP_{\bsy{11}}
       \cP_{\bsy{12}}   \cP_{\bsy{13}}   \cP_{\bsy{14}}   \cP_{\bsy{15}}
       { \color{brown} \cR_{\hat{2},\bsy{16}}   \cR_{\hat{1},\bsy{22}} }   .
\end{gather*}
Also see Fig.~\ref{fig:7simplex_eq_polyhedron}.
In terms of $\hat{\cR} = \cR   \cP_{\bsy{35}}   \cP_{\bsy{26}}   \cP_{\bsy{17}}$, it collapses to
\begin{gather*}
   \hat{\cR}_{\hat{1},\bsy{1},\bsy{2},\bsy{3},\bsy{4},\bsy{5},\bsy{6},\bsy{7} }
   \hat{\cR}_{\hat{2},\bsy{1},\bsy{8},\bsy{9},\bsy{10},\bsy{11}, \bsy{12}, \bsy{13}}
   \hat{\cR}_{\hat{3},\bsy{2},\bsy{8},\bsy{14},\bsy{15},\bsy{16},\bsy{17},\bsy{18}}
   \hat{\cR}_{\hat{4},\bsy{3},\bsy{9},\bsy{14},\bsy{19},\bsy{20},\bsy{21},\bsy{22} } \nonumber \\
   \hat{\cR}_{\hat{5},\bsy{4},\bsy{10},\bsy{15},\bsy{19},\bsy{23},\bsy{24},\bsy{25}  }
  \hat{\cR}_{\hat{6},\bsy{5},\bsy{11},\bsy{16},\bsy{20},\bsy{23},\bsy{26},\bsy{27} }
   \hat{\cR}_{\hat{7},\bsy{6},\bsy{12},\bsy{17},\bsy{21},\bsy{24},\bsy{26},\bsy{28} }
   \hat{\cR}_{\hat{8},\bsy{7},\bsy{13},\bsy{18},\bsy{22},\bsy{25},\bsy{27},\bsy{28} }  \nonumber \\
\qquad {} =  \hat{\cR}_{\hat{8},\bsy{7},\bsy{13},\bsy{18},\bsy{22},\bsy{25},\bsy{27},\bsy{28} }
     \hat{\cR}_{\hat{7},\bsy{6},\bsy{12},\bsy{17},\bsy{21},\bsy{24},\bsy{26},\bsy{28} }
     \hat{\cR}_{\hat{6},\bsy{5},\bsy{11},\bsy{16},\bsy{20},\bsy{23},\bsy{26},\bsy{27} }
     \hat{\cR}_{\hat{5},\bsy{4},\bsy{10},\bsy{15},\bsy{19},\bsy{23},\bsy{24},\bsy{25}  } \nonumber \\
\qquad \quad \ \hat{\cR}_{\hat{4},\bsy{3},\bsy{9},\bsy{14},\bsy{19},\bsy{20},\bsy{21},\bsy{22} }
     \hat{\cR}_{\hat{3},\bsy{2},\bsy{8},\bsy{14},\bsy{15},\bsy{16},\bsy{17},\bsy{18}}
     \hat{\cR}_{\hat{2},\bsy{1},\bsy{8},\bsy{9},\bsy{10},\bsy{11}, \bsy{12}, \bsy{13}}
     \hat{\cR}_{\hat{1},\bsy{1},\bsy{2},\bsy{3},\bsy{4},\bsy{5},\bsy{6},\bsy{7} }   .
\end{gather*}

\begin{rem}
The form in which the above zonohedra appear, i.e., decomposed in two complementary parts, reveals
an interesting feature. If we identify antipodal edges (carrying the same label) of the boundaries,
in each of the two parts, we obtain the same projective polyhedron.\footnote{Corresponding resolutions
of small cubes have to be chosen.}
From the cube, associated with the Yang--Baxter equation, we obtain in this way two copies of the hemicube.
 From the permutahedron, associated with the Zamolodchikov equation, we obtain two copies of
a ``hemi-permutahedron''. Here the identif\/ication of antipodal edges means
identif\/ication of a permutation with the corresponding reversed permutation, e.g., $1234 \cong 4321$,
$1324 \cong 4231$.
\end{rem}

\subsection{Lax systems for simplex equations}
\label{subsec:simplex_Lax}

We promote the maps $\cR_J\colon  \cU_{\vec{P}(J)} \rightarrow \cU_{\cev{P}(J)}$,
$J \in {[N+2] \choose N}$, to ``localized'' maps
\begin{gather*}
    \cL_J \colon \ \cU_J  \longrightarrow \mathrm{Map}(\cU_{\vec{P}(J)}, \cU_{\cev{P}(J)}), \\
    \hphantom{\cL_J \colon}{} \ u_J  \longmapsto \cL_J(u_J)\colon \  \cU_{\vec{P}(J)} \rightarrow \cU_{\cev{P}(J)}   .
\end{gather*}
In $B(N+2,N-1)$, counterparts of the two maximal chains, of which $B(N+1,N-1)$ consists, appear as
chains for all $\hat{k} = [N+2] \setminus \{k\} = \{ k_1,\ldots,k_{N+1} \} \in {[N+2] \choose N+1}$:
\begin{gather*}
    \cC_{\hat{k}, \mathrm{lex} }\colon \  [\alpha_{\hat{k}}] \stackrel{ \widehat{k k_{N+1}} }{\longrightarrow} [\rho_{\hat{k},1}]
            \stackrel{ \widehat{k k_N} }{\longrightarrow} \cdots
            \stackrel{\widehat{k k_2}}{\longrightarrow} [\rho_{\hat{k},N}]
            \stackrel{\widehat{k k_1}}{\longrightarrow} [\omega_{\hat{k}}]   , \nonumber \\
    \cC_{\hat{k}, \mathrm{rev} } \colon \  [\alpha_{\hat{k}}] \stackrel{\widehat{k k_1}}{\longrightarrow} [\sigma_{\hat{k},1}]
            \stackrel{\widehat{k k_2}}{\longrightarrow} \cdots
            \stackrel{\widehat{k k_N}}{\longrightarrow} [\sigma_{\hat{k},N}]
            \stackrel{ \widehat{k k_{N+1}} }{\longrightarrow} [\omega_{\hat{k}}]   ,
\end{gather*}
where $\alpha_{\hat{k}}$, $\omega_{\hat{k}}$, $\rho_{\hat{k},i}$, $\sigma_{\hat{k},i}$ are admissible linear
orders of ${\hat{k} \choose N-1}$, the $k_i$ are assumed to be in natural order, and
$\widehat{kk_i} = \hat{k} \setminus \{k_i\} = [N+2] \setminus \{k,k_i\}$.
For each $k \in [N+2]$, we then impose the localized $N$-simplex equation,
\begin{gather*}
     \cL_{\tilde{\cC}_{\hat{k},\mathrm{lex}}}(\bsy{u}_{\vec{P}(\hat{k})})
   = \cL_{\tilde{\cC}_{\hat{k},\mathrm{rev}}}(\bsy{v}_{\cev{P}(\hat{k})})   ,
      %\label{simplex_eq_Lax_eq}
\end{gather*}
with $\bsy{u}_{\vec{P}(\hat{k})} = (u_{\widehat{k k_{N+1}}}, \ldots, u_{\widehat{k k_1}})$ and
$\bsy{v}_{\cev{P}(\hat{k})} = (v_{\widehat{k k_1}}, \ldots, v_{\widehat{k k_{N+1}}})$. We assume that,
for each $k \in [N+2]$, this equation uniquely determines a map
$\cR_{\hat{k}}\colon \bsy{u}_{\vec{P}(\hat{k})} \mapsto \bsy{v}_{\cev{P}(\hat{k})}$.

In terms of
\begin{gather*}
      \hat{\cL}_{\widehat{k k_i}} := \cL_{\widehat{k k_i}}   \cP_{\widehat{k k_i}}   ,
\end{gather*}
the above equation has the form
\begin{gather}
     \hat{\cL}_{\widehat{k k_1},\bsy{X}_{\bsy{A}_1}}   \hat{\cL}_{\widehat{k k_2},\bsy{X}_{\bsy{A}_2}}\!
       \cdots   \hat{\cL}_{\widehat{k k_{N+1}},\bsy{X}_{\bsy{A}_{N+1}}}\!
  = \Big( \hat{\cL}_{\widehat{k k_{N+1}},\bsy{X}_{\bsy{A}_{N+1}}}\!  \cdots
       \hat{\cL}_{\widehat{k k_2},\bsy{X}_{\bsy{A}_2}}   \hat{\cL}_{\widehat{k k_1},\bsy{X}_{\bsy{A}_1}} \Big)
         \circ \cR_{\hat{k}}   . \!\! \label{simplex_eq_hat_Lax_eq}
\end{gather}
Here $\bsy{X}_{\bsy{A}_i} = (\bsy{x}_{\bsy{a}_{i,1}},\ldots,\bsy{x}_{\bsy{a}_{i,N+1}})$, where
$1 \leq \bsy{x}_{\bsy{a}_{i,j}} \leq c(N+2,N-1)$,
are increasing sequences of positive
integers and $\bsy{A}_i = (\bsy{a}_{i,1}, \ldots, \bsy{a}_{i,N+1})$, where
$1 \leq \bsy{a}_{i,j} \leq c(N+1,N-1)$, are the multi-indices introduced in
Remark~\ref{rem:simplex_eq_structure}.

With $\rho = (J_1,\ldots,J_{c(N+2,N)}) \in A(N+2,N)$, we associate the composition
\begin{gather*}
 \hat{\cL}_\rho = \hat{\cL}_{J_{c(N+2,N)},\bsy{A}_{c(N+2,N)}} \cdots \hat{\cL}_{J_1,\bsy{A}_1} \colon \
 \cU_\eta \rightarrow \cU_\eta
\end{gather*}
of the corresponding maps. Here $\eta \in A(N+2,N-1)$ is the reverse lexicographical order of ${[N+2] \choose N-1}$, and
the multi-indices~$\bsy{A}_i$ specify the positions of the elements of~$P(J_i)$ in~$\eta$.
The next observation will be important in the following.

\begin{lem}
\label{lemma:L-exchanges}
If $J,J' \in {[N+2] \choose N}$ satisfy $E(J) \cap E(J') = \varnothing$,
then $\hat{\cL}_{J,\bsy{A}}   \hat{\cL}_{J',\bsy{A}'} = \hat{\cL}_{J',\bsy{A}'}    \hat{\cL}_{J,\bsy{A}}$
$($acting on some $\cU_\mu$, $\mu \in A(N+2,N-1))$.
\end{lem}

\begin{proof}
It is easily verif\/ied that
$ E(J) \cap E(J') = \varnothing \Longleftrightarrow P(J) \cap P(J') = \varnothing$.
Hence, if $E(J) \cap E(J') = \varnothing$, then $\hat{\cL}_{J,\bsy{A}}$ and $\hat{\cL}_{J',\bsy{A}'}$ must
act on distinct positions in $\cU_\mu$, hence they commute.
\end{proof}

Now we sketch a proof of the claim that the $(N+1)$-simplex equation
\begin{gather*}
      \cR_{\tilde{\cC}_{\mathrm{lex}}} = \cR_{\tilde{\cC}_{\mathrm{rev}}}   ,
\end{gather*}
where $\tilde{\cC}_{\mathrm{lex}}$ and $\tilde{\cC}_{\mathrm{rev}}$ constitute a resolution
of $B(N+2,N)$,
arises as a consistency condition of the above Lax system.
We start with $\hat{\cL}_\alpha$, where $\alpha \in A(N+2,N)$ is the lexicographical order of~${[N+2] \choose N}$, and proceed according to the resolution $\tilde{\cC}_{\mathrm{lex}}$.
The above Lemma guarantees that there is a permutation of $\hat{\cL}$'s, corresponding to the
resolution of $[\alpha]$ that leads to $\cU_{\rho_0} = \cP_{\rho_0,\alpha}  \cU_\alpha$, which arranges
that $\hat{\cL}_{\vec{P}(\widehat{N+2})}$ acts on $\cU_{\rho_0}$ at consecutive positions.
This yields $\hat{\cL}_{\rho_0} \circ \cP_{\rho_0,\alpha}$.
Next we apply the respective Lax equation~(\ref{simplex_eq_hat_Lax_eq}), which results in
$\hat{\cL}_{\rho_0'} \circ (\cR_{\widehat{N+2},\bsy{a}_0} \cP_{\rho_0,\alpha})$.
Proceeding in this way, we f\/inally arrive at $\hat{\cL}_\omega \circ \cR_{\tilde{\cC}_{\mathrm{lex}}}$.
Starting again with $\hat{\cL}_\alpha$, but now following the resolu\-tion~$\tilde{\cC}_{\mathrm{rev}}$,
we f\/inally obtain $\hat{\cL}_\omega \circ \cR_{\tilde{\cC}_{\mathrm{rev}}}$.
Since we assumed that the Lax equations uniquely determine the respective maps, we can conclude
that the $(N+1)$-simplex equation holds.

\begin{exam}
\label{ex:simplex_N=3_Lax_again}
Let $N=3$. Then $\alpha = (\widehat{45},\widehat{35},\widehat{34},\widehat{25},
\widehat{24},\widehat{23},\widehat{15},\widehat{14},\widehat{13},\widehat{12}) \in A(5,3)$ and
$\eta = (45,35,34,25,24,23,15,14,13,12) \in A(5,2)$, from which we can read of\/f the position indices
to obtain
\begin{gather*}
   \hat{\cL}_\alpha = \hat{\cL}_{\widehat{12},\bsy{123}}   \hat{\cL}_{\widehat{13},\bsy{145}}
                      \hat{\cL}_{\widehat{14},\bsy{246}}   \hat{\cL}_{\widehat{15},\bsy{356}}
                      \hat{\cL}_{\widehat{23},\bsy{178}}   \hat{\cL}_{\widehat{24},\bsy{279}}
                      \hat{\cL}_{\widehat{25},\bsy{389}}   \hat{\cL}_{\widehat{34},\bsy{470}}
                      \hat{\cL}_{\widehat{35},\bsy{580}}   \hat{\cL}_{\widehat{45},\bsy{690}}   .
\end{gather*}
The Lax system takes the form
\begin{gather*}
     \hat{\cL}_{ \widehat{kk_1},\bsy{x}_{\bsy{1}},\bsy{x}_{\bsy{2}},\bsy{x}_{\bsy{3}} }
      \hat{\cL}_{ \widehat{kk_2},\bsy{x}_{\bsy{1}},\bsy{x}_{\bsy{4}},\bsy{x}_{\bsy{5}} }
      \hat{\cL}_{ \widehat{kk_3},\bsy{x}_{\bsy{2}},\bsy{x}_{\bsy{4}},\bsy{x}_{\bsy{6}} }
      \hat{\cL}_{ \widehat{kk_4},\bsy{x}_{\bsy{3}},\bsy{x}_{\bsy{5}},\bsy{x}_{\bsy{6}} } \\
\qquad =  \Big( \hat{\cL}_{ \widehat{kk_4},\bsy{x}_{\bsy{3}},\bsy{x}_{\bsy{5}},\bsy{x}_{\bsy{6}} }
            \hat{\cL}_{ \widehat{kk_3},\bsy{x}_{\bsy{2}},\bsy{x}_{\bsy{4}},\bsy{x}_{\bsy{6}} }
            \hat{\cL}_{\widehat{kk_2},\bsy{x}_{\bsy{1}},\bsy{x}_{\bsy{4}},\bsy{x}_{\bsy{5}} }
            \hat{\cL}_{\widehat{kk_1},\bsy{x}_{\bsy{1}},\bsy{x}_{\bsy{2}},\bsy{x}_{\bsy{3}} } \Big)
      \circ \cR_{\hat{k}}   ,
\end{gather*}
where $1 \leq \bsy{x}_1 < \bsy{x}_2 < \cdots < \bsy{x}_6 \leq 10$.
Now we have
\begin{gather*}
 \hat{\cL}_\alpha  =  \hat{\cL}_{\widehat{12},\bsy{123}}   \hat{\cL}_{\widehat{13},\bsy{145}}
                      \hat{\cL}_{\widehat{14},\bsy{246}}   {\color{brown} \hat{\cL}_{\widehat{15},\bsy{356}} }
                      \hat{\cL}_{\widehat{23},\bsy{178}}   \hat{\cL}_{\widehat{24},\bsy{279}}
                      {\color{brown} \hat{\cL}_{\widehat{25},\bsy{389}} }   \hat{\cL}_{\widehat{34},\bsy{470}}
                      {\color{brown} \hat{\cL}_{\widehat{35},\bsy{580}}   \hat{\cL}_{\widehat{45},\bsy{690}} } \\
 \hphantom{\hat{\cL}_\alpha}{}  \stackrel{\cP_{\bsy{4}} \cP_{\bsy{5}} \cP_{\bsy{6}} \cP_{\bsy{3}}}{=}
                      \hat{\cL}_{\widehat{12},\bsy{123}}   \hat{\cL}_{\widehat{13},\bsy{145}}
                      \hat{\cL}_{\widehat{14},\bsy{246}}   \hat{\cL}_{\widehat{23},\bsy{178}}
                      \hat{\cL}_{\widehat{24},\bsy{279}}   \hat{\cL}_{\widehat{34},\bsy{470}}
                      {\color{brown} \hat{\cL}_{\widehat{15},\bsy{356}}   \hat{\cL}_{\widehat{25},\bsy{389}}
                      \hat{\cL}_{\widehat{35},\bsy{580}}   \hat{\cL}_{\widehat{45},\bsy{690}} } \\
\hphantom{\hat{\cL}_\alpha}{}   \stackrel{\cR_{\hat{5},\bsy{1}}}{=}
                      \hat{\cL}_{\widehat{12},\bsy{123}}   \hat{\cL}_{\widehat{13},\bsy{145}}
                      {\color{brown} \hat{\cL}_{\widehat{14},\bsy{246}} }  \hat{\cL}_{\widehat{23},\bsy{178}}
                      {\color{brown} \hat{\cL}_{\widehat{24},\bsy{279}}   \hat{\cL}_{\widehat{34},\bsy{470}}
                      \hat{\cL}_{\widehat{45},\bsy{690}} }   \hat{\cL}_{\widehat{35},\bsy{580}}
                      \hat{\cL}_{\widehat{25},\bsy{389}}   \hat{\cL}_{\widehat{15},\bsy{356}} \\
\hphantom{\hat{\cL}_\alpha}{}   \stackrel{\cP_{\bsy{7}}}{=}
                      \hat{\cL}_{\widehat{12},\bsy{123}}   \hat{\cL}_{\widehat{13},\bsy{145}}
                      \hat{\cL}_{\widehat{23},\bsy{178}}   {\color{brown} \hat{\cL}_{\widehat{14},\bsy{246}}
                      \hat{\cL}_{\widehat{24},\bsy{279}}   \hat{\cL}_{\widehat{34},\bsy{470}}
                      \hat{\cL}_{\widehat{45},\bsy{690}} }   \hat{\cL}_{\widehat{35},\bsy{580}}
                      \hat{\cL}_{\widehat{25},\bsy{389}}   \hat{\cL}_{\widehat{15},\bsy{356}} \\
\hphantom{\hat{\cL}_\alpha}{}   \stackrel{\cR_{\hat{4},\bsy{4}}}{=}
                      \hat{\cL}_{\widehat{12},\bsy{123}}   {\color{brown} \hat{\cL}_{\widehat{13},\bsy{145}}
                      \hat{\cL}_{\widehat{23},\bsy{178}} }   \hat{\cL}_{\widehat{45},\bsy{690}}
                      {\color{brown} \hat{\cL}_{\widehat{34},\bsy{470}} }   \hat{\cL}_{\widehat{24},\bsy{279}}
                      \hat{\cL}_{\widehat{14},\bsy{246}}   {\color{brown} \hat{\cL}_{\widehat{35},\bsy{580}} }
                      \hat{\cL}_{\widehat{25},\bsy{389}}   \hat{\cL}_{\widehat{15},\bsy{356}} \\
\hphantom{\hat{\cL}_\alpha}{}    \stackrel{\cP_{\bsy{8}} \cP_{\bsy{7}} \cP_{\bsy{4}} \cP_{\bsy{3}}}{=}
                      \hat{\cL}_{\widehat{12},\bsy{123}}   \hat{\cL}_{\widehat{45},\bsy{690}}
                      {\color{brown} \hat{\cL}_{\widehat{13},\bsy{145}}   \hat{\cL}_{\widehat{23},\bsy{178}}
                      \hat{\cL}_{\widehat{34},\bsy{470}}   \hat{\cL}_{\widehat{35},\bsy{580}} }
                      \hat{\cL}_{\widehat{24},\bsy{279}}   \hat{\cL}_{\widehat{14},\bsy{246}}
                      \hat{\cL}_{\widehat{25},\bsy{389}}   \hat{\cL}_{\widehat{15},\bsy{356}} \\
\hphantom{\hat{\cL}_\alpha}{}   \stackrel{\cR_{\hat{3},\bsy{5}}}{=}
                      {\color{brown} \hat{\cL}_{\widehat{12},\bsy{123}} }  \hat{\cL}_{\widehat{45},\bsy{690}}
                      \hat{\cL}_{\widehat{35},\bsy{580}}   \hat{\cL}_{\widehat{34},\bsy{470}}
                      {\color{brown} \hat{\cL}_{\widehat{23},\bsy{178}} }   \hat{\cL}_{\widehat{13},\bsy{145}}
                      {\color{brown} \hat{\cL}_{\widehat{24},\bsy{279}} }   \hat{\cL}_{\widehat{14},\bsy{246}}
                      {\color{brown} \hat{\cL}_{\widehat{25},\bsy{389}} }   \hat{\cL}_{\widehat{15},\bsy{356}} \\
\hphantom{\hat{\cL}_\alpha}{}   \stackrel{\cP_{\bsy{7}} \cP_{\bsy{8}} \cP_{\bsy{9}} \cP_{\bsy{3}} \cP_{\bsy{2}} \cP_{\bsy{4}} }{=}
                      \hat{\cL}_{\widehat{45},\bsy{690}}   \hat{\cL}_{\widehat{35},\bsy{580}}
                      \hat{\cL}_{\widehat{34},\bsy{470}}   {\color{brown} \hat{\cL}_{\widehat{12},\bsy{123}}
                      \hat{\cL}_{\widehat{23},\bsy{178}}   \hat{\cL}_{\widehat{24},\bsy{279}}
                      \hat{\cL}_{\widehat{25},\bsy{389}} }   \hat{\cL}_{\widehat{13},\bsy{145}}
                      \hat{\cL}_{\widehat{14},\bsy{246}}   \hat{\cL}_{\widehat{15},\bsy{356}} \\
\hphantom{\hat{\cL}_\alpha}{}   \stackrel{\cR_{\hat{2},\bsy{4}}}{=}
                      \hat{\cL}_{\widehat{45},\bsy{690}}   \hat{\cL}_{\widehat{35},\bsy{580}}
                      \hat{\cL}_{\widehat{34},\bsy{470}}   \hat{\cL}_{\widehat{25},\bsy{389}}
                      \hat{\cL}_{\widehat{24},\bsy{279}}   \hat{\cL}_{\widehat{23},\bsy{178}}
                      {\color{brown} \hat{\cL}_{\widehat{12},\bsy{123}}   \hat{\cL}_{\widehat{13},\bsy{145}}
                      \hat{\cL}_{\widehat{14},\bsy{246}}   \hat{\cL}_{\widehat{15},\bsy{356}} } \\
\hphantom{\hat{\cL}_\alpha}{}   \stackrel{\cR_{\hat{1},\bsy{1}}}{=}
                      \hat{\cL}_{\widehat{45},\bsy{690}}   \hat{\cL}_{\widehat{35},\bsy{580}}
                      \hat{\cL}_{\widehat{34},\bsy{470}}   \hat{\cL}_{\widehat{25},\bsy{389}}
                      \hat{\cL}_{\widehat{24},\bsy{279}}   \hat{\cL}_{\widehat{23},\bsy{178}}
                      \hat{\cL}_{\widehat{15},\bsy{356}}   \hat{\cL}_{\widehat{14},\bsy{246}}
                      \hat{\cL}_{\widehat{13},\bsy{145}}   \hat{\cL}_{\widehat{12},\bsy{123}}   ,
\end{gather*}
where an index $\bsy{0}$ stands for $\bsy{10}$ (ten). Here we indicated
over the equality signs the maps that act on the arguments of the $\hat{\cL}$'s in the respective transformation step.
Returning to the proper notation, the result is $\hat{\cL}_\alpha = \hat{\cL}_\omega \circ \cR_{\tilde{\cC}_{\mathrm{lex}}}$,
which determines the left hand side of the $4$-simplex equation~(\ref{4-simplex_eq}).
We marked in brown those $\hat{\cL}$'s that have
to be brought together in order to allow for an application of a Lax equation. We could have omitted
all the position indices in the above computation, since they are automatically compatible according to
Lemma~\ref{lemma:L-exchanges}. But we kept them for comparison with corresponding computations
in the literature (see, e.g.,~\cite{Hiet+Nijh97, Mail+Nijh89pre,Mail+Nijh89PLB,Mail+Nijh89GTMP,Mail+Nijh90,Michi+Nijh93}),
where only these position indices appear, but not the ``combinatorial indices'' that nicely guided us through the
above computation.
\end{exam}

\subsection{Reductions of simplex equations}
\label{subsec:simplex_red}
The relation between the Bruhat order $B(N+1,N-1)$ and the $N$-simplex equation,
together with the projection of Bruhat orders, def\/ined in Remark~\ref{rem:Bruhat_reduction}, induces
a relation between neighboring simplex equations:
\begin{gather*}
    \begin{array}{@{}ccc} B(N+2,N) & \longleftrightarrow & \mbox{$(N+1)$-simplex equation} \\
                           \downarrow &     &   \downarrow  \\
                       B(N+1,N-1) & \longleftrightarrow & \mbox{$N$-simplex equation}
    \end{array}
\end{gather*}
Since the structure of the $N$-simplex equation can be read of\/f in full detail from $B(N+1,N-2)$, we shall
consider the projection $B(N+2,N-1) \rightarrow B(N+1,N-2)$. Let us choose $k=N+2$
in Remark~\ref{rem:Bruhat_reduction}.
Then all vertices of $B(N+2,N-1)$ connected by edges
labeled by $\widehat{j,\mbox{$N$+2}}$ (in complementary notation), with some $j<N+2$, are
identif\/ied under the projection.

For the example of $B(4,1)$ and $k=4$, the projection is shown in Fig.~\ref{fig:B41_red}.
The $3$-simplex equation
\begin{gather*}
     \hat{\cR}^{(3)}_{\hat{1},\bsy{123}}   \hat{\cR}^{(3)}_{\hat{2},\bsy{145}}   \hat{\cR}^{(3)}_{\hat{3},\bsy{246}}   \hat{\cR}^{(3)}_{\hat{4},\bsy{356}}
   = \hat{\cR}^{(3)}_{\hat{4},\bsy{356}}   \hat{\cR}^{(3)}_{\hat{3},\bsy{246}}   \hat{\cR}^{(3)}_{\hat{2},\bsy{145}}   \hat{\cR}^{(3)}_{\hat{1},\bsy{123}}
\end{gather*}
acts on $\cU_{12} \times \cU_{13} \times \cU_{14} \times \cU_{23} \times \cU_{24} \times \cU_{34}$.
The projection formally reduces this to the $2$-simplex equation
\begin{gather*}
     \hat{\cR}^{(2)}_{\hat{1},\bsy{12}}   \hat{\cR}^{(2)}_{\hat{2},\bsy{13}}   \hat{\cR}^{(2)}_{\hat{3},\bsy{23}}
   = \hat{\cR}^{(2)}_{\hat{3},\bsy{23}}   \hat{\cR}^{(2)}_{\hat{2},\bsy{13}}   \hat{\cR}^{(2)}_{\hat{1},\bsy{12}}
       ,
\end{gather*}
acting on $\cU_{12} \times \cU_{13} \times \cU_{23}$. Of course, in the two equations we are dealing
with dif\/ferent types of maps (indicated by a superscript ${}^{(2)}$, respectively ${}^{(3)}$).
The following relation holds.

\begin{prop}
Let $\hat{\cR}^{(N)}$ satisfy the $N$-simplex equation. Let $\cU_{\widehat{j,N+2}}$,
$j=1,2,\ldots,N+1$, be sets,
$f_j\colon  \cU_{\widehat{j,N+2}} \rightarrow \cU_{\widehat{j,N+2}}$ and
$\hat{\cR}^{(N+1)}_{\widehat{N+2}}\colon
\cU_{\widehat{1,N+2}} \times \cU_{\widehat{2,N+2}} \times \cdots \times \cU_{\widehat{N+1,N+2}} \rightarrow
\cU_{\widehat{1,N+2}} \times \cU_{\widehat{2,N+2}} \times \cdots \times \cU_{\widehat{N+1,N+2}}$
any maps such that
\begin{gather*}
    (f_1 \times f_2 \times \cdots \times f_{N+1}) \, \hat{\cR}^{(N+1)}_{\widehat{N+2}}
      = \hat{\cR}^{(N+1)}_{\widehat{N+2}}   (f_1 \times f_2 \times \cdots \times f_{N+1})
\end{gather*}
holds. Setting
\begin{gather*}
    \hat{\cR}^{(N+1)}_{\hat{j}} := \hat{\cR}^{(N)}_{\hat{j}} \times f_j, \qquad    j=1,2,\ldots,N+1   ,
\end{gather*}
then yields a solution $\hat{\cR}^{(N+1)}_{\hat{j}}$, $j=1,\ldots,N+2$, of the $(N+1)$-simplex equation.
\end{prop}
\begin{proof}
The statement can be easily verif\/ied by a direct computation.
\end{proof}

 In particular, if the maps $f_j$, $j=1,2,\ldots,N+1$, are identity functions, then the condition
in the proposition is trivially satisf\/ied and the new map def\/ined in terms of the $N$-simplex map
solves the $(N+1)$-simplex equation.

\section{Polygon equations}
\label{sec:polygon_eqs}
In this section we address realizations of Tamari orders $T(N,n)$ in terms of sets and maps between
Cartesian\footnote{If the sets are supplied with a linear structure, we may
as well consider tensor products or direct sums.}
products of these sets. After some preparations in Section~\ref{subsec:res_of_T}, polygon equations
will be introduced in Section~\ref{subsec:polygon_eqs}, which contains explicit expressions
up to the $11$-gon equation, and the associated polyhedra.
Section~\ref{subsec:polygon_Lax} discusses the integrability of polygon equations.
Reductions of polygon equations associated with reductions of Tamari orders are
the subject of Section~\ref{subsec:polygon_red}.
In the following, we write~$\bar{\rho}$ instead of~$\rho^{(b)}$, for an admissible linear order~$\rho$.

\subsection[Resolutions of $T(N,N-2)$ and polyhedra]{Resolutions of $\boldsymbol{T(N,N-2)}$ and polyhedra}
\label{subsec:res_of_T}
The Tamari order $T(N,N-2)$ consists of the two maximal chains
\begin{gather*}
    \cC_o \colon \  [\bar{\alpha}] \stackrel{ \hat{N} }{\longrightarrow} [\bar{\rho}_1]
            \stackrel{ \widehat{N-2} }{\longrightarrow} [\bar{\rho}_3] \longrightarrow  \cdots \longrightarrow
            [\bar{\rho}_{N+m-3}] \stackrel{ \widehat{2-m} }{\longrightarrow} [\bar{\omega}]
                         , \nonumber \\
    \cC_e \colon \  [\bar{\alpha}] \stackrel{ \widehat{1+m} }{\longrightarrow} [\bar{\sigma}_{1+m}]
            \stackrel{ \widehat{3+m} }{\longrightarrow} [\bar{\sigma}_{3+m}] \longrightarrow
            \cdots \longrightarrow [\bar{\sigma}_{N-5}]
            \stackrel{ \widehat{N-3} }{\longrightarrow} [\bar{\sigma}_{N-3}]
            \stackrel{ \widehat{N-1} }{\longrightarrow} [\bar{\omega}]   ,
                    %\label{T(N,N-2)-chains}
\end{gather*}
where $m = N \, \mathrm{mod} \, 2$. There are resolutions $\tilde{\cC}_o$ and $\tilde{\cC}_e$ in
$A^{(b)}(N,N-2)$, both starting with $\bar{\alpha}$ and both ending with $\bar{\omega}$,
\begin{gather*}
\begin{tikzpicture}
  \node (A) at (-2,0) {$\bar{\alpha}$};
  \node (B) at (2,0) {$\bar{\omega}$};
  \path[->,font=\scriptsize,>=angle 90]
  (A) edge [bend left] node[above] {$\tilde{\cC}_o$} (B)
      edge [bend right] node[below] {$\tilde{\cC}_e$} (B);
\end{tikzpicture}
\end{gather*}
Using the correspondence between elements of $A^{(b)}(N,N-2)$ and maximal chains of $T(N,N-3)$ (see
Remark~\ref{rem:A(N,n+1)-B(N,n)_for_b,r}),
each of the two resolutions corresponds to a sequence of maximal chains of~$T(N,N-3)$.
For $\bar{\rho} \in A^{(b)}(N,N-2)$, let $\cC_{\bar{\rho}}$ be the corresponding maximal chain of $T(N,N-3)$.
The resolution $\tilde{\cC}_o$, respectively $\tilde{\cC}_e$, is then a rule for deforming
$\cC_{\bar{\alpha}}$ stepwise into $\cC_{\bar{\omega}}$.

The resolution of $T(N,N-2)$, given by $\tilde{\cC}_o$ and $\tilde{\cC}_e$, can be regarded as
a rule to construct a polyhedron. If $N$ is odd, i.e., $N=2n+1$, then $\bar{\alpha}$ and $\bar{\omega}$
have both $n(n+1)/2$ elements. The construction rules are then exactly the same as in the case treated in
Section~\ref{subsec:res_of_B(N+1,N-1)}. Each appearance of an inversion corresponds to a $2n$-gon.

If $N$ is even, i.e., $N=2n$, then $\bar{\alpha}$ has $n(n+1)/2$ elements and $\bar{\omega}$
has $n(n-1)/2$ elements. Starting from the top vertex, the chain corresponding to $\bar{\alpha}$
($\bar{\omega}$) is drawn counterclockwise (clockwise). The two chains $\bar{\alpha}$ and $\bar{\omega}$
then join to form an $n^2$-gon. Again, the two sides of the $N$-gon equation correspond to sequences
of maximal chains that deform $\bar{\alpha}$ into $\bar{\omega}$.
But in this case we do not obtain a zonohedron, since an inversion is represented by a $(2n-1)$-gon,
hence an odd polygon.

\subsection{Polygon equations and associated polyhedra}
\label{subsec:polygon_eqs}
Let $N \in \bbN$, $N>1$, and $0 \leq n \leq N-1$.
With $\bar{\rho} \in A^{(b)}(N,n)$ we associate the corresponding Cartesian product $\cU_{\bar{\rho}}$
of sets $\cU_J$, $J \in \bar{\rho}$.
For each $K \in {[N] \choose n+1}$, let there be a map
\begin{gather*}
     \cT_K \colon \ \cU_{\vec{P_o}(K)} \rightarrow \cU_{\cev{P_e}(K)}   ,
\end{gather*}
where $\vec{P_o}(K)$ and $\cev{P_e}(K)$ have been def\/ined in Section~\ref{subsec:Bruhat_decomp},
and $\cU_{\vec{P_o}(K)}$, $\cU_{\cev{P_e}(K)}$ are the corresponding Cartesian products, i.e.,
\begin{gather*}
    \cU_{\vec{P_o}(K)}
   :=  \cU_{ K \setminus \{k_{n+1}\} } \times \cU_{ K \setminus \{k_{n-1}\} } \times \cdots \times
    \cU_{ K \setminus \{k_{1 + (n \, \mathrm{mod}\, 2)}\} }    , \\
    \cU_{\cev{P_e}(K)}
   := \cU_{ K \setminus \{k_{2 - (n \, \mathrm{mod}\, 2)}\} } \times \cdots \times
    \cU_{ K \setminus \{k_{n-2}\} } \times \cU_{ K \setminus \{k_n\} }   .
\end{gather*}

\begin{rem}
In case of the \emph{dual} $B^{(r)}(N,n)$ of the Tamari order $T(N,n)$, we are dealing instead with maps
$\cS_K \colon  \cU_{\vec{P_e}(K)} \longrightarrow \cU_{\cev{P_o}(K)}$.
\end{rem}

Let $[\bar{\rho}] \stackrel{K}{\longrightarrow} [\bar{\rho}']$, $K \in {[N] \choose n+1}$,
be an inversion in $T(N,n)$. Hence $\vec{P_o}(K) \subset \bar{\rho}$ and
$\cev{P_e}(K) \subset \bar{\rho}'$ (where $\subset$ means subsequence).
If $\vec{P_o}(K)$ appears in $\bar{\rho}$ at consecutive positions,
starting at position $\bsy{a}$, we extend $\cT_K$ to a map
$\cT_{K,\bsy{a}}\colon  \cU_{\bar{\rho}} \rightarrow \cU_{\bar{\rho}'}$, which acts
non-trivially only on the sets labeled by the elements of $P_o(K)$.

For a maximal chain $\cC\colon  [\bar{\alpha}] \stackrel{K_1}{\longrightarrow} [\bar{\rho}_1]
\stackrel{K_2}{\longrightarrow} [\bar{\rho}_2] \longrightarrow \cdots \stackrel{K_r}{\longrightarrow}
[\bar{\omega}]$ of $T(N,n)$,
let $\tilde{\cC}$ be a resolution of $\cC$ in $A^{(b)}(N,n)$.
We write $\cT_{\tilde{\cC}}\colon \cU_{\bar{\alpha}} \rightarrow  \cU_{\bar{\omega}}$
for the corresponding composition of maps~$\cT_{K_i,\bsy{a}_i}$ and~$\cP_{\bsy{a}}$.\footnote{$\cT_{\tilde{\cC}}$ has the form
$
    \cP_{\bar{\omega},\bar{\rho}'_r}   \cT_{K_r,\bsy{a}_r}
                       \cP_{\bar{\rho}_{r-1},\bar{\rho}_{r-1}'} \cdots \cP_{\bar{\rho}_2,\bar{\rho}'_2}
                       \cT_{K_2,\bsy{a}_2}   \cP_{\bar{\rho}_1,\bar{\rho}'_1}
                       \cT_{K_1,\bsy{a}_1}   \cP_{\bar{\rho}_0,\bar{\alpha}}   .
$
Here, for $i=1,\dots, r$, $\bar{\rho}_{i-1} \in [\bar{\rho}_{i-1}]$
(where $\bar{\rho}_0 \in [\bar{\alpha}]$) such
that $\vec{P_o}(K_i)$ appears in it at consecutive positions, starting at $\bsy{a}_i$.
$\bar{\rho}'_i \in [\bar{\rho}_i]$ is the result of applying $\cT_{K_i,\bsy{a}_i}$
to $\bar{\rho}_i$
(so that $\bar{\rho}'_i$ contains $\cev{P_e}(K_i)$ at consecutive positions).
}
Turning to $T(N,N-2)$ and choosing $\alpha$ as the lexicographically ordered set ${[N] \choose N-2}$
and $\omega$ as $\alpha$ in reverse order, the $N$-\emph{gon equation} is def\/ined by
\begin{gather*}
      \cT_{\tilde{\cC}_o} = \cT_{\tilde{\cC}_e}   ,
\end{gather*}
which is independent of the choice of resolutions.

For odd $N$, i.e., $N=2n+1$, $\bar{\alpha}$ is the lexicographically ordered sequence of elements
$\widehat{(2j)(2k+1)}$, with $j=1,\ldots, n$ and $k=j,\ldots,n$. $\bar{\omega}$ is
the reverse lexicographically ordered sequence of elements $\widehat{(2j-1)(2k)}$, where
$j=1,\ldots, n$ and $k=j,\ldots,n$. Here
$\cT_{\hat{k}}\colon  \cU_{\vec{P_o}(\hat{k})} \rightarrow \cU_{\cev{P_e}(\hat{k})}$ acts between Cartesian products
having $n$ factors. In terms of\footnote{The permutation map $\cP_{\hat{k}}$ achieves a reversion.}
\begin{gather*}
     \hat{\cT}_{\hat{k}} = \cT_{\hat{k}} \cP_{\hat{k}} \colon \
     \cU_{\cev{P_o}(\hat{k})} \rightarrow \cU_{\cev{P_e}(\hat{k})}   ,
\end{gather*}
the two sides of the $(2n+1)$-gon equation
become maps $\cU_{\mathrm{rev}(\bar{\alpha})} \rightarrow \cU_{\bar{\omega}}$, where
$\mathrm{rev}(\bar{\alpha})$ is $\bar{\alpha}$ reverse lexicographically ordered.
$\bar{\alpha}$ and $\bar{\omega}$ have both $c(n+1,2) = \frac{1}{2} n (n+1)$ elements.
The ``hatted polygon equation''
can be obtained either by substituting $\cT_{\hat{k}} = \hat{\cT}_{\hat{k}} \cP_{\hat{k}}$
in the original polygon equation, or by starting with $\cU_{\mathrm{rev}(\bar{\alpha})}$
and stepwise mapping it to $\cU_{\bar{\omega}}$, following $\cC_o$ and~$\cC_e$. It has the form
\begin{gather*}
 \hat{\cT}_{\hat{1},\bsy{B}_1} \hat{\cT}_{\hat{3},\bsy{B}_3} \cdots \hat{\cT}_{\widehat{2n+1},\bsy{B}_{2n+1}}
 = \hat{\cT}_{\widehat{2n},\bsy{B}_{2n}}  \cdots \hat{\cT}_{\hat{4},\bsy{B}_4}
   \hat{\cT}_{\hat{2},\bsy{B}_2}   ,
\end{gather*}
where $\bsy{B}_k = (\bsy{b}_{k,1},\ldots,\bsy{b}_{k,n})$, with $1 \leq \bsy{b}_{k,i} \leq c(n+1,2)$,
is the multi-index (increasing sequence of positive integers) specifying the positions, in the
respectice active linear order, which take part in the action of the map~$\hat{\cT}_{\hat{k}}$.
Examples will be presented below.

For even $N$, i.e., $N=2n$, $\bar{\alpha}$ is the lexicographically ordered sequence
$\widehat{(2j-1)(2k)}$, where $j=1,\ldots, n$ and $k=j,\ldots,n$, and $\bar{\omega}$ is
the reverse lexicographically ordered sequence $\widehat{(2j)(2k+1)}$,
$j=1,\ldots, n-1$ and $k=j,\ldots,n-1$. Thus, $\bar{\alpha}$ has $c(n+1,2)$ and
$\bar{\omega}$ has $c(n,2)$ elements. Now~$\cT_{\hat{k}}$ maps $\cU_{\vec{P_o}(\hat{k})}$,
which has $n$ factors, to $\cU_{\cev{P_e}(\hat{k})}$, which has $n-1$ factors. Also in this
case the polygon equation can be expressed in compact form without the need of permutation maps.
But in order to achieve this, we have to modify the range to
$\cU_{\widehat{0k}} \times \cU_{\cev{P_e}(\hat{k})}$,
where the sets $\cU_{\widehat{0k}}$ play the role
of placeholders, they are irrelevant for the process of evaluation of the polygon
equation. We def\/ine
\begin{gather*}
     \hat{\cT}_{\hat{k}} := (u_{\widehat{0k}},\cT_{\hat{k}} \cP_{\hat{k}}) \colon \
     \cU_{\cev{P_o}(\hat{k})} \rightarrow \cU_{\widehat{0k}} \times \cU_{\cev{P_e}(\hat{k})}   ,
\end{gather*}
choosing f\/ixed elements $u_{\widehat{0k}} \in \cU_{\widehat{0k}}$.
The $2n$-gon equation then takes the form
\begin{gather}
   \hat{\cT}_{\hat{2},\bsy{B}_2} \hat{\cT}_{\hat{4},\bsy{B}_4} \cdots
     \hat{\cT}_{\widehat{2n},\bsy{B}_{2n}}
 = \hat{\cT}_{\widehat{2n-1},\bsy{B}_{2n-1}} \cdots \hat{\cT}_{\hat{3},\bsy{B}_3} \hat{\cT}_{\hat{1},\bsy{B}_1}
     ,   \label{even_polygon_eq_hatted_version}
\end{gather}
where $\bsy{B}_k = (\bsy{b}_{k,1}, \ldots, \bsy{b}_{k,n})$, $1 \leq \bsy{b}_{k,i} \leq c(n+1,2)$.
This requires setting
$\cU_{\widehat{0 \, (2l)}} = \cU_{\widehat{0 \, (2l-1)}}$ for $l=1,\ldots,n$. Then both sides of
(\ref{even_polygon_eq_hatted_version}) are maps
$\cU_{\mathrm{rev}(\bar{\alpha})} \rightarrow \cU_{\widehat{01}} \times \cdots \times \cU_{\widehat{0 \, (2l-1)}}
\times \cU_{\bar{\omega}} = \cU_{\widehat{02}} \times \cdots \times \cU_{\widehat{0 \, (2l)}}
\times \cU_{\bar{\omega}}$.

\paragraph{Digon equation.}
The two maximal chains of $T(2,0)$ lead to the \emph{digon equation} $\cT_1 = \cT_2$ for the two maps
$\cT_i\colon \cU_\varnothing \rightarrow \cU_\varnothing$.

\paragraph{Trigon equation.}
The two maximal chains of $T(3,1)$ are $1 \stackrel{12}{\rightarrow} 2 \stackrel{23}{\rightarrow} 3$
and $1 \stackrel{13}{\rightarrow} 3$. The maps $\cT_{ij}\colon \cU_i \rightarrow \cU_j$, $i<j$,
have to satisfy the \emph{trigon equation} $\cT_{23}   \cT_{12} = \cT_{13}$.

\begin{figure}[t]\centering
\includegraphics[width=.25\linewidth]{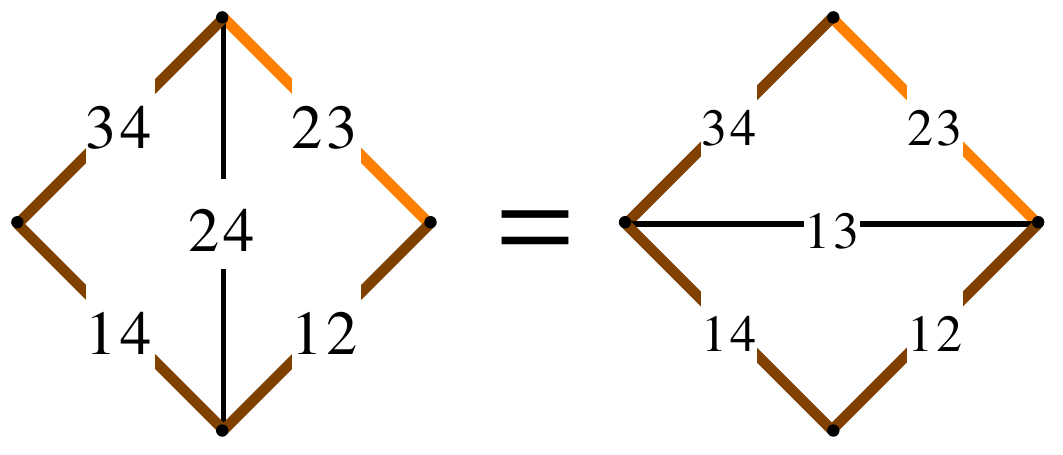}
\caption{Graphical representation of the tetragon equation (using complementary notation for the edge labels, with hats omitted).
\label{fig:tetragon_eq_on_tetrahedron} }
\end{figure}

\paragraph{Tetragon equation.}
The two maximal chains of $T(4,2)$ are
\begin{gather*}
 \begin{array}{@{}c@{\;\;}c@{\;\;}c@{\;\;}c@{\;\;}c@{\;\;}c@{}}
 \begin{minipage}{.3cm} 12 \\ 23 \\ 34 \end{minipage} & \stackrel{123}{\longrightarrow} &
   \begin{minipage}{.3cm} 13 \\ 34 \end{minipage} & \stackrel{134}{\longrightarrow} &
      \begin{minipage}{.3cm} 14 \end{minipage}
 \end{array}
 \qquad
  \begin{array}{@{}c@{\;\;}c@{\;\;}c@{\;\;}c@{\;\;}c@{\;\;}c@{}}
 \begin{minipage}{.3cm} 12 \\ 23 \\ 34 \end{minipage} & \stackrel{234}{\longrightarrow} &
   \begin{minipage}{.3cm} 12 \\ 14 \end{minipage} & \stackrel{124}{\longrightarrow} &
      \begin{minipage}{.3cm} 14 \end{minipage}
 \end{array}
\end{gather*}
The \emph{tetragon equation} is thus
\begin{gather*}
     \cT_{134}   \cT_{123,\bsy{1}} = \cT_{124}   \cT_{234,\bsy{2}}  ,  %\label{4-gon_eq}
\end{gather*}
for maps $\cT_{ijk}\colon \cU_{ij} \times \cU_{jk} \rightarrow \cU_{ik}$, $i<j<k$.
Using complementary notation, the tetragon equation reads
\begin{gather*}
    \cT_{\hat{2}}   \cT_{\hat{4},\bsy{1}} = \cT_{\hat{3}}   \cT_{\hat{1},\bsy{2}}    .
\end{gather*}
Also see Fig.~\ref{fig:tetragon_eq_on_tetrahedron}.
The hatted version of the tetragon equation is
\begin{gather*}
    \hat{\cT}_{\hat{2},\bsy{13}}   \hat{\cT}_{\hat{4},\bsy{23}}
  = \hat{\cT}_{\hat{3},\bsy{23}}   \hat{\cT}_{\hat{1},\bsy{12}}   ,
\end{gather*}
which can be read of\/f from
\begin{gather*}
 \begin{array}{@{}c@{\;\;}c@{\;\;}c@{\;\;}c@{\;\;}c@{\;\;}c@{}}
 \begin{minipage}{.3cm} $\widehat{12}$ \\ $\widehat{14}$ \\ $\widehat{34}$ \end{minipage} &
          \xrightarrow[\bsy{23}]{\hat{4}} &
   \begin{minipage}{.3cm} $\widehat{12}$ \\ $\widehat{04}$ \\ $\widehat{24}$ \end{minipage} &
          \xrightarrow[\bsy{13}]{\hat{2}} &
      \begin{minipage}{.3cm} $\widehat{02}$ \\ $\widehat{04}$ \\ $\widehat{23}$ \end{minipage}
 \end{array}
 \qquad
  \begin{array}{@{}c@{\;\;}c@{\;\;}c@{\;\;}c@{\;\;}c@{\;\;}c@{}}
 \begin{minipage}{.3cm} $\widehat{12}$ \\ $\widehat{14}$ \\ $\widehat{34}$ \end{minipage} &
        \xrightarrow[\bsy{12}]{\hat{1}} &
   \begin{minipage}{.3cm} $\widehat{01}$ \\ $\widehat{13}$ \\ $\widehat{34}$ \end{minipage} &
        \xrightarrow[\bsy{23}]{\hat{3}} &
      \begin{minipage}{.3cm} $\widehat{01}$ \\ $\widehat{03}$ \\ $\widehat{23}$ \end{minipage}
 \end{array}
\end{gather*}
As here, also in the following we will sometimes superf\/luously display read-of\/f position indices under the arrows.

\begin{figure}[t]
\centering
\includegraphics[width=.3\linewidth]{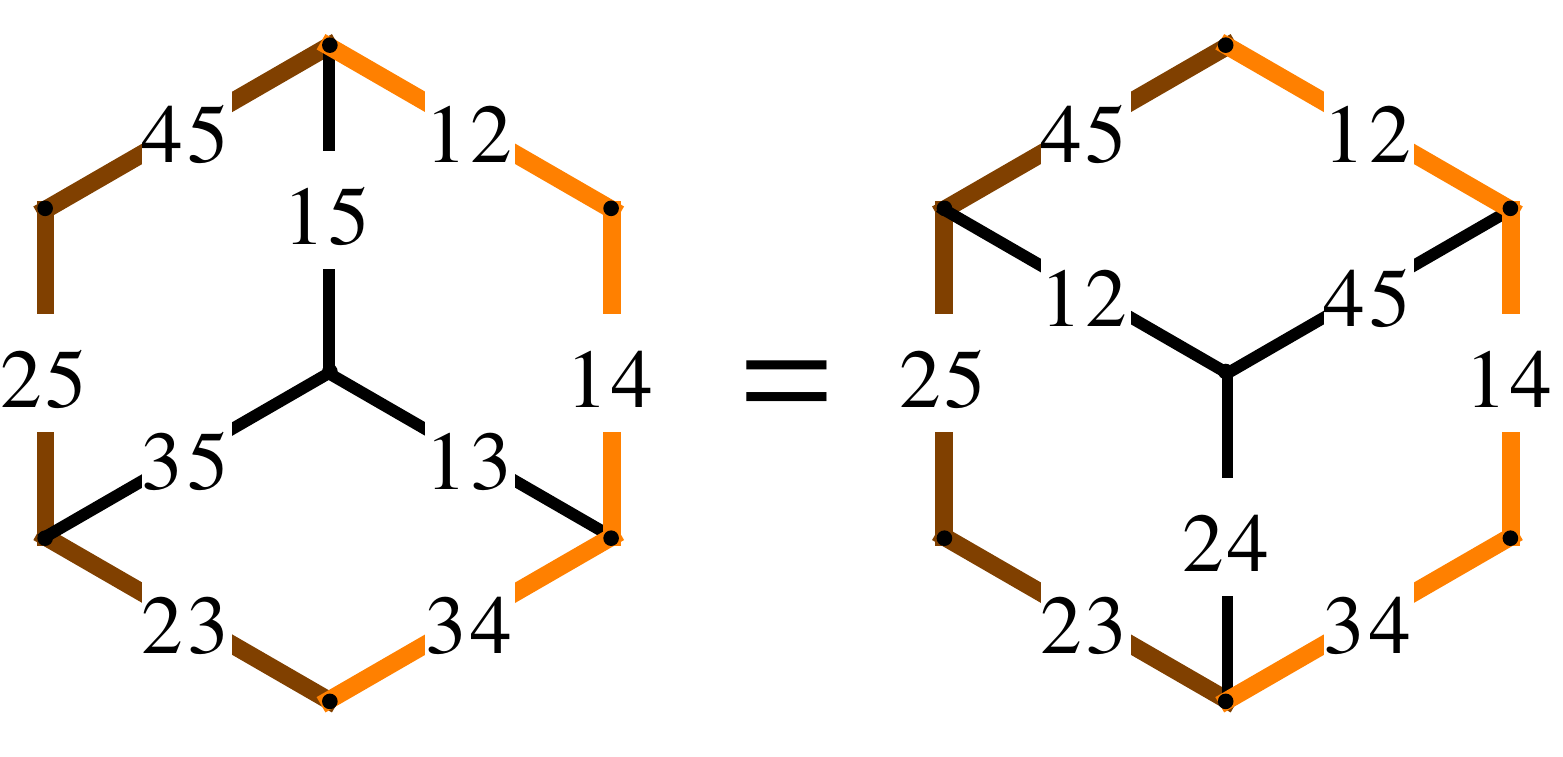}
\caption{Two complementary sides of the cube formed by~$T(5,2)$. We use complementary notation for
the labels. The equality presents the pentagon equation in shorthand form. See
Fig.~\ref{fig:pentagon_eq_on_cube} for the expanded form (using original labeling).
\label{fig:pentagon_equation} }
\end{figure}

\paragraph{Pentagon equation.}
The two maximal chains of $T(5,3)$ can be resolved to
\begin{gather*}
 \begin{array}{@{}c@{\;\;}c@{\;\;}c@{\;\;}c@{\;\;}c@{\;\;}c@{\;\;}c@{}}
  \begin{minipage}{.5cm} 123 \\ 134 \\ 145 \end{minipage} & \stackrel{1234}{\longrightarrow} &
   \begin{minipage}{.5cm} 234 \\ 124 \\ 145 \end{minipage} & \stackrel{1245}{\longrightarrow} &
    \begin{minipage}{.5cm} 234 \\ 245 \\ 125 \end{minipage} & \stackrel{2345}{\longrightarrow} &
     \begin{minipage}{.5cm} 345 \\ 235 \\ 125 \end{minipage}
 \end{array}
 \qquad
  \begin{array}{@{}c@{\;\;}c@{\;\;}c@{\;\;}c@{\;\;}c@{\;\;}c@{\;\;}c@{\;\;}c@{}}
  \begin{minipage}{.5cm} 123 \\ 134 \\ 145 \end{minipage} & \stackrel{1345}{\longrightarrow} &
   \begin{minipage}{.5cm} 123 \\ 345 \\ 135 \end{minipage} & \stackrel{ \bsy{\sim} }{\longrightarrow} &
    \begin{minipage}{.5cm} 345 \\ 123 \\ 135 \end{minipage} & \stackrel{1235}{\longrightarrow} &
     \begin{minipage}{.5cm} 345 \\ 235 \\ 125 \end{minipage}
 \end{array}
\end{gather*}
They describe deformations of maximal chains of $T(5,2)$, see Fig.~\ref{fig:pentagon_eq_on_cube}.
Here we consider maps $\cT_{ijkl}\colon \cU_{ijk} \times \cU_{ikl} \rightarrow \cU_{jkl} \times \cU_{ijl}$,
$i<j<k<l$. Using complementary notation, the \emph{pentagon equation} is thus
\begin{gather}
    \cT_{\hat{1},\bsy{1}}   \cT_{\hat{3},\bsy{2}}   \cT_{\hat{5},\bsy{1}}
  = \cT_{\hat{4},\bsy{2}}   \cP_{\bsy{1}}  \cT_{\hat{2},\bsy{2}}   ,
                \label{5-gon_eq}
\end{gather}
also see Fig.~\ref{fig:pentagon_equation}.
In terms of $\hat{\cT} := \cT   \cP$, it takes the form
\begin{gather}
     \hat{\cT}_{\hat{1},\bsy{1}\bsy{2}}   \hat{\cT}_{\hat{3},\bsy{1}\bsy{3}}   \hat{\cT}_{\hat{5},\bsy{2}\bsy{3}}
  =  \hat{\cT}_{\hat{4},\bsy{2}\bsy{3}}   \hat{\cT}_{\hat{2},\bsy{1}\bsy{2}}   .
             \label{5-gon_eq_mod_cn}
\end{gather}

\begin{figure}[t]
\centering
\includegraphics[width=.9\linewidth]{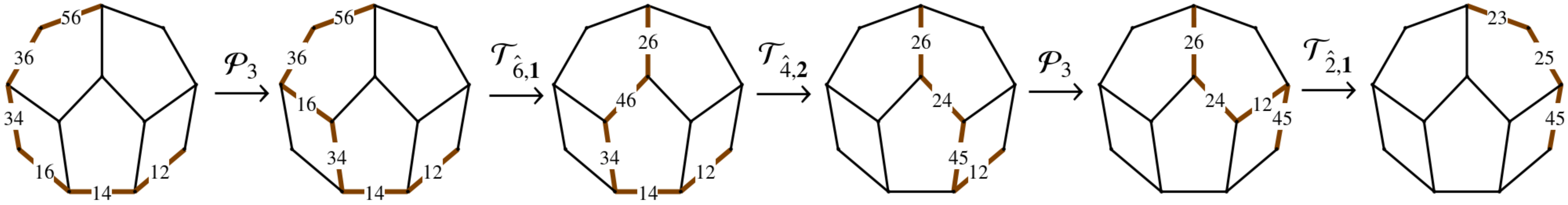}
\includegraphics[width=.75\linewidth]{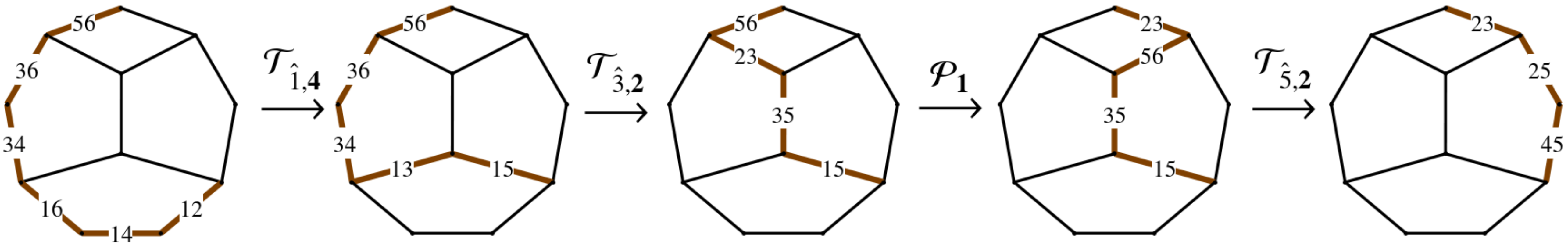}
\caption{The left-hand side of the hexagon equation (\ref{6-gon_eq}) corresponds to a sequence
of maximal chains (f\/irst row) on one side of the associahedron in three dimensions,
formed by the Tamari lattice $T(6,3)$. The right-hand side corresponds to a sequence
of maximal chains (second row) on the complementary side.
\label{fig:hexagon_eq_on_associahedron} }
\end{figure}

\begin{figure}[t]\centering
\includegraphics[width=.35\linewidth]{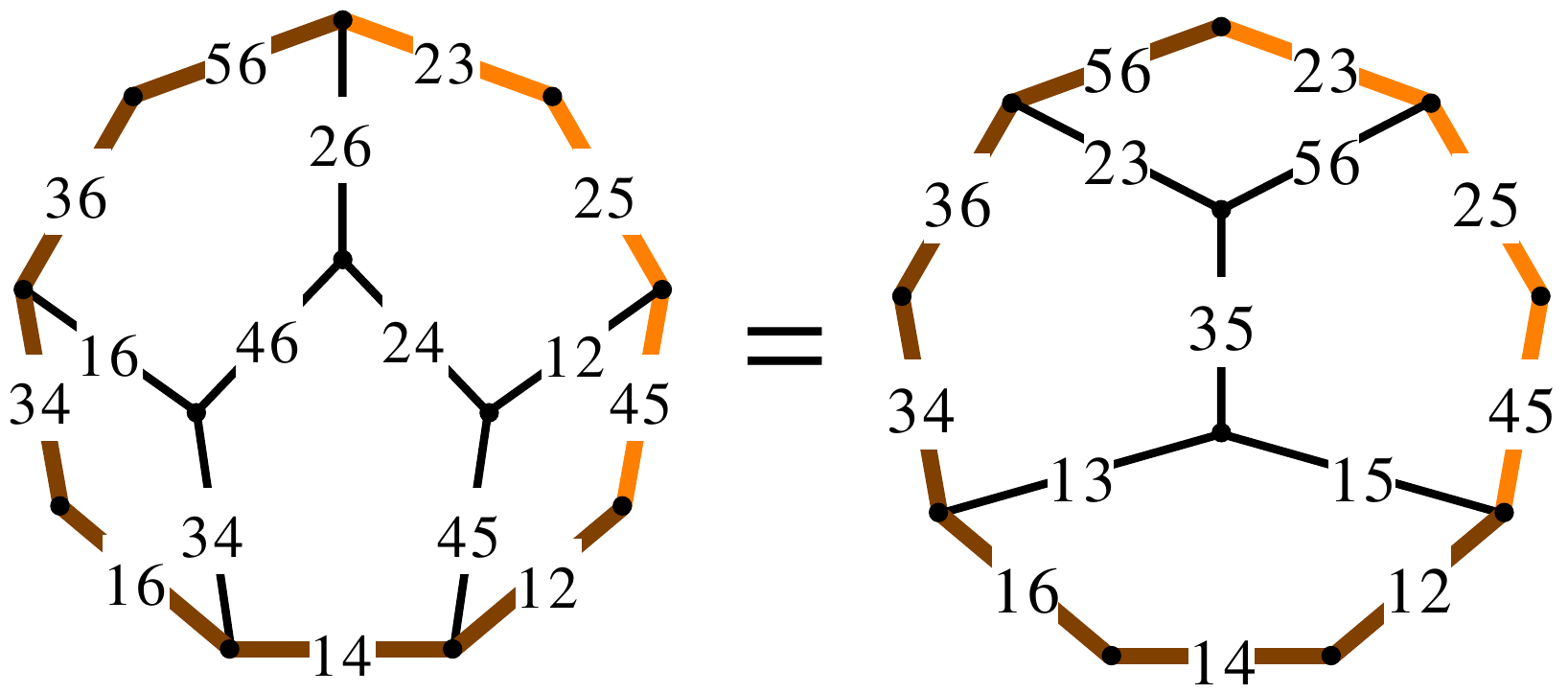}
\caption{Two complementary sides of the associahedron. The equality expresses the hexagon equation.
\label{fig:hexagon_eq_on_associahedron_2} }
\end{figure}

\paragraph{Hexagon equation.}
We will treat this case in some more detail. The two maximal chains of $T(6,4)$ are
\begin{gather*}
    \cC_o\colon \  [\bar{\alpha}] \stackrel{12345}{\longrightarrow} [\bar{\rho}_1]
            \stackrel{12356}{\longrightarrow} [\bar{\rho}_3] \stackrel{13456}{\longrightarrow} [\bar{\omega}]
                         , \qquad
    \cC_e \colon \  [\bar{\alpha}] \stackrel{23456}{\longrightarrow}  [\bar{\sigma}_1]
            \stackrel{12456}{\longrightarrow}  [\bar{\sigma}_3]
            \stackrel{12346}{\longrightarrow} [\bar{\omega}]  ,
\end{gather*}
with $\bar{\alpha} = (1234,1245,1256,2345,2356,3456)$. The equivalence class $[\bar{\alpha}]$
contains another linear order, which is $(1234,1245,2345,1256,2356,3456)$. We obtain
\begin{gather*}
 \tilde{\cC}_o\colon \quad \begin{array}{@{}c@{\;\;}c@{\;\;}c@{\;\;}c@{\;\;}c@{\;\;}c@{\;\;}c@{\;\;}c@{\;\;}c@{\;\;}c@{\;\;}c@{}}
  \begin{minipage}{.7cm} 1234 \\ 1245 \\ 1256 \\ 2345 \\ 2356 \\ 3456 \end{minipage}
          & \stackrel{ \bsy{\sim} }{\longrightarrow} &
  \begin{minipage}{.7cm} 1234 \\ 1245 \\ 2345 \\ 1256 \\ 2356 \\ 3456 \end{minipage}
      & \stackrel{12345}{\longrightarrow} &
  \begin{minipage}{.7cm} 1345 \\ 1235 \\ 1256 \\ 2356 \\ 3456 \end{minipage}
      & \stackrel{12356}{\longrightarrow} &
  \begin{minipage}{.7cm} 1345 \\ 1356 \\ 1236 \\ 3456 \end{minipage}
      & \stackrel{ \bsy{\sim} }{\longrightarrow} &
  \begin{minipage}{.7cm} 1345 \\ 1356 \\ 3456 \\ 1236 \end{minipage}
      & \stackrel{13456}{\longrightarrow} &
  \begin{minipage}{.7cm} 1456 \\ 1346 \\ 1236 \end{minipage}
 \end{array}          % \label{T(6,4)_chain_1_resolved}
 \\[1ex]
 \tilde{\cC}_e\colon \quad \begin{array}{@{}c@{\;\;}c@{\;\;}c@{\;\;}c@{\;\;}c@{\;\;}c@{\;\;}c@{\;\;}c@{\;\;}c@{}}
  \begin{minipage}{.7cm} 1234 \\ 1245 \\ 1256 \\ 2345 \\ 2356 \\ 3456 \end{minipage}
          & \stackrel{23456}{\longrightarrow} &
  \begin{minipage}{.7cm} 1234 \\ 1245 \\ 1256 \\ 2456 \\ 2346 \end{minipage}
          & \stackrel{12456}{\longrightarrow} &
  \begin{minipage}{.7cm} 1234 \\ 1456 \\ 1246 \\ 2346 \end{minipage}
          & \stackrel{ \bsy{\sim} }{\longrightarrow} &
  \begin{minipage}{.7cm} 1456 \\ 1234 \\ 1246 \\ 2346 \end{minipage}
          & \stackrel{12346}{\longrightarrow} &
  \begin{minipage}{.7cm} 1456 \\ 1346 \\ 1236 \end{minipage}
 \end{array}
      %\label{T(6,4)_chain_2_resolved}
\end{gather*}
In this case, we consider maps $\cT_{ijklm}\colon \cU_{ijkl} \times \cU_{ijlm} \times \cU_{jklm}
\rightarrow \cU_{iklm} \times \cU_{ijkm}$, $i<j<k<l<m$. Using complementary notation, the
chains $\tilde{\cC}_o$ and $\tilde{\cC}_e$ read\footnote{Again, the boldface digits are positions, counted from
top to bottom in a column, on which the corresponding action takes place. Here they are always consecutive and can thus be
abbreviated to the f\/irst, as done in~(\ref{6-gon_eq}). }
\begin{gather*}
 \begin{array}{@{}c@{\,}c@{\,}c@{\,}c@{\,}c@{\,}c@{\,}c@{\,}c@{\,}c@{\,}c@{\,}c@{}}
 \begin{minipage}{.4cm} $\widehat{56}$ \\[.5ex] $\widehat{36}$ \\[.5ex] $\widehat{34}$ \\[.5ex] $\widehat{16}$ \\[.5ex]
          $\widehat{14}$ \\[.5ex] $\widehat{12}$ \end{minipage}
          \xrightarrow[\bsy{34}]{ \bsy{\sim} } &
  \begin{minipage}{.4cm} $\widehat{56}$ \\[.5ex] $\widehat{36}$ \\[.5ex] $\widehat{16}$ \\[.5ex] $\widehat{34}$ \\[.5ex]
           $\widehat{14}$ \\[.5ex] $\widehat{12}$ \end{minipage}
      \xrightarrow[\bsy{123}]{ \hat{6} } &
  \begin{minipage}{.4cm} $\widehat{26}$ \\[.5ex] $\widehat{46}$ \\[.5ex] $\widehat{34}$ \\[.5ex] $\widehat{14}$ \\[.5ex] $\widehat{12}$ \end{minipage}
      \xrightarrow[\bsy{234}]{ \hat{4} } &
  \begin{minipage}{.4cm} $\widehat{26}$ \\[.5ex] $\widehat{24}$ \\[.5ex] $\widehat{45}$ \\[.5ex] $\widehat{12}$ \end{minipage}
       \xrightarrow[\bsy{34}]{ \bsy{\sim} } &
  \begin{minipage}{.4cm} $\widehat{26}$ \\[.5ex] $\widehat{24}$ \\[.5ex] $\widehat{12}$ \\[.5ex] $\widehat{45}$ \end{minipage}
       \xrightarrow[\bsy{123}]{ \hat{2} } &
  \begin{minipage}{.4cm} $\widehat{23}$ \\[.5ex] $\widehat{25}$ \\[.5ex] $\widehat{45}$ \end{minipage}
 \end{array}
   \qquad
 \begin{array}{@{}c@{\;}c@{\;}c@{\;}c@{\;}c@{\;}c@{\;}c@{\;}c@{\;}c@{}}
  \begin{minipage}{.4cm} $\widehat{56}$ \\[.5ex] $\widehat{36}$ \\[.5ex] $\widehat{34}$ \\[.5ex] $\widehat{16}$ \\[.5ex]
        $\widehat{14}$ \\[.5ex] $\widehat{12}$ \end{minipage}
         \xrightarrow[\bsy{456}]{ \hat{1} } &
  \begin{minipage}{.4cm} $\widehat{56}$ \\[.5ex] $\widehat{36}$ \\[.5ex] $\widehat{34}$ \\[.5ex] $\widehat{13}$ \\[.5ex] $\widehat{15}$ \end{minipage}
          \xrightarrow[\bsy{234}]{ \hat{3} } &
  \begin{minipage}{.4cm} $\widehat{56}$ \\[.5ex] $\widehat{23}$ \\[.5ex] $\widehat{35}$ \\[.5ex] $\widehat{15}$ \end{minipage}
          \xrightarrow[\bsy{12}]{ \bsy{\sim} } &
  \begin{minipage}{.4cm} $\widehat{23}$ \\[.5ex] $\widehat{56}$ \\[.5ex] $\widehat{35}$ \\[.5ex] $\widehat{15}$ \end{minipage}
          \xrightarrow[\bsy{234}]{ \hat{5} } &
  \begin{minipage}{.4cm} $\widehat{23}$ \\[.5ex] $\widehat{25}$ \\[.5ex] $\widehat{45}$ \end{minipage}
 \end{array}
\end{gather*}
This corresponds to the two sequences of graphs in Fig.~\ref{fig:hexagon_eq_on_associahedron}.
We read of\/f the
\emph{hexagon equation}
\begin{gather}
    \cT_{\hat{2},\bsy{1}}   \cP_{\bsy{3}}   \cT_{\hat{4},\bsy{2}}   \cT_{\hat{6},\bsy{1}}
    \cP_{\bsy{3}}
 = \cT_{\hat{5},\bsy{2}}   \cP_{\bsy{1}}  \cT_{\hat{3},\bsy{2}}   \cT_{\hat{1},\bsy{4}}   .
   \label{6-gon_eq}
\end{gather}
Fig.~\ref{fig:hexagon_eq_on_associahedron_2}
is a short-hand form of Fig.~\ref{fig:hexagon_eq_on_associahedron}.
In a categorical setting, a similar diagram appeared in~\cite[p.~218]{Kapr+Voev94PSPM},
and in~\cite[p.~189]{Stre98} as a $4$-cycle condition.
According to the prescription given for even polygon equations in the beginning of this subsection, we obtain
the following hatted version of the hexagon equation,
\begin{gather}
    \hat{\cT}_{\hat{2},\bsy{145}} \hat{\cT}_{\hat{4},\bsy{246}} \hat{\cT}_{\hat{6},\bsy{356}}
  = \hat{\cT}_{\hat{5},\bsy{356}} \hat{\cT}_{\hat{3},\bsy{245}} \hat{\cT}_{\hat{1},\bsy{123}}   ,
       \label{hexagon_eq_hatted}
\end{gather}
which can be read of\/f from
\begin{gather*}
 \begin{array}{@{}c@{\,}c@{\,}c@{\,}c@{\,}c@{\,}c@{\,}c@{}}
 \begin{minipage}{.6cm} $\widehat{12}$ \\[.5ex] $\widehat{14}$ \\[.5ex] $\widehat{16}$ \\[.5ex] $\widehat{34}$ \\[.5ex]
          $\widehat{36}$ \\[.5ex] $\widehat{56}$ \end{minipage}
          \xrightarrow[\bsy{356}]{ \hat{6} } &
  \begin{minipage}{.6cm} $\widehat{12}$ \\[.5ex] $\widehat{14}$ \\[.5ex] $\widehat{06}$ \\[.5ex] $\widehat{34}$ \\[.5ex]
           $\widehat{26}$ \\[.5ex] $\widehat{46}$ \end{minipage}
          \xrightarrow[\bsy{246}]{ \hat{4} } &
  \begin{minipage}{.6cm} $\widehat{12}$ \\[.5ex] $\widehat{04}$ \\[.5ex] $\widehat{06}$ \\[.5ex] $\widehat{24}$ \\[.5ex]
          $\widehat{26}$ \\[.5ex] $\widehat{45}$ \end{minipage}
          \xrightarrow[\bsy{145}]{ \hat{2} } &
  \begin{minipage}{.6cm} $\widehat{02}$ \\[.5ex] $\widehat{04}$ \\[.5ex] $\widehat{06}$ \\[.5ex] $\widehat{23}$ \\[.5ex]
        $\widehat{25}$ \\[.5ex] $\widehat{45}$ \end{minipage}
 \end{array}
   \qquad
 \begin{array}{@{}c@{\;}c@{\;}c@{\;}c@{\;}c@{\;}c@{\;}c@{}}
  \begin{minipage}{.6cm} $\widehat{12}$ \\[.5ex] $\widehat{14}$ \\[.5ex] $\widehat{16}$ \\[.5ex] $\widehat{34}$ \\[.5ex]
        $\widehat{36}$ \\[.5ex] $\widehat{56}$ \end{minipage}
         \xrightarrow[\bsy{123}]{ \hat{1} } &
  \begin{minipage}{.6cm} $\widehat{01}$ \\[.5ex] $\widehat{13}$ \\[.5ex] $\widehat{15}$ \\[.5ex] $\widehat{34}$ \\[.5ex]
        $\widehat{36}$ \\[.5ex] $\widehat{56}$ \end{minipage}
         \xrightarrow[\bsy{245}]{ \hat{3} } &
  \begin{minipage}{.6cm} $\widehat{01}$ \\[.5ex] $\widehat{03}$ \\[.5ex] $\widehat{15}$ \\[.5ex] $\widehat{23}$
          \\[.5ex] $\widehat{35}$ \\[.5ex] $\widehat{56}$     \end{minipage}
         \xrightarrow[\bsy{356}]{ \hat{5} } &
  \begin{minipage}{.6cm} $\widehat{01}$ \\[.5ex] $\widehat{03}$ \\[.5ex] $\widehat{05}$ \\[.5ex] $\widehat{23}$
          \\[.5ex] $\widehat{25}$ \\[.5ex] $\widehat{45}$ \end{minipage}
 \end{array}
\end{gather*}
Some versions of (\ref{hexagon_eq_hatted}) appeared in \cite{Kash14,Kore11,Kore12,Kore13,Kore14,Kore15,Kore+Sady13}
as a ``Pachner relation'' for a map realizing Pachner moves of triangulations of a
four-dimensional manifold.

\begin{figure}[t]\centering
\includegraphics[width=.6\linewidth]{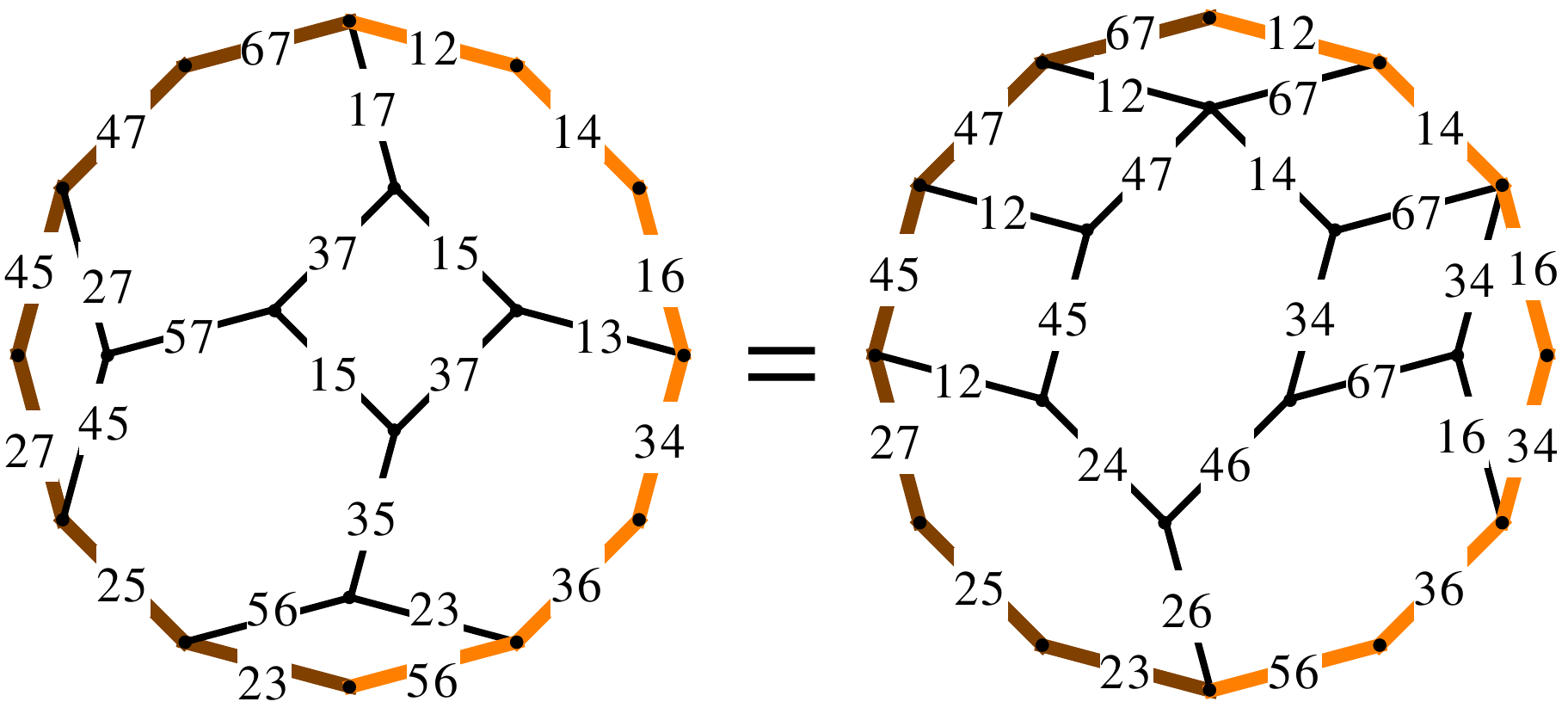}
\caption{Two complementary sides of the Edelman--Reiner polyhedron, formed by $T(7,4)$.
The equality represents the heptagon equation.
\label{fig:heptagon_eq_on_ER-polyhedron} }
\end{figure}

\paragraph{Heptagon equation.}
Here we consider maps $\cT_{ijklmp} \colon \cU_{ijklm} \times \cU_{ijkmp} \times \cU_{iklmp}
\rightarrow \cU_{jklmp} \times \cU_{ijlmp} \times \cU_{ijklp}$, $i<j<k<l<m<p$.
The two maximal chains of $T(7,5)$ lead to the \emph{heptagon equation}
\begin{gather}
   \cT_{\hat{1},\bsy{1}}    \cT_{\hat{3},\bsy{3}}   \cP_{\bsy{5}}   \cP_{\bsy{2}}
     \cT_{\hat{5},\bsy{3}}   \cT_{\hat{7},\bsy{1}}   \cP_{\bsy{3}}
 = \cP_{\bsy{3}}   \cT_{\hat{6},\bsy{4}}   \cP_{\bsy{3}}   \cP_{\bsy{2}}   \cP_{\bsy{1}}
    \cT_{\hat{4},\bsy{3}}   \cP_{\bsy{2}}   \cP_{\bsy{3}}
    \cT_{\hat{2},\bsy{4}}   ,
    \label{7-gon_eq}
\end{gather}
using complementary notation. Fig.~\ref{fig:heptagon_eq_on_ER-polyhedron}
shows the two sides of the Edelman--Reiner polyhedron~\cite{Edel+Rein96}, formed by $T(7,4)$,
on which the respective sides of this equation correspond to sequences of maximal chains.
In terms of $\hat{\cT} := \cT  \cP_{\bsy{1} \bsy{3}}$, the heptagon equation takes the form
\begin{gather}
    \hat{\cT}_{\hat{1},\bsy{1}\bsy{2}\bsy{3}}   \hat{\cT}_{\hat{3},\bsy{1}\bsy{4}\bsy{5}}
    \hat{\cT}_{\hat{5},\bsy{2}\bsy{4}\bsy{6}}   \hat{\cT}_{\hat{7},\bsy{3}\bsy{5}\bsy{6}}
  = \hat{\cT}_{\hat{6},\bsy{3}\bsy{5}\bsy{6}}   \hat{\cT}_{\hat{4},\bsy{2}\bsy{4}\bsy{5}}
    \hat{\cT}_{\hat{2},\bsy{1}\bsy{2}\bsy{3}}   .
       \label{7-gon_eq_mod}
\end{gather}
An equation with this structure appeared in~\cite{Volk97}.

\begin{figure}[t]\centering
\includegraphics[width=.7\linewidth]{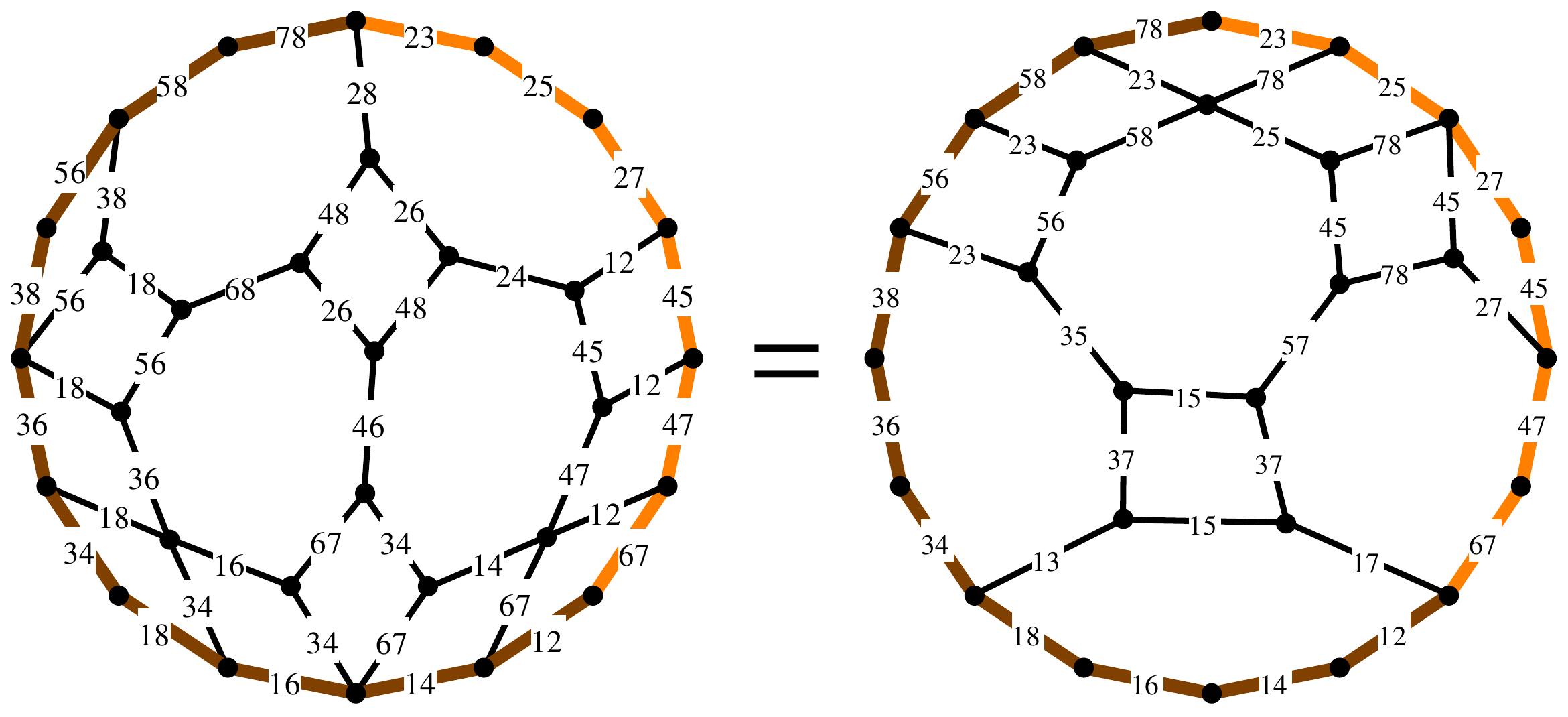}
\caption{Two complementary sides of the polyhedron formed by~$T(8,5)$. Equality represents the octagon equation.
\label{fig:octagon_eq_on_polyhedron} }
\end{figure}

\paragraph{Octagon equation.}
In case of $T(8,6)$, we consider maps
$\cT_{ijklmpq} \colon \cU_{ijklmp} \times \cU_{ijklpq} \times \cU_{ijlmpq} \times \cU_{jklmpq}
  \rightarrow  \cU_{iklmpq} \times \cU_{ijkmpq} \times \cU_{ijklmp}$, $i<j<k<l<m<p<q$, subject to
the \emph{octagon equation}
\begin{gather*}
   \cT_{\hat{2},\bsy{1}}   \cP_{\bsy{4}}   \cP_{\bsy{5}}   \cP_{\bsy{6}}
   \cT_{\hat{4},\bsy{3}}   \cP_{\bsy{6}}   \cP_{\bsy{5}}   \cP_{\bsy{2}}
   \cT_{\hat{6},\bsy{3}}   \cP_{\bsy{6}}   \cT_{\hat{8},\bsy{1}}   \cP_{\bsy{4}}
     \cP_{\bsy{5}}   \cP_{\bsy{6}}   \cP_{\bsy{3}}
   =  \cP_{\bsy{3}}   \cT_{\hat{7},\bsy{4}}    \cP_{\bsy{3}}   \cP_{\bsy{2}}   \cP_{\bsy{1}}
   \cT_{\hat{5},\bsy{3}}   \cP_{\bsy{6}}   \cP_{\bsy{2}}   \cP_{\bsy{3}}
   \cT_{\hat{3},\bsy{4}}   \cT_{\hat{1},\bsy{7}}   .   %\label{8-gon_eq}
\end{gather*}
See Fig.~\ref{fig:octagon_eq_on_polyhedron}.
The hatted version of the octagon equation is
\begin{gather*}
    \hat{\cT}_{\hat{2},\bsy{1},\bsy{5},\bsy{6},\bsy{7}} \hat{\cT}_{\hat{4},\bsy{2},\bsy{5},\bsy{8},\bsy{9}}
    \hat{\cT}_{\hat{6},\bsy{3},\bsy{6},\bsy{8},\bsy{10}} \hat{\cT}_{\hat{8},\bsy{4},\bsy{7},\bsy{9},\bsy{10}}
  = \hat{\cT}_{\hat{7},\bsy{4},\bsy{7},\bsy{9},\bsy{10}} \hat{\cT}_{\hat{5},\bsy{3},\bsy{6},\bsy{8},\bsy{9}}
    \hat{\cT}_{\hat{3},\bsy{2},\bsy{5},\bsy{6},\bsy{7}} \hat{\cT}_{\hat{1},\bsy{1},\bsy{2},\bsy{3},\bsy{4}}  .
\end{gather*}
The position indices can be read of\/f from
\begin{gather*}
 \begin{array}{@{}c@{\,}c@{\,}c@{\,}c@{\,}c@{\,}c@{\,}c@{\,}c@{\,}c@{}}
 \begin{minipage}{.5cm} $\widehat{12}$ \\[.5ex] $\widehat{14}$ \\[.5ex] $\widehat{16}$ \\[.5ex] $\widehat{18}$ \\[.5ex]
          $\widehat{34}$ \\[.5ex] $\widehat{36}$ \\[.5ex] $\widehat{38}$ \\[.5ex] $\widehat{56}$
          \\[.5ex] $\widehat{58}$ \\[.5ex] $\widehat{78}$ \end{minipage}
           \xrightarrow[\bsy{4},\bsy{7},\bsy{9},\bsy{10}]{ \hat{8} } &
 \begin{minipage}{.5cm} $\widehat{12}$ \\[.5ex] $\widehat{14}$ \\[.5ex] $\widehat{16}$ \\[.5ex] $\widehat{08}$ \\[.5ex]
          $\widehat{34}$ \\[.5ex] $\widehat{36}$ \\[.5ex] $\widehat{28}$ \\[.5ex] $\widehat{56}$
          \\[.5ex] $\widehat{48}$ \\[.5ex] $\widehat{68}$ \end{minipage}
           \xrightarrow[\bsy{3},\bsy{6},\bsy{8},\bsy{10}]{ \hat{6} } &
 \begin{minipage}{.5cm} $\widehat{12}$ \\[.5ex] $\widehat{14}$ \\[.5ex] $\widehat{06}$ \\[.5ex] $\widehat{08}$ \\[.5ex]
          $\widehat{34}$ \\[.5ex] $\widehat{26}$ \\[.5ex] $\widehat{28}$ \\[.5ex] $\widehat{46}$
          \\[.5ex] $\widehat{48}$ \\[.5ex] $\widehat{67}$ \end{minipage}
           \xrightarrow[\bsy{2},\bsy{5},\bsy{8},\bsy{9}]{ \hat{4} } &
 \begin{minipage}{.5cm} $\widehat{12}$ \\[.5ex] $\widehat{04}$ \\[.5ex] $\widehat{06}$ \\[.5ex] $\widehat{08}$ \\[.5ex]
          $\widehat{24}$ \\[.5ex] $\widehat{26}$ \\[.5ex] $\widehat{28}$ \\[.5ex] $\widehat{45}$
          \\[.5ex] $\widehat{47}$ \\[.5ex] $\widehat{67}$ \end{minipage}
           \xrightarrow[\bsy{1},\bsy{5},\bsy{6},\bsy{7}]{ \hat{2} } &
 \begin{minipage}{.5cm} $\widehat{02}$ \\[.5ex] $\widehat{04}$ \\[.5ex] $\widehat{06}$ \\[.5ex] $\widehat{08}$ \\[.5ex]
          $\widehat{23}$ \\[.5ex] $\widehat{25}$ \\[.5ex] $\widehat{27}$ \\[.5ex] $\widehat{45}$
          \\[.5ex] $\widehat{47}$ \\[.5ex] $\widehat{67}$ \end{minipage}
 \end{array}
           \qquad
 \begin{array}{@{}c@{\,}c@{\,}c@{\,}c@{\,}c@{\,}c@{\,}c@{\,}c@{\,}c@{}}
 \begin{minipage}{.5cm} $\widehat{12}$ \\[.5ex] $\widehat{14}$ \\[.5ex] $\widehat{16}$ \\[.5ex] $\widehat{18}$ \\[.5ex]
          $\widehat{34}$ \\[.5ex] $\widehat{36}$ \\[.5ex] $\widehat{38}$ \\[.5ex] $\widehat{56}$
          \\[.5ex] $\widehat{58}$ \\[.5ex] $\widehat{78}$ \end{minipage}
           & \stackrel{ \hat{1} }{\longrightarrow} &
 \begin{minipage}{.5cm} $\widehat{02}$ \\[.5ex] $\widehat{13}$ \\[.5ex] $\widehat{15}$ \\[.5ex] $\widehat{17}$ \\[.5ex]
          $\widehat{34}$ \\[.5ex] $\widehat{36}$ \\[.5ex] $\widehat{38}$ \\[.5ex] $\widehat{56}$
          \\[.5ex] $\widehat{58}$ \\[.5ex] $\widehat{78}$ \end{minipage}
          & \stackrel{ \hat{3} }{\longrightarrow} &
 \begin{minipage}{.5cm} $\widehat{01}$ \\[.5ex] $\widehat{03}$ \\[.5ex] $\widehat{15}$ \\[.5ex] $\widehat{17}$ \\[.5ex]
          $\widehat{23}$ \\[.5ex] $\widehat{35}$ \\[.5ex] $\widehat{37}$ \\[.5ex] $\widehat{56}$
          \\[.5ex] $\widehat{58}$ \\[.5ex] $\widehat{78}$ \end{minipage}
          & \stackrel{ \hat{5} }{\longrightarrow} &
 \begin{minipage}{.5cm} $\widehat{01}$ \\[.5ex] $\widehat{03}$ \\[.5ex] $\widehat{05}$ \\[.5ex] $\widehat{17}$ \\[.5ex]
          $\widehat{23}$ \\[.5ex] $\widehat{25}$ \\[.5ex] $\widehat{37}$ \\[.5ex] $\widehat{45}$
          \\[.5ex] $\widehat{57}$ \\[.5ex] $\widehat{78}$ \end{minipage}
          & \stackrel{ \hat{7} }{\longrightarrow} &
 \begin{minipage}{.5cm} $\widehat{01}$ \\[.5ex] $\widehat{03}$ \\[.5ex] $\widehat{05}$ \\[.5ex] $\widehat{07}$ \\[.5ex]
          $\widehat{23}$ \\[.5ex] $\widehat{25}$ \\[.5ex] $\widehat{27}$ \\[.5ex] $\widehat{45}$
          \\[.5ex] $\widehat{47}$ \\[.5ex] $\widehat{67}$ \end{minipage}
 \end{array}
\end{gather*}

\begin{figure}[t]\centering
\includegraphics[width=.8\linewidth]{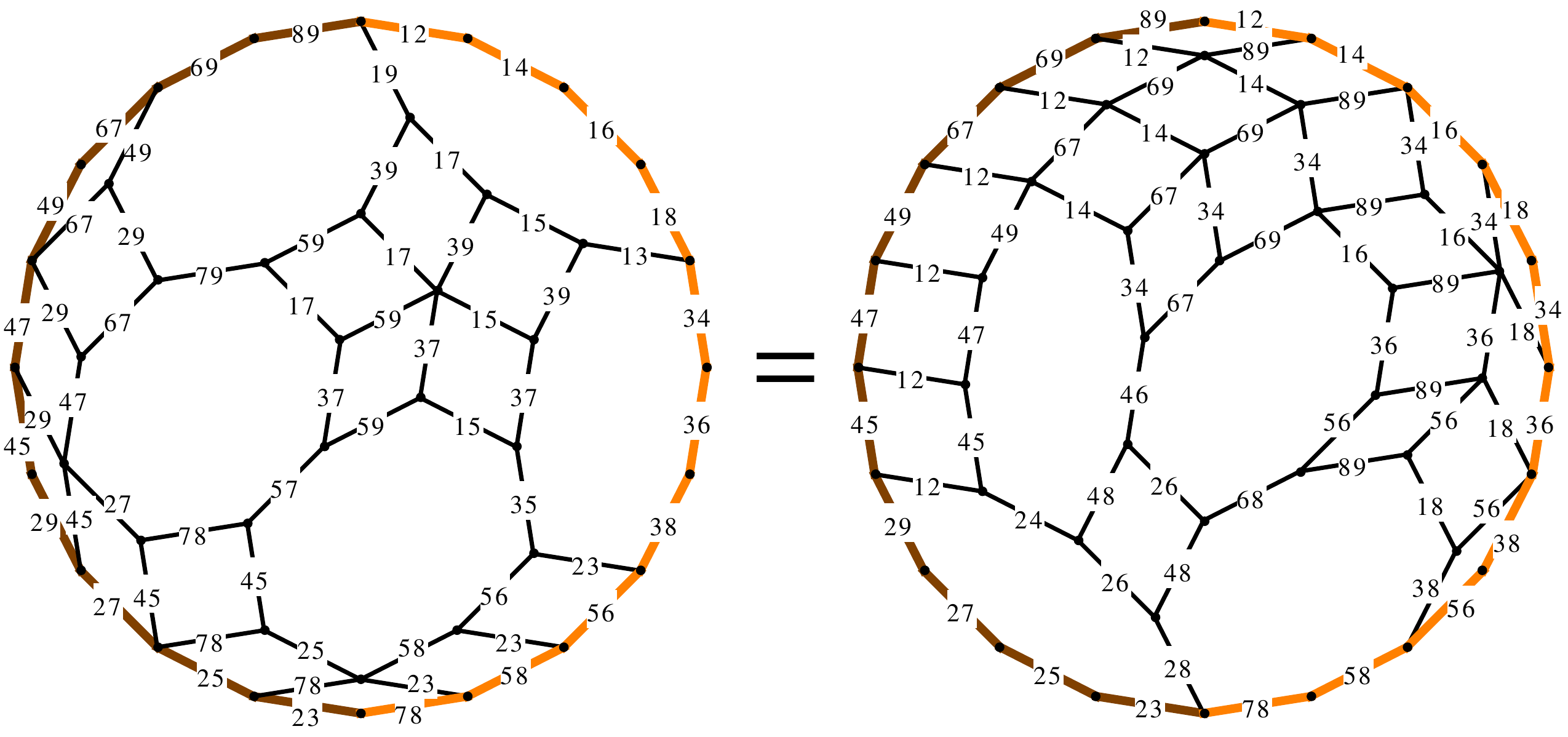}
\caption{Two complementary sides of the polyhedral part of $T(9,6)$ (which is non-polyhedral due to the
existence of small cubes). Equality represents the enneagon equation.
\label{fig:enneagon_eq_on_polyhedron} }
\end{figure}

\paragraph{Enneagon equation.}
For $T(9,7)$ we f\/ind
\begin{gather*}
  \cT_{\hat{1},\bsy{1}}   \cT_{\hat{3},\bsy{4}}
    \cP_{\bsy{7}}   \cP_{\bsy{8}}   \cP_{\bsy{9}}   \cP_{\bsy{3}}   \cP_{\bsy{2}}   \cP_{\bsy{4}}
    \cT_{\hat{5},\bsy{5}}
    \cP_{\bsy{8}}   \cP_{\bsy{7}}   \cP_{\bsy{4}}   \cP_{\bsy{3}}
    \cT_{\hat{7},\bsy{4}}
    \cP_{\bsy{7}}
    \cT_{\hat{9},\bsy{1}}
    \cP_{\bsy{4}}    \cP_{\bsy{5}}    \cP_{\bsy{6}}   \cP_{\bsy{3}}   \nonumber         \\
 \qquad{} =
    \cP_{\bsy{7}}   \cP_{\bsy{4}}   \cP_{\bsy{5}}   \cP_{\bsy{6}}   \cP_{\bsy{3}}
    \cT_{\hat{8},\bsy{7}}
    \cP_{\bsy{6}}   \cP_{\bsy{5}}   \cP_{\bsy{4}}   \cP_{\bsy{3}}   \cP_{\bsy{2}}   \cP_{\bsy{1}}
    \cT_{\hat{6},\bsy{5}}
    \cP_{\bsy{8}}   \cP_{\bsy{4}}   \cP_{\bsy{3}}   \cP_{\bsy{2}}   \cP_{\bsy{5}}   \cP_{\bsy{4}}   \cP_{\bsy{3}}
    \cT_{\hat{4},\bsy{5}}
    \cP_{\bsy{4}}   \cP_{\bsy{5}}   \cP_{\bsy{6}}
    \cT_{\hat{2},\bsy{7}}   ,
            %\label{9-gon_eq}
\end{gather*}
which can be visualized on $T(9,6)$, see Fig.~\ref{fig:enneagon_eq_on_polyhedron}.
In terms of $\hat{\cT} := \cT  \cP_{\bsy{14}}  \cP_{\bsy{23}}$,
the \emph{enneagon} (or \emph{nonagon}) \emph{equation} takes the compact form
\begin{gather*}
    \hat{\cT}_{\hat{1},\bsy{1},\bsy{2},\bsy{3},\bsy{4}}   \hat{\cT}_{\hat{3},\bsy{1},\bsy{5},\bsy{6},\bsy{7}}
    \hat{\cT}_{\hat{5},\bsy{2},\bsy{5},\bsy{8},\bsy{9}}
    \hat{\cT}_{\hat{7},\bsy{3},\bsy{6},\bsy{8},\bsy{10}}   \hat{\cT}_{\hat{9},\bsy{4},\bsy{7},\bsy{9},\bsy{10}}
 =  \hat{\cT}_{\hat{8},\bsy{4},\bsy{7},\bsy{9},\bsy{10}}   \hat{\cT}_{\hat{6},\bsy{3},\bsy{6},\bsy{8},\bsy{9}}
    \hat{\cT}_{\hat{4},\bsy{2},\bsy{5},\bsy{6},\bsy{7}}   \hat{\cT}_{\hat{2},\bsy{1},\bsy{2},\bsy{3},\bsy{4}}   .
\end{gather*}

\begin{figure}[t]\centering
\includegraphics[width=.9\linewidth]{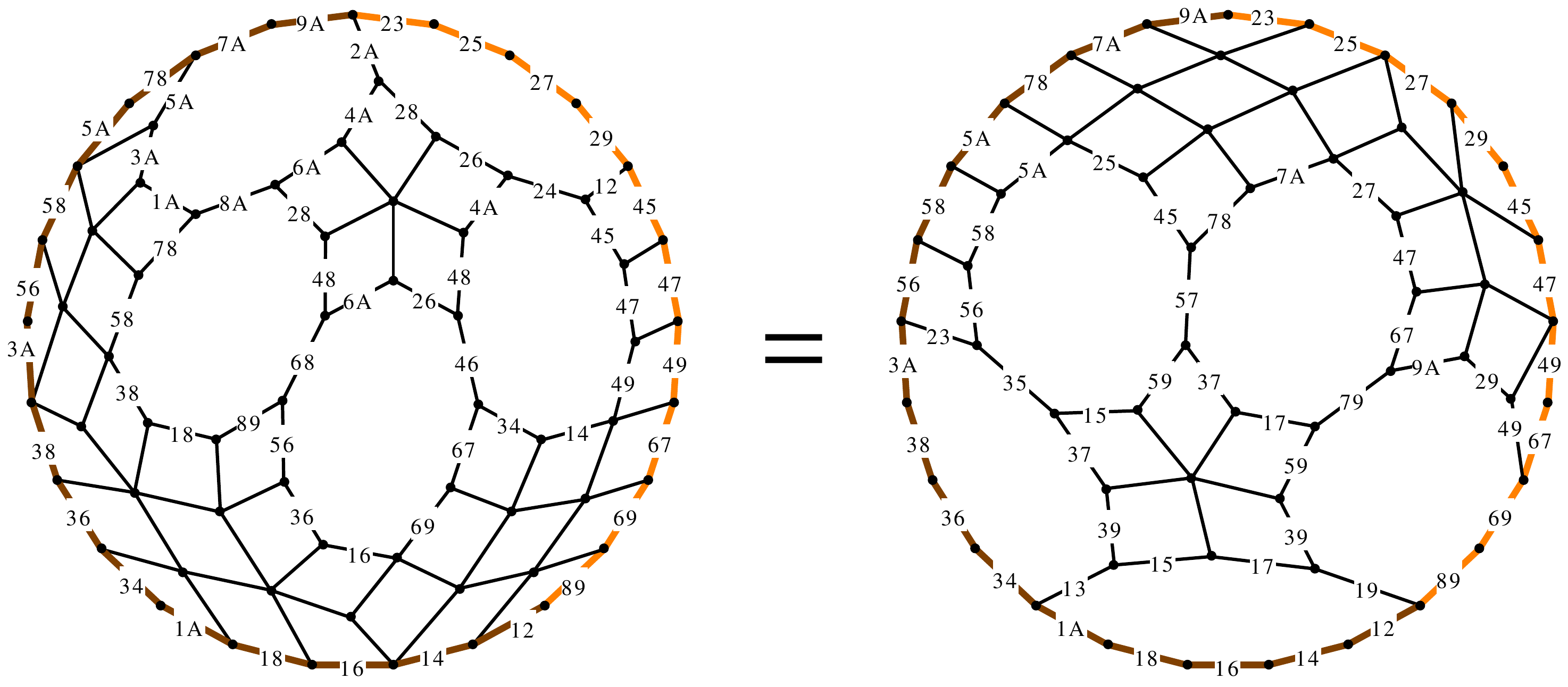}
\caption{Two complementary sides of the polyhedral part of $T(10,7)$ (small cubes are resolved).
Equality represents the decagon equation. Here $A$ stands for $10$.
\label{fig:decagon_eq_on_polyhedron} }
\end{figure}

\paragraph{Decagon equation.}
For $T(10,8)$ we obtain the equation
\begin{gather*}
   \cT_{\hat{2},\bsy{1}}
    \cP_{\bsy{5}}   \cP_{\bsy{6}}   \cP_{\bsy{7}}   \cP_{\bsy{8}}   \cP_{\bsy{9}}
    \cP_{\bsy{10}}
    \cT_{\hat{4},\bsy{4}}
    \cP_{\bsy{8}}   \cP_{\bsy{9}}   \cP_{\bsy{10}}
    \cP_{\bsy{7}}   \cP_{\bsy{8}}   \cP_{\bsy{9}}
    \cP_{\bsy{3}}   \cP_{\bsy{2}}   \cP_{\bsy{4}}
    \cT_{\hat{6},\bsy{5}}
    \cP_{\bsy{9}}   \cP_{\bsy{10}}
    \cP_{\bsy{8}}   \cP_{\bsy{7}}   \\
   \cP_{\bsy{4}}   \cP_{\bsy{3}}
    \cT_{\hat{8},\bsy{4}}
    \cP_{\bsy{8}}   \cP_{\bsy{9}}   \cP_{\bsy{10}}   \cP_{\bsy{7}}
    \cT_{\widehat{10},\bsy{1}}
    \cP_{\bsy{5}}   \cP_{\bsy{6}}   \cP_{\bsy{7}}   \cP_{\bsy{8}}   \cP_{\bsy{9}}
    \cP_{\bsy{10}}   \cP_{\bsy{4}}   \cP_{\bsy{5}}   \cP_{\bsy{6}}   \cP_{\bsy{3}} \\
 \qquad{} =
    \cP_{\bsy{7}}   \cP_{\bsy{4}}   \cP_{\bsy{5}}   \cP_{\bsy{6}}   \cP_{\bsy{3}}
    \cT_{\hat{9},\bsy{7}}
    \cP_{\bsy{6}}   \cP_{\bsy{5}}   \cP_{\bsy{4}}   \cP_{\bsy{3}}
    \cP_{\bsy{2}}   \cP_{\bsy{1}}
    \cT_{\hat{7},\bsy{5}}
    \cP_{\bsy{9}}   \cP_{\bsy{10}}   \cP_{\bsy{8}}   \cP_{\bsy{4}}   \cP_{\bsy{3}}
    \cP_{\bsy{2}}   \cP_{\bsy{5}}   \cP_{\bsy{4}}   \cP_{\bsy{3}}
    \cT_{\hat{5},\bsy{5}}    \\
\qquad\quad \   \cP_{\bsy{9}}   \cP_{\bsy{10}}   \cP_{\bsy{4}}   \cP_{\bsy{5}}   \cP_{\bsy{6}}
    \cT_{\hat{3},\bsy{7}}   \cT_{\hat{1},\bsy{11}}   ,
\end{gather*}
and Fig.~\ref{fig:decagon_eq_on_polyhedron} shows the corresponding polyhedral representation
obtained from $T(10,7)$.
The hatted version of the decagon equation is
\begin{gather*}
     \hat{\cT}_{\hat{2},\bsy{1},\bsy{6},\bsy{7},\bsy{8},\bsy{9}}
    \hat{\cT}_{\hat{4},\bsy{2},\bsy{6},\bsy{10},\bsy{11},\bsy{12}}
    \hat{\cT}_{\hat{6},\bsy{3},\bsy{7},\bsy{10},\bsy{13},\bsy{14}}
    \hat{\cT}_{\hat{8},\bsy{4},\bsy{8},\bsy{11},\bsy{13},\bsy{15}}
    \hat{\cT}_{\hat{10},\bsy{5},\bsy{9},\bsy{12},\bsy{14},\bsy{15}} \\
  \qquad =  \hat{\cT}_{\hat{9},\bsy{5},\bsy{9},\bsy{12},\bsy{14},\bsy{15}}
    \hat{\cT}_{\hat{7},\bsy{4},\bsy{8},\bsy{11},\bsy{13},\bsy{14}}
    \hat{\cT}_{\hat{5},\bsy{3},\bsy{7},\bsy{10},\bsy{11},\bsy{12}}
    \hat{\cT}_{\hat{3},\bsy{2},\bsy{6},\bsy{7},\bsy{8},\bsy{9}}
    \hat{\cT}_{\hat{1},\bsy{1},\bsy{2},\bsy{3},\bsy{4},\bsy{5}}   .
\end{gather*}

\begin{figure}[t]\centering
\includegraphics[width=1.\linewidth]{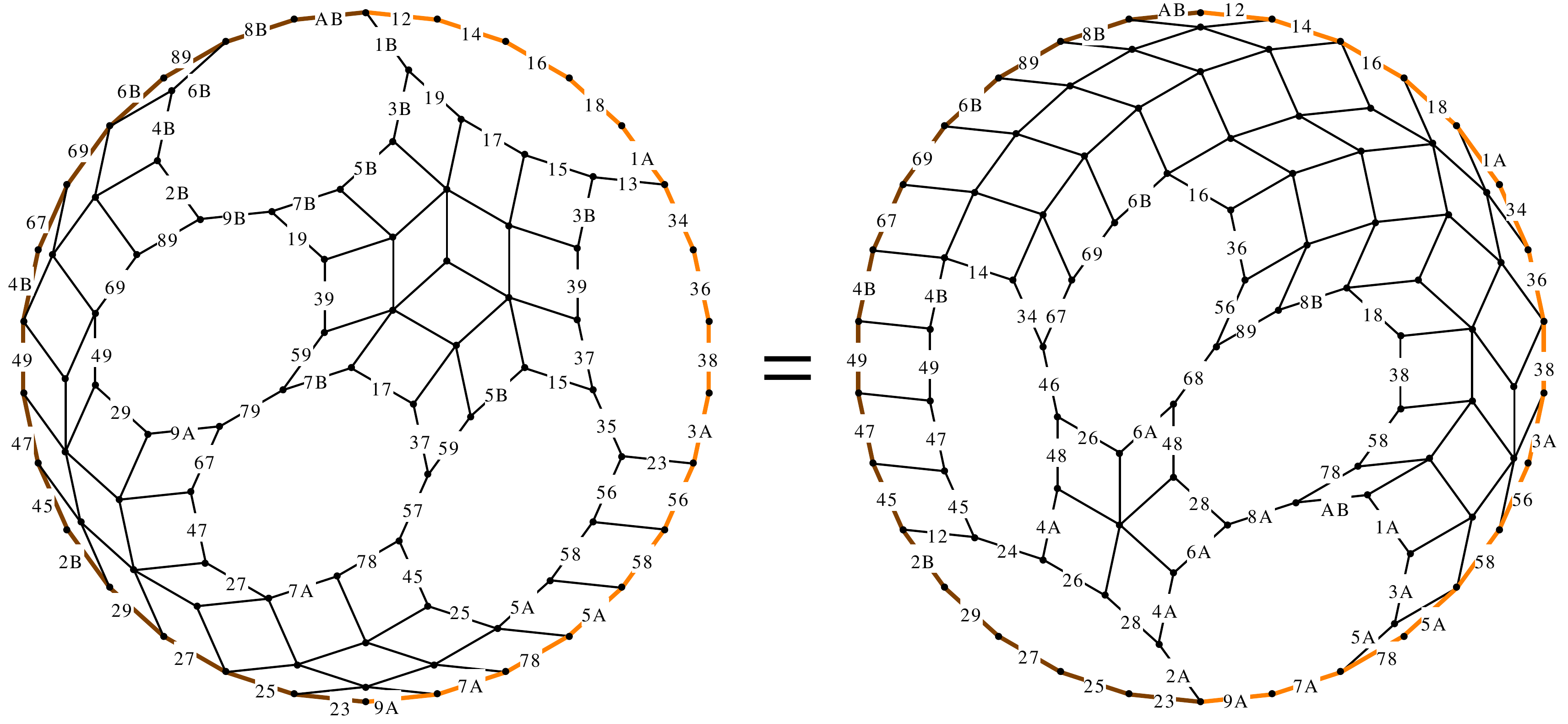}
\caption{Two complementary sides of the polyhedral part of $T(11,8)$ (small cubes are resolved).
Equality represents the hendecagon equation. Here we set $A:=10$, $B:=11$.
\label{fig:hendecagon_eq_on_polyhedron} }
\end{figure}

\paragraph{Hendecagon equation.}
For $T(11,9)$, the associated equation reads
\begin{gather*}
  \cT_{\hat{1},\bsy{1}}   \cT_{\hat{3},\bsy{5}}
    \cP_{\bsy{9}}   \cP_{\bsy{10}}   \cP_{\bsy{11}}   \cP_{\bsy{12}}   \cP_{\bsy{13}}
    \cP_{\bsy{14}}
    \cP_{\bsy{4}}   \cP_{\bsy{3}}   \cP_{\bsy{2}}   \cP_{\bsy{5}}   \cP_{\bsy{4}}
    \cP_{\bsy{6}}
    \cT_{\hat{5},\bsy{7}}
    \cP_{\bsy{11}}   \cP_{\bsy{12}}   \cP_{\bsy{13}}   \cP_{\bsy{10}}   \cP_{\bsy{11}}
    \cP_{\bsy{12}}   \nonumber \\
   \cP_{\bsy{6}}
    \cP_{\bsy{5}}   \cP_{\bsy{4}}   \cP_{\bsy{3}}   \cP_{\bsy{7}}   \cP_{\bsy{6}}
    \cT_{\hat{7},\bsy{7}}
    \cP_{\bsy{11}}   \cP_{\bsy{12}}   \cP_{\bsy{10}}   \cP_{\bsy{9}}
    \cP_{\bsy{6}}   \cP_{\bsy{5}}   \cP_{\bsy{4}}
    \cT_{\hat{9},\bsy{5}}
    \cP_{\bsy{9}}   \cP_{\bsy{10}}   \cP_{\bsy{11}}    \cP_{\bsy{8}}
    \cT_{\widehat{11},\bsy{1}}               \nonumber  \\
  \cP_{\bsy{5}}   \cP_{\bsy{6}}   \cP_{\bsy{7}}   \cP_{\bsy{8}}   \cP_{\bsy{9}}
    \cP_{\bsy{10}}   \cP_{\bsy{4}}   \cP_{\bsy{5}}   \cP_{\bsy{6}}   \cP_{\bsy{3}}  \nonumber \\
\qquad{} =
    \cP_{\bsy{12}}   \cP_{\bsy{9}}   \cP_{\bsy{10}}   \cP_{\bsy{11}}   \cP_{\bsy{8}}
    \cP_{\bsy{5}}   \cP_{\bsy{6}}   \cP_{\bsy{7}}   \cP_{\bsy{8}}   \cP_{\bsy{9}}
    \cP_{\bsy{10}}   \cP_{\bsy{4}}   \cP_{\bsy{5}}   \cP_{\bsy{6}}   \cP_{\bsy{3}}
    \cT_{\widehat{10},\bsy{11}}
    \cP_{\bsy{10}}   \cP_{\bsy{9}}   \cP_{\bsy{8}}   \cP_{\bsy{7}}
    \cP_{\bsy{6}}   \cP_{\bsy{5}}   \cP_{\bsy{4}}   \cP_{\bsy{3}}   \cP_{\bsy{2}}
    \cP_{\bsy{1}}      \nonumber \\
\qquad \quad \    \cT_{\hat{8},\bsy{8}}
    \cP_{\bsy{12}}   \cP_{\bsy{13}}   \cP_{\bsy{11}}   \cP_{\bsy{7}}   \cP_{\bsy{6}}
    \cP_{\bsy{5}}   \cP_{\bsy{4}}   \cP_{\bsy{3}}   \cP_{\bsy{2}}
    \cP_{\bsy{8}}   \cP_{\bsy{7}}   \cP_{\bsy{6}}   \cP_{\bsy{5}}   \cP_{\bsy{4}}
    \cP_{\bsy{3}}
    \cT_{\hat{6},\bsy{7}}
    \cP_{\bsy{11}}   \cP_{\bsy{12}}   \cP_{\bsy{6}}   \cP_{\bsy{5}}   \cP_{\bsy{4}}
    \cP_{\bsy{7}}   \cP_{\bsy{6}}   \cP_{\bsy{5}}    \nonumber  \\
\qquad \quad \   \cP_{\bsy{8}}   \cP_{\bsy{7}}   \cP_{\bsy{6}}   \cT_{\hat{4},\bsy{8}}
    \cP_{\bsy{7}}   \cP_{\bsy{8}}   \cP_{\bsy{9}}   \cP_{\bsy{10}}
    \cT_{\hat{2},\bsy{11}}   ,
     %\label{11-gon_eq}
\end{gather*}
also see Fig.~\ref{fig:hendecagon_eq_on_polyhedron}.
In terms of $\hat{\cT} = \cT   \cP_{\bsy{15}}   \cP_{\bsy{24}}$, it takes the form
\begin{gather*}
 \hat{\cT}_{\hat{1},\bsy{1},\bsy{2},\bsy{3},\bsy{4},\bsy{5}}
    \hat{\cT}_{\hat{3},\bsy{1},\bsy{6},\bsy{7},\bsy{8},\bsy{9}}
    \hat{\cT}_{\hat{5},\bsy{2},\bsy{6},\bsy{10},\bsy{11},\bsy{12}}
    \hat{\cT}_{\hat{7},\bsy{3},\bsy{7},\bsy{10},\bsy{13},\bsy{14}}
    \hat{\cT}_{\hat{9},\bsy{4},\bsy{8},\bsy{11},\bsy{13},\bsy{15}}
    \hat{\cT}_{\widehat{11},\bsy{5},\bsy{9},\bsy{12},\bsy{14},\bsy{15}} \nonumber \\
\qquad {} =  \hat{\cT}_{\widehat{10},\bsy{5},\bsy{9},\bsy{12},\bsy{14},\bsy{15}}
    \hat{\cT}_{\hat{8},\bsy{4},\bsy{8},\bsy{11},\bsy{13},\bsy{14}}
    \hat{\cT}_{\hat{6},\bsy{3},\bsy{7},\bsy{10},\bsy{11},\bsy{12}}
    \hat{\cT}_{\hat{4},\bsy{2},\bsy{6},\bsy{7},\bsy{8},\bsy{9}}
    \hat{\cT}_{\hat{2},\bsy{1},\bsy{2},\bsy{3},\bsy{4},\bsy{5}}   .
    %\label{11-gon_eq_mod}
\end{gather*}

\begin{rem}
Disregarding the indices that specify on which sets the maps act, the left-hand side
of the $N$-simplex equation has the same structure as the left-hand side of the $(2N+1)$-gon
equation (but there is no such relation between the right-hand sides, of course).
Accordingly, the corresponding halfs of the associated polyhedra coincide up to the labeling.
\end{rem}

\subsection{Lax systems for polygon equations}
\label{subsec:polygon_Lax}
In this subsection we consider the case where the maps $\cT_J$, $J \in {[N+1] \choose N-1}$, are
``localized'' to maps
\begin{gather*}
    \cL_J \colon \ \cU_J \longrightarrow \mathrm{Map}(\cU_{\vec{P_o}(J)}, \cU_{\cev{P_e}(J)}),  \\
    \hphantom{\cL_J \colon}{} \ u_J \longmapsto \cL_J(u_J): \quad \cU_{\vec{P_o}(J)} \rightarrow  \cU_{\cev{P_e}(J)}  .
\end{gather*}
In $T(N+1,N-2)$, counterparts of the two maximal chains, of which $T(N,N-2)$ consists,
appear as chains for all $\hat{k} \in {[N+1] \choose N}$, $k \in [N+1]$,
\begin{gather*}
    \cC_{\hat{k},o} \colon \  [\bar{\alpha}_{\hat{k}}] \stackrel{ \widehat{k k_N} }{\longrightarrow} [\bar{\rho}_{\hat{k},1}]
            \stackrel{ \widehat{k k_{N-2}} }{\longrightarrow} [\bar{\rho}_{\hat{k},2}]
            \longrightarrow \cdots \longrightarrow [\bar{\rho}_{\hat{k},N+m-3}]
            \stackrel{ \widehat{k k_{2-m}} }{\longrightarrow} [\bar{\omega}_{\hat{k}}]
              , \nonumber \\
    \cC_{\hat{k},e} \colon \ [\bar{\alpha}_{\hat{k}}] \stackrel{ \widehat{k k_{1+m}} }{\longrightarrow} [\bar{\sigma}_{\hat{k},1}]
            \stackrel{ \widehat{k k_{3+m}} }{\longrightarrow} [\bar{\sigma}_{\hat{k},2}]
            \longrightarrow \cdots
            \stackrel{ \widehat{k k_{N-1}} }{\longrightarrow} [\bar{\omega}_{\hat{k}}] \, ,
\end{gather*}
where we wrote $\hat{k} = (k_1,\ldots,k_N)$, $k_1 < k_2 < \cdots < k_N$. Here
$\bar{\alpha}_{\hat{k}},\bar{\omega}_{\hat{k}},\bar{\rho}_{\hat{k},i}$ and $\bar{\sigma}_{\hat{k},i}$ are reduced
admissible linear orders of ${\hat{k} \choose N-2}$, and $m := N \, \mathrm{mod} \, 2$.
Let $\tilde{\cC}_{\hat{k},o}$ and $\tilde{\cC}_{\hat{k},e}$ be resolutions of the above chains.
We consider the following system of localized $N$-gon equations,
\begin{gather}
     \cL_{\tilde{\cC}_{\hat{k},o}}(\bsy{u}_{\vec{P_o}(\hat{k})})
   = \cL_{\tilde{\cC}_{\hat{k},e}}(\bsy{v}_{\cev{P_e}(\hat{k})}),
     \qquad    \hat{k} \in {[N+1] \choose N}   ,    \label{polygon_Lax_system}
\end{gather}
where
\begin{gather}
     \bsy{u}_{\vec{P_o}(\hat{k})} = \big(u_{ \widehat{k k_N} }, u_{ \widehat{k k_{N-2}} }, \ldots,
          u_{ \widehat{k k_{2-m}} }\big)   , \qquad
     \bsy{v}_{\cev{P_e}(\hat{k})} = \big(v_{\widehat{k k_{1+m}}}, \ldots,
          v_{\widehat{k k_{N-3}}}, \ldots, v_{\widehat{k k_{N-1}}}\big)   .
\end{gather}
We shall assume that each of these equations uniquely determines a map
$\cT_{\hat{k}}$ via $\bsy{u}_{\vec{P_o}(\hat{k})} \mapsto \bsy{v}_{\cev{P_e}(\hat{k})}$.

A hatted version of $\cL_J$ is def\/ined in the same way as the hatted version of $\cT_J$ (see the beginning
of Section~\ref{sec:polygon_eqs}), dif\/ferently for even and odd polygon equations.
The above system (\ref{polygon_Lax_system}) then has the form
\begin{gather*}
    \hat{\cL}_{ \widehat{k k_{2-m}},\bsy{Y}_{\bsy{B}_{2-m}} } \cdots \hat{\cL}_{ \widehat{k k_{N-2}},\bsy{Y}_{\bsy{B}_{N-2}}}
    \hat{\cL}_{ \widehat{k k_N},\bsy{Y}_{\bsy{B}_N} }\\
    \qquad{}
  = \Big( \hat{\cL}_{ \widehat{k k_{N-1}},\bsy{Y}_{\bsy{B}_{N-1}}} \hat{\cL}_{ \widehat{k k_{N-3}},\bsy{Y}_{\bsy{B}_{N-3}} }
    \cdots \hat{\cL}_{ \widehat{k k_{m+1}},\bsy{Y}_{\bsy{B}_{m+1}} } \Big) \circ \cT_{\hat{k}}   ,
\end{gather*}
where $\hat{k} \in {[N+1] \choose N}$. Here $\bsy{Y}_{\bsy{B}_i} = (\bsy{y}_{\bsy{b}_{i,1}}, \ldots,
\bsy{y}_{\bsy{b}_{i,n}})$, where $n=\lfloor N/2 \rfloor = (N-m)/2$,
is an increasing sequence of integers, and
$\bsy{B}_i = (\bsy{b}_{i,1}, \ldots, \bsy{b}_{i,n})$ is a multi-index, as def\/ined previously. We have
$1 \leq \bsy{b}_{i,j} \leq c(n+1,2)$ and $1 \leq \bsy{y}_{\bsy{b}_{i,j}} \leq c(n+m+1,3)$.

With $\bar{\rho} = (J_1,\ldots, J_r) \in A^{(b)}(N+1,N-1)$ we associate the composition of maps
$\hat{\cL}_{\bar{\rho}} = \hat{\cL}_{J_r, \bsy{B}_r} \cdots \hat{\cL}_{J_1,\bsy{B}_1}$. The domain
of $\hat{\cL}_{\bar{\rho}}$ is $\cU_{\mathrm{rev}(\bar{\eta})}$, where $\eta \in A(N+1,N-1)$ is the
lexicographical order of ${[N+1] \choose N-2}$. $\bsy{B}_1$ is the multi-index of the positions of the
elements of $P(J_1)$ in $\mathrm{rev}(\bar{\eta})$ (which has $c(n+m+1,3)$ elements). It seems to be a
dif\/f\/icult task to f\/ind a general formula that determines the other multi-indices.

In a similar way as in the case of simplex equations, one can show that the $(N+1)$-gon equation
$\cT_{\tilde{\cC}_o} = \cT_{\tilde{\cC}_e}$, where $\tilde{\cC}_o$ and $\tilde{\cC}_e$
constitute a resolution of $T(N+1,N-1)$, arises as a~consistency condition of the above Lax
system. Here one starts with $\hat{\cL}_{\bar{\alpha}}$, $\bar{\alpha} \in A^{(b)}(N+1,N-1)$, where
$\alpha$ is the lexicographical order of ${[N+1] \choose N-1}$, and follows the two resolutions.

\begin{exam}
For $N=6$, we have
\begin{gather*}
   \mathrm{rev}(\bar{\eta}) = (\widehat{123},\widehat{125},\widehat{127},\widehat{145},
           \widehat{147},\widehat{167},\widehat{345},\widehat{347},\widehat{367},\widehat{567})
\end{gather*}
and $\bar{\alpha} = (\widehat{67},\widehat{47},\widehat{45},\widehat{27},\widehat{25},\widehat{23}) \in A^{(b)}(7,5)$, so that
\begin{gather*}
 \begin{array}{@{}c@{\,}c@{\,}c@{\,}c@{\,}c@{\,}c@{\,}c@{\,}c@{\,}c@{\,}c@{\,}c@{\,}c@{\,}c@{}}
 \begin{minipage}{.7cm} $\widehat{123}$ \\[.5ex] $\widehat{125}$ \\[.5ex] $\widehat{127}$ \\[.5ex] $\widehat{145}$ \\[.5ex]
          $\widehat{147}$ \\[.5ex] $\widehat{167}$ \\[.5ex] $\widehat{345}$ \\[.5ex] $\widehat{347}$
          \\[.5ex] $\widehat{367}$ \\[.5ex] $\widehat{567}$
       \end{minipage}
          \xrightarrow[\bsy{690}]{\widehat{67}} &
 \begin{minipage}{.7cm} $\widehat{123}$ \\[.5ex] $\widehat{125}$ \\[.5ex] $\widehat{127}$ \\[.5ex] $\widehat{145}$ \\[.5ex]
          $\widehat{147}$ \\[.5ex] $\widehat{067}$ \\[.5ex] $\widehat{345}$ \\[.5ex] $\widehat{347}$
          \\[.5ex] $\widehat{267}$ \\[.5ex] $\widehat{467}$
       \end{minipage}
          \xrightarrow[\bsy{580}]{\widehat{47}} &
 \begin{minipage}{.7cm} $\widehat{123}$ \\[.5ex] $\widehat{125}$ \\[.5ex] $\widehat{127}$ \\[.5ex] $\widehat{145}$ \\[.5ex]
          $\widehat{047}$ \\[.5ex] $\widehat{067}$ \\[.5ex] $\widehat{345}$ \\[.5ex] $\widehat{247}$
          \\[.5ex] $\widehat{267}$ \\[.5ex] $\widehat{457}$
       \end{minipage}
          \xrightarrow[\bsy{470}]{\widehat{45}} &
 \begin{minipage}{.7cm} $\widehat{123}$ \\[.5ex] $\widehat{125}$ \\[.5ex] $\widehat{127}$ \\[.5ex] $\widehat{045}$ \\[.5ex]
          $\widehat{047}$ \\[.5ex] $\widehat{067}$ \\[.5ex] $\widehat{245}$ \\[.5ex] $\widehat{247}$
          \\[.5ex] $\widehat{267}$ \\[.5ex] $\widehat{456}$
       \end{minipage}
          \xrightarrow[\bsy{389}]{\widehat{27}} &
 \begin{minipage}{.7cm} $\widehat{123}$ \\[.5ex] $\widehat{125}$ \\[.5ex] $\widehat{027}$ \\[.5ex] $\widehat{045}$ \\[.5ex]
          $\widehat{047}$ \\[.5ex] $\widehat{067}$ \\[.5ex] $\widehat{245}$ \\[.5ex] $\widehat{237}$
          \\[.5ex] $\widehat{257}$ \\[.5ex] $\widehat{456}$
       \end{minipage}
           \xrightarrow[\bsy{279}]{\widehat{25}} &
  \begin{minipage}{.7cm} $\widehat{123}$ \\[.5ex] $\widehat{025}$ \\[.5ex] $\widehat{027}$ \\[.5ex] $\widehat{045}$ \\[.5ex]
          $\widehat{047}$ \\[.5ex] $\widehat{067}$ \\[.5ex] $\widehat{235}$ \\[.5ex] $\widehat{237}$
          \\[.5ex] $\widehat{256}$ \\[.5ex] $\widehat{456}$
       \end{minipage}
          \xrightarrow[\bsy{178}]{\widehat{23}} &
  \begin{minipage}{.7cm} $\widehat{023}$ \\[.5ex] $\widehat{025}$ \\[.5ex] $\widehat{027}$ \\[.5ex] $\widehat{045}$ \\[.5ex]
          $\widehat{047}$ \\[.5ex] $\widehat{067}$ \\[.5ex] $\widehat{234}$ \\[.5ex] $\widehat{236}$
          \\[.5ex] $\widehat{256}$ \\[.5ex] $\widehat{456}$
       \end{minipage}
 \end{array}
\end{gather*}
from which we can read of\/f the position (i.e., boldface) indices of $\hat{\cL}_{\bar{\alpha}}$. The Lax system reads
\begin{gather*}
   \hat{\cL}_{ \hat{k k_2},\bsy{y}_{\bsy{1}},\bsy{y}_{\bsy{4}},\bsy{y}_{\bsy{5}} }
   \hat{\cL}_{ \hat{kk_4},\bsy{y}_{\bsy{2}},\bsy{y}_{\bsy{4}},\bsy{y}_{\bsy{6}} }
   \hat{\cL}_{ \hat{kk_6},\bsy{y}_{\bsy{3}},\bsy{y}_{\bsy{5}},\bsy{y}_{\bsy{6}} }
 = \hat{\cL}_{ \hat{kk_5},\bsy{y}_{\bsy{3}},\bsy{y}_{\bsy{5}},\bsy{y}_{\bsy{6}} }
   \hat{\cL}_{ \hat{kk_3},\bsy{y}_{\bsy{2}},\bsy{y}_{\bsy{4}},\bsy{y}_{\bsy{5}} }
   \hat{\cL}_{ \hat{kk_1},\bsy{y}_{\bsy{1}},\bsy{y}_{\bsy{2}},\bsy{y}_{\bsy{3}} }  ,
\end{gather*}
where $1 \leq \bsy{y}_{\bsy{b}} \leq 10$.
The consistency condition is now obtained from\footnote{Here we depart from our notation and indicate over an
equality sign the maps that act on the arguments of the $\hat{\cL}$'s in the respective transformation step.
Furthermore, we write $\bsy{0}$ instead of $\bsy{10}$. }
\begin{gather*}
   \hat{\cL}_{\bar{\alpha}}  =  \hat{\cL}_{\widehat{23},\bsy{178}} \hat{\cL}_{\widehat{25},\bsy{279}} {\color{brown} \hat{\cL}_{\widehat{27},\bsy{389}} }
                                \hat{\cL}_{\widehat{45},\bsy{470}} {\color{brown} \hat{\cL}_{\widehat{47},\bsy{580}} \hat{\cL}_{\widehat{67},\bsy{690}} } \\
 \hphantom{\hat{\cL}_{\bar{\alpha}}}{}    \stackrel{\cP_{\bsy{3}}}{=}
                                \hat{\cL}_{\widehat{23},\bsy{178}} \hat{\cL}_{\widehat{25},\bsy{279}} \hat{\cL}_{\widehat{45},\bsy{470}}
                                {\color{brown} \hat{\cL}_{\widehat{27},\bsy{389}} \hat{\cL}_{\widehat{47},\bsy{580}} \hat{\cL}_{\widehat{67},\bsy{690}} } \\
 \hphantom{\hat{\cL}_{\bar{\alpha}}}{} \stackrel{\cT_{\hat{7},\bsy{1}}}{=}
                                \hat{\cL}_{\widehat{23},\bsy{178}} {\color{brown} \hat{\cL}_{\widehat{25},\bsy{279}} \hat{\cL}_{\widehat{45},\bsy{470}}
                                \hat{\cL}_{\widehat{57},\bsy{690}} } \hat{\cL}_{\widehat{37},\bsy{589}} \hat{\cL}_{\widehat{17},\bsy{356}} \\
 \hphantom{\hat{\cL}_{\bar{\alpha}}}{} \stackrel{\cT_{\hat{5},\bsy{3}}}{=}
                                {\color{brown} \hat{\cL}_{\widehat{23},\bsy{178}} } \hat{\cL}_{\widehat{56},\bsy{690}} {\color{brown} \hat{\cL}_{\widehat{35},\bsy{479}} }
                                \hat{\cL}_{\widehat{15},\bsy{246}} {\color{brown} \hat{\cL}_{\widehat{37},\bsy{589}} } \hat{\cL}_{\widehat{17},\bsy{356}} \\
 \hphantom{\hat{\cL}_{\bar{\alpha}}}{} \stackrel{\cP_{\bsy{2}} \cP_{\bsy{5}}}{=}
                                \hat{\cL}_{\widehat{56},\bsy{690}} {\color{brown} \hat{\cL}_{\widehat{23},\bsy{178}} \hat{\cL}_{\widehat{35},\bsy{479}}
                                \hat{\cL}_{\widehat{37},\bsy{589}} } \hat{\cL}_{\widehat{15},\bsy{246}} \hat{\cL}_{\widehat{17},\bsy{356}} \\
 \hphantom{\hat{\cL}_{\bar{\alpha}}}{} \stackrel{\cT_{\hat{3},\bsy{3}}}{=}
                                \hat{\cL}_{\widehat{56},\bsy{690}} \hat{\cL}_{\widehat{36},\bsy{589}} \hat{\cL}_{\widehat{34},\bsy{478}}
                                {\color{brown} \hat{\cL}_{\widehat{13},\bsy{145}} \hat{\cL}_{\widehat{15},\bsy{246}} \hat{\cL}_{\widehat{17},\bsy{356}} } \\
 \hphantom{\hat{\cL}_{\bar{\alpha}}}{} \stackrel{\cT_{\hat{1},\bsy{1}}}{=}
                                \hat{\cL}_{\widehat{56},\bsy{690}} \hat{\cL}_{\widehat{36},\bsy{589}} \hat{\cL}_{\widehat{34},\bsy{478}}
                                \hat{\cL}_{\widehat{16},\bsy{356}} \hat{\cL}_{\widehat{14},\bsy{245}} \hat{\cL}_{\widehat{12},\bsy{123}}  ,
\end{gather*}
which is $\hat{\cL}_{\bar{\alpha}} = \hat{\cL}_{\bar{\omega}} \circ \cT_{\tilde{\cC}_o}$, and
\begin{gather*}
   \hat{\cL}_{\bar{\alpha}} = {\color{brown} \hat{\cL}_{\widehat{23},\bsy{178}} \hat{\cL}_{\widehat{25},\bsy{279}} \hat{\cL}_{\widehat{27},\bsy{389}} }
                                \hat{\cL}_{\widehat{45},\bsy{470}} \hat{\cL}_{\widehat{47},\bsy{580}} \hat{\cL}_{\widehat{67},\bsy{690}} \\
\hphantom{\hat{\cL}_{\bar{\alpha}}}{} \stackrel{\cT_{\hat{2},\bsy{4}}}{=}
                                \hat{\cL}_{\widehat{26},\bsy{389}} {\color{brown} \hat{\cL}_{\widehat{24},\bsy{278}} } \hat{\cL}_{\widehat{12},\bsy{123}}
                                {\color{brown} \hat{\cL}_{\widehat{45},\bsy{470}} \hat{\cL}_{\widehat{47},\bsy{580}} } \hat{\cL}_{\widehat{67},\bsy{690}} \\
\hphantom{\hat{\cL}_{\bar{\alpha}}}{} \stackrel{\cP_{\bsy{2}} \cP_{\bsy{3}}}{=}
                                \hat{\cL}_{\widehat{26},\bsy{389}} {\color{brown} \hat{\cL}_{\widehat{24},\bsy{278}} \hat{\cL}_{\widehat{45},\bsy{470}}
                                \hat{\cL}_{\widehat{47},\bsy{580}} } \hat{\cL}_{\widehat{12},\bsy{123}} \hat{\cL}_{\widehat{67},\bsy{690}} \\
\hphantom{\hat{\cL}_{\bar{\alpha}}}{} \stackrel{\cT_{\hat{4},\bsy{3}}}{=}
                                {\color{brown} \hat{\cL}_{\widehat{26},\bsy{389}} \hat{\cL}_{\widehat{46},\bsy{580}} } \hat{\cL}_{\widehat{34},\bsy{478}}
                                \hat{\cL}_{\widehat{14},\bsy{245}} \hat{\cL}_{\widehat{12},\bsy{123}} {\color{brown} \hat{\cL}_{\widehat{67},\bsy{690}} } \\
\hphantom{\hat{\cL}_{\bar{\alpha}}}{} \stackrel{\cP_{\bsy{3}} \cP_{\bsy{2}} \cP_{\bsy{1}}}{=}
                                {\color{brown} \hat{\cL}_{\widehat{26},\bsy{389}} \hat{\cL}_{\widehat{46},\bsy{580}} \hat{\cL}_{\widehat{67},\bsy{690}} }
                                \hat{\cL}_{\widehat{34},\bsy{478}} \hat{\cL}_{\widehat{14},\bsy{245}} \hat{\cL}_{\widehat{12},\bsy{123}} \\
\hphantom{\hat{\cL}_{\bar{\alpha}}}{} \stackrel{\cT_{\hat{6},\bsy{4}}}{=}
                                \hat{\cL}_{\widehat{56},\bsy{690}} \hat{\cL}_{\widehat{36},\bsy{589}} \hat{\cL}_{\widehat{16},\bsy{356}}
                                {\color{brown} \hat{\cL}_{\widehat{34},\bsy{478}} } \hat{\cL}_{\widehat{14},\bsy{245}} \hat{\cL}_{\widehat{12},\bsy{123}} \\
\hphantom{\hat{\cL}_{\bar{\alpha}}}{} \stackrel{\cP_{\bsy{3}}}{=}
                                \hat{\cL}_{\widehat{56},\bsy{690}} \hat{\cL}_{\widehat{36},\bsy{589}} \hat{\cL}_{\widehat{34},\bsy{478}}
                                \hat{\cL}_{\widehat{16},\bsy{356}} \hat{\cL}_{\widehat{14},\bsy{245}} \hat{\cL}_{\widehat{12},\bsy{123}} ,
\end{gather*}
which is $\hat{\cL}_{\bar{\alpha}} = \hat{\cL}_{\bar{\omega}} \circ \cT_{\tilde{\cC}_e}$. Hence
\begin{gather*}
      \cT_{\tilde{\cC}_o}
    = \cT_{\hat{1},\bsy{1}} \cT_{\hat{3},\bsy{3}} \cP_{\bsy{2}} \cP_{\bsy{5}} \cT_{\hat{5},\bsy{3}} \cT_{\hat{7},\bsy{1}} \cP_{\bsy{3}}
    = \cP_{\bsy{3}} \cT_{\hat{6},\bsy{4}} \cP_{\bsy{3}} \cP_{\bsy{2}} \cP_{\bsy{1}} \cT_{\hat{4},\bsy{3}} \cP_{\bsy{2}} \cP_{\bsy{3}} \cT_{\hat{2},\bsy{4}}
    = \cT_{\tilde{\cC}_e} ,
\end{gather*}
which is the heptagon equation.
\end{exam}

\begin{exam}
\label{ex:pentagon-Lax->hexagon}
Let $N=5$. Then $\bar{\alpha} = (\widehat{56},\widehat{36},\widehat{34},\widehat{16},\widehat{14}, \widehat{12}) \in A^{(b)}(6,4)$
and $\bar{\eta} = (\widehat{456}, \widehat{256}, \widehat{236}$, $\widehat{234}) \in A^{(b)}(6,3)$.
We thus have the chain
\begin{gather*}
 \begin{array}{@{}c@{\,}c@{\,}c@{\,}c@{\,}c@{\,}c@{\,}c@{\,}c@{\,}c@{\,}c@{\,}c@{\,}c@{\,}c@{}}
 \begin{minipage}{.7cm} $\widehat{456}$ \\[.5ex] $\widehat{256}$ \\[.5ex] $\widehat{236}$ \\[.5ex] $\widehat{234}$
       \end{minipage}
          \xrightarrow[\bsy{12}]{\widehat{56}} &
 \begin{minipage}{.7cm} $\widehat{156}$ \\[.5ex] $\widehat{356}$ \\[.5ex] $\widehat{236}$ \\[.5ex] $\widehat{234}$
       \end{minipage}
          \xrightarrow[\bsy{23}]{\widehat{36}} &
 \begin{minipage}{.7cm} $\widehat{156}$ \\[.5ex] $\widehat{136}$ \\[.5ex] $\widehat{346}$ \\[.5ex] $\widehat{234}$
       \end{minipage}
          \xrightarrow[\bsy{34}]{\widehat{34}} &
 \begin{minipage}{.7cm} $\widehat{156}$ \\[.5ex] $\widehat{136}$ \\[.5ex] $\widehat{134}$ \\[.5ex] $\widehat{345}$
       \end{minipage}
          \xrightarrow[\bsy{12}]{\widehat{16}} &
 \begin{minipage}{.7cm} $\widehat{126}$ \\[.5ex] $\widehat{146}$ \\[.5ex] $\widehat{134}$ \\[.5ex] $\widehat{345}$
       \end{minipage}
          \xrightarrow[\bsy{23}]{\widehat{14}} &
  \begin{minipage}{.7cm} $\widehat{126}$ \\[.5ex] $\widehat{124}$ \\[.5ex] $\widehat{145}$ \\[.5ex] $\widehat{345}$
       \end{minipage}
          \xrightarrow[\bsy{12}]{\widehat{12}} &
  \begin{minipage}{.7cm} $\widehat{123}$ \\[.5ex] $\widehat{125}$ \\[.5ex] $\widehat{145}$ \\[.5ex] $\widehat{345}$
       \end{minipage}
 \end{array}
\end{gather*}
from which we can read of\/f the position indices in the expression
\begin{gather*}
   \cL_{\bar{\alpha}} = \cL_{\widehat{12},\bsy{1}} \cL_{\widehat{14},\bsy{2}} \cL_{\widehat{16},\bsy{1}}
                        \cL_{\widehat{34},\bsy{3}} \cL_{\widehat{36},\bsy{2}} \cL_{\widehat{56},\bsy{1}}   .
\end{gather*}
The Lax system (\ref{polygon_Lax_system}) consists of the localized pentagon equations
\begin{gather*}
    \cL_{\widehat{kk_1},\bsy{a}} \cL_{\widehat{kk_3},\bsy{a+1}} \cL_{\widehat{kk_5},\bsy{a}}
    = \Big( \cL_{\widehat{kk_4},\bsy{a}+1} \cP_{\bsy{a}} \cL_{\widehat{kk_2},\bsy{a}+1} \Big) \circ \cT_{\hat{k}}   ,
\end{gather*}
where $k \in [6]$, $\hat{k} = \{k_1,\ldots,k_5\}$ with $k_1 < \cdots < k_5$.
The consistency condition is now obtained as follows. We have
\begin{gather*}
       \cL_{\bar{\alpha}}
    =  \cL_{\widehat{12},\bsy{1}} \cL_{\widehat{14},\bsy{2}} {\color{brown} \cL_{\widehat{16},\bsy{1}} }
       \cL_{\widehat{34},\bsy{3}} {\color{brown} \cL_{\widehat{36},\bsy{2}} \cL_{\widehat{56},\bsy{1}} }
   \stackrel{\cP_{\bsy{3}}}{=}
       \cL_{\widehat{12},\bsy{1}} \cL_{\widehat{14},\bsy{2}} \cL_{\widehat{34},\bsy{3}}
       {\color{brown} \cL_{\widehat{16},\bsy{1}} \cL_{\widehat{36},\bsy{2}} \cL_{\widehat{56},\bsy{1}} } \\
\hphantom{\cL_{\bar{\alpha}}}{} \stackrel{\cT_{\hat{6},\bsy{1}}}{=}
       \cL_{\widehat{12},\bsy{1}} {\color{brown} \cL_{\widehat{14},\bsy{2}} \cL_{\widehat{34},\bsy{3}}
       \cL_{\widehat{46},\bsy{2}} } \cP_{\bsy{1}} \cL_{\widehat{26},\bsy{2}}
    \stackrel{\cT_{\hat{4},\bsy{2}}}{=}
       {\color{brown} \cL_{\widehat{12},\bsy{1}} } \cL_{\widehat{45},\bsy{3}}  \cP_{\bsy{2}}
       {\color{brown} \cL_{\widehat{24},\bsy{3}} } \cP_{\bsy{1}}
       {\color{brown} \cL_{\widehat{26},\bsy{2}} } \\
\hphantom{\cL_{\bar{\alpha}}}{} \stackrel{\cP_{\bsy{3}}}{=}
       \cL_{\widehat{45},\bsy{3}} {\color{brown} \cL_{\widehat{12},\bsy{1}} } \cP_{\bsy{2}}
       {\color{brown} \cL_{\widehat{24},\bsy{3}} } \cP_{\bsy{1}}
       {\color{brown} \cL_{\widehat{26},\bsy{2}} }
    = \cL_{\widehat{45},\bsy{3}} {\color{brown} \cL_{\widehat{12},\bsy{1}} } \cP_{\bsy{2}} \cP_{\bsy{1}}
       {\color{brown} \cL_{\widehat{24},\bsy{3}} \cL_{\widehat{26},\bsy{2}} } \\
\hphantom{\cL_{\bar{\alpha}}}{} =  \cL_{\widehat{45},\bsy{3}} \cP_{\bsy{2}} \cP_{\bsy{1}} {\color{brown} \cL_{\widehat{12},\bsy{2}}
       \cL_{\widehat{24},\bsy{3}} \cL_{\widehat{26},\bsy{2}} }
   \stackrel{\cT_{\hat{2},\bsy{1}}}{=}
       \cL_{\widehat{45},\bsy{3}} \cP_{\bsy{2}} \cP_{\bsy{1}} \cL_{\widehat{25},\bsy{3}} \cP_{\bsy{2}}
       \cL_{\widehat{23},\bsy{3}}   ,
\end{gather*}
which is $\cL_{\bar{\alpha}} = \cL_{\bar{\omega}} \circ \cT_{\tilde{\cC}_o}$.
This yields the left hand side of the hexagon equation~(\ref{6-gon_eq}). The right hand side of~(\ref{6-gon_eq})
is obtained if we proceed according to the chain $\cC_e$. The full compositions of maps appearing in this computation
can be visualized as maximal chains of~$T(6,2)$, which forms a $4$-hypercube.
For larger values of~$N$, the corresponding computation, based on (\ref{polygon_Lax_system}), becomes very complicated.
This is in contrast to the derivation of the consistency condition using the hatted version of the Lax system.
\end{exam}

\begin{rem}
\label{rem:extended_braid_group}
The derivation of the hexagon equation for the maps $\cT_{\hat{j}}$ in Example~\ref{ex:pentagon-Lax->hexagon}
still works if $\cP$ is \emph{any} braiding map (solution of the Yang--Baxter equation), provided that the
following relations hold,
\begin{gather*}
  \cL_{\bsy{a}}   \cL_{\bsy{a}+1}   \cL_{\bsy{a}} = \cL_{\bsy{a}+1}   \cP_{\bsy{a}}   \cL_{\bsy{a}+1}   , \qquad
      \cP_{\bsy{a}}   \cP_{\bsy{a}+1}   \cP_{\bsy{a}} = \cP_{\bsy{a}+1}   \cP_{\bsy{a}}   \cP_{\bsy{a}+1}   , \\
  \cL_{\bsy{a}}   \cP_{\bsy{a}+1}   \cP_{\bsy{a}} = \cP_{\bsy{a}+1}   \cP_{\bsy{a}}   \cL_{\bsy{a}+1}   , \qquad
      \cP_{\bsy{a}}   \cP_{\bsy{a}+1}   \cL_{\bsy{a}} = \cL_{\bsy{a}+1}   \cP_{\bsy{a}}   \cP_{\bsy{a}+1}   , \\
  \cP_{\bsy{a}}   \cL_{\bsy{b}} = \cL_{\bsy{b}}   \cP_{\bsy{a}} \qquad \mbox{for} \quad |\bsy{a}-\bsy{b}| > 1    ,
\end{gather*}
which determine an extension of the braid group. If~$\cP$ is the transposition map, as assumed in this work
outside of this remark, these relations become identities, with the exception of the f\/irst, the pentagon equation.
\end{rem}

\subsection{Reductions of polygon equations}
\label{subsec:polygon_red}
\begin{figure}[t]\centering
\includegraphics[width=.4\linewidth]{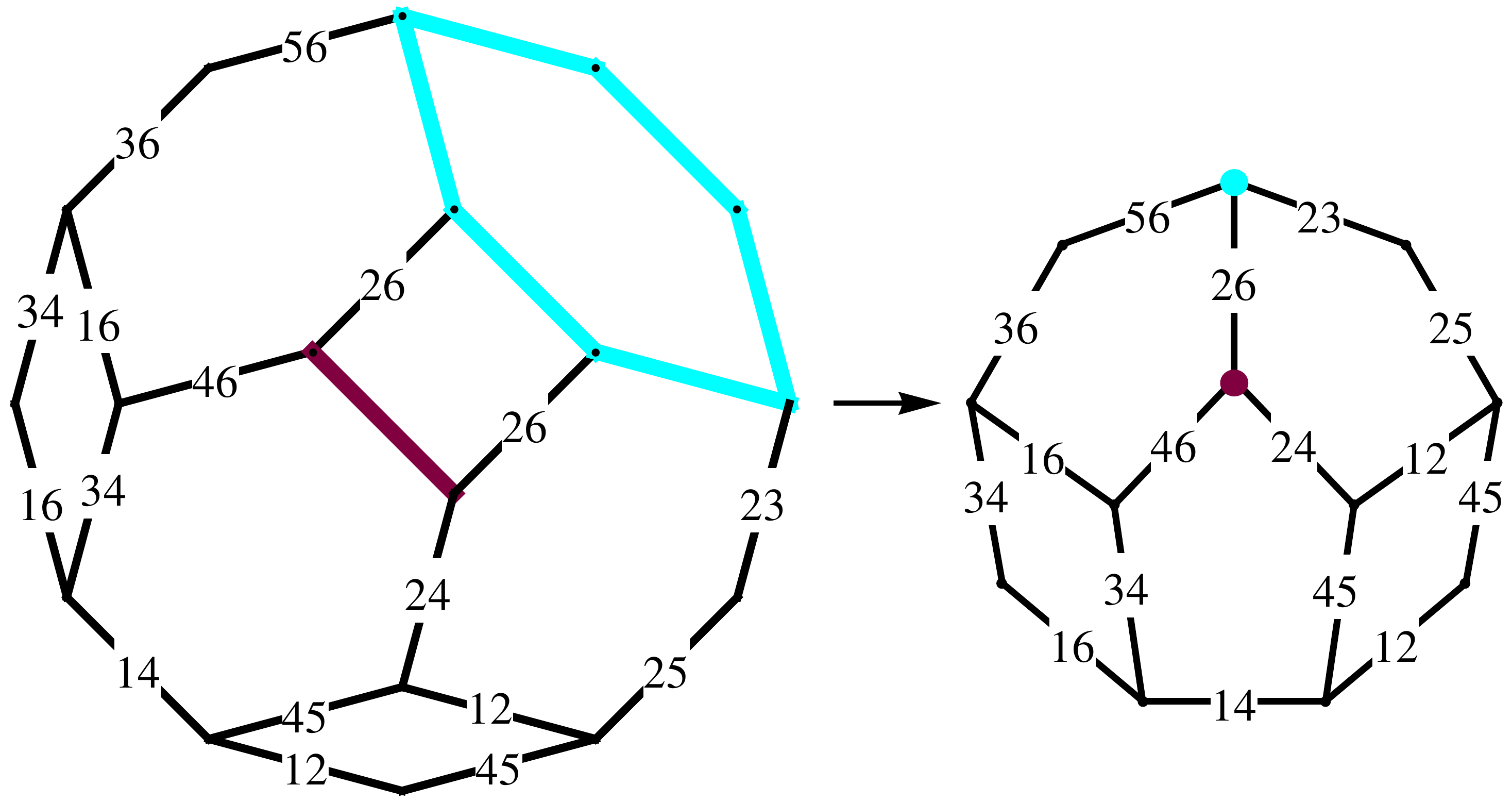}
\hspace{1cm}
\includegraphics[width=.4\linewidth]{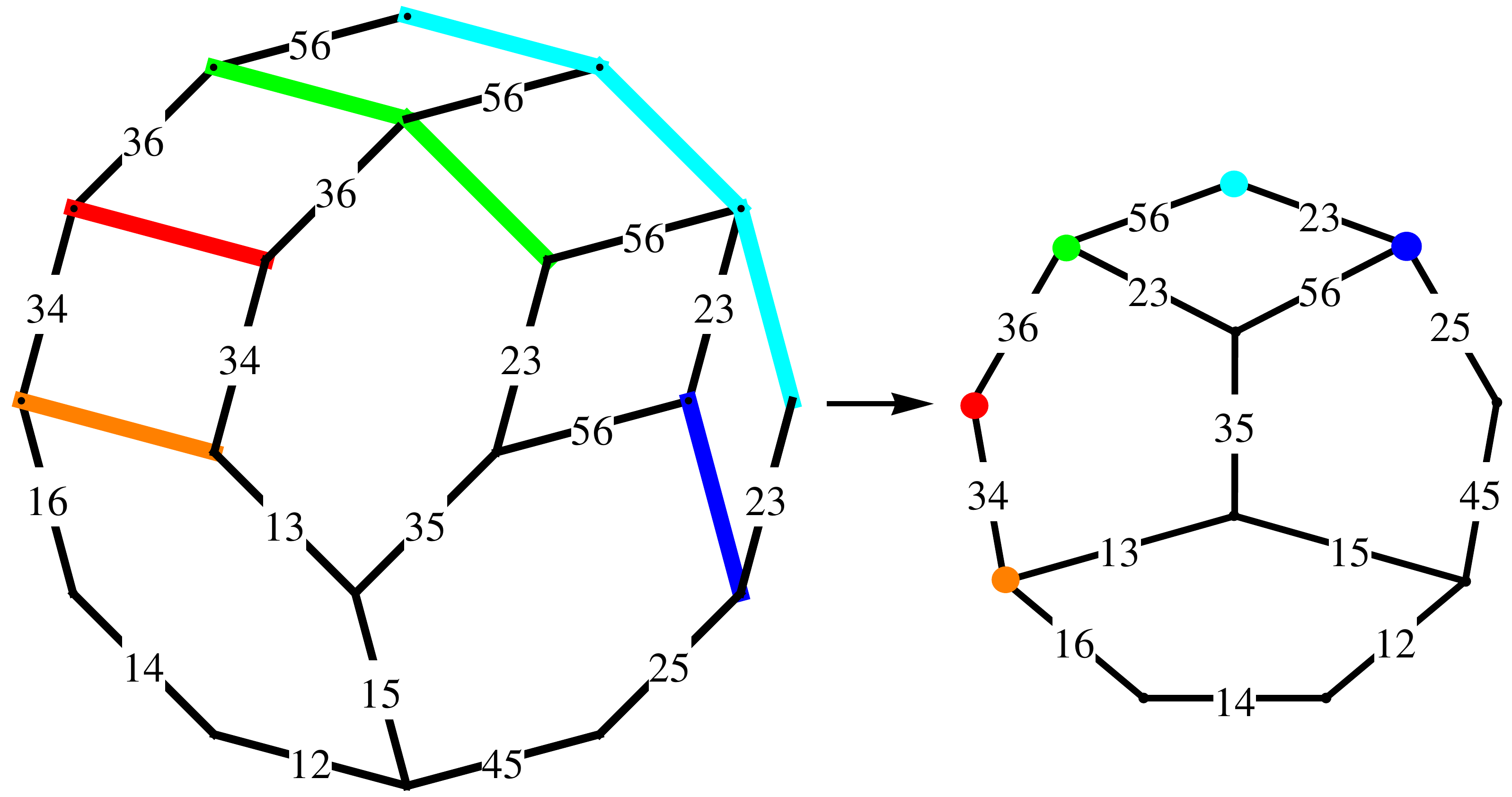}
\caption{Projection of $T(7,4)$ (Edelman--Reiner polyhedron) to $T(6,3)$ (associahedron). We use complementary labeling.
\label{fig:T74_red} }
\end{figure}

\begin{figure}[t]\centering
\includegraphics[width=.45\linewidth]{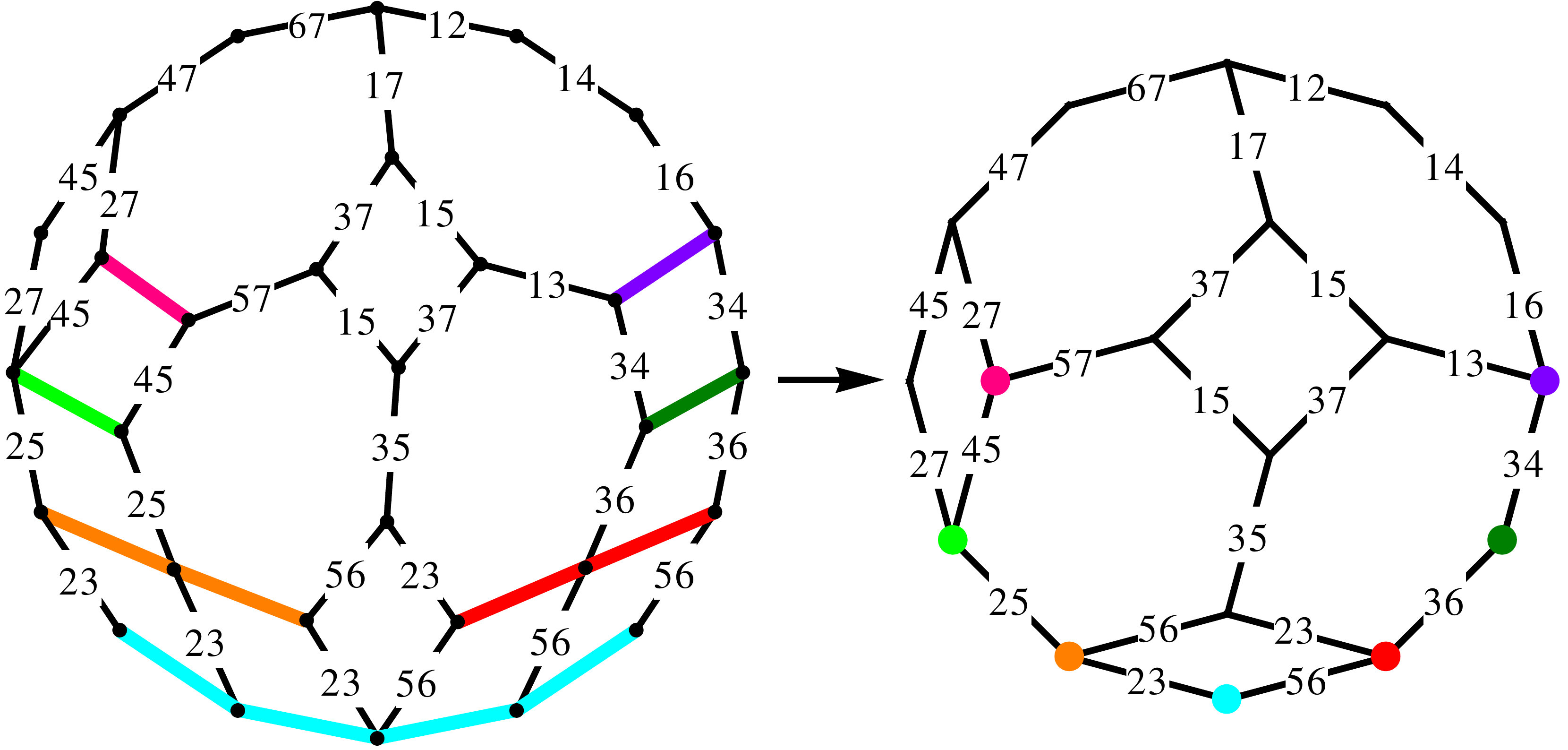}
\hspace{.5cm}
\includegraphics[width=.45\linewidth]{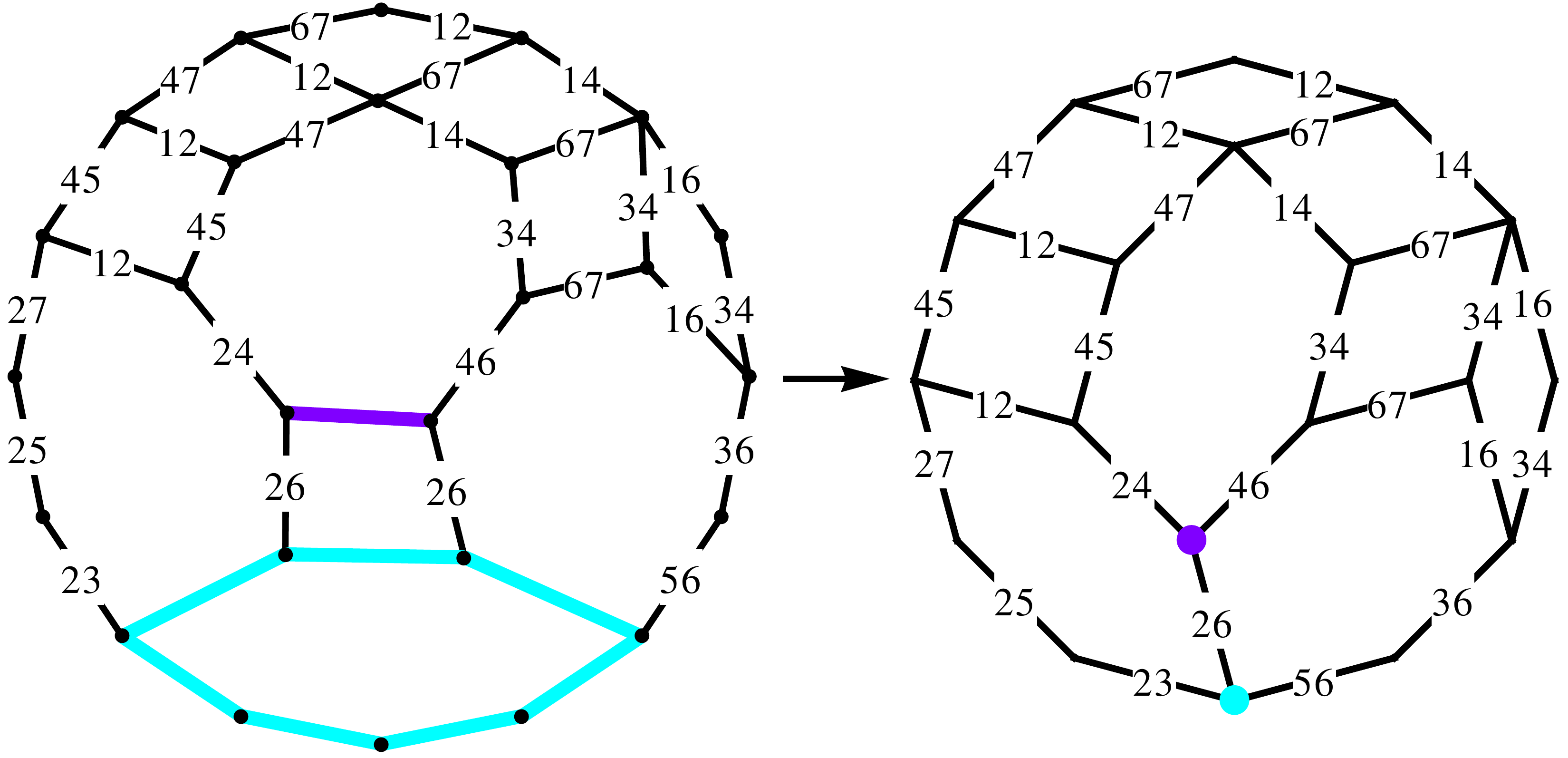}
\caption{Projection of $T(8,5)$ to $T(7,4)$. As in Fig.~\ref{fig:T74_red}, the coloring
marks those parts of the two Tamari orders that are related by the projection.
\label{fig:T85_red} }
\end{figure}

The relation between the Tamari order $T(N,N-2)$ and the $N$-gon equation,
together with the projection of Tamari orders def\/ined in Remark~\ref{rem:Tamari_red}, induces
a relation between neighboring polygon equations:
\begin{gather*}
    \begin{array}{@{}ccc@{}} T(N,N-2) & \longleftrightarrow & \mbox{$N$-gon equation} \\
                           \downarrow &     &   \downarrow  \\
                       T(N-1,N-3) & \longleftrightarrow & \mbox{$(N-1)$-gon equation}
    \end{array}
\end{gather*}
But we have to consider the projection $T(N,N-3) \rightarrow T(N-1,N-4)$ (for $N>4$) in order to
display the full structure of the corresponding polygon equations. Examples are shown in
Figs.~\ref{fig:T74_red} and \ref{fig:T85_red}.
We use the set $\{0,1,2,3,4,5,N-1\}$ instead of $\{1,2,3,4,5,6,N\}$.
Unlike the case of simplex equations, there is a substantial dif\/ference between odd and
even polygon equations.

Let $N$ be odd, i.e., $N=2n+1$. Setting
\begin{gather*}
    \hat{\cT}^{(2n+1)}_{\hat{j}} := \hat{\cT}^{(2n)}_{\hat{j}}, \qquad   j=1,\ldots,2n   ,
\end{gather*}
and choosing for
\begin{gather*}
    \hat{\cT}^{(2n+1)}_{\hat{0}} \colon \
    \cU_{\widehat{02}} \times \cU_{\widehat{04}} \times \cdots \times \cU_{\widehat{0 (2n)}}
   \longrightarrow
   \cU_{\widehat{01}} \times \cU_{\widehat{03}} \times \cdots \cU_{\widehat{0 (2n-1)}}
\end{gather*}
the identity map\footnote{Recall the identif\/ications made in the def\/inition of $\hat{\cT}^{(2n)}$.},
it follows that $\hat{\cT}^{(2n+1)}_{\hat{j}}$, $j=0,\ldots,2n$, satisfy the $(2n+1)$-gon equation.

\begin{exam}
The heptagon equation (with labels $0,1,\ldots,6$) reads
\begin{gather*}
    \hat{\cT}^{(7)}_{\hat{0},\bsy{123}} \hat{\cT}^{(7)}_{\hat{2},\bsy{145}} \hat{\cT}^{(7)}_{\hat{4},\bsy{246}} \hat{\cT}^{(7)}_{\hat{6},\bsy{356}}
  = \hat{\cT}^{(7)}_{\hat{5},\bsy{356}} \hat{\cT}^{(7)}_{\hat{3},\bsy{245}} \hat{\cT}^{(7)}_{\hat{1},\bsy{123}}   .
\end{gather*}
If $\hat{\cT}^{(7)}_{\hat{0}}$ is the identity map, this reduces to the hexagon equation
\begin{gather*}
    \hat{\cT}^{(6)}_{\hat{2},\bsy{145}} \hat{\cT}^{(6)}_{\hat{4},\bsy{246}} \hat{\cT}^{(6)}_{\hat{6},\bsy{356}}
  = \hat{\cT}^{(6)}_{\hat{5},\bsy{356}} \hat{\cT}^{(6)}_{\hat{3},\bsy{245}} \hat{\cT}^{(6)}_{\hat{1},\bsy{123}} \, .
\end{gather*}
\end{exam}

Let now $N$ be even, i.e., $N=2n$. For $j \in [2n]$, $j \neq 1$, we have
\begin{gather*}
    \hat{\cT}^{(2n)}_{\hat{j}} = \big(u_{\widehat{0j}}, \cT^{(2n)}_{\hat{j}}   \cP_{\hat{j}}\big)   ,
\end{gather*}
where $\hat{j} = \{ 1,j_2,\ldots,j_{2n-1} \}$ with $1 < j_2 < \cdots < j_{2n-1}$, and $u_{\widehat{0j}} \in \cU_{\widehat{0j}}$.
These are maps
\begin{gather*}
   \hat{\cT}^{(2n)}_{\hat{j}} \colon \
   \cU_{\widehat{1j}} \times \cU_{\widehat{jj_3}} \times \cdots \times \cU_{\widehat{j j_{2n-1}}}
   \longrightarrow
   \cU_{\widehat{0j}} \times \cU_{\widehat{jj_2}} \times \cU_{\widehat{jj_4}} \times \cdots \times \cU_{\widehat{j j_{2n-2}}}   .
\end{gather*}
Ignoring the f\/irst argument of these maps, the $2n$-gon equation implies that the resulting maps have to satisfy
the $(2n-1)$-gon equation.
Conversely, let $\hat{\cT}^{(2n-1)}_{\hat{j}}$, $j=2,3,\ldots,2n$, satisfy the $(2n-1)$-gon equation. We extend these maps
trivially via
\begin{gather*}
    \hat{\cT}^{(2n)}_{\hat{j}}\big(u_{\widehat{1j}},u_{\widehat{jj_3}},\ldots,u_{\widehat{jj_{2n-1}}}\big) :=
    \hat{\cT}^{(2n-1)}_{\hat{j}}\big(u_{\widehat{jj_3}},\ldots,u_{\widehat{jj_{2n-1}}}\big)   .
\end{gather*}
Furthermore, we assume that $\cU_{\widehat{01}} = \cU_{\widehat{12}}$ and $\cU_{\widehat{1(2j-1)}} = \cU_{\widehat{1(2j)}}$,
$j=2,3,\ldots,n$, and we choose for
$\hat{\cT}^{(2n)}_{\hat{1}}$ the identity map. Then, after dropping the f\/irst position index in all terms, we f\/ind that
$\hat{\cT}^{(2n)}_{\hat{j}}$, $j=1,\ldots,2n$, solve the $2n$-gon equation, with hatted indices shifted by 1, and
with position indices shifted by~$n$.
In this way, the $2n$-gon equation reduces to the $(2n-1)$-gon equation.

\begin{exam}
The octagon equation
\begin{gather*}
    \hat{\cT}^{(8)}_{\hat{2},\bsy{1},\bsy{5},\bsy{6},\bsy{7}} \hat{\cT}^{(8)}_{\hat{4},\bsy{2},\bsy{5},\bsy{8},\bsy{9}}
    \hat{\cT}^{(8)}_{\hat{6},\bsy{3},\bsy{6},\bsy{8},\bsy{10}} \hat{\cT}^{(8)}_{\hat{8},\bsy{4},\bsy{7},\bsy{9},\bsy{10}}
  = \hat{\cT}^{(8)}_{\hat{7},\bsy{4},\bsy{7},\bsy{9},\bsy{10}} \hat{\cT}^{(8)}_{\hat{5},\bsy{3},\bsy{6},\bsy{8},\bsy{9}}
    \hat{\cT}^{(8)}_{\hat{3},\bsy{2},\bsy{5},\bsy{6},\bsy{7}} \hat{\cT}^{(8)}_{\hat{1},\bsy{1},\bsy{2},\bsy{3},\bsy{4}}
\end{gather*}
reduces in the way described above to the heptagon equation (with shifted labels)
\begin{gather*}
    \hat{\cT}^{(7)}_{\hat{2},\bsy{5},\bsy{6},\bsy{7}} \hat{\cT}^{(7)}_{\hat{4},\bsy{5},\bsy{8},\bsy{9}}
    \hat{\cT}^{(7)}_{\hat{6},\bsy{6},\bsy{8},\bsy{10}} \hat{\cT}^{(7)}_{\hat{8},\bsy{7},\bsy{9},\bsy{10}}
  = \hat{\cT}^{(7)}_{\hat{7},\bsy{7},\bsy{9},\bsy{10}} \hat{\cT}^{(7)}_{\hat{5},\bsy{6},\bsy{8},\bsy{9}}
    \hat{\cT}^{(7)}_{\hat{3},\bsy{5},\bsy{6},\bsy{7}}    .
\end{gather*}
\end{exam}

\section{Three color decomposition of simplex equations}
\label{sec:simplex_polygon_rel}
The existence of a decomposition of a Bruhat order into a Tamari order, the corresponding
dual Tamari order, and a mixed order, suggests that there should be a way to construct solutions of a
simplex equation from solutions of the respective polygon equation and its dual, provided a
compatibility condition, associated with the mixed order, is fulf\/illed.
As in Section~\ref{sec:polygon_eqs}, we associate with $K \in {[N] \choose N-1}$ a map
$\cT_K$ and a dual map $\cS_K$. We set
\begin{gather}
      \cR_K := \cP''   (\cT_K \times \cS_K)   \cP'   ,   \label{R<-T,S}
\end{gather}
where $\cP'$, $\cP''$ are compositions of transposition maps achieving the necessary shuf\/f\/ling
of $\cU_{\vec{P_o}(K)}$ and $\cU_{\vec{P_e}(K)}$, respectively of $\cU_{\cev{P_o}(K)}$
and $\cU_{\cev{P_e}(K)}$, so that $\cR_K$ has the correct structure
of a simplex map $\cU_{\vec{P}(K)} \rightarrow \cU_{\cev{P}(K)}$.
The $(N-1)$-simplex equation then indeed reduces to the $N$-gon equation for~$\cT_K$ and the dual
$N$-gon equation for $\cS_K$, and an additional compatibility condition. This includes
one of the results in~\cite{Kash+Serg98}: special solutions of the $4$-simplex equation
can be constructed from solutions of the pentagon equation and its dual.
The corresponding compatibility condition is~(1.7) in~\cite{Kash+Serg98}.

\begin{figure}[t]\centering
\includegraphics[width=.4\linewidth]{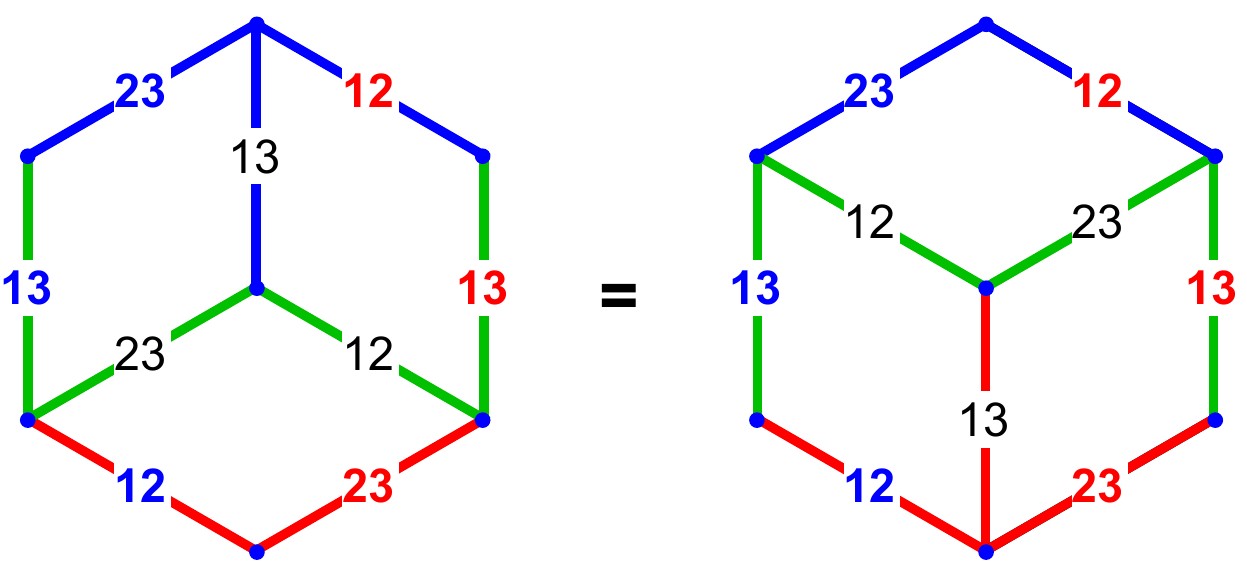}

\vspace{.2cm}

\includegraphics[width=.2\linewidth]{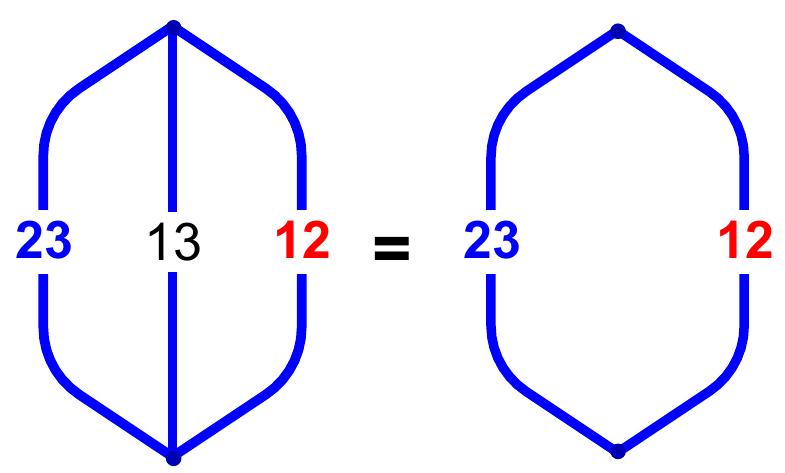}
\hspace{.4cm}
\includegraphics[width=.2\linewidth]{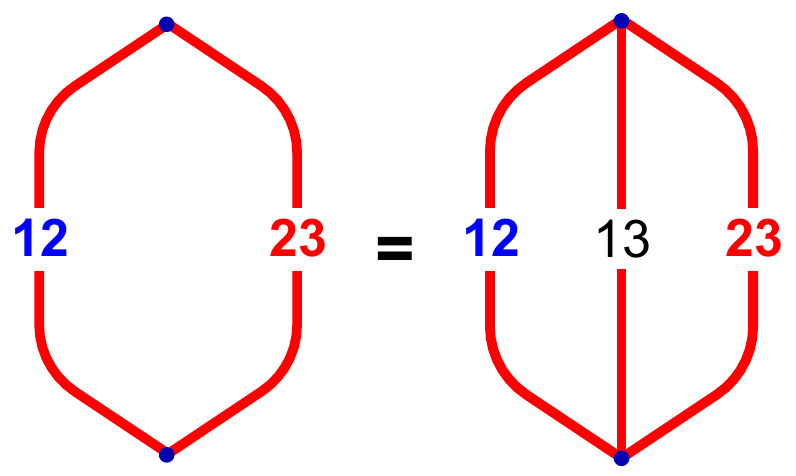}
\hspace{.4cm}
\includegraphics[width=.2\linewidth]{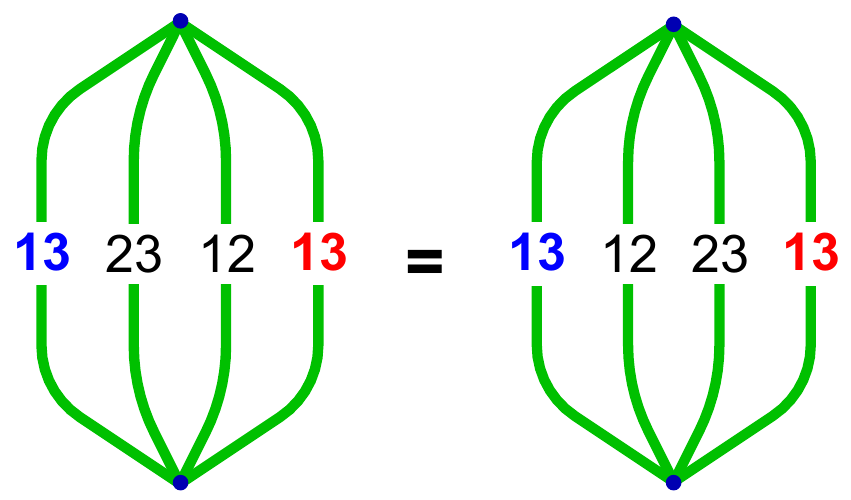}
\caption{Decomposition of the Yang--Baxter equation viewed on the cube $B(3,0)$ (with
complementary edge labeling, but hats omitted). Following the action of the composition of
Yang--Baxter maps on the left and the right-hand side of the Yang--Baxter equation, we observe
that the action splits as indicated by the three dif\/ferent colors.
A set of edges having the same color corresponds to one of three equations, represented by
the graphs in the second row (where all vertices connected by edges having a dif\/ferent color have
been identif\/ied). Blue: trigon equation, red: dual trigon equation, green: compatibility condition.
Here, and also in some of the following f\/igures, labels of edges of an initial (f\/inal)
maximal chain are marked blue (red), which has no further meaning.
\label{fig:2simplex_to_trigon_reduction} }
\end{figure}

\paragraph{2-simplex and trigon equation.}
If $N=3$, we consider maps $\cR_{ij} \colon \cU_i \times \cU_j \rightarrow \cU_j \times \cU_i$,
$\cT_{ij}\colon \cU_i \rightarrow \cU_j$, $\cS_{ij} \colon \cU_j \rightarrow \cU_i$.
In complementary notation, we have, for example,
$\cR_{\hat{3}} \colon \cU_{\widehat{23}} \times \cU_{\widehat{13}} \rightarrow \cU_{\widehat{13}} \times \cU_{\widehat{23}}$,
$\cT_{\hat{3}} \colon \cU_{\widehat{23}} \rightarrow \cU_{\widehat{13}}$, and
$\cS_{\hat{3}} \colon \cU_{\widehat{13}} \rightarrow \cU_{\widehat{23}}$.
We set $\cR_{\hat{k}} = \cT_{\hat{k}} \times \cS_{\hat{k}}$, hence
$\cR_{\hat{k},\bsy{a}} = \cT_{\hat{k},\bsy{a}}   \cS_{\hat{k},\bsy{a}+1}$.
The $2$-simplex equation
$\cR_{\hat{1},\bsy{1}}   \cR_{\hat{2},\bsy{2}}   \cR_{\hat{3},\bsy{1}}
 = \cR_{\hat{3},\bsy{2}}   \cR_{\hat{2},\bsy{1}}   \cR_{\hat{1},\bsy{2}}$ then
becomes
\begin{gather*}
    \cT_{\hat{1},\bsy{1}}   \cS_{\hat{1},\bsy{2}}  \cT_{\hat{2},\bsy{2}}   \cS_{\hat{2},\bsy{3}}
       \cT_{\hat{3},\bsy{1}}   \cS_{\hat{3},\bsy{2}}
  = \cT_{\hat{3},\bsy{2}}   \cS_{\hat{3},\bsy{3}}  \cT_{\hat{2},\bsy{1}}   \cS_{\hat{2},\bsy{2}}
       \cT_{\hat{1},\bsy{2}}   \cS_{\hat{1},\bsy{3}}   ,
\end{gather*}
which splits into
\begin{gather*}
    \cT_{\hat{1}}   \cT_{\hat{3}} = \cT_{\hat{2}}   , \qquad
    \cS_{\hat{2}} = \cS_{\hat{3}}   \cS_{\hat{1}}    , \qquad
    \cS_{\hat{1}}   \cT_{\hat{2}}   \cS_{\hat{3}} = \cT_{\hat{3}}   \cS_{\hat{2}}   \cT_{\hat{1}}   .
\end{gather*}
The f\/irst two are the trigon equation and its dual. The last equation is an additional condition.
A graphical representation of this ``decomposition'' of the Yang--Baxter equation is shown in
Fig.~\ref{fig:2simplex_to_trigon_reduction}.

\begin{figure}[t]\centering
\includegraphics[width=.6\linewidth]{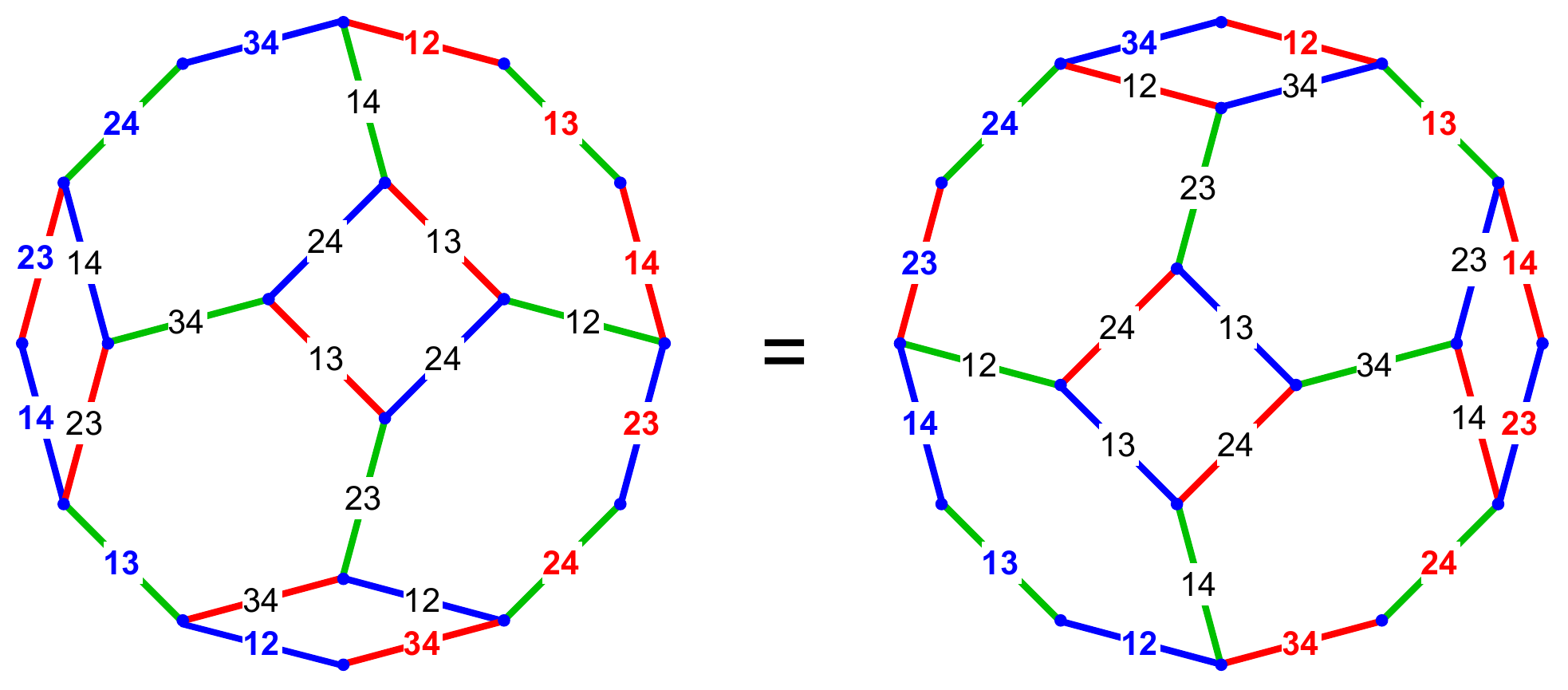}

\vspace{.2cm}

\includegraphics[width=.28\linewidth]{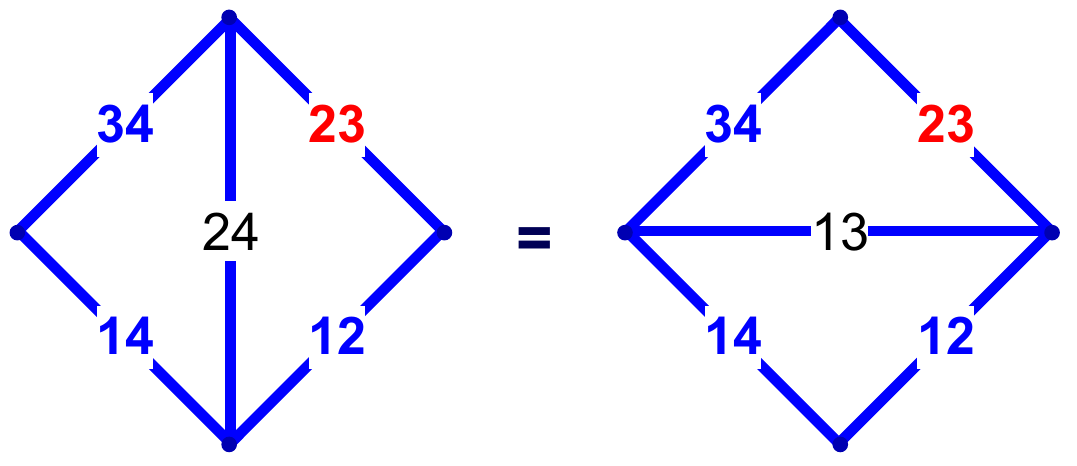}
\hspace{.2cm}
\includegraphics[width=.28\linewidth]{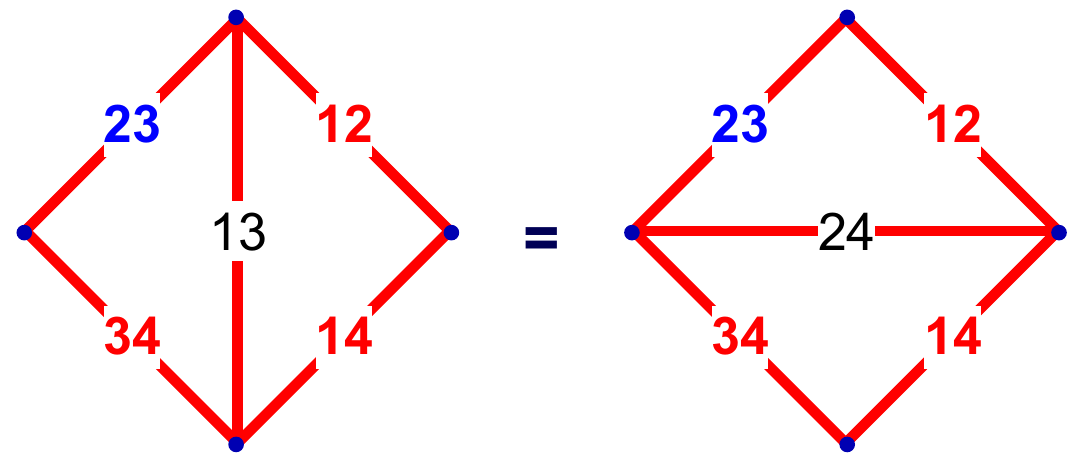}
\hspace{.2cm}
\includegraphics[width=.28\linewidth]{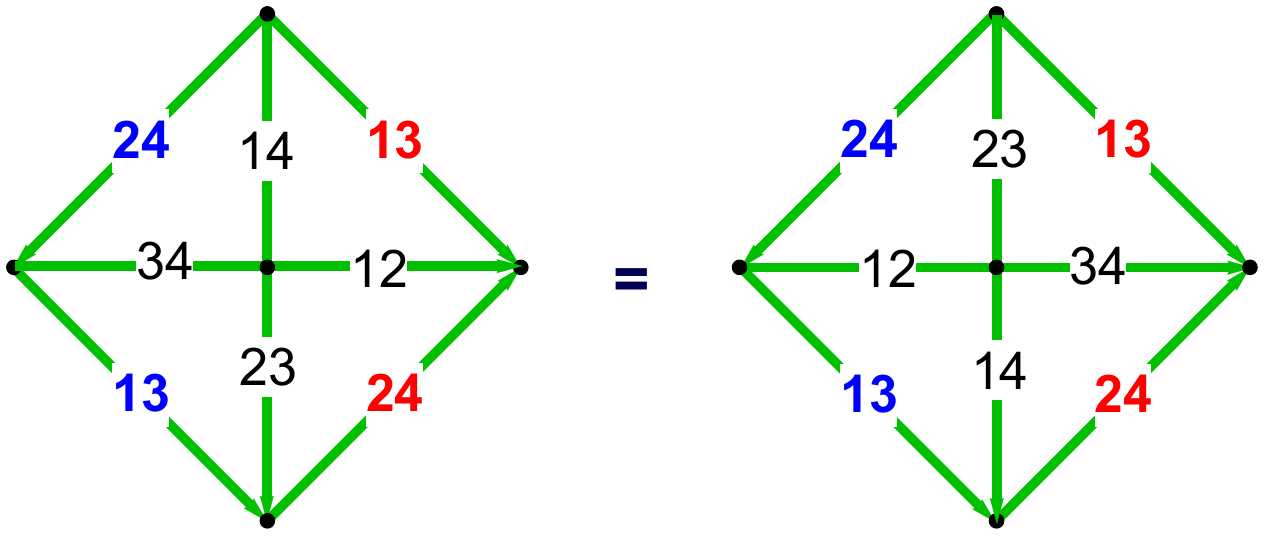}

\caption{Decomposition of the $3$-simplex (tetrahedron, Zamolodchikov) equation, viewed on
the permutahedron formed by $B(4,1)$.
The second row of f\/igures graphically represents the resulting tetragon equation,
dual tetragon equation, and the compatibility condition.
\label{fig:3simplex_to_tetragon_reduction} }
\end{figure}

\paragraph{3-simplex and tetragon equation.}
 For $N=4$, we have
$\cR_{\hat{1}}\colon \cU_{\widehat{14}} \times \cU_{\widehat{13}} \times \cU_{\widehat{12}}
\rightarrow \cU_{\widehat{12}} \times \cU_{\widehat{13}} \times \cU_{\widehat{14}}$,
$\cT_{\hat{1}}\colon \cU_{\widehat{14}} \times \cU_{\widehat{12}} \rightarrow \cU_{\widehat{13}}$,
$\cS_{\hat{1}}\colon \cU_{\widehat{13}} \rightarrow \cU_{\widehat{12}} \times \cU_{\widehat{14}}$,
etc.\footnote{In the setting of linear spaces, $\cT_{\hat{k}}$ will be a product and $\cS_{\hat{k}}$
a coproduct. }
According to (\ref{R<-T,S}), we have to set
$\cR_{\hat{k}} = \cP_{\bsy{1}}   (\cT_{\hat{k}} \times \cS_{\hat{k}})   \cP_{\bsy{2}}$.
This means $\cR_{\hat{k}}(u,v,w) = (\cS_{\hat{k}}(v)_1 , \cT_{\hat{k}}(u,w), \cS_{\hat{k}}(v)_2)$,
where $\cS_{\hat{k}}(v) =: (\cS_{\hat{k}}(v)_1 , \cS_{\hat{k}}(v)_2)$.
The $3$-simplex equation (\ref{3-simplex_eq_cn}) then leads to
\begin{gather*}
    \cT_{\hat{2}}   \cT_{\hat{4},\bsy{1}} = \cT_{\hat{3}}   \cT_{\hat{1},\bsy{2}}   , \qquad
    \cS_{\hat{1},\bsy{1}}   \cS_{\hat{3}} = \cS_{\hat{4},\bsy{2}}   \cS_{\hat{2}}   , \qquad
    \cT_{\hat{1},\bsy{1}}   \cS_{\hat{2},\bsy{2}}   \cT_{\hat{3},\bsy{2}}   \cS_{\hat{4},\bsy{1}}
     = \cT_{\hat{4},\bsy{2}}   \cS_{\hat{3},\bsy{1}}   \cT_{\hat{2},\bsy{1}}   \cS_{\hat{1},\bsy{2}}   .
\end{gather*}
The f\/irst two equations are the tetragon equation and its dual. The three equations correspond to
$B^{(b)}(4,2)$, $B^{(r)}(4,2)$ and $B^{(g)}(4,2)$, respectively, cf. Example~\ref{ex:B(4,2)_decomp}.
Also see Fig.~\ref{fig:3simplex_to_tetragon_reduction}.

\begin{rem}
\label{rem:line_diagram_from_polyhedron}
By drawing a line through the midpoints of parallel edges,
the half-polytopes of $B(4,1)$ in Fig.~\ref{fig:3simplex_to_tetragon_reduction}
are mapped to the diagrams in Fig.~\ref{fig:B42_line_diagram}, presented in
Fig.~\ref{fig:3simplex_decomp_line_diagram} in a slightly dif\/ferent way.
A tetragon is mapped in this way to a crossing, a hexagon to a node with six legs and a number
$\hat{k}$ that associates it with the map $\hat{\cT}_{\hat{k}}$. The tetragon equation, its dual
and the compatibility equation are then represented by the further graphs in
Fig.~\ref{fig:3simplex_decomp_line_diagram}.
\begin{figure}[t]\centering
\includegraphics[width=.5\linewidth]{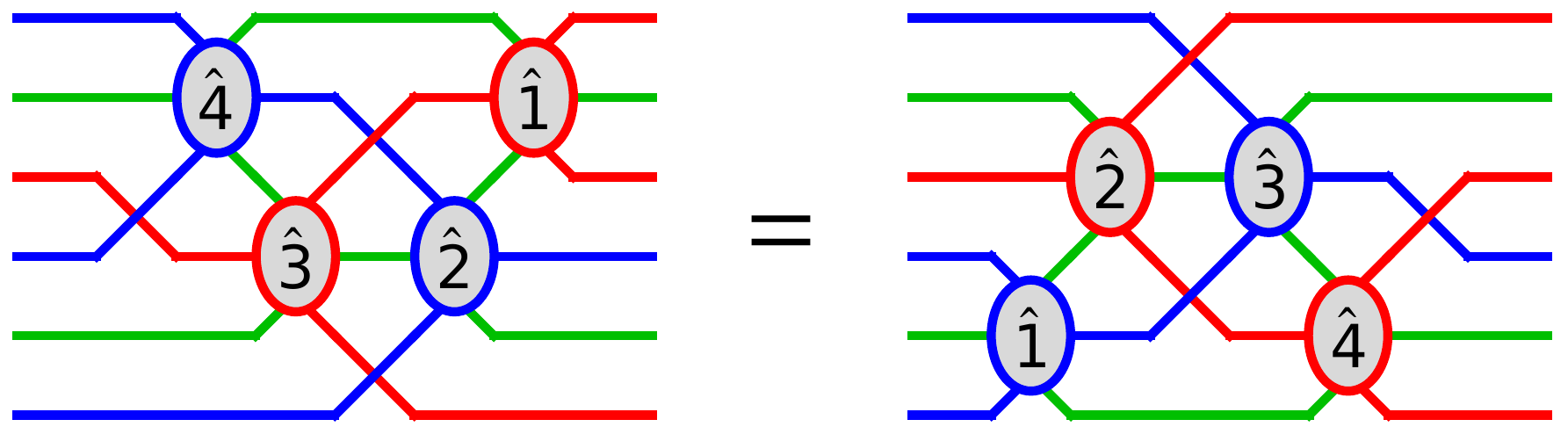}

\vspace{.3cm}

\includegraphics[width=.4\linewidth]{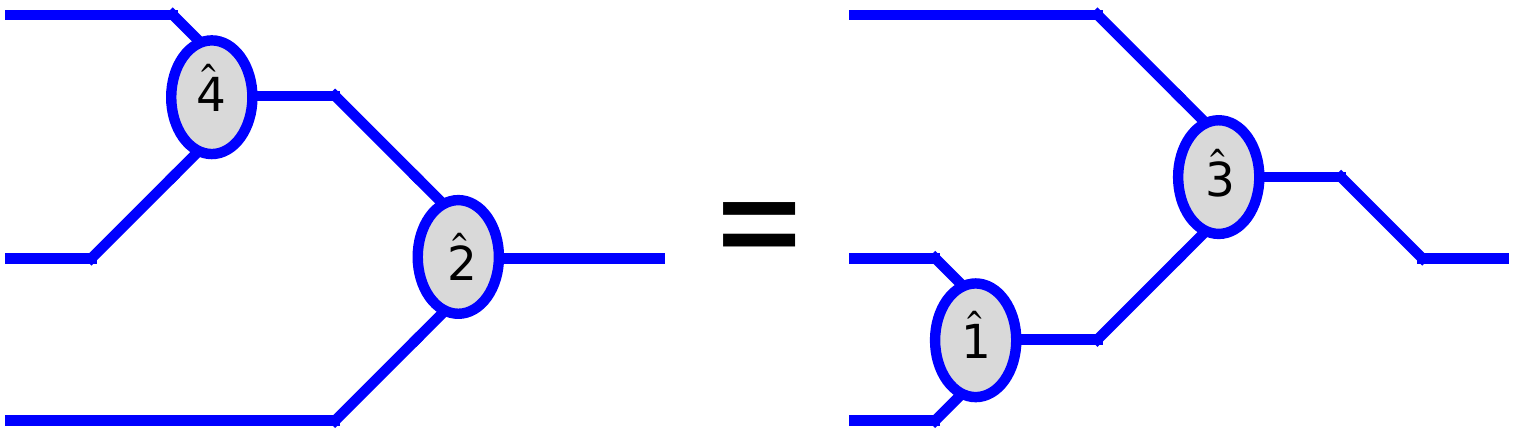}
\quad
\includegraphics[width=.4\linewidth]{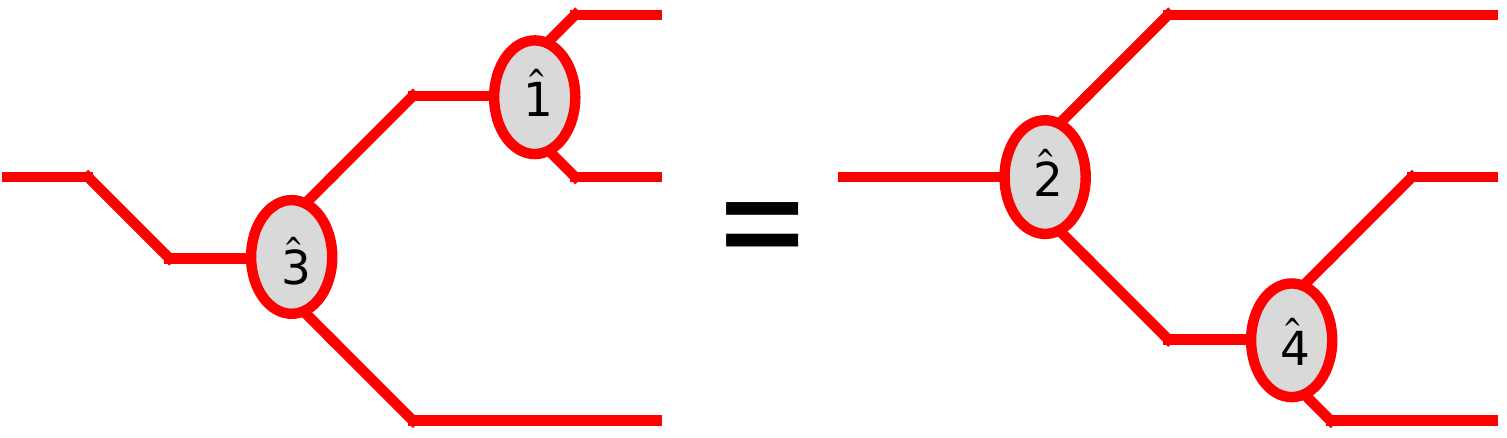}
\vspace{.3cm}

\includegraphics[width=.4\linewidth]{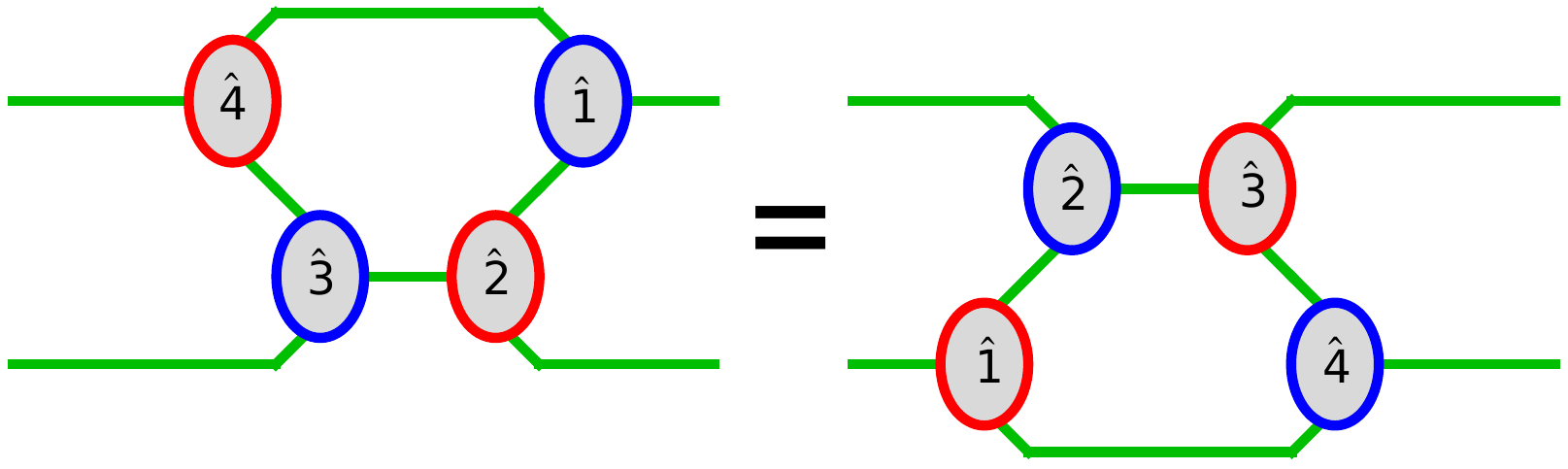}

\caption{The f\/irst line shows the dual of the permutahedron equality in
Fig.~\ref{fig:3simplex_to_tetragon_reduction},
in the sense of Remark~\ref{rem:line_diagram_from_polyhedron}.
The further diagrams represent the three parts that arise from the
``decomposition'' of the $3$-simplex equation~(\ref{3-simplex_eq_cn}).
\label{fig:3simplex_decomp_line_diagram} }
\end{figure}
\end{rem}

\begin{figure}[t]\centering
\includegraphics[width=.7\linewidth]{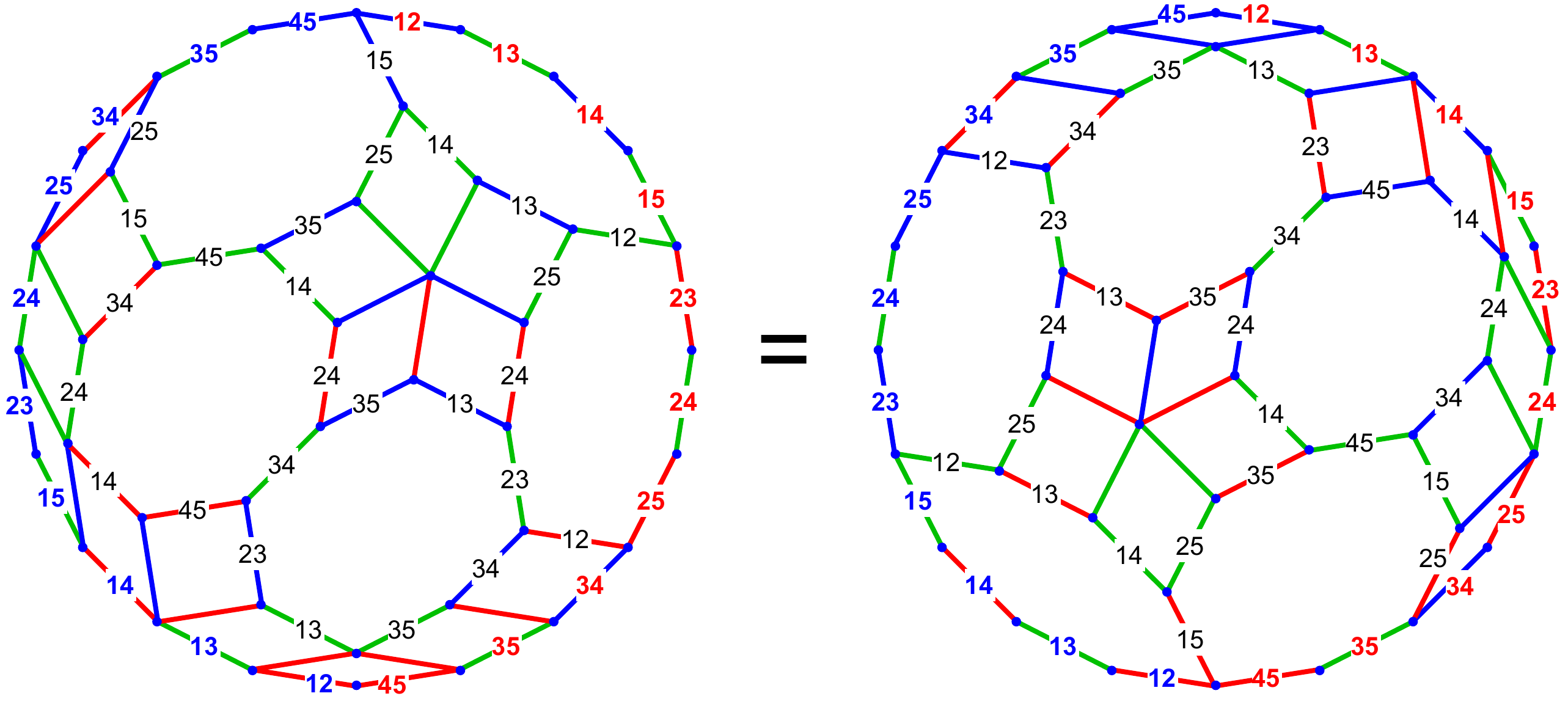}

\vspace{.2cm}

\includegraphics[width=.27\linewidth]{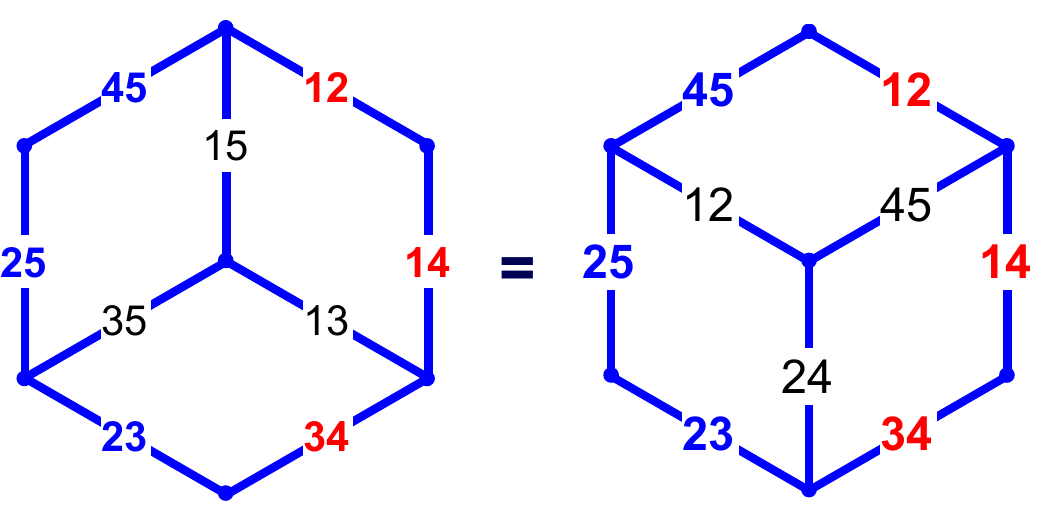}
\hspace{.4cm}
\includegraphics[width=.27\linewidth]{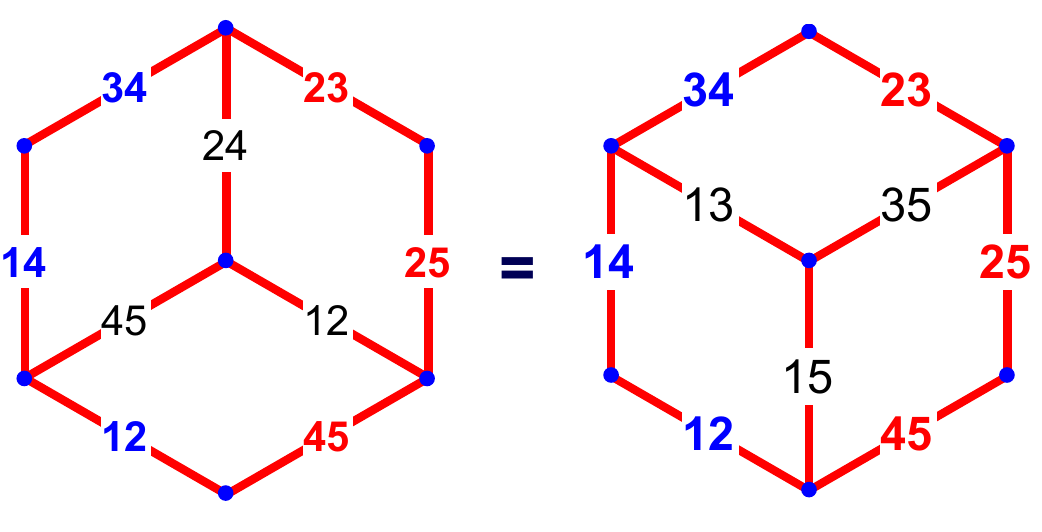}
\hspace{.4cm}
\includegraphics[width=.31\linewidth]{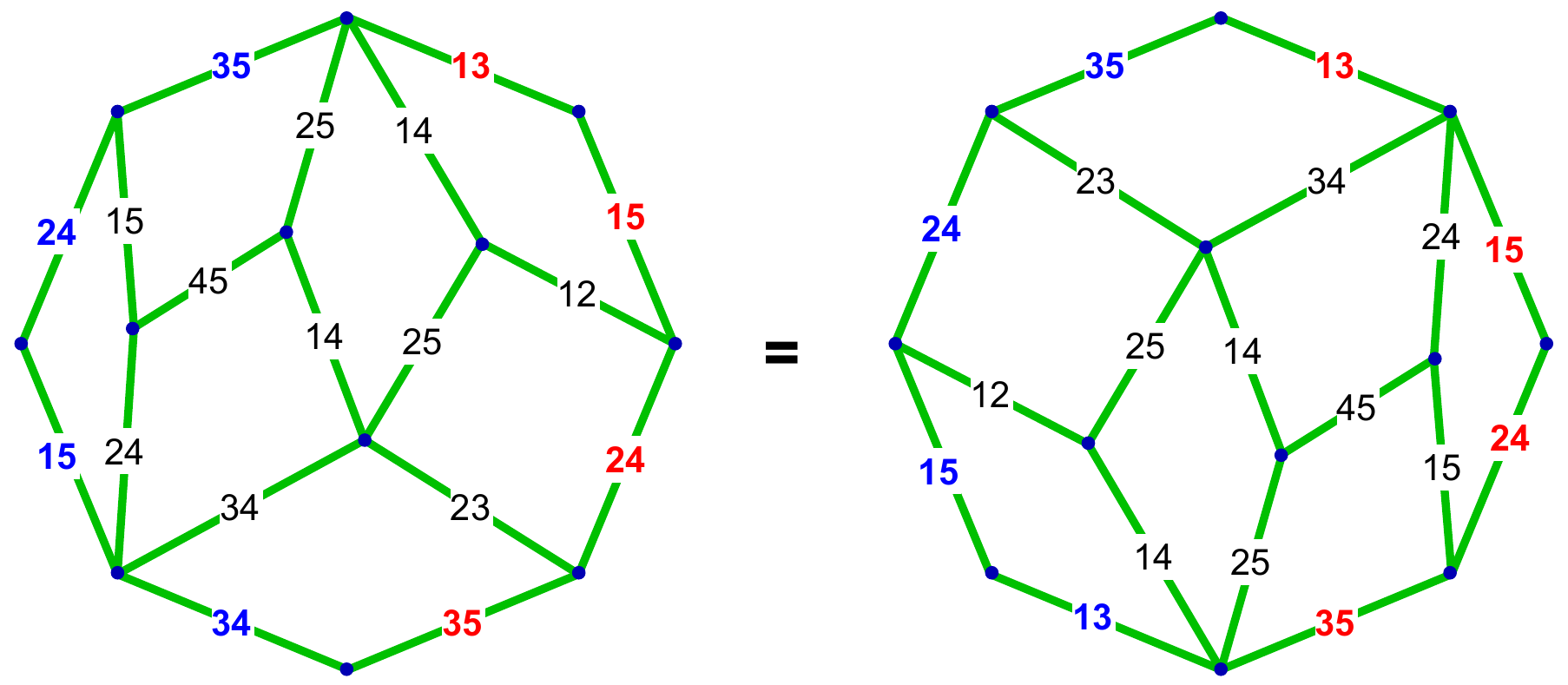}

\caption{Decomposition of the $4$-simplex (Bazhanov--Stroganov) equation, viewed on
the Felsner--Ziegler polyhedron formed by~$B(5,2)$.
The resulting pentagon, dual pentagon and compatibility equations are
represented by the graphs in the second row.
\label{fig:4simplex_to_pentagon_reduction} }
\end{figure}

\paragraph{4-simplex and pentagon equation.}
For $N=5$, we have
$\cR_{\hat{1}} \colon \cU_{\widehat{15}} \times \cU_{\widehat{14}} \times \cU_{\widehat{13}} \times \cU_{\widehat{12}}
\rightarrow \cU_{\widehat{12}} \times \cU_{\widehat{13}} \times \cU_{\widehat{14}} \times \cU_{\widehat{15}}$,
$\cT_{\hat{1}} \colon \cU_{\widehat{15}} \times \cU_{\widehat{13}} \rightarrow \cU_{\widehat{12}} \times \cU_{\widehat{14}}$,
$\cS_{\hat{1}} \colon \cU_{\widehat{14}}  \times \cU_{\widehat{12}} \rightarrow \cU_{\widehat{13}} \times \cU_{\widehat{15}}$,
etc.
We have to set $\cR_{\hat{1}} = \cP_{\bsy{2}}   \cT_{\hat{1},\bsy{3}}   \cS_{\hat{1},\bsy{1}}   \cP_{\bsy{2}}$, etc.,
hence
$\hat{\cR}_{\hat{k},\bsy{1}\bsy{2}\bsy{3}\bsy{4}}
   = \hat{\cT}_{\hat{k},\bsy{1}\bsy{3}}  \hat{\cS}_{\hat{k},\bsy{2}\bsy{4}}
     \cP_{\bsy{1},\bsy{2}}  \cP_{\bsy{3}\bsy{4}}
   = \hat{\cT}_{\hat{k},\bsy{1}\bsy{3}}  \cP_{\bsy{1},\bsy{2}}  \cP_{\bsy{3}\bsy{4}}
     \hat{\cS}_{\hat{k},\bsy{1}\bsy{3}}$.
The $4$-simplex equation (\ref{4-simplex_eq_hatR_cn}) then splits into the pentagon equation
(\ref{5-gon_eq_mod_cn}), its dual
\begin{gather*}
    \hat{\cS}_{\hat{2},\bsy{1}\bsy{2}} \hat{\cS}_{\hat{4},\bsy{2}\bsy{3}}
    = \hat{\cS}_{\hat{5},\bsy{2}\bsy{3}} \hat{\cS}_{\hat{3},\bsy{1}\bsy{3}}
         \hat{\cS}_{\hat{1},\bsy{1}\bsy{2}} ,
\end{gather*}
and the additional condition
\begin{gather*}
  \hat{\cS}_{\hat{1},\bsy{1}\bsy{2}}   \hat{\cT}_{\hat{2},\bsy{1}\bsy{3}}   \hat{\cS}_{\hat{3},\bsy{1}\bsy{4}}
    \, \hat{\cT}_{\hat{4},\bsy{2}\bsy{4}}   \hat{\cS}_{\hat{5},\bsy{3}\bsy{4}}
  = \hat{\cT}_{\hat{5},\bsy{2}\bsy{4}}   \hat{\cS}_{\hat{4},\bsy{3}\bsy{4}}   \hat{\cT}_{\hat{3},\bsy{1}\bsy{4}}
      \hat{\cS}_{\hat{2},\bsy{1}\bsy{2}}   \hat{\cT}_{\hat{1},\bsy{1}\bsy{3}}
\end{gather*}
(cf.~(1.7) in~\cite{Kash+Serg98}). In terms of $\cT$ and $\cS$, we have (\ref{5-gon_eq}) and{\samepage
\begin{gather*}
    \cS_{\hat{2},\bsy{1}}   \cP_{\bsy{2}}   \cS_{\hat{4},\bsy{1}}
    = \cS_{\hat{5},\bsy{2}}   \cS_{\hat{3},\bsy{1}}   \cS_{\hat{1},\bsy{2}}   , \qquad
   \cS_{\hat{1},\bsy{1}}   \cT_{\hat{2},\bsy{2}}   \cS_{\hat{3},\bsy{3}}   \cP_{\bsy{1}}
      \cT_{\hat{4},\bsy{2}}   \cS_{\hat{5},\bsy{1}}   \cP_{\bsy{2}}
  = \cP_{\bsy{2}}   \cT_{\hat{5},\bsy{3}}   \cS_{\hat{4},\bsy{2}}   \cP_{\bsy{3}}   \cT_{\hat{3},\bsy{1}}
      \cS_{\hat{2},\bsy{2}}   \cT_{\hat{1},\bsy{3}}   .
\end{gather*}
Also see Fig.~\ref{fig:4simplex_to_pentagon_reduction}.}

\begin{figure}[t]\centering
\includegraphics[width=.8\linewidth]{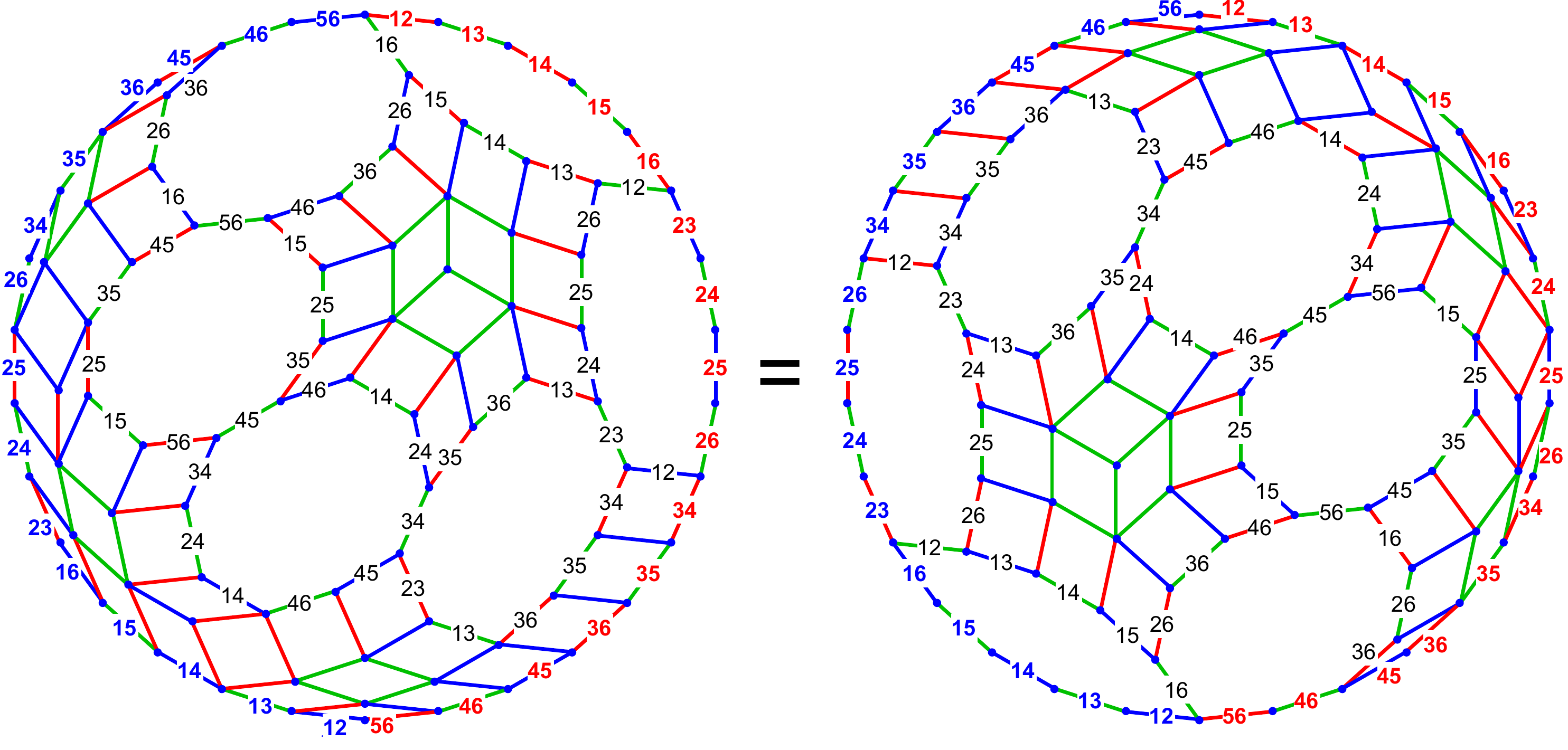}

\vspace{.2cm}

\includegraphics[width=.29\linewidth]{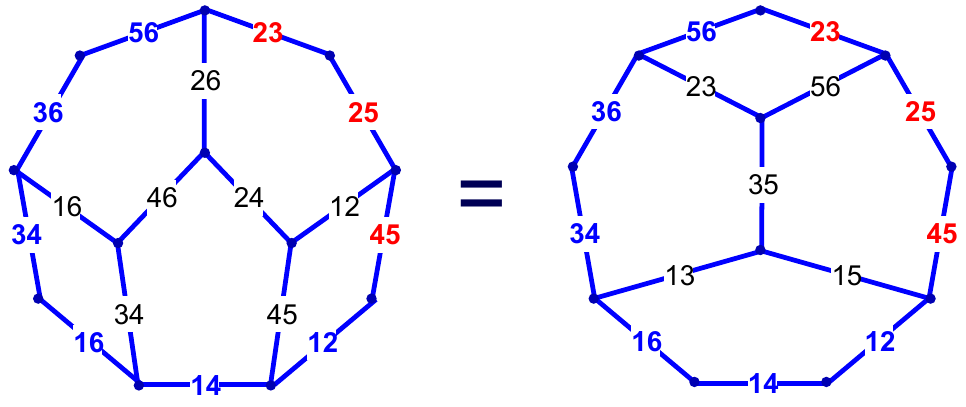}
\hspace{.1cm}
\includegraphics[width=.29\linewidth]{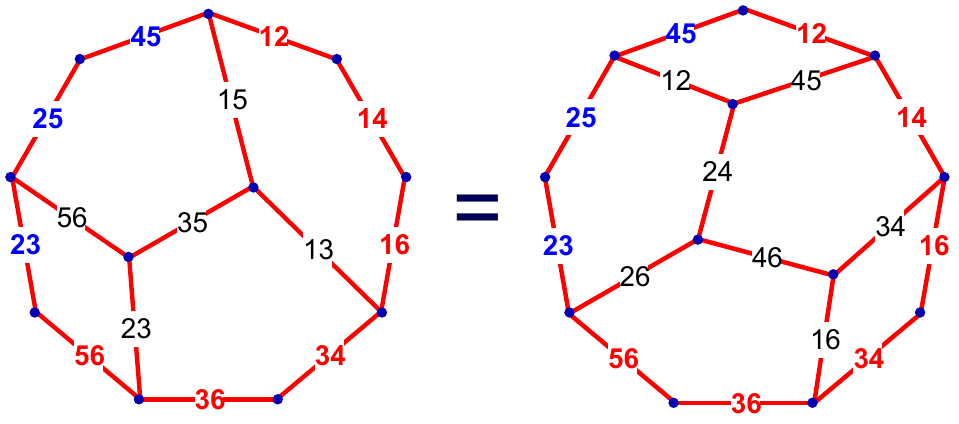}
\hspace{.1cm}
\includegraphics[width=.34\linewidth]{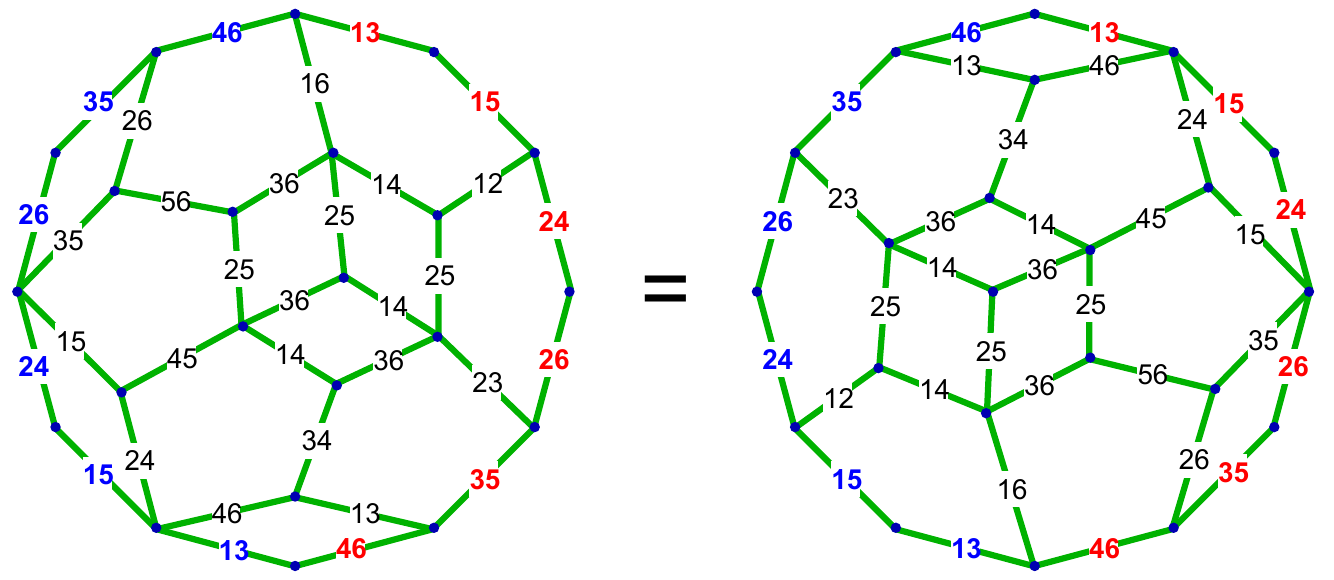}

\caption{Decomposition of the $5$-simplex equation on the polyhedron formed by~$B(6,3)$.
The resulting hexagon, dual hexagon and compatibility equations are graphically represented in the second row.
\label{fig:5simplex_to_hexagon_reduction} }
\end{figure}

\paragraph{5-simplex and hexagon equation.}
For $N=6$, we have
$\cR_{\hat{1}} \colon \cU_{\widehat{16}} \times \cU_{\widehat{15}} \times \cU_{\widehat{14}}
 \times \cU_{\widehat{13}} \times \cU_{\widehat{12}}
\rightarrow \cU_{\widehat{12}} \times \cU_{\widehat{13}} \times \cU_{\widehat{14}}
 \times \cU_{\widehat{15}} \times \cU_{\widehat{16}}$,
$\cT_{\hat{1}} \colon \cU_{\widehat{16}} \times \cU_{\widehat{14}} \times \cU_{\widehat{12}}
 \rightarrow \cU_{\widehat{13}} \times \cU_{\widehat{15}}$,
$\cS_{\hat{1}} \colon \cU_{\widehat{15}} \times \cU_{\widehat{13}} \rightarrow
  \cU_{\widehat{12}} \times \cU_{\widehat{14}} \times \cU_{\widehat{16}}$,
etc.
We set
$\cR_{\hat{k}} = \cP_{\bsy{3}} \cP_{\bsy{1}} \cP_{\bsy{2}}   (\cT_{\hat{k}} \times \cS_{\hat{k}} )
\cP_{\bsy{4}} \cP_{\bsy{2}} \cP_{\bsy{3}}$. The $5$-simplex equation (\ref{5simplex_eq})
then decomposes into the hexagon equation (\ref{6-gon_eq}) for $\cT$, the dual hexagon equation
\begin{gather*}
    \cS_{\hat{1},\bsy{1}}   \cS_{\hat{3},\bsy{2}}   \cP_{\bsy{3}}   \cS_{\hat{5},\bsy{1}}
  = \cP_{\bsy{3}}   \cS_{\hat{6},\bsy{4}}   \cS_{\hat{4},\bsy{2}}   \cP_{\bsy{1}}
    \cS_{\hat{2},\bsy{2}}   ,
\end{gather*}
and
\begin{gather*}
    \cT_{\hat{1},\bsy{1}}   \cS_{\hat{2},\bsy{3}}   \cT_{\hat{3},\bsy{4}}   \cP_{\bsy{6}}
      \cP_{\bsy{2}}   \cP_{\bsy{3}}   \cP_{\bsy{2}}   \cS_{\hat{4},\bsy{4}}   \cT_{\hat{5},\bsy{3}}
    \cS_{\hat{6},\bsy{1}}   \cP_{\bsy{4}}   \cP_{\bsy{2}}
  = \cP_{\bsy{4}}   \cP_{\bsy{2}}   \cT_{\hat{6},\bsy{5}}   \cS_{\hat{5},\bsy{3}}   \cT_{\hat{4},\bsy{2}}
    \cP_{\bsy{5}}   \cP_{\bsy{4}}   \cP_{\bsy{5}}   \cP_{\bsy{1}}   \cS_{\hat{3},\bsy{2}}
       \cT_{\hat{2},\bsy{3}}   \cS_{\hat{1},\bsy{5}}    .
\end{gather*}
See Fig.~\ref{fig:5simplex_to_hexagon_reduction}.

\begin{figure}[t]\centering
\includegraphics[width=.9\linewidth]{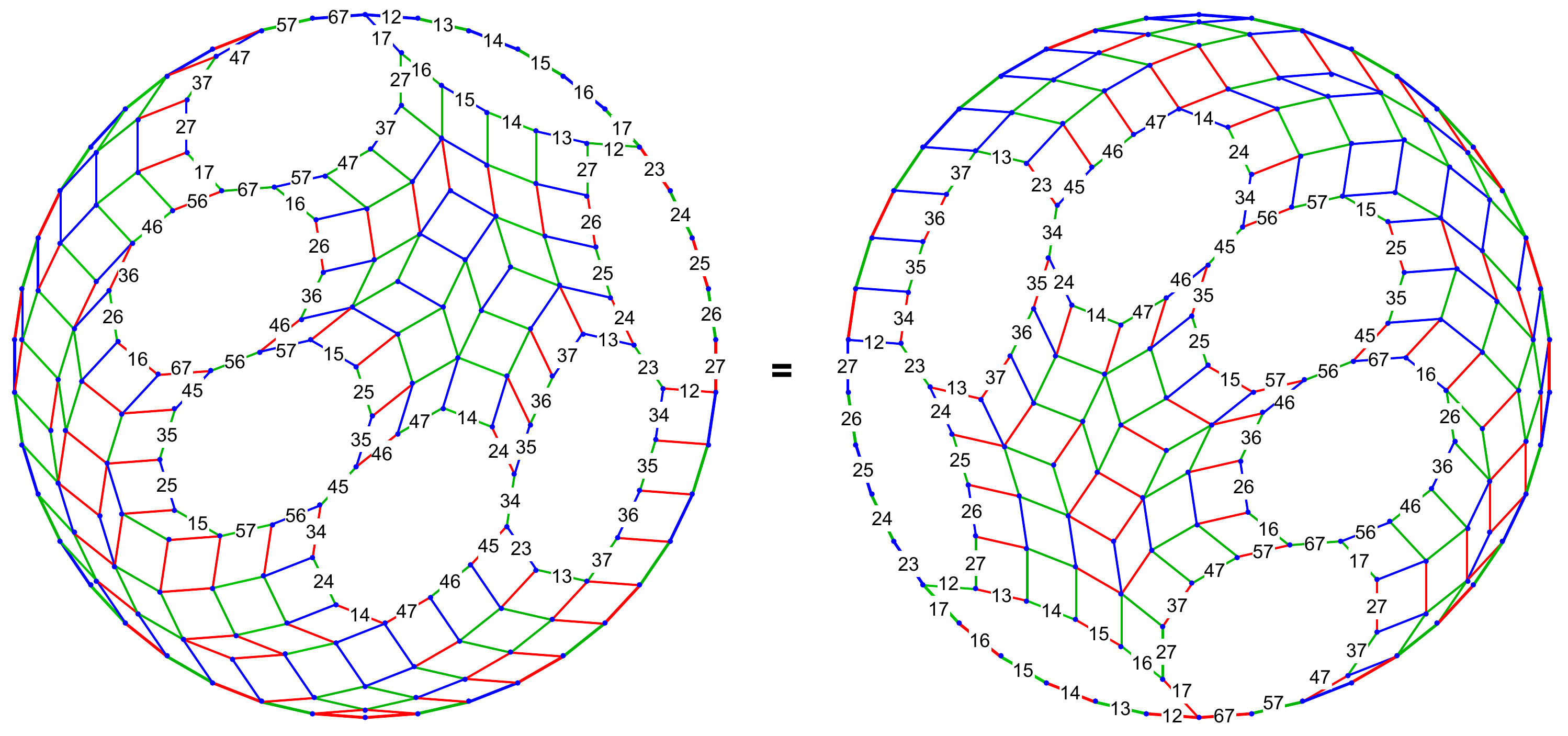}
\vspace{.2cm}

\includegraphics[width=.42\linewidth]{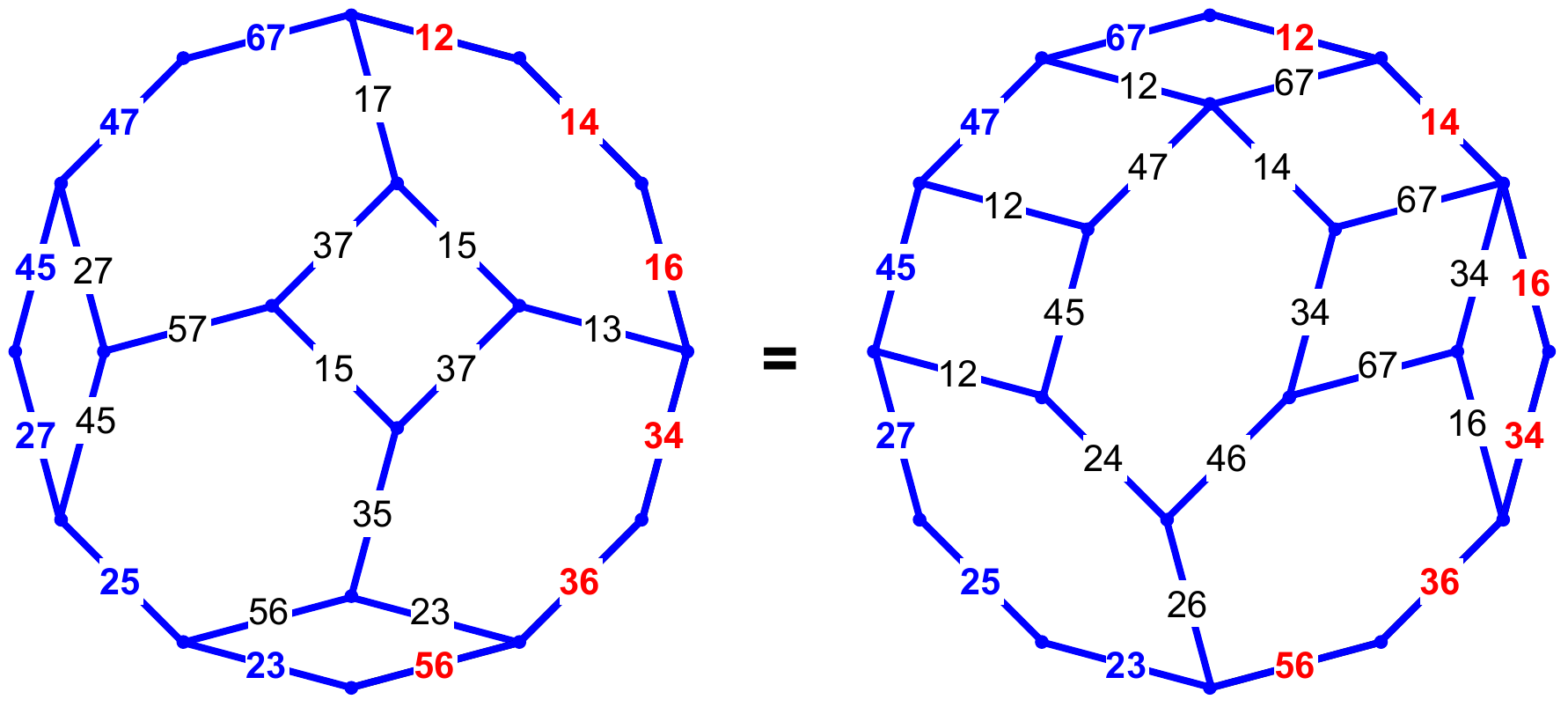}
\hspace{.5cm}
\includegraphics[width=.42\linewidth]{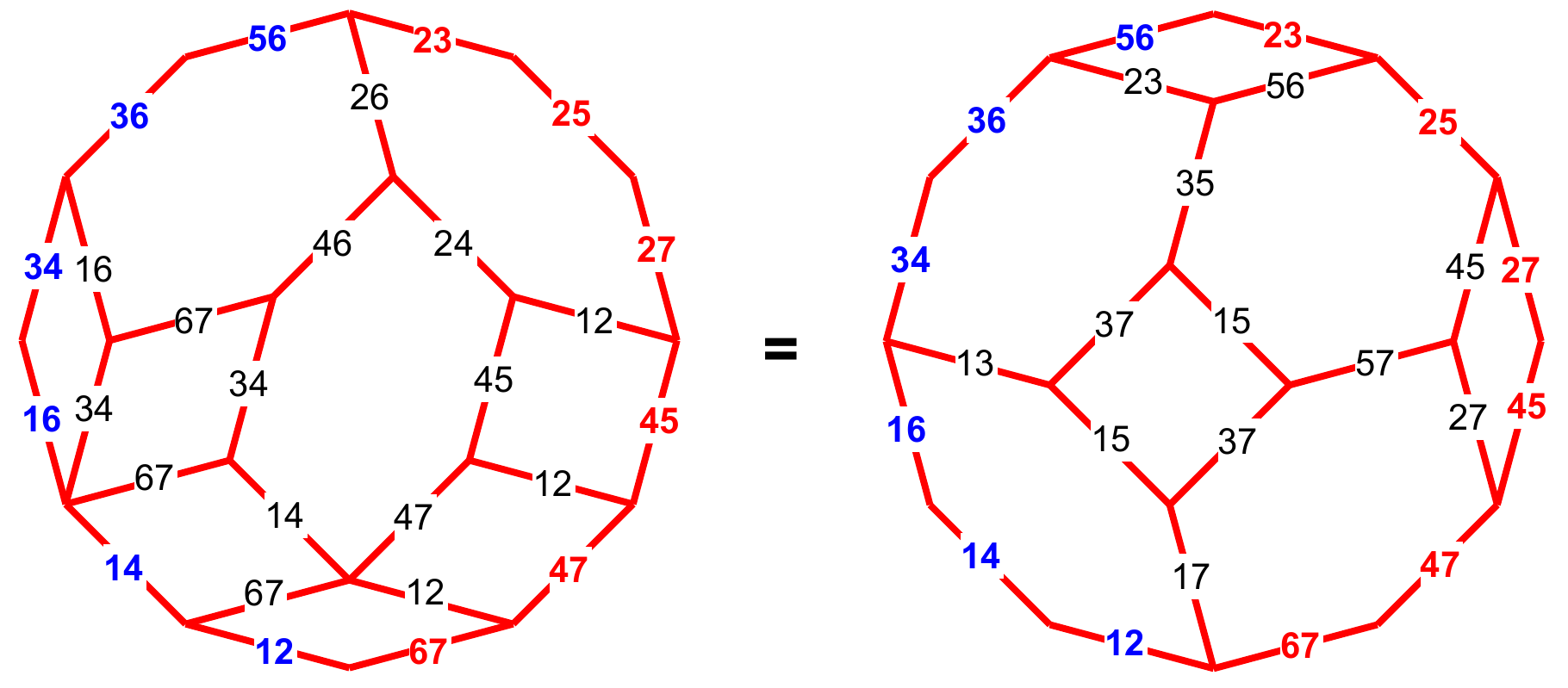}
\vspace{.2cm}

\includegraphics[width=.9\linewidth]{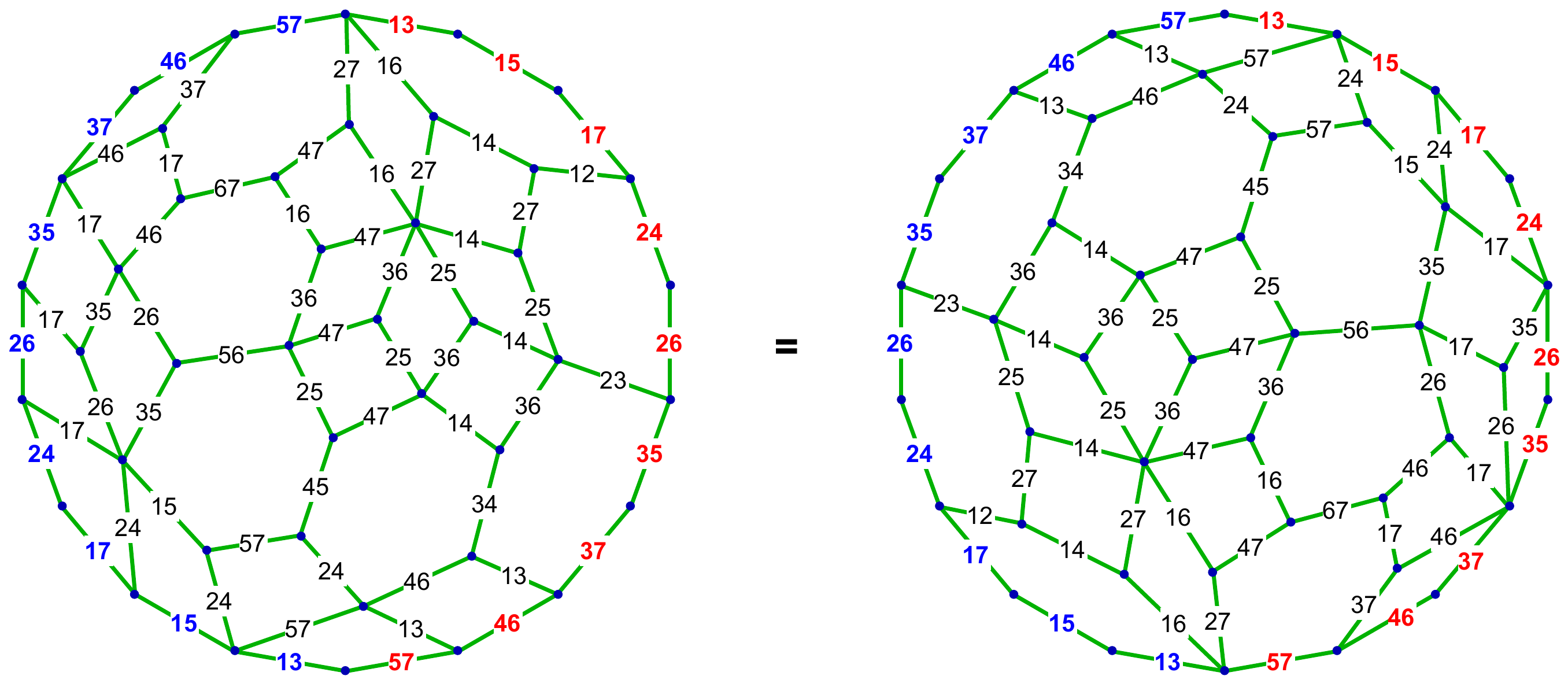}

\caption{Decomposition of the $6$-simplex equation on the polyhedron formed by~$B(7,4)$.
The resulting heptagon, dual heptagon and compatibility equations are
represented by the graphs in the last two rows.
\label{fig:6simplex_to_heptagon_reduction} }
\end{figure}

\paragraph{6-simplex and heptagon equation.}
For $N=7$, we have
$\cR_{\hat{1}}\colon \cU_{\widehat{17}} \times \cU_{\widehat{16}} \times \cU_{\widehat{15}} \times \cU_{\widehat{14}}
 \times \cU_{\widehat{13}} \times \cU_{\widehat{12}}
\rightarrow \cU_{\widehat{12}} \times \cU_{\widehat{13}} \times \cU_{\widehat{14}}
 \times \cU_{\widehat{15}} \times \cU_{\widehat{16}} \times \cU_{\widehat{17}}$,
$\cT_{\hat{1}}\colon \cU_{\widehat{17}} \times \cU_{\widehat{15}} \times \cU_{\widehat{13}}
 \rightarrow \cU_{\widehat{12}} \times \cU_{\widehat{14}} \times \cU_{\widehat{16}}$,
$\cS_{\hat{1}}\colon \cU_{\widehat{16}} \times \cU_{\widehat{14}} \times \cU_{\widehat{12}} \rightarrow
  \cU_{\widehat{13}} \times \cU_{\widehat{15}} \times \cU_{\widehat{17}}$,
etc.
We set
$\cR_{\hat{k}}
 = \cP_{\bsy{4}} \cP_{\bsy{2}} \cP_{\bsy{3}}  (\cT_{\hat{k}} \times \cS_{\hat{k}} )
        \cP_{\bsy{3}} \cP_{\bsy{2}} \cP_{\bsy{4}}
 = \cP_{\bsy{4}} \cP_{\bsy{2}} \cP_{\bsy{3}}  \cT_{\hat{k},\bsy{1}} \cS_{\hat{k},\bsy{4}}
        \cP_{\bsy{3}} \cP_{\bsy{2}} \cP_{\bsy{4}}$.
The $6$-simplex equation (\ref{6simplex_eq}) then decomposes into the heptagon equation~(\ref{7-gon_eq}), respectively~(\ref{7-gon_eq_mod}),
and the dual heptagon equation
\begin{gather*}
    \cS_{\hat{2},\bsy{1}}   \cP_{\bsy{3}}   \cP_{\bsy{4}}   \cP_{\bsy{5}}   \cS_{\hat{4},\bsy{2}}
    \cP_{\bsy{4}}   \cP_{\bsy{3}}   \cS_{\hat{6},\bsy{1}}   \cP_{\bsy{3}}
  = \cP_{\bsy{3}}   \cS_{\hat{7},\bsy{4}}   \cS_{\hat{5},\bsy{2}}   \cP_{\bsy{4}}   \cP_{\bsy{1}}
    \cS_{\hat{3},\bsy{2}}   \cS_{\hat{1},\bsy{4}}   ,
\end{gather*}
respectively,
\begin{gather*}
    \hat{\cS}_{\hat{2},\bsy{123}}   \hat{\cS}_{\hat{4},\bsy{245}}   \hat{\cS}_{\hat{6},\bsy{356}}
  = \hat{\cS}_{\hat{7},\bsy{356}}   \hat{\cS}_{\hat{5},\bsy{246}}   \hat{\cS}_{\hat{3},\bsy{145}}   \hat{\cS}_{\hat{1},\bsy{123}}   ,
\end{gather*}
and the compatibility condition
\begin{gather*}
   \cT_{\hat{1},\bsy{1}}   \cS_{\hat{2},\bsy{3}}   \cP_{\bsy{2}}   \cP_{\bsy{1}}   \cP_{\bsy{3}}
    \cT_{\hat{3},\bsy{5}}   \cP_{\bsy{7}}   \cP_{\bsy{8}}   \cP_{\bsy{4}}   \cP_{\bsy{3}}   \cS_{\hat{4},\bsy{5}}
    \cP_{\bsy{7}}   \cP_{\bsy{4}}   \cP_{\bsy{3}}   \cP_{\bsy{2}}  \cT_{\hat{5},\bsy{5}}   \cP_{\bsy{7}}
    \cS_{\hat{6},\bsy{3}}   \cP_{\bsy{5}}   \cT_{\hat{7},\bsy{1}}   \cP_{\bsy{3}} \cP_{\bsy{4}}   \cP_{\bsy{5}}
    \cP_{\bsy{6}}   \cP_{\bsy{2}}  \nonumber \\
  =  \cP_{\bsy{7}}   \cP_{\bsy{5}}   \cP_{\bsy{3}}   \cP_{\bsy{4}}   \cP_{\bsy{5}}   \cP_{\bsy{6}}   \cP_{\bsy{2}}
    \cS_{\hat{7},\bsy{7}}   \cT_{\hat{6},\bsy{5}}   \cP_{\bsy{7}}   \cP_{\bsy{8}}   \cP_{\bsy{6}}   \cS_{\hat{5},\bsy{3}}
    \cP_{\bsy{5}}   \cP_{\bsy{6}}   \cP_{\bsy{2}}   \cP_{\bsy{1}}   \cT_{\hat{4},\bsy{3}}   \cP_{\bsy{5}}   \cP_{\bsy{6}}
    \cP_{\bsy{7}}   \cP_{\bsy{2}}   \cS_{\hat{3},\bsy{3}}   \cT_{\hat{2},\bsy{5}}   \cS_{\hat{1},\bsy{7}}    ,
\end{gather*}
respectively
\begin{gather*}
    \hat{\cT}_{\hat{1},\bsy{123}}   \hat{\cS}_{\hat{2},\bsy{145}}   \hat{\cT}_{\hat{3},\bsy{167}}
    \hat{\cS}_{\hat{4},\bsy{268}}
    \hat{\cT}_{\hat{5},\bsy{469}}   \hat{\cS}_{\hat{6},\bsy{379}}   \hat{\cT}_{\hat{7},\bsy{589}}
  = \hat{\cS}_{\hat{7},\bsy{379}}   \hat{\cT}_{\hat{6},\bsy{589}}   \hat{\cS}_{\hat{5},\bsy{269}}
    \hat{\cT}_{\hat{4},\bsy{467}}   \hat{\cS}_{\hat{3},\bsy{168}}   \hat{\cT}_{\hat{2},\bsy{123}}
    \hat{\cS}_{\hat{1},\bsy{145}}    .
\end{gather*}
See Fig.~\ref{fig:6simplex_to_heptagon_reduction}.

\section{Conclusions}
\label{sec:concl}
The main result of this work is the existence of an inf\/inite family of ``polygon equations'' that generalize
the pentagon equation in very much the same way as the simplex equations generalize the Yang--Baxter
equation. The underlying combinatorial structure in case of simplex equations is given by (higher)
Bruhat orders~\cite{Manin+Schecht86a,Manin+Shekhtman86b,Manin+Schecht89}. Underlying
the polygon equations are (higher) Tamari orders.

We also introduced a visualization of simplex as well as polygon equations as deformations of maximal chains
of posets forming 1-skeletons of polyhedra.
This geometrical representation revealed various deep relations between such equations.

An intermediate result, worth to highlight, is the (three color) decomposition in
Section~\ref{subsec:Bruhat_decomp} of any (higher) Bruhat order into a (higher) Tamari order,
the corresponding dual Tamari order, and a ``mixed order''. From this we recovered a
relation between the pentagon and the $4$-simplex equation, observed in~\cite{Kash+Serg98}.
Moreover, we showed that this is just a special case of a relation between any simplex equation
and its associated polygon equation.
Another (more profound) observation made in~\cite{Kash+Serg98} concerns a relation between
the pentagon equation and the $3$-simplex equation. This seems not to have a corresponding
generalization.

Further exploration of the higher polygon equations is required.
We expect that they will play a role in similarly diverse problems as the pentagon equation does.
A major task will be the search for relevant solutions in suitable frameworks. Such a framework could be
the KP hierarchy, since a subclass of its soliton solutions realizes higher Tamari orders
\cite{DMH11KPT,DMH12KPBT}.

\renewcommand{\theequation} {\Alph{section}.\arabic{equation}}
\renewcommand{\thesection} {\Alph{section}}

\makeatletter
\newcommand\appendix@section[1]{
  \refstepcounter{section}
  \orig@section*{Appendix \@Alph\c@section: #1}
  \addcontentsline{toc}{section}{Appendix \@Alph\c@section: #1}
}
\let\orig@section\section
\g@addto@macro\appendix{\let\section\appendix@section}
\makeatother

\begin{appendix}

\section{A dif\/ferent view of simplex and polygon equations}
\label{app:alg}
Let $\cB$ be a monoid and $N > 2$ an integer. With each $J \in {[N] \choose N-2}$, we associate
an invertible element $L_J$ of $\cB$, and with each $K \in {[N] \choose N-1}$, we associate elements
$R_K$, $R_K'$. They shall be subject to the conditions~(\cite{DMH12KPBT}, also see~\cite{Byts+Volk13} for
a related structure)
\begin{gather}
  L_J   L_{J'} = L_{J'}   L_J \qquad \mbox{if} \quad E(J) \cap E(J') = \varnothing ,
            \label{monoid_cond_1} \\
 L_J   R_K = R_K   L_J   , \qquad
       L_J   R'_K = R'_K   L_J
       \qquad  \mbox{if} \quad J \notin P(K)   , \label{monoid_cond_2} \\
  L_{\vec{P}(K)}   R'_K = R_K   L_{\cev{P}(K)}   ,  \label{monoid_cond_3}
\end{gather}
where
\begin{gather*}
    L_{\vec{P}(K)} = L_{K \setminus \{k_{N-1}\}} \cdots L_{K \setminus \{k_1\}}  , \qquad
    L_{\cev{P}(K)} = L_{K \setminus \{k_1\}} \cdots L_{K \setminus \{k_{N-1}\}}  ,
\end{gather*}
and $K = \{k_1,\ldots,k_{N-1}\}$ with $k_1 < k_2 < \cdots < k_{N-1}$.
For any sequence $\rho = (J_1,\ldots,J_r)$, let $L_\rho = L_{J_1} \cdots L_{J_r}$.
If $\rho \in A(N,N-2)$, then $L_{\rho'} = L_\rho$ for any $\rho' \in [\rho]$, according
to~(\ref{monoid_cond_1}). Hence~$L_\rho$ represents $[\rho]$.

\begin{prop}
\label{prop:monoid_simplex_eq}
\begin{gather*}
   R_{\vec{P}([N])} = R_{\cev{P}([N])} \quad \Longleftrightarrow \quad
   R'_{\vec{P}([N])} = R'_{\cev{P}([N])}   .
\end{gather*}
\end{prop}

\begin{proof}\sloppy
We follow the two maximal chains~(\ref{B(N,N-2)_chains}) of $B(N,N-2)$.
Let $\alpha$ be the lexicographical order on ${[N] \choose N-2}$, and $\omega$ the reverse
lexicographical order. Let us start
with $L_\alpha R'_{\vec{P}([N])} = L_\alpha R'_{\hat{N}} \cdots R'_{\hat{1}}$ and move
all $L_J$, $J \in P(\hat{N})$, to consecutive positions in $L_\alpha$, using~(\ref{monoid_cond_1}).
\mbox{Using~(\ref{monoid_cond_2})},
we commute $R'_{\hat{N}}$ to the left until we have the substring $L_{\vec{P}(\hat{N})} R'_{\hat{N}}$.
Then we use (\ref{monoid_cond_3}) to replace this by $R_{\hat{N}} \, L_{\cev{P}(\hat{N})}$. Using
(\ref{monoid_cond_2}) again, we commute $R_{\hat{N}}$ to the left of all $L$'s, thus obtaining~$R_{\hat{N}} L_{\rho_1} R'_{\widehat{N-1}} \cdots R'_{\hat{1}}$. Continuing in this way,
we f\/inally get $L_\alpha R'_{\vec{P}([N])} = R_{\vec{P}([N])} L_\omega$. For the second
maximal chain of $B(N,N-2)$, we obtain $L_\alpha R'_{\cev{P}([N])} = R_{\cev{P}([N])} L_\omega$
correspon\-dingly. Since $L_\alpha$ and $L_\omega$ are invertible (since we assume that the $L$'s
are invertible), the statement of the proposition follows.
\end{proof}

The proposition says that the elements $R_K$, $K \in {[N] \choose N-1}$, satisfy (the algebraic
version of) the $(N-1)$-simplex
equation if and only if this is so for $R'_K$, $K \in {[N] \choose N-1}$.
Choosing for all~$R'_K$ the identity
element of~$\cB$, (\ref{monoid_cond_3}) becomes the Lax system $L_{\vec{P}(K)} = R_K   L_{\cev{P}(K)}$.

\begin{exam}
For $N=3$, the Lax system reads $L_i L_j = R_{ij} \, L_j L_i$, $1 \leq i<j \leq 3$, so that
$R_{ij} = L_i L_j L_i^{-1} L_j^{-1} =: [L_i,L_j]$ is a commutator in a group. The condition~(\ref{monoid_cond_1}) is empty and~(\ref{monoid_cond_2}) requires $[[L_i,L_j],L_k] = e$ for $i<j$, $k \neq i,j$, where $e$ is
the identity element. Hence, if $G$ is the group
$\langle g_1,g_2,g_3 \, | \, [[g_1,g_2],g_3] = [[g_2,g_3],g_1] = [[g_1,g_3],g_2]=e \rangle$,
then Proposition~\ref{prop:monoid_simplex_eq} implies that
$ R_{ij} := [g_i,g_j] $, $i<j$, satisfy the Yang--Baxter equation.
If $G$ is abelian, then $R_{ij} = e$.
\end{exam}

We are led to the following by the three color decomposition.
Let us keep (\ref{monoid_cond_1}), but replace (\ref{monoid_cond_2}) and (\ref{monoid_cond_3}) by
\begin{gather}
  L_J T_K = T_K L_J   , \qquad L_J T'_K = T'_K L_J
     \qquad  \mbox{if} \quad J \notin P(K)   , \label{monoid_T_cond2} \\
  L_{\vec{P_o}(K)}   T'_K = T_K   L_{\cev{P_e}(K)}    .\nonumber
\end{gather}
Then we have
\begin{gather*}
   T_{\vec{P_o}([N])} = T_{\cev{P_e}([N])} \quad \Longleftrightarrow \quad
   T'_{\vec{P_o}([N])} = T'_{\cev{P_e}([N])}   .
\end{gather*}
The proof is analogous to that of Proposition~\ref{prop:monoid_simplex_eq}, but here we
start with $L_{\alpha^{(b)}} T'_{\vec{P_o}([N])}$.
Choosing for~$T'_K$ the identity element, we have the Lax system $L_{\vec{P_o}(K)} = T_K   L_{\cev{P_e}(K)}$
for (an algebraic version of) the $N$-gon equation.

Let us now keep (\ref{monoid_cond_1}), but replace (\ref{monoid_cond_2}) and (\ref{monoid_cond_3}) by
\begin{gather}
  L_J S_K = S_K L_J   , \qquad L_J S'_K = S'_K L_J
     \qquad  \mbox{if} \quad J \notin P(K)   , \label{monoid_dT_cond2} \\
  L_{\vec{P_e}(K)}   S'_K = S_K   L_{\cev{P_o}(K)}   .\nonumber
\end{gather}
Then we have
\begin{gather*}
   S_{\vec{P_e}([N])} = S_{\cev{P_o}([N])} \quad \Longleftrightarrow \quad
   S'_{\vec{P_e}([N])} = S'_{\cev{P_o}([N])}   .
\end{gather*}
Here the proof starts with $L_{\alpha^{(r)}} S'_{\vec{P_e}([N])}$.
Choosing for $T'_K$ the identity element, we have the Lax system $L_{\vec{P_e}(K)} = S_K   L_{\cev{P_o}(K)}$
for (an algebraic version of) the dual $N$-gon equation.

Next, let (\ref{monoid_cond_1}), (\ref{monoid_T_cond2}) and (\ref{monoid_dT_cond2}) hold, and in addition
\begin{gather*}
  L_{\vec{P_e}(K)}  T'_K = T_K   L_{\cev{P_o}(K)}    , \qquad
       L_{\vec{P_o}(K)}   S'_K = S_K   L_{\cev{P_e}(K)}   .
\end{gather*}
For odd $N$, the mixed equation reads
\begin{gather*}
    S_{\hat{N}} T_{\widehat{N-1}} \cdots T_{\hat{2}} S_{\hat{1}}
    = T_{\hat{1}} S_{\hat{2}} \cdots S_{\widehat{N-1}} T_{\hat{N}}   ,
\end{gather*}
while for even $N$ is has the form
\begin{gather*}
    S_{\hat{N}} T_{\widehat{N-1}} \cdots S_{\hat{2}} T_{\hat{1}}
    = S_{\hat{1}} T_{\hat{2}} \cdots S_{\widehat{N-1}} T_{\hat{N}}   .
\end{gather*}
We f\/ind that the mixed equation holds for $S_K$, $T_K$ if and only if it holds for $S'_K$, $T'_K$.
The proof starts with
$L_{\alpha^{(g)}} S'_{\hat{N}} T'_{\widehat{N-1}} \cdots T'_{\hat{2}} S'_{\hat{1}}$ for odd~$N$,
and correspondingly for even~$N$.

\begin{rem}
In the present framework, the pentagon equation reads\footnote{This corresponds to
the pentagon equation (\ref{5-gon_eq}) without the position indices and with
reversion of the order on both sides. }
\begin{gather*}
    T_{1,2,3,4}    T_{1,2,4,5}    T_{2,3,4,5} =  T_{1,3,4,5}    T_{1,2,3,5}   .
\end{gather*}
Here we inserted commas, which we mostly omitted before.
We translate the labels as follows. If a label $i_1$, $i_2$, $i_3$, $i_4$ contains a pair with
$i_{r+1} = i_r +2$ (higher shifts do not appear),
then we make the substitution $i_r,i_{r+1} \mapsto i_r(i_r+1),i_{r+1}$,
where~$i_r$ $(i_r+1)$ is understood as a two-digit expression.
Finally we drop the very last index of the resulting new label.
If there is no index pair of the above kind in a label, we keep the label, but also drop
the very last index. This translates the above pentagon equation to
\begin{gather*}
    T_{1,2,3}   T_{1,23,4}   T_{2,3,4} = T_{12,3,4}   T_{1,2,34}   .
\end{gather*}
In this form the pentagon equation shows up in Drinfeld's theory of associators (see, e.g.,
\cite{Alek+Toro12, Etin+Schi98}).
In the same way, the associated (tetragon) Lax equation
$L_{1,2,3} L_{1,3,4}   T'_{1,2,3,4} = L_{2,3,4} L_{1,2,4}$ becomes
\begin{gather*}
     L_{1,2} L_{12,3}   T'_{1,2,3} = L_{2,3} L_{1,23}   ,
\end{gather*}
which becomes the \emph{twist equation} in the context of associators (see, e.g., equation~(2)
in~\cite{Alek+Toro12}). Furthermore, the $3$-simplex equation in the present framework is
$R_{1,2,3} R_{1,2,4} R_{1,3,4} R_{2,3,4} = R_{2,3,4} R_{1,3,4} R_{1,2,4} R_{1,2,3}$, which
translates to
\begin{gather*}
     R_{1,2}   R_{1,23}   R_{12,3}   R_{2,3} =  R_{2,3}   R_{12,3}   R_{1,23}   R_{1,2}   ,
\end{gather*}
and the hexagon equation
$T_{1,2,3,4,5} T_{1,2,3,5,6} T_{1,3,4,5,6} = T_{2,3,4,5,6} T_{1,2,4,5,6} T_{1,2,3,4,6}$
becomes
\begin{gather*}
    T_{1,2,3,4}   T_{1,2,34,5}   T_{12,3,4,5} = T_{2,3,4,5}   T_{1,23,4,5}   T_{1,2,3,45}   ,
\end{gather*}
and so forth.
\end{rem}

\end{appendix}

\subsection*{Acknowledgments}
We have to thank an anonymous referee for comments that led to some
corrections in our previous version of Section~\ref{subsec:Bruhat_decomp}.

\pdfbookmark[1]{References}{ref}
\LastPageEnding

\end{document}